\documentstyle[epsfig,natbib2,natbibmnfix]{mn}

\newcommand{\bea}{\begin{eqnarray}}
\newcommand{\eea}{\end{eqnarray}}
\newcommand{\bc}{\begin{center}}
\newcommand{\ec}{\end{center}}

\renewcommand{\vec}[1]{ {\bmath #1} }

\renewcommand{\thefootnote}{\fnsymbol{footnote}}

\title[Galiliean-invariant cosmological hydrodynamical simulations on
  a moving mesh]{{\em E pur si muove:} Galiliean-invariant cosmological
  hydrodynamical simulations on a moving mesh}

\author[V.~Springel]{\parbox{18cm}{Volker
Springel\footnotemark[1]\vspace{0.3cm}}\\ Max-Planck-Institut
f\"{u}r Astrophysik, Karl-Schwarzschild-Stra\ss{}e 1, 85740 Garching
bei M\"{u}nchen, Germany}

\begin{document}
\maketitle
\begin{abstract}
  Hydrodynamic cosmological simulations at present usually employ either the
  Lagrangian smoothed particle hydrodynamics (SPH) technique, or Eulerian
  hydrodynamics on a Cartesian mesh with (optional) adaptive mesh refinement
  (AMR).  Both of these methods have disadvantages that negatively impact
  their accuracy in certain situations, for example the suppression of fluid
  instabilities in the case of SPH, and the lack of Galilean-invariance and
  the presence of overmixing in the case of AMR.  We here propose a novel
  scheme which largely eliminates these weaknesses. It is based on a moving
  unstructured mesh defined by the Voronoi tessellation of a set of discrete
  points. The mesh is used to solve the hyperbolic conservation laws of ideal
  hydrodynamics with a finite volume approach, based on a second-order unsplit
  Godunov scheme with an exact Riemann solver. The mesh-generating points can
  in principle be moved arbitrarily. If they are chosen to be stationary, the
  scheme is equivalent to an ordinary Eulerian method with second order
  accuracy. If they instead move with the velocity of the local flow, one
  obtains a Lagrangian formulation of continuum hydrodynamics that does not
  suffer from the mesh distortion limitations inherent in other mesh-based
  Lagrangian schemes. In this mode, our new method is fully
  Galilean-invariant, unlike ordinary Eulerian codes, a property that is of
  significant importance for cosmological simulations where highly
  supersonic bulk flows are common.  In addition, the new scheme can adjust
  its spatial resolution automatically and continuously, and hence inherits
  the principal advantage of SPH for simulations of cosmological structure
  growth.  The high accuracy of Eulerian methods in the treatment of shocks is
  also retained, while the treatment of contact discontinuities improves. We
  discuss how this approach is implemented in our new code {\small AREPO},
  both in 2D and 3D, and is parallelized for distributed memory computers. We
  also discuss techniques for adaptive refinement or derefinement of the
  unstructured mesh.  We introduce an individual time-step approach for finite
  volume hydrodynamics, and present a high-accuracy treatment of self-gravity
  for the gas that allows the new method to be seamlessly combined with a
  high-resolution treatment of collisionless dark matter.  We use a suite of
  test problems to examine the performance of the new code and argue that the
  hydrodynamic moving-mesh scheme proposed here provides an attractive and
  competitive alternative to current SPH and Eulerian techniques.
\end{abstract}

\section{Introduction}
\renewcommand{\thefootnote}{\fnsymbol{footnote}}
\footnotetext[1]{E-mail: volker@mpa-garching.mpg.de}

Numerical simulations have become an indispensable tool to study astrophysical
problems of structure formation. They are the method of choice to predict the
fully non-linear outcome of the well-specified initial conditions of the
standard $\Lambda$CDM cosmology. In fact, they have played an instrumental
role to establish the viability of the standard cosmogony, and continue to be
of crucial importance for theoretical research on galaxy formation.

When only dark matter is considered, the current generation of
cosmological codes have reached a high-degree of accuracy, allowing an
impressive dynamic range in high-resolution studies of dark matter
clustering. There is now a consensus emerging in the field about
important key results, such as the central dark matter density profile
of collapsed halos \citep{Navarro2008,Stadel2008}. This is important
progress, which is in part due to the fact that there is little doubt
about what is required to achieve high accuracy in collisionless
simulations; this is simply an accurate gravitational force
calculation (which can be easily and objectively tested), accurate
time integration (also easy to check) and use of a large number of
particles (to make the collisionless dynamics more faithful, and
resolve smaller scales).

However, the situation is different for hydrodynamic cosmological
simulations. Here a variety of fundamentally quite different numerical
methods are in use, the most prominent ones are Lagrangian smoothed
particle hydrodynamics \citep[SPH;][]{Lucy1977,Gingold1977,Monaghan1992}
and Eulerian mesh-based hydrodynamics \citep[e.g.][]{Stone1992} with or
without adaptive mesh refinement \citep[AMR;][]{Berger1989}, but also
more exotic schemes have been proposed, such as treating hydrodynamics
through an approximation of the {\em collisional} Boltzmann equation
\citep{Xu1997,Slyz1999}.  An issue of great concern is that these
methods sometimes yield conflicting results even for basic calculations
that only consider non-radiative hydrodynamics
\citep[e.g.][]{Agertz2007,Tasker2008,Mitchell2008}.  Perhaps the most
famous example is the Santa Barbara cluster comparison project
\citep{Frenk1999}, and the systematic offsets in the core entropy that
are apparently produced between SPH and AMR codes. The right answer to
this problem is presently still unclear \citep[but see][for some
  hints]{Mitchell2008}. This uncertainty compromises the trust one would
like to have in the predictive power of ab-initio hydrodynamical
cosmological simulation, especially when applied to the full problem of
galaxy formation, where additional processes such as radiative cooling,
star formation and feedback must be included. The latter bring about
significant additional complexity, and further extend the dynamic range
that needs to be addressed.

It has become clear over recent years that both SPH and AMR suffer from
fundamental problems that make them inaccurate in certain regimes. SPH codes
have comparatively poor shock resolution, and offer only low-order accuracy
for the treatment of contact discontinuities. Worse, they appear to suppress
fluid instabilities under certain conditions \citep{Agertz2007}, as a result
of a spurious surface tension and inaccurate gradient estimates across density
jumps. While it is possible to alleviate these effects by introducing
artificial heat conduction or mixing terms \citep{Price2007KH,Wadsley2008}, or
a modified treatment of the artificial viscosity \citep{Dolag2005}, it is
still unclear whether any of these suggestions provides a universal solution
that generally improves the results without introducing significant problems
in other situations. In any case, the absence of any entropy production
through mixing in SPH, as particularly apparent in the entropy-formulation of
SPH \citep{Springel2002}, is an important conceptual difference to Eulerian
codes, where entropy is implicitly produced when fluxes with different
thermodynamic state are mixed together in a single cell.

Eulerian methods are the traditional method to solve the system of
hyperbolic partial differential equations that constitute ideal
hydrodynamics. There are decades of experience with these methods in
computational fluid dynamics, and accurate Godunov schemes exist which
offer high-order spatial accuracy, have negligible postshock
oscillations, and low numerical diffusivity. However, fundamental
problems remain with these methods as well. Perhaps the most serious
one is their lack of Galilean-invariance, making the results sensitive
to the presence of bulk velocities
\citep[e.g.][]{Wadsley2008,Tasker2008}. This is a source of
substantial concern in simulations of galaxy formation, where galaxies
move with large speeds relative to each other, speeds that are often
orders of magnitude larger than the sound speed of the dense
interstellar medium that one wants to follow
hydrodynamically. Similarly, it is also challenging with AMR to follow
a highly refined region that moves with large velocity relative to the
reference frame adopted for the calculation as a whole, because
refinement criteria that correctly `anticipate' the motion of a system
across a grid are difficult to construct.

Another concern lies in the mixing inherent in multi-dimensional
Eulerian hydrodynamics. This provides for an implicit source of
entropy, with sometimes unclear consequences, a situation that prompted
\citet{Wadsley2008} to propose an explicit modeling of the mixing
through additional terms in the fluid equations. Even though it is
clear that some mixing helps and provides a dissipation scale for the
finite resolution, there may well be overmixing if the resolution is
limited or the bulk velocities are large. Also, it is rather unclear
whether the turbulent cascades that actually happen in nature are
correctly captured if the AMR hierarchy is truncated at a certain
maximum refinement level \citep{Iapichino2008a,Iapichino2008b}.  It
has been suggested that this can lead to unphysical solutions for
fluid instabilities like the Rayleigh-Taylor instability, and that
recovery of the correct behaviour requires subresolution models for
turbulence \citep{Scannapieco2008}.  In any case, the different
treatment of mixing is arguably the most fundamental difference
between SPH and AMR \citep[see also][]{Trac2007,Mitchell2008}.

It has also become clear that current cosmological AMR codes presently in use
have problems to accurately treat structure formation driven by gravitational
instability \citep{Shea2005,Heitmann2007}. This happens because it is quite
difficult to refine `early enough' on {\em all} the many small density
fluctuations that grow at high redshift, and if a refinement is placed, the
resolution increases {\em discontinuously} by a factor of 2 per dimension.  In
typical calculations, this introduces a subtle suppression of the growth of
small halos, such that the halo mass functions show a deficit of small halos
at late times. The AMR approach is therefore not ideal for a high accuracy
treatment of the N-body problem posed by cosmic structure; only when very fine
base meshes and conservative refinement criteria are adopted, do AMR results
approximatively recover those obtained comparatively easily by SPH codes,
which treat self-gravity in a Lagrangian fashion, and do not have
discontinuous jumps in resolution.

As has long been recognized, Eulerian methods have also problems to
properly resolve flows where the kinetic energy is much larger than
the thermal energy, and both the pre- and postshock gas move
supersonically with respect to the grid. This situation is ubiquitous
in cosmological applications, and prompted the development of schemes
that try to circumvent the problem when necessary, such as the `dual
energy formalism' \citep{Bryan1995} or schemes that evolve a
conservation law for the entropy outside of shocks
\citep{Ryu1993}. Usage of such schemes usually means that exact energy
conservation is sacrificed in favour of a more accurate treatment of
the gas entropy.  The need for such fixes is in part a consequence of
the choice of a fixed reference frame for describing the flow, and
hence is related to the Galilean non-invariance of the Eulerian
treatment. Indeed, there have been attempts to solve the bulk-flow
problem by formulating the equations such that a more natural
reference frame can be adopted. In particular, \citet{Trac2004}
developed a special method where a frame change is introduced when the
gas-dynamical equations are coupled to self-gravity. The frame
velocity is estimated based on a smoothed large-scale velocity
field. This relatively simple approach can reduce the artefacts
stemming from large bulk flows, but it does not really render the
results invariant of the original reference frame and therefore does
not provide a complete solution for the non-Galilean invariance of the
underlying Eulerian approach.

A more radical approach is to let the mesh itself move. This is an obvious and
old idea, but one fraught with many practical difficulties that have so far
prevented any widespread use in astrophysics and cosmology. There have been a
number of attempts that seemed promising however. In particular,
\citet{Whitehurst1995} presented his first order accurate code {\small FLAME}
for hydrodynamics based on Delaunay and Voronoi tessellations and his `signal
method' which was able to perform quite well on a number of test problems.
Unfortunately no practical applications followed.

\citet{Gnedin1995} and \cite{Pen1998} have presented moving mesh hydrodynamic
algorithms that have successfully been applied to a range of cosmological
problems.  Their methods rely on the continuous deformation of a Cartesian
grid. However, the need to limit the maximum grid distortions severely limits
the flexibility of the codes for situations in which the mesh becomes heavily
distorted, and special measures were required to let the codes evolve
cosmological density fields into a highly clustered state. For example,
\citet{Gnedin1995} addressed this by letting an Eulerian solver take over in
regions where the Lagrangian approach fails due to severe mesh distortions.
In general, mesh tangling (manifested in `bow-tie' cells and hourglass like
mesh motions) is the traditional problem of multi-dimensional Lagrangian
hydrodynamics. In arbitrary Lagrange-Eulerian (ALE) approaches, remapping
techniques to more regular meshes are used to counteract the deteriorating
influence of mesh distortions, allowing the calculation to continue past the
point where it would otherwise be stopped by mesh twisting.  The remapping is a
diffusive operation, however, and the task to automatically construct `good' new
regularized meshes is very challenging in general.  This appears to have impaired
wide-spread adoption of ALE techniques in astronomy thus far, apart from
notable exceptions in stellar astrophysics \citep{Murphy2008}.

Another interesting study directly related to our approach was that of
\citet{Xu1997}, who presented an N-body and hydro-solver on an {\em
  unstructured}, fixed mesh. This work used a Delaunay tessellation, and the
hydrodynamic scheme was formulated based on a gas-kinetic approach, with the
goal to apply it to cosmological simulations of structure formation. However,
the method appears to have not been investigated much further afterwards
(except for an unpublished master thesis by M.~Ruetalo, U.~of Toronto,
privately communicated to us by J.~R.~Bond).  We note that unstructured
triangular meshes are regularly used in engineering applications, however
often in the context of stationary flows, for example around airplane foils
\citep[see][for a review]{Mavripilis1997}.

We here propose a new formulation of continuum hydrodynamics based on an
unstructured mesh. The mesh is defined as the Voronoi tessellation of a set of
discrete mesh-generating points, which are in principle allowed to move
freely. We show how a finite-volume hydrodynamic scheme with the Voronoi cells
as principle control volumes can be consistently defined. Most importantly,
due to the mathematical properties of the Voronoi tessellation, the mesh
continuously deforms and changes its topology as a result of the point motion,
without ever leading to the dreaded mesh-tangling effects that are the curse
of traditional ALE methods. Our method therefore retains the principal
advantage of the mesh-free SPH approach: It offers free and unrestricted,
continuous adjustment of its resolution to local clustering. In addition, we
show that our new method is Galilean invariant when the mesh is moved along
with the flow. There are no preferred directions in it, unlike in Cartesian
grids. Thanks to its Lagrangian nature, mesh refinement is normally not needed
when one wants to maintain roughly constant mass resolution, but if desired,
the Voronoi mesh may also be adaptively refined or derefined.

With these properties, the moving-mesh approach represents a
compromise between SPH and AMR. It inherits the automatic adaptivity,
geometric flexibility and Galilean invariance of SPH, while it shares
the high-accuracy treatment of shocks, shear waves, and fluid
instabilities, as well as the low noise and the absence of artificial
viscosity, with AMR. A further advantage of the method lies in its
ability to easily handle boundary conditions at curved surfaces that
can be stationary, move with the flow, or are governed by a prescribed
velocity field.  We also show how the method can be made adaptive in
time by means of individual timesteps, and how it can be coupled to a
high-resolution gravitational solver (a TreePM scheme) that gets
around the problems experienced by the current generation of AMR codes
in cosmological structure formation calculations. We think this makes
the new code {\small AREPO}\footnote{\parbox[t]{5.5cm}{Named after the
    enigmatic word AREPO in the Latin palindromic sentence {\em sator
      arepo tenet opera rotas}, the `Sator
    Square'.}\hfill \parbox[b]{1.6cm}{\resizebox{1.6cm}{!}{\includegraphics{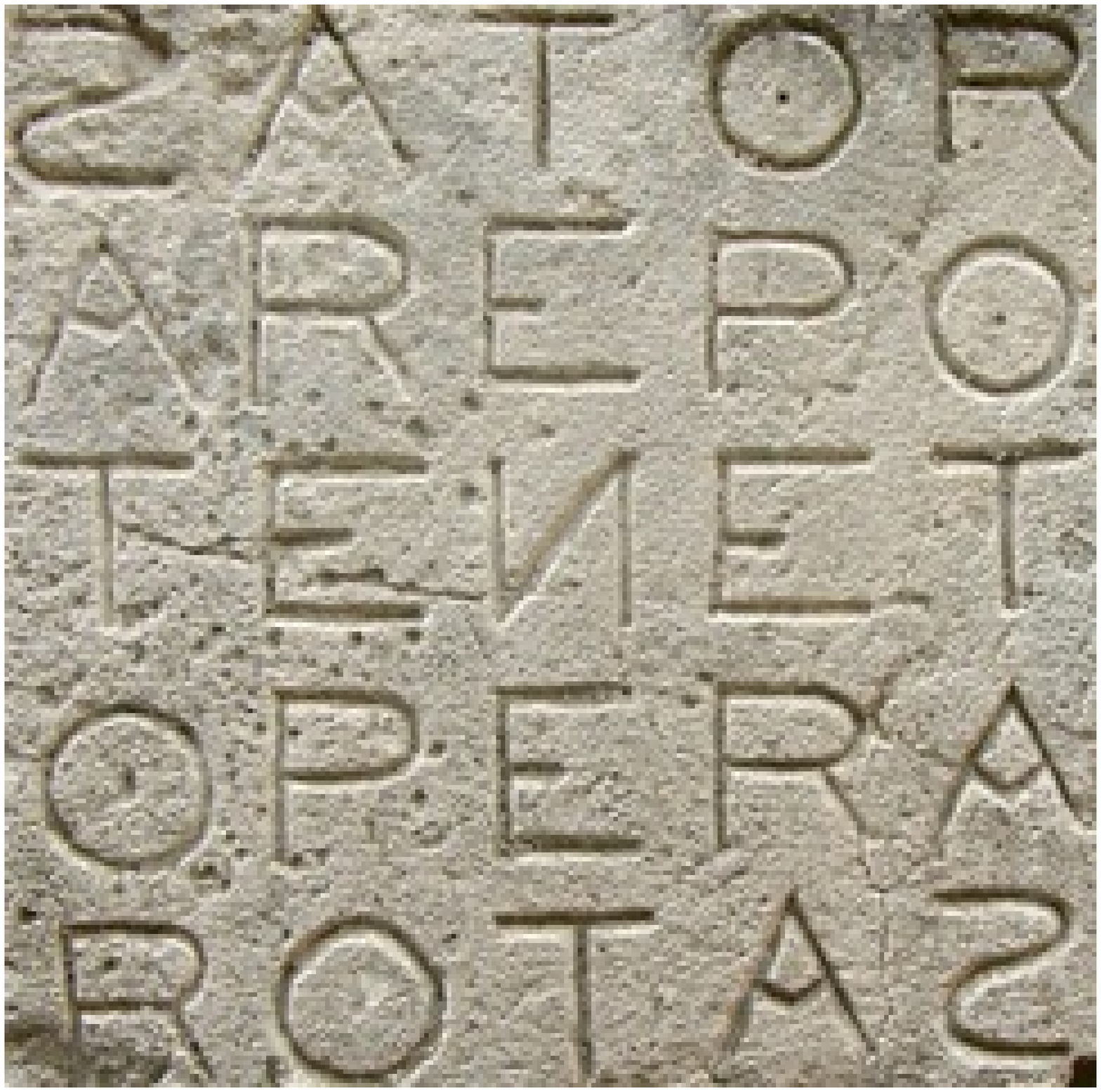}}
\vspace*{-1.1cm}}}
that we built with this approach a very interesting and competitive method for
future applications in cosmology, as well as in other fields.

We demonstrate the performance of {\small AREPO} in a number of test
problems, which include purely hydrodynamical tests in 1D, 2D, and 3D,
as well as simulations where self-gravity is included. We also present
comparisons with the state-of-the-art Eulerian code {\small ATHENA}
\citep{Stone2008}, both to validate our hydrodynamic algorithms, and
to discuss issues of Galilean (non)invariance.  Because our scheme
relies on Voronoi meshes, it is very important to develop algorithms
that are able to construct the mesh rapidly and robustly on
distributed memory platforms.  We will therefore discuss in some
detail the solutions we have developed for this problem.

This paper is structured as follows. In Section~\ref{SecMesh}, we discuss our
mesh generation algorithms, both in 2D and 3D. In Section~\ref{SecHydro}, we
then formulate continuum hydrodynamics on the Voronoi mesh, based on a
finite-volume ansatz and a second-order accurate extension of Godunov's
method. In Section~\ref{SecMeshRegularity}, we discuss how the mesh motion can
be steered to maintain constant mass or volume per cell, or to improve mesh
regularity. Our treatment of self-gravity is described in
Section~\ref{SecGravity}, and the refinement or de-refinement of the
unstructured mesh in Section~\ref{SecRefinement}. In
Section~\ref{SecTimeintegration}, we outline our methods for time-integration,
in particular the use of individual timesteps, and we describe the basic
architecture of our new simulation code. We then turn to an extensive
discussion of test problems, including pure hydrodynamical tests in
Section~\ref{SecHydroTests}, and tests that include the gravitational effects
from the gas itself and from a collisionless dark matter component in
Section~\ref{SecGravityTests}. Finally, we summarize and discuss our findings
in Section~\ref{SecDiscussion}.

\section{Generating Delaunay and Voronoi meshes} \label{SecMesh}

\begin{figure*}
\bc
\resizebox{17cm}{!}{\includegraphics{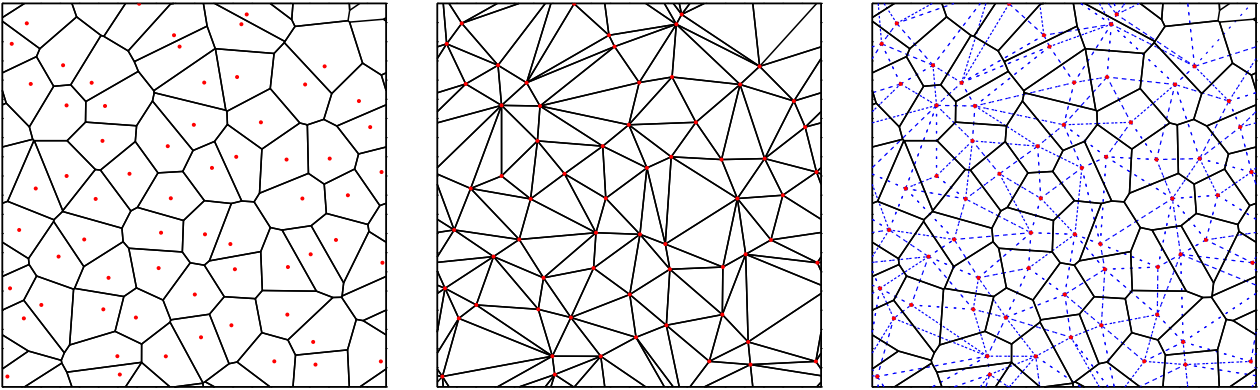}}\\
\caption{Example of a Voronoi and Delaunay tessellation in 2D, with periodic
  boundary conditions. The panel on the left shows the Voronoi tessellation
  for $N=64$ points (shown as red circles), the panel in the middle gives the
  corresponding Delaunay tessellation, while the panel on the right shows both
  simultaneously (solid lines show the Voronoi, dashed lines the Delaunay
  tessellation).
  \label{FigVoronoiExample}}
\ec
\end{figure*}

For a given set of points, a Voronoi tessellation of space consists of
non-overlapping cells around each of the sites such that each cell contains
the region of space closer to it than any of the other sites.  This
definition holds both in 2D and 3D, and can be readily extended to higher
dimensions if desired. A direct consequence of this definition is that the
cells are polygons in 2D and polyhedra in 3D, with faces that are equidistant
to the mesh-generating points of each pair of neighbouring cells.

Closely related to the Voronoi tessellation is the Delaunay tessellation, which
is in fact the topological dual of the Voronoi diagram. In 2D, the Delaunay
tessellation for a given set of points is a triangulation of the plane, where
the points serve as vertices of the triangles.  The defining property of the
Delaunay triangulation is that each circumcircle around one of the triangles
of the tessellation is not allowed to contain any of the other mesh-generating
points in its interior. This {\em empty circumcircle} property distinguishes
the Delaunay triangulation from the many other triangulations of the plane
that are possible for the point set. Furthermore, this condition {\em
  uniquely} determines the triangulation for points in general position.
Similarly, in three dimensions, the Delaunay tessellation is formed by
tetrahedra that are not allowed to contain any of the points inside their
circumspheres.

As an example, we show in Figure~\ref{FigVoronoiExample} the Delaunay and
Voronoi tessellations for a small set of points in 2D, enclosed in a box with
imposed periodic boundary conditions. The midpoints of the circumcircles
around each Delaunay triangle form the vertices of the Voronoi cells, and for
each line in the Delaunay diagram, there is an orthogonal face in the Voronoi
tessellation. This topological duality also holds in 3D, where each edge of a
tetrahedron lies orthogonal to a face of a Voronoi polyhedron.

Delaunay and Voronoi tessellations are basic constructions in computational
geometry, and numerous mathematical properties are known for them
\citep{Okabe2000}. For example, the Delaunay triangulation maximises the minimum
angle among all possible triangulations for a given point set.  For points
in general location, the Delaunay and Voronoi tessellations are unique. If
there exist circles with more than 3 points on them (or spheres with more than
4 points in 3D), the Delaunay triangulation contains degenerate cases where
the triangulation may flip by an infinitesimal motion of one of the
points. Note however that the Voronoi tessellation is still unique in this
case. In fact, an edge between two degenerate points of the Delaunay
triangulation has a dual Voronoi area of zero size. Nevertheless, degeneracies
can be a significant problem for the robustness of mesh-construction
algorithms, an issue we will discuss in more detail later on.

There is a sizable body of literature in computational geometry on algorithms
for constructing the Delaunay and Voronoi tessellations. It is in general much
easier to construct the Delaunay tessellation and obtain the Voronoi
tessellation from it, instead of trying to directly construct the Voronoi
tessellation. The Voronoi construction hence effectively reduces to the problem
of constructing the Delaunay triangulation, an approach we will also follow
here.

The different construction algorithms for the Delaunay triangulation include:
\begin{enumerate}
\item incremental insertion,
\item projection of the convex hull of a higher dimensional embedding,
\item recursive subdivision (divide \& conquer),
\item direct incremental construction,
\item improving an arbitrary triangulation by flipping.
\end{enumerate}

Incremental insertion due to \citet{Bowyer1981} and \citet{Watson1981} is
conceptionally the simplest approach.  Here one starts with a valid
tessellation, inserts an additional point, and then repairs the mesh locally by
`flipping' triangles/tetrahedra to restore Delaunayhood (see below).  It can
be shown that the worst case behaviour for this method (for unfavourable input
particle sets) scales as $N^2$, but in practice, the observed scaling is 
much better. In fact, for point sets in general location which are added to
the tessellation in random order, a scaling of $N\log N$ is reached.

\begin{figure*}
\bc
\resizebox{14cm}{!}{\includegraphics{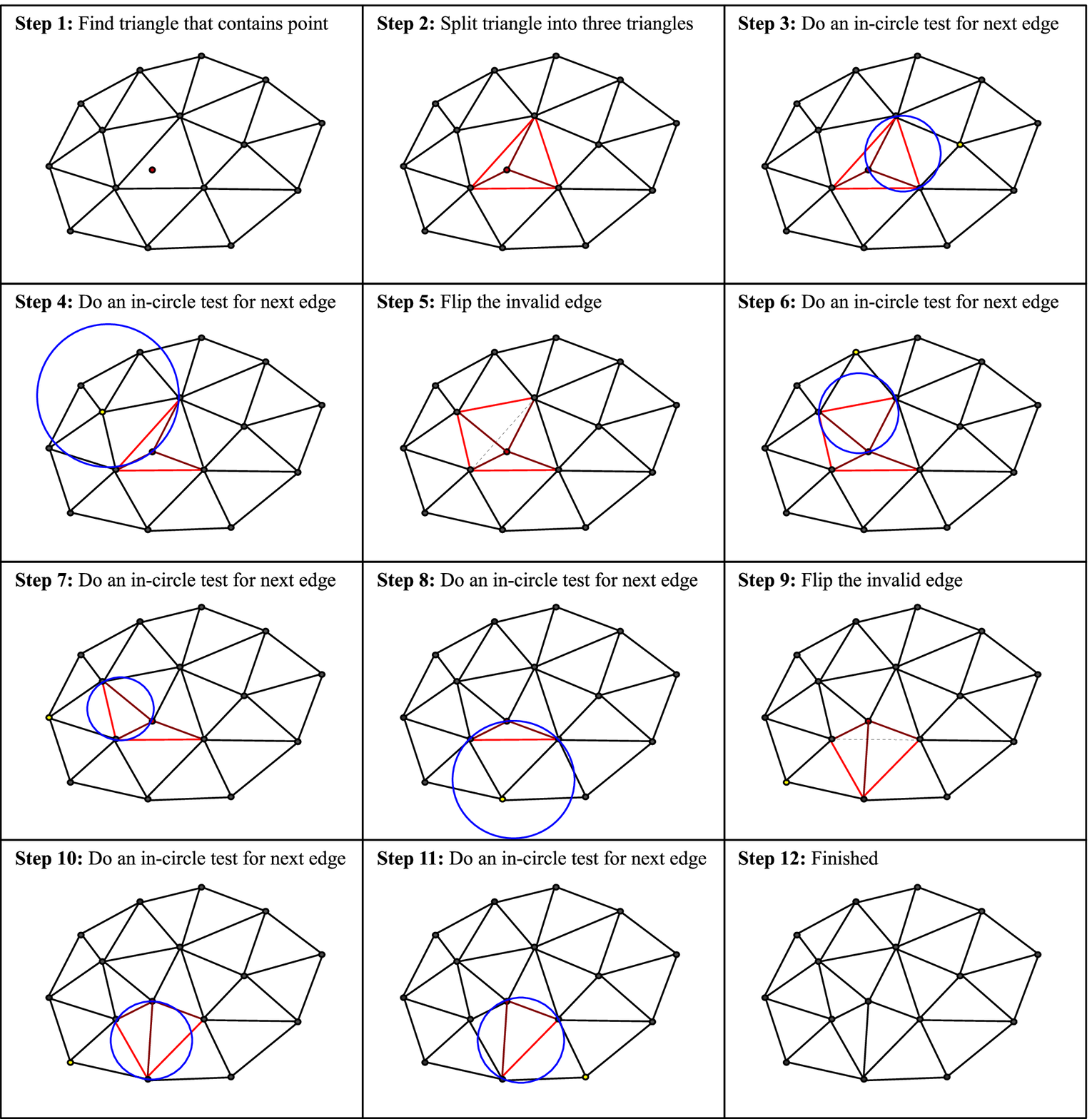}}
\caption{The point insertion algorithm in 2D. 
We start with a valid Delaunay
triangulation in which we want to insert an additional point. 
We first locate the triangle containing the point (step 1), 
then split it into three triangles (step 2). 
The edges (drawn in red) in the  new triangles opposite of the inserted point  may violate the in-circle 
criterion and need to be tested individually. If an edge is Delaunay (step 3),
it is part of the final tessellation, but if it violates the in-circle
criterion (step 4), the edge needs to be flipped in the quadrilateral formed
by the adjacent triangles (step 5). The flip generates additional edges that
need to be tested (steps 6 and 7). Any violating edge found (e.g. step 9)
needs to be corrected by flips. Once all remaining new edges are validated
(steps 10 and 11), we arrive again at a valid Delaunay tessellation (step 12).
\label{FigInsertionAlgorithmus}}
\ec
\end{figure*}

Another interesting method is obtained by adding an additional coordinate to
the point set, $r^2=x^2+y^2+z^2$, which effectively produces a higher
dimensional embedding of the form of a paraboloid. The convex hull of this
lifted point set yields the Delaunay triangulation when projected down onto
the original lower dimensional space. This method hence reduces the Delaunay
triangulation to the problem of finding the convex hull in $n$-dimensional
space, for which the {\em quickhull} algorithm can be used.

In two dimensions, the fastest algorithm is based on a divide and conquer
strategy, as proposed by \citet{Guibas1985} and refined by
\citet{Dwyer1987}. Here the point set is recursively subdivided, until a
single triangle can be constructed. These sets are then merged along the
dividing lines. Unfortunately, this elegant approach is difficult to implement
in three dimensions, primarily because of the difficulty of constructing a
two-dimensional merging phase along the dividing planes. \citet{Cignoni1998}
overcame this problem in the {\em Dewall} algorithm, essentially by reversing
the order of the split and merge steps. These authors first construct a
``wall'' of Delaunay triangles directly, which splits the tessellation into two
halfs; those can then be processed recursively in turn.

Direct incremental construction techniques start out from one Delaunay
edge, and then find the correct point that completes it to form a
Delaunay triangle. This has been used by \citet{Weygaert1994}, for
example, who applied Voronoi tessellations for a statistical analysis
of cosmic structures \citep[a comprehensive dicussion and overview
  about this topic is given by][]{Weygaert2007}.

Finally, the flipping method starts from an arbitrary triangulation, and then
tries to give it the Delaunay property by local changes in the triangulation
(``flips''). In 2D, it can be shown that this can always succeed through
simple flips of edges between two adjacent triangles. However, in 3D, one may
get get stuck with tetrahedralizations that are not flipable into the correct
Delaunay triangulation. While this may appear as a show stopper for
incremental insertion algorithms in 3D, \citet{Edelsbrunner1996} have shown
that this is not the case. Provided one starts with a valid Delaunay
triangulation, local flips can always restore Delaunayhood after a further
point has been inserted into the mesh, so that the incremental insertion
strategy is actually a robust algorithm also for the three-dimensional case.

We use the incremental insertion strategy in our new hydrodynamical code. It
is among the fastest known algorithms, and most importantly for us, it allows
implementing our particular parallelization strategy for distributed memory
machines, which requires that additional points from other processors can be
easily added to an existing local tessellation. This task can not be readily
accomplished with the other tessellation approaches, where normally the full
point set needs to be known already at the start of the tessellation
procedure.

We illustrate the sequential insertion algorithm in
Figure~\ref{FigInsertionAlgorithmus}.  Starting from a valid Delaunay
tessellation, the new point first needs to be located in one of the triangles
(or tetrahedra in 3D), a problem we shall discuss further below. After this
first step, the identified triangle is then subdivided into 3 triangles by
inserting the point, yielding a new triangulation. However, one or several of
the new triangles may now violate the {\em empty circumcircle} criterion. We
note that the latter can also be formulated for individual edges; we say an
edge is a {\em Delaunay edge} if there exists a circle through both of its
endpoints which does not contain any other point in its interior. It can be
shown that if an edge is Delaunay, it is part of the correct Delaunay
triangulation. It is easy to show that the three edges around the newly
inserted point are Delaunay, but the opposite edges may have lost this
property as a result of the insertion (marked in red in `Step 2' of
Fig.~\ref{FigInsertionAlgorithmus}). These edges must be tested in turn using
the in-circle criterion. If a violating edge is found (Step 4), it is flipped
in the quadrilateral formed by the two adjacent triangles. This produces two
more edges that may now have lost the Delaunay property, and which lie again
opposite of the inserted point. These edges are added to the list of edges
that need to be tested with the in-circle criterion. The algorithm continues
until this list is exhausted, at which point the new site has been
successfully inserted, and a new valid Delaunay triangulation has been
obtained.

To make sure that every point that needs to be inserted always lies in a
triangle to begin with, we start the tessellation procedure with a fiducial
large triangle enclosing the whole system. Especially in 3D dimensions, this
simplifies the algorithms enormously, as the difficult case of an insertion of
a point outside of the convex hull of the current tessellation does not have
to be dealt with. 

In practice, we will always use periodic or reflecting boundaries that are
realized with a layer of {\em ghost cells} (see below). The enclosing triangle
is chosen large enough that both the primary simulation domain and the ghost
region are enclosed in its interior, such that the enclosing triangle's shape
or orientation does not influence the used part of the final tessellation in
any way.

The geometric in-circle test
can be formulated compactly in terms of an evaluation of a
determinant. For example, in 2D, the in-circle test is given by
\begin{eqnarray}
T_{\rm InCircle}(\vec{a},\vec{b},\vec{c},\vec{d})=
\left|
\begin{array}{cccc}
1& a_x & a_y & a_x^2+a_y^2\\
1& b_x & b_y & b_x^2+b_y^2\\
1& c_x & c_y & c_x^2+c_y^2\\
1& d_x & d_y & d_x^2+d_y^2\\
\end{array}
\right| =  \nonumber \\
\left|
\begin{array}{ccc}
b_x - a_x & b_y - a_y & (b_x-a_x)^2+(b_y-a_y)^2\\
c_x - a_x & c_y - a_y & (c_x-a_x)^2+(c_y-a_y)^2\\
d_x - a_x & d_y - a_y & (d_x-a_x)^2+(d_y-a_y)^2\\
\end{array}
\right|.
\end{eqnarray} 
Provided the triangle $(\vec{a}, \vec{b}, \vec{c})$ is positively oriented,
this gives $T_{\rm InCircle}(\vec{a},\vec{b},\vec{c},\vec{d})<0$ if the point
$\vec{d}$ lies inside the circumsphere of the triangle, and $T_{\rm
  InCircle}(\vec{a},\vec{b},\vec{c},\vec{d})>0$ if the point is outside.
$T_{\rm InCircle}(\vec{a},\vec{b},\vec{c},\vec{d})=0$ corresponds to the
interesting case that $\vec{d}$ lies exactly on the circumsphere of the
triangle. It turns out that correct detection of this degenerate case is
problematic in the light of finite floating point precision on a computer, but
crucial for the stability of the mesh-generating algorithm, an issue which we
shall discuss further below.

The orientation of a triangle can also be established with a determinant,
through
\begin{eqnarray}
T_{\rm Orient2D}(\vec{a},\vec{b},\vec{c})=
\left|
\begin{array}{ccc}
1& a_x & a_y \\
1& b_x & b_y \\
1& c_x & c_y \\
\end{array}
\right|
=
\left|
\begin{array}{cc}
b_x - a_x & b_y-a_y \\
c_x - a_x & c_y -a_y \\
\end{array}
\right|.
\end{eqnarray} 
A positive value indicates positive orientation.  Internally, we always store
the triangles/tetrahedra of our Delaunay triangulation such that they are
positively oriented, which minimizes the required number of orientation tests.

\begin{figure}
\bc
\resizebox{3.5cm}{!}{\includegraphics{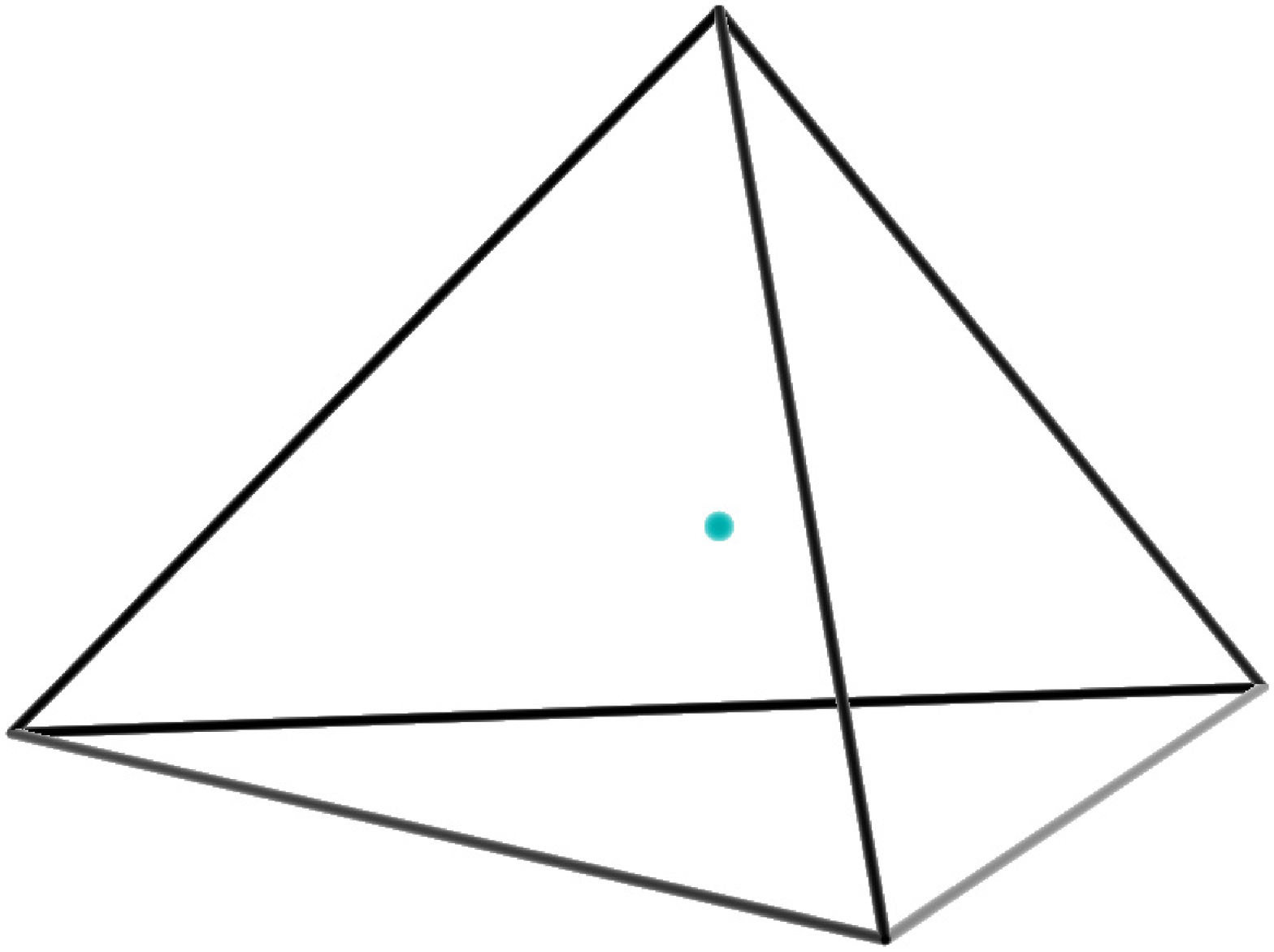}}\ \ %
\resizebox{3.5cm}{!}{\includegraphics{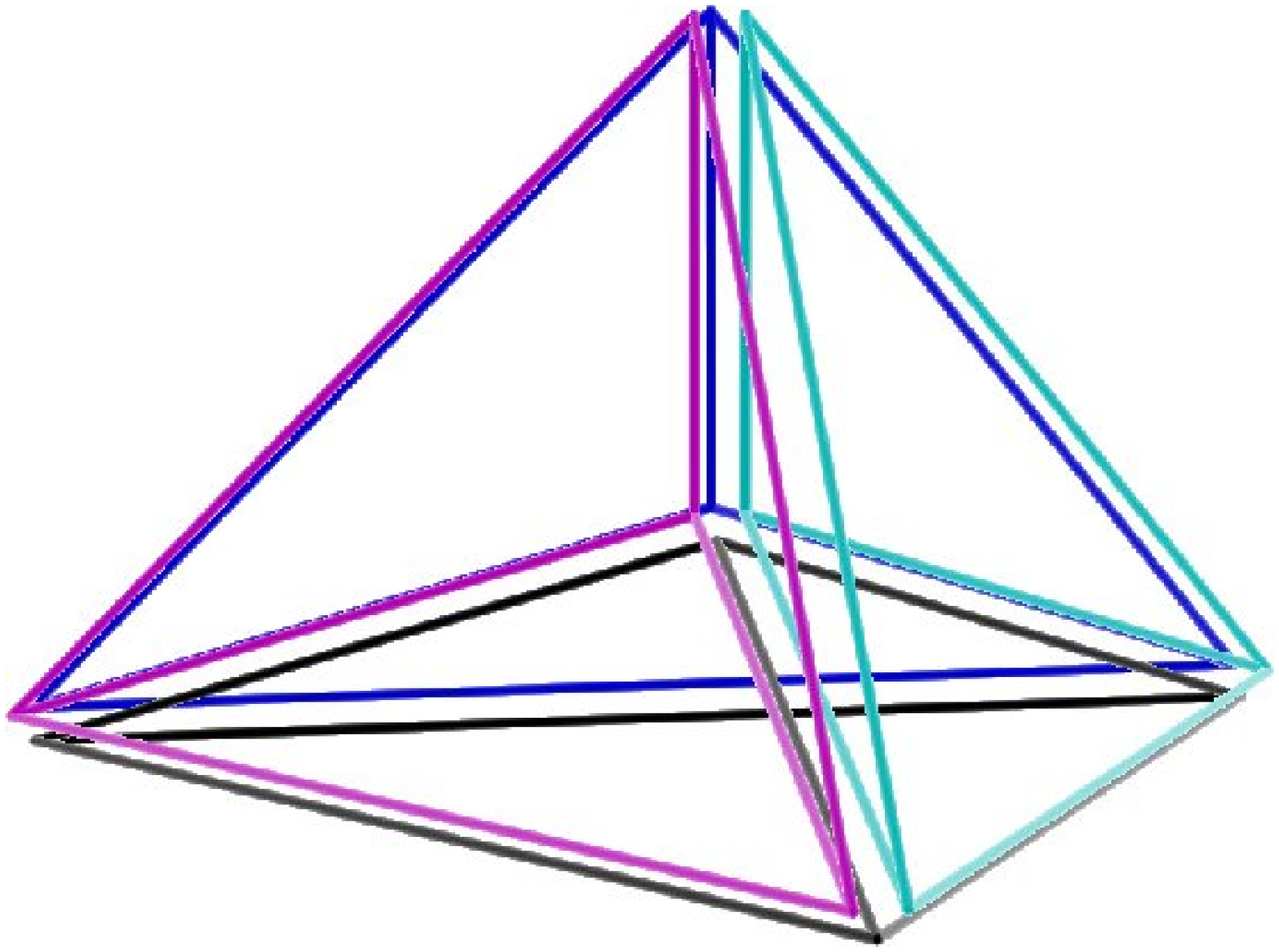}}\\
\caption{A `1-to-4' flip. A newly inserted point splits its insertion
 tetrahedron into 4 daughter tetrahedra.
\label{Fig1-4-Flip}}
\ec
\end{figure}

In three dimensions, the incremental construction algorithm works very
similarly, apart from a few additional complications. Briefly, when a point is
inserted, we now need to carry out a `1-to-4 flip', i.e.~we replace the
insertion tetrahedron by four new tetrahedra, as illustrated in
Figure~\ref{Fig1-4-Flip}.

\begin{figure}
\bc
\resizebox{3.5cm}{!}{\includegraphics{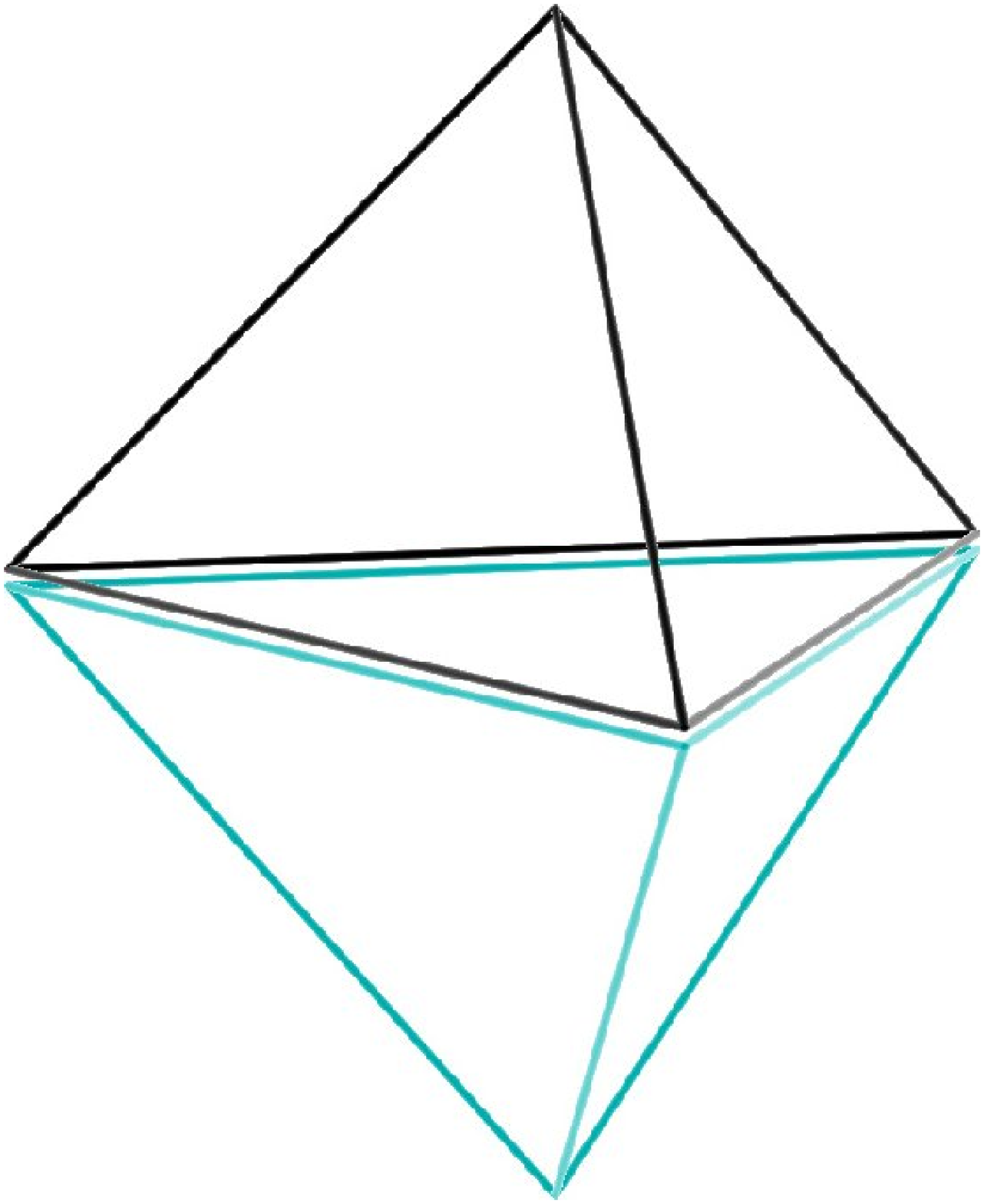}}\ \ %
\resizebox{3.5cm}{!}{\includegraphics{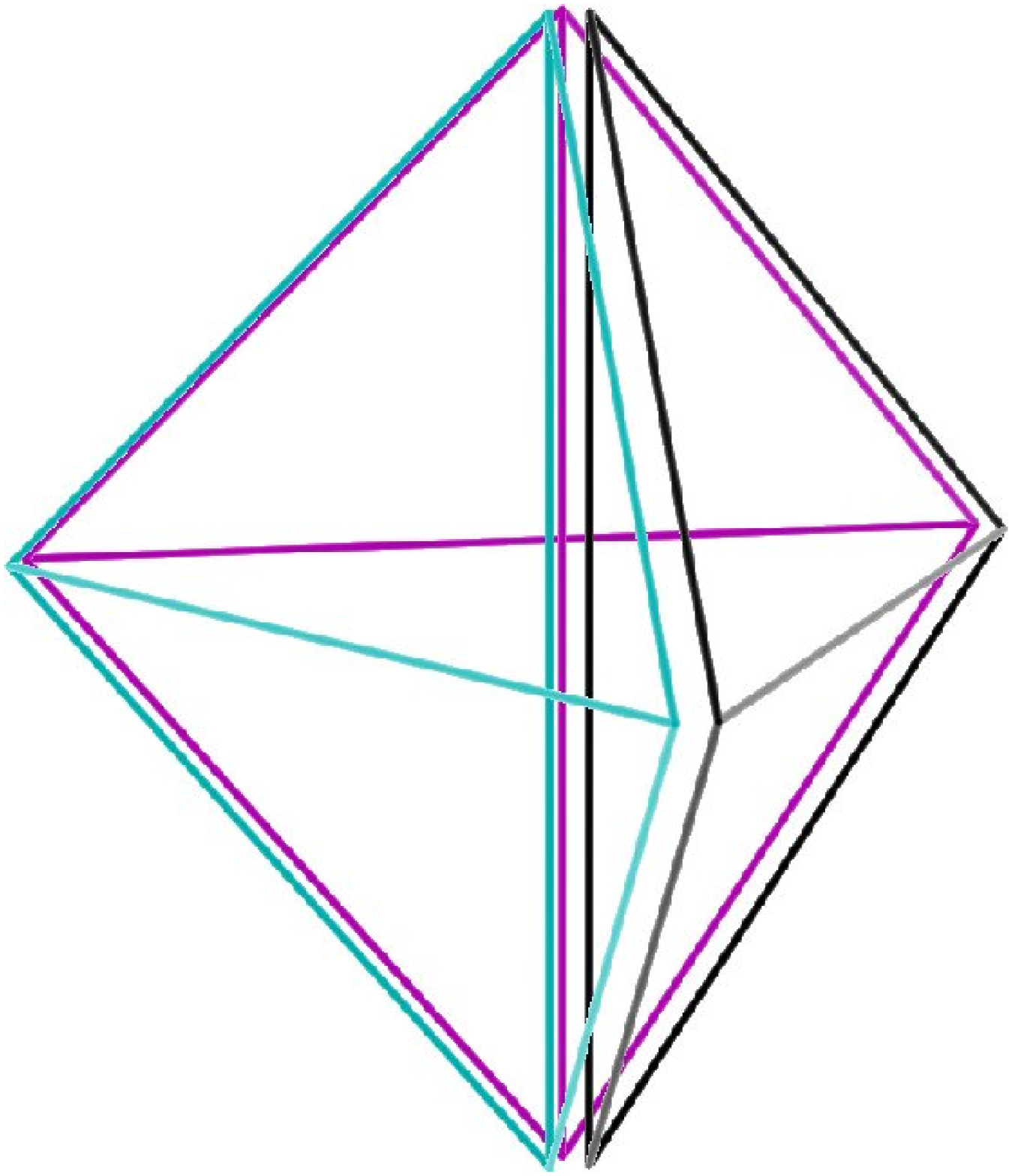}}\\
\caption{The standard replacement operation in 3D required for restoring
Delaunayhood. It consists of 2-to-3 (from left to right) or 3-to-2 (from right
to left) flips of tetrahedra. Note that the 2-to-3 
flip is only possible if the 
line connecting the two
points opposite of the common face intersects the interior of this face.
 Conversely, the 3-to-2 flip is only possible 
if an edge is shared by exactly three tetrahedra.
\label{Fig2-3-Flip}}
\ec
\end{figure}

Just as in 2D, this can render tetrahedra that share a face with the four new
tetrahedra invalid. These tetrahedra have to be subjected to the {\em
  in-sphere test} with the inserted point. We store the faces that need to be
tested on a stack, where we specify the face that needs to be tested for
Delaunayhood with a reference to a tetrahedron and the face's opposite point
(which is always the inserted point). If a face that is pulled from the stack
fails the in-sphere test, we need to check how we can replace the two adjacent
tetrahedra.  Unlike in 2D, we cannot simply replace two tetrahedra with two
other tetrahedra. Instead, we may be able to replace the two tetrahedra with
three tetrahedra, in a `2-to-3 flip', provided the line connecting the two
tips opposite of the common triangle of the two tetrahedra intersects this
triangle in its interior. This is illustrated in Figure~\ref{Fig2-3-Flip}. If
on the other hand the intersection point lies outside {\em one} of the edges
of the common triangle, then there is a tetrahedron formed by this edge and
the two tips which needs to be included in the replacement operation. We can
then replace these three tetrahedra with two, in a `3-to-2 flip', which is
just the reverse of the `2-to-3 flip' shown in Figure~\ref{Fig2-3-Flip}. We
note that the intersection point may also lie outside {\em two} of the edges
of the common triangle; in this case the violating face is not flipable and
can be skipped. It can be shown that the algorithm nevertheless finishes
successfully thanks to the flips that can be carried out for other violating
faces.  Depending on the type of the flip that has been performed, either two
or three new faces need to be put onto the test stack. The tests and flips are
then continued until the stack is empty, at which point the Delaunay
tessellation is valid again.

In three dimensions, the relevant determinant for the in-sphere test is given
by
\begin{equation}
T_{\rm InSphere}(\vec{a},\vec{b},\vec{c},\vec{d},\vec{e})=
\left|
\begin{array}{ccccc}
1& a_x & a_y & a_z & a_x^2+a_y^2+a_z^2\\
1& b_x & b_y & b_z & b_x^2+b_y^2+b_z^2\\
1& c_x & c_y & c_z & c_x^2+c_y^2+c_z^2\\
1& d_x & d_y & d_z & d_x^2+d_y^2+d_z^2\\
1& e_x & e_y & e_z & e_x^2+e_y^2+e_z^2\\
\end{array}
\right|.
\end{equation}
This is negative if the point $\vec{e}$ lies inside the circumsphere
around the positively oriented tetrahedron
$(\vec{a},\vec{b},\vec{c},\vec{d}$), it is positive if the point lies outside,
and zero if it is exactly on the circumsphere. The orientation of a
tetrahedron can be established by testing
the sign of 
\begin{eqnarray}
T_{\rm Orient3D}(\vec{a},\vec{b},\vec{c},\vec{d})=
\left|
\begin{array}{cccc}
1& a_x & a_y & a_z \\
1& b_x & b_y & b_z \\
1& c_x & c_y & c_z \\
1& d_x & d_y & d_z \\
\end{array}
\right|,
\end{eqnarray} 
which is positive for positive orientation.\footnote{Note that the value of
  the determinant is equal to six times the volume of the tetrahedron spanned
  by the four points.}

Once the Delaunay triangulation is generated, we calculate the areas and
centres of the Voronoi faces, and the volumes of all Voronoi cells, as well as
their centres-of-mass. To this end, we first calculate the midpoints of the
circumspheres around each tetrahedron; these points form the vertices of the
Voronoi cells. We then introduce a new data structure for each Voronoi face,
storing the face area and references to the two adjacent cells, information
that is later needed to determine the hydrodynamic fluxes across the face. To
calculate the area of a Voronoi face, we circle in clockwise fashion around
all tetrahedra that share the same Delaunay edge between the two
mesh-generating points that belong to the face.  Note that the line connecting
these two points need not necessarily intersect the face.  Once we have
determined the area of the face, we can also easily obtain the volumes of the
two equally-sized pyramids formed by the face and its two associated
mesh-generating points. The volume of each Voronoi cell is then obtained as a
sum over the pyramid volumes of all the cell's surface polyhedra.

\subsection{Data structures for the tessellation}

From the above it is clear that an important practical consideration for
working with an unstructured polyhedral mesh is the use of efficient data
structures to represent the tessellation. Ideally, the data structure should
allow rapid and convenient access to the topological objects of the
tessellation, such as individual triangles, the surfaces of Voronoi polyhedra,
and neighbourhood relations, while at the same time not requiring too much
memory.  In 2D, \citet{Guibas1985} introduced an elegant {\em quad-edge} data
structure which can encode both the Delaunay triangulation and its dual at the
same time. Besides storing references to the points of an edge, this
edge-based structure stores links to the first adjacent edges in a clockwise
or anticlockwise direction around the end points. While being very elegant, it
is difficult to extend this structure to three dimensions. The `face-edge'
structure of \citet{Dobkin1989} is one such possibility, but it produces
substantial memory overhead. We therefore follow the approach taken in most
codes for 3D Delaunay triangulation and adopt full tetrahedra as basic data
structures for describing the mesh. In 2D, we correspondingly use full
triangles.

For each tetrahedron, we store references to its four points. These are
oriented such that the fourth point lies above the oriented triangle formed by
the first three points.  We also store along with each point of a tetrahedron
a reference to the adjacent tetrahedron that lies opposite of the point and
shares a face with the present tetrahedron. In addition, we store the index
location of the point in this adjacent tetrahedron that lies opposite to the
common face. This simplifies various types of construction and navigation
tasks in the tessellation. Note that using this data structure we can also
easily specify individual faces of tetrahedra in terms of a reference to the
tetrahedron and to the point that lies {\em opposite} to the face in question.

A disadvantage of our data structure is its comparatively large memory
requirement. If pointers are used for references to the 4 vertices and 4
adjacent tetrahedra of each tetrahedron, 36 bytes are required on 32-bit
architectures per tetrahedron, or 68 bytes on 64-bit machines (using
integer indices instead of pointers can however easily reduce this back
to 36 bytes, which is completely sufficient if there are less than $\sim
200\,{\rm GB}$ or so available per MPI task, which is well above the parameters
of current supercomputers).  For a random point set, there are on
average $\sim 6.77$ Delaunay tetrahedra per point \citep{Weygaert1994},
giving $\sim 244$ bytes (or $\sim$460 bytes if pointers on 64-bit
architectures are used) per point for storing the mesh.  Additional
storage is required to hold a list of faces for the flux calculation.
This sums up to a relatively hefty cost in terms of memory, a factor $3$
to $4$ or so larger than what is needed for the tree construction in a
Tree-SPH code like {\small GADGET-2}, but not something that
prohibitively restricts the sizes of possible simulations.  However, we
note that by exploiting the adjacency relations to label nearby
tetrahedra, the memory cost could in principle be reduced to just about
7.5 bytes per tetrahedron in 3D \citep{Blandford2005}. We leave such
memory optimizations for future work.

\subsection{Point location}

The point location in the above insertion algorithm can be a limiting factor,
as a simple search through all triangles/tetrahedra would produce an
$N^2$-scaling of the algorithm. But there are different possible approaches
for speeding up the point location. One idea is to store the past insertion
history of Delaunay triangles in a directed acyclic graph
\citep{Edelsbrunner1996}, such that the insertion triangle can be localized
through a tree-walk. However, the manipulation of the history graph requires
complex bookkeeping and large amounts of memory.

Another method is the `jump and walk' procedure for point location first
proposed by \cite{Muecke1996}. Here one walks through the tessellation from a
random triangle in the direction of the point that is to be inserted.  We will
adopt this strategy.  However, instead of starting at a random triangle or
using a search grid for rapid location of an initial triangle, we order the
points that are to be inserted along a space-filling Peano-Hilbert curve
\citep{Springel2005}, a trick that has also been employed by the tessellation
code {\em tess3} \citep{Liu2005}.  This guarantees that the next point that is
inserted is always spatially close to the previous one. If we hence remember a
pointer to the last processed triangle/tetrahedron, we can start the search in
the immediate neighbourhood of the insertion triangle, and only a very small
number of steps are required to arrive at the correct triangle. An additional
advantage of this scheme lies in cache utilization benefits; thanks to the
spatial proximity of subsequent insertion points, much of the required
memory has been accessed recently and may hence still be resident in the processor's
cache, which increases performance.

To test whether a point lies inside a given tetrahedron, we calculate whether
it lies above all four of the planes defined by its oriented triangles. If the
point does not lie in the current tetrahedron, we determine which of its faces
is intersected by a line from the centre of mass of the tetrahedron to the
insertion point, and then change to the adjacent tetrahedron on the other side
of the selected face.

The above requires an efficient way to test whether a given point lies inside
a tetrahedron (or triangle in 2D). Since all our tetrahedra are positively
oriented, one way to do this is to use four orientation tests: if the point in
question lies above all four oriented triangles of the tetrahedron, it must be
inside. However, this is slow due to the required evaluation of four
determinants. A faster method is to expand the coordinates of the given point
in terms of the three linearly independent vectors spanned by the four points
of the tetrahedron. This involves a linear system of equations which can be
quickly solved with Gauss elimination. The values of the expansion
coefficients $\alpha$, $\beta$ and $\gamma$ then directly indicate whether
the point lies inside the tetrahedron. This is the case if we simultaneously
have $\alpha>0$, $\beta>0$, $\gamma>0$ and $\alpha+\beta+\gamma<1$.  Only when
there is a danger of obtaining the incorrect result with this method due to
numerical round-off, we use instead an {\em exact} evaluation of the four
orientation tests, which we discuss in more detail below.

\subsection{Treatment of degenerate cases}

\begin{figure}
\bc
\resizebox{7cm}{!}{\includegraphics{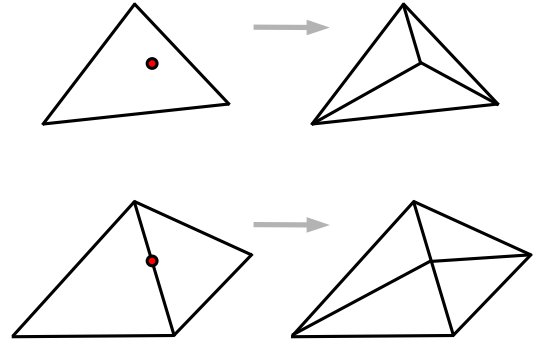}}
\caption{Point insertion in 2D in the normal case (top, via a 1-to-3 flip) and
  the degenerate case (bottom), where the point lies {\em exactly} on an edge of the current
  tessellation.  In the latter case, the two triangles need to be replaced
  with four triangles (a 2-to-4 flip).
  \label{Fig2Ddegeneracy}}
\ec
\end{figure}

\begin{figure}
\bc
\resizebox{3.5cm}{!}{\includegraphics{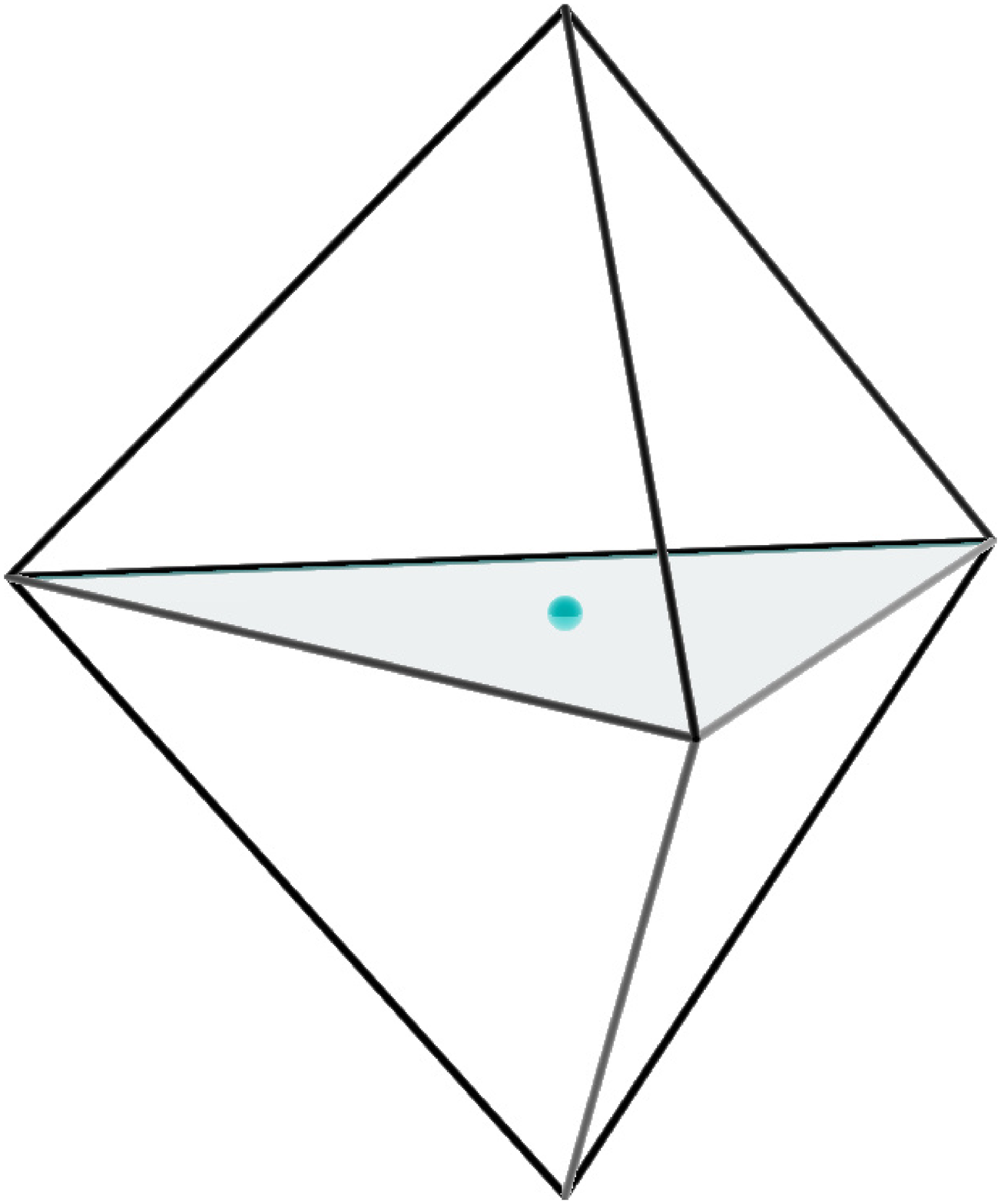}}\ \ %
\resizebox{3.5cm}{!}{\includegraphics{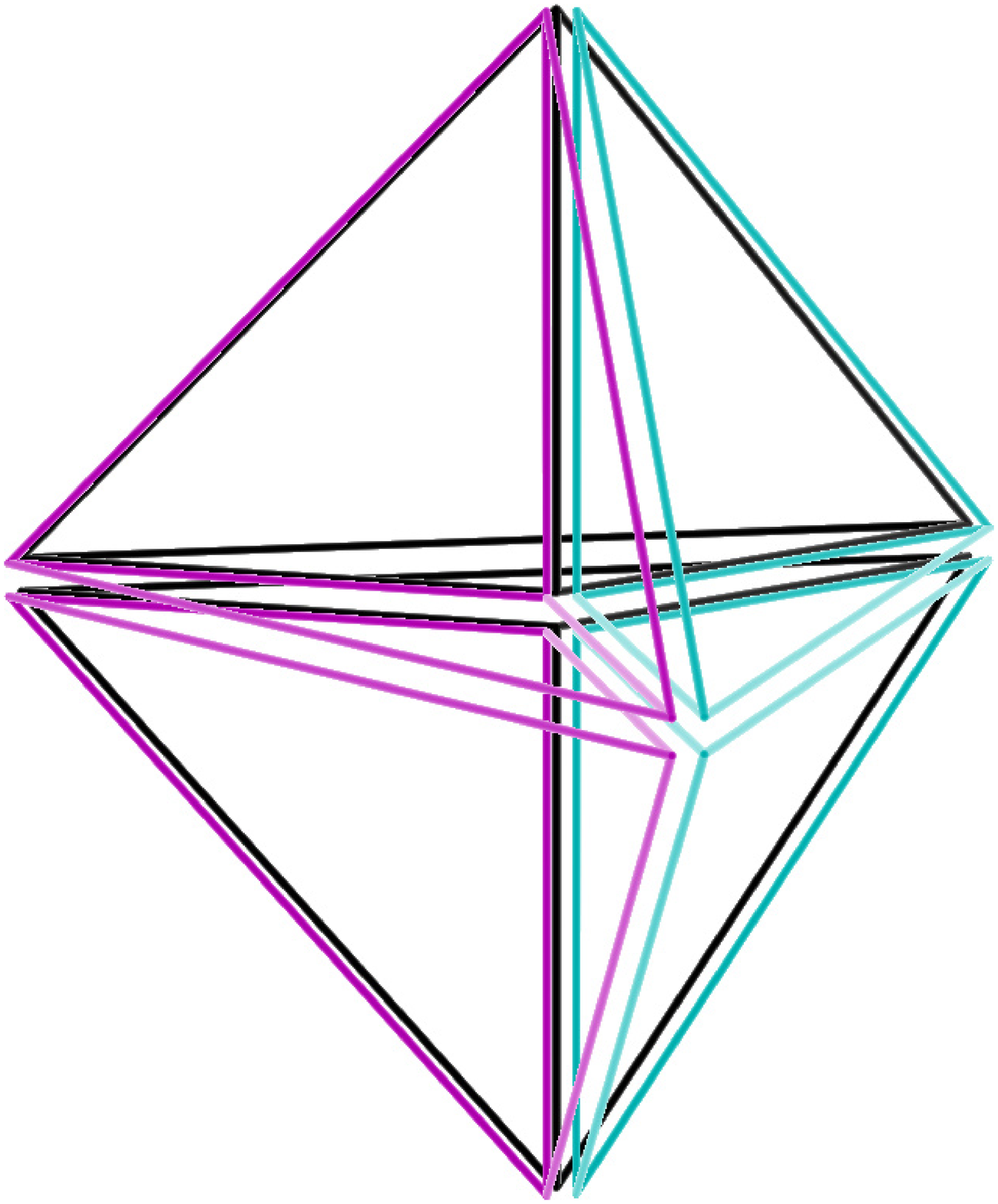}}\\
\caption{Degeneracy during point insertion in 3D. Here the point 
that is to be inserted falls onto a face of the current 
tessellation. The 
  two tetrahedra involved need to be replaced 
by six tetrahedra, constituting a 2-to-6 flip.
\label{Fig2-6-Flip}}
\ec
\end{figure}

\begin{figure}
\bc
\resizebox{3.5cm}{!}{\includegraphics{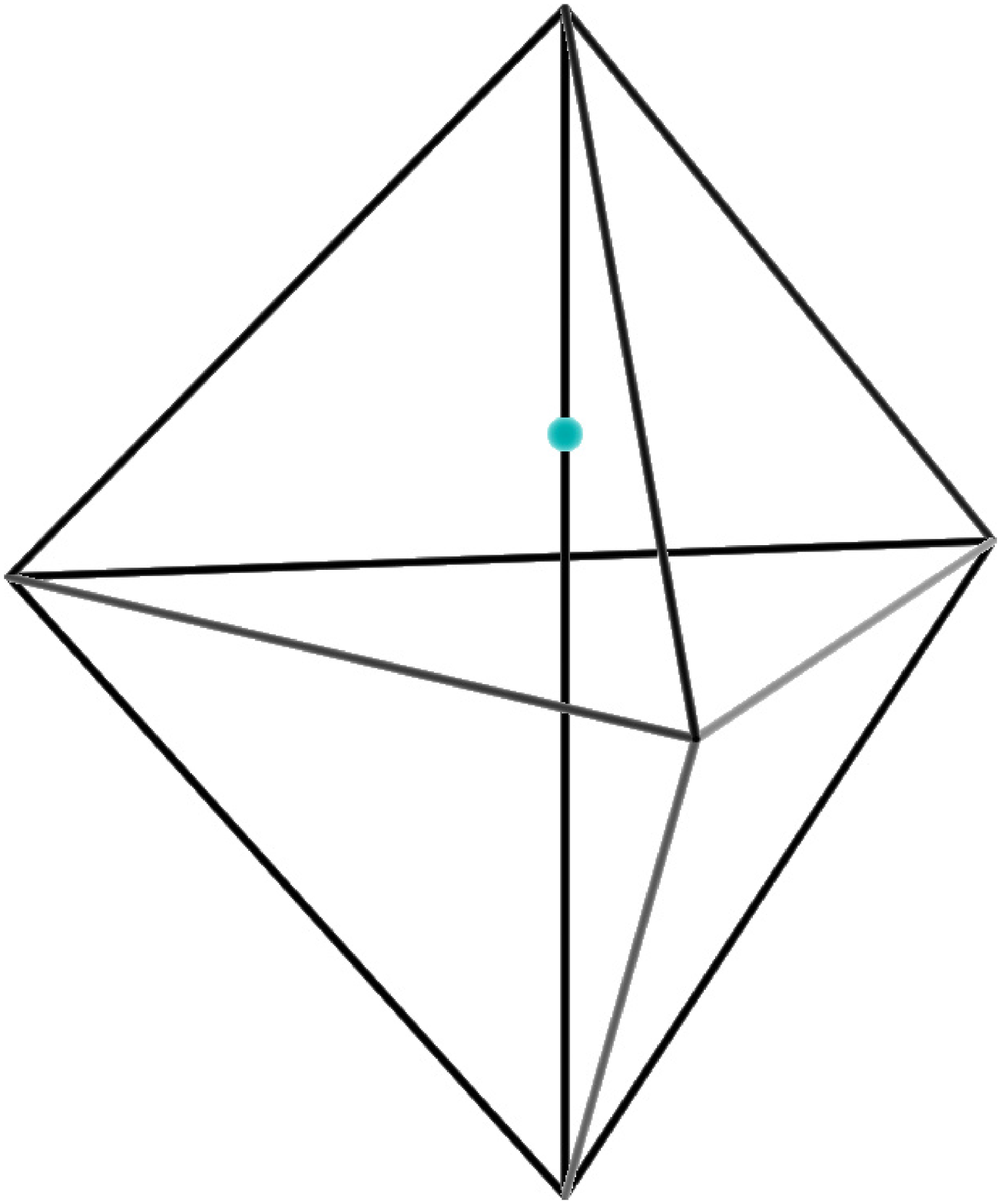}}\ \ %
\resizebox{3.5cm}{!}{\includegraphics{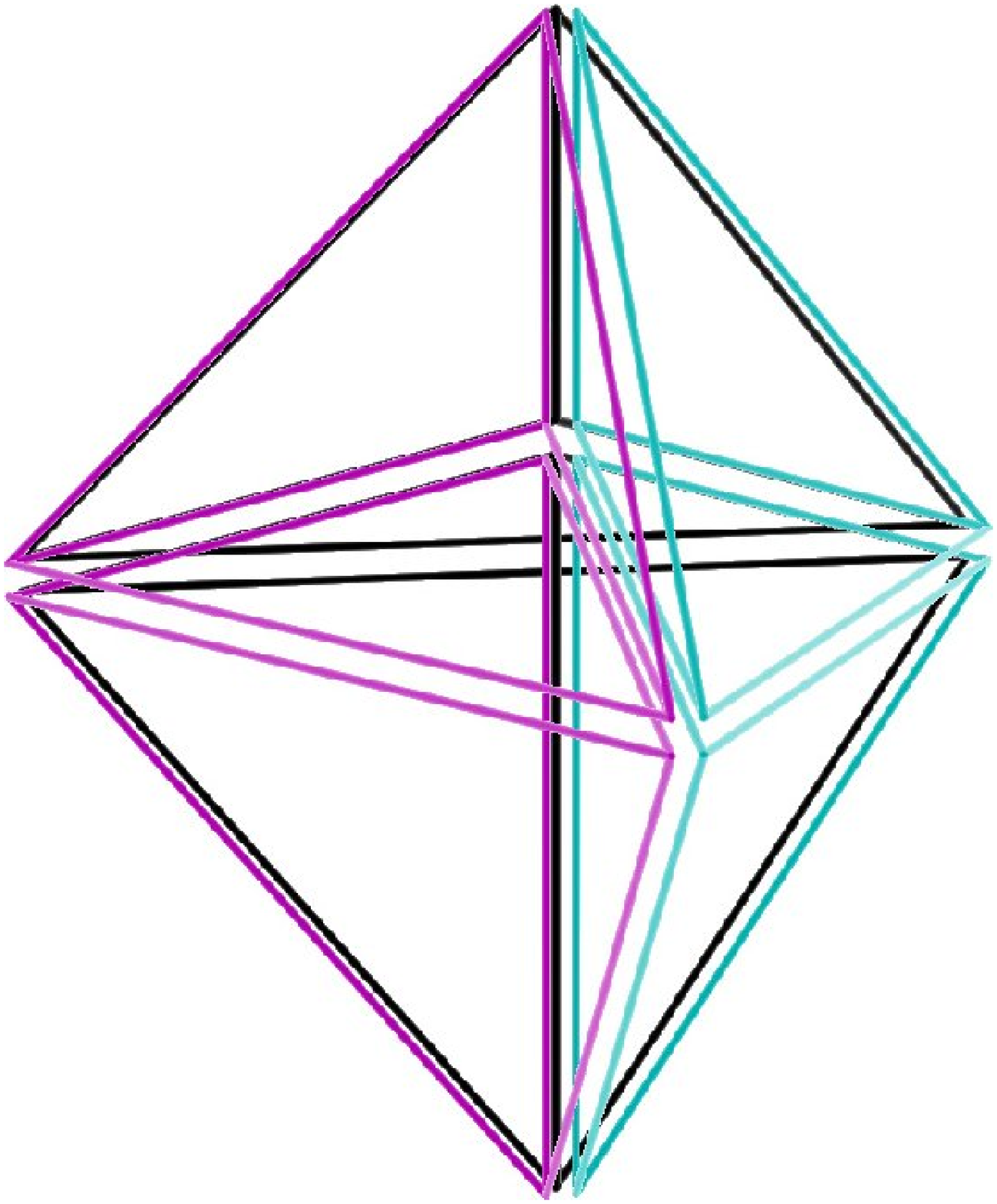}}\vspace*{0.4cm}\\
\resizebox{3.5cm}{!}{\includegraphics{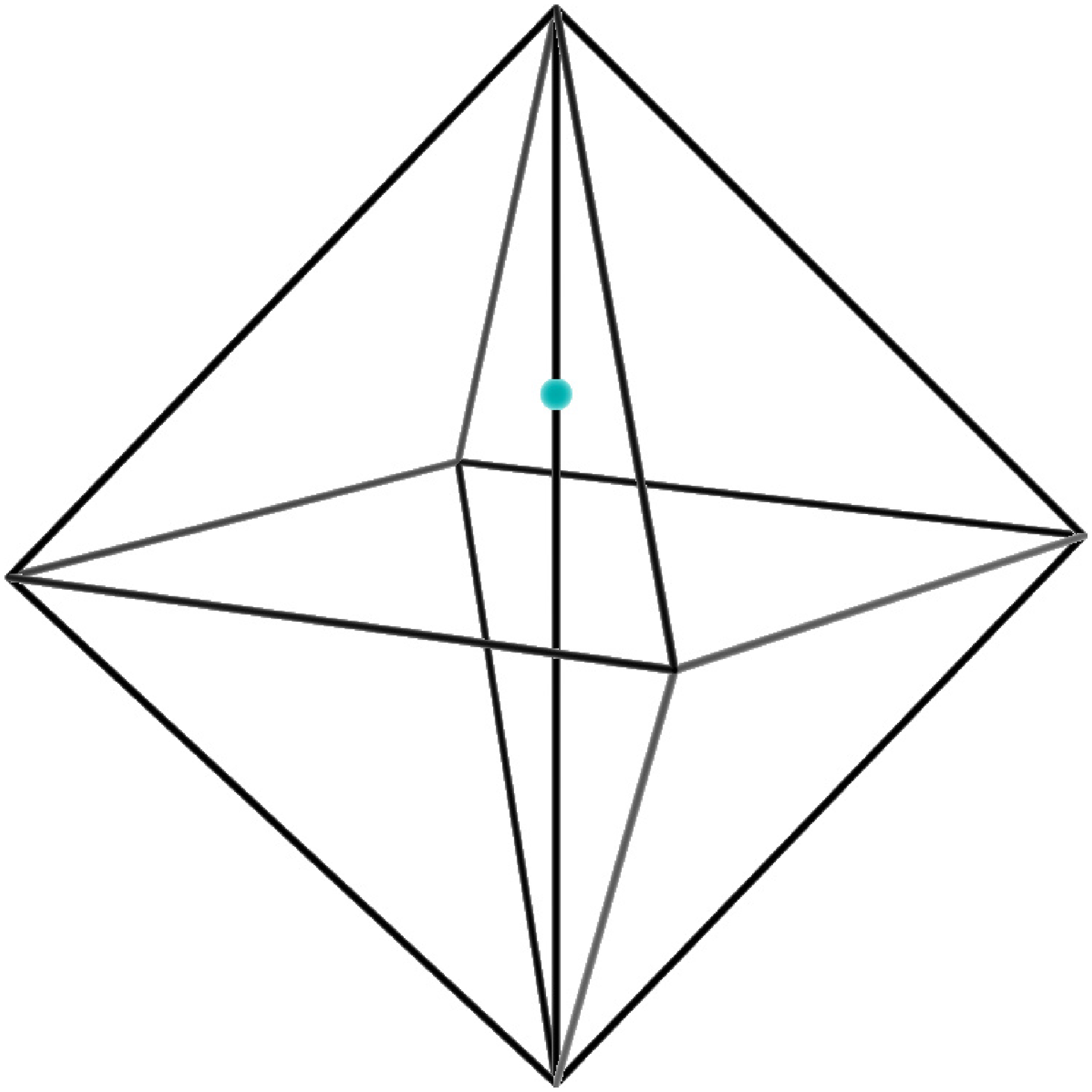}}\ \ %
\resizebox{3.5cm}{!}{\includegraphics{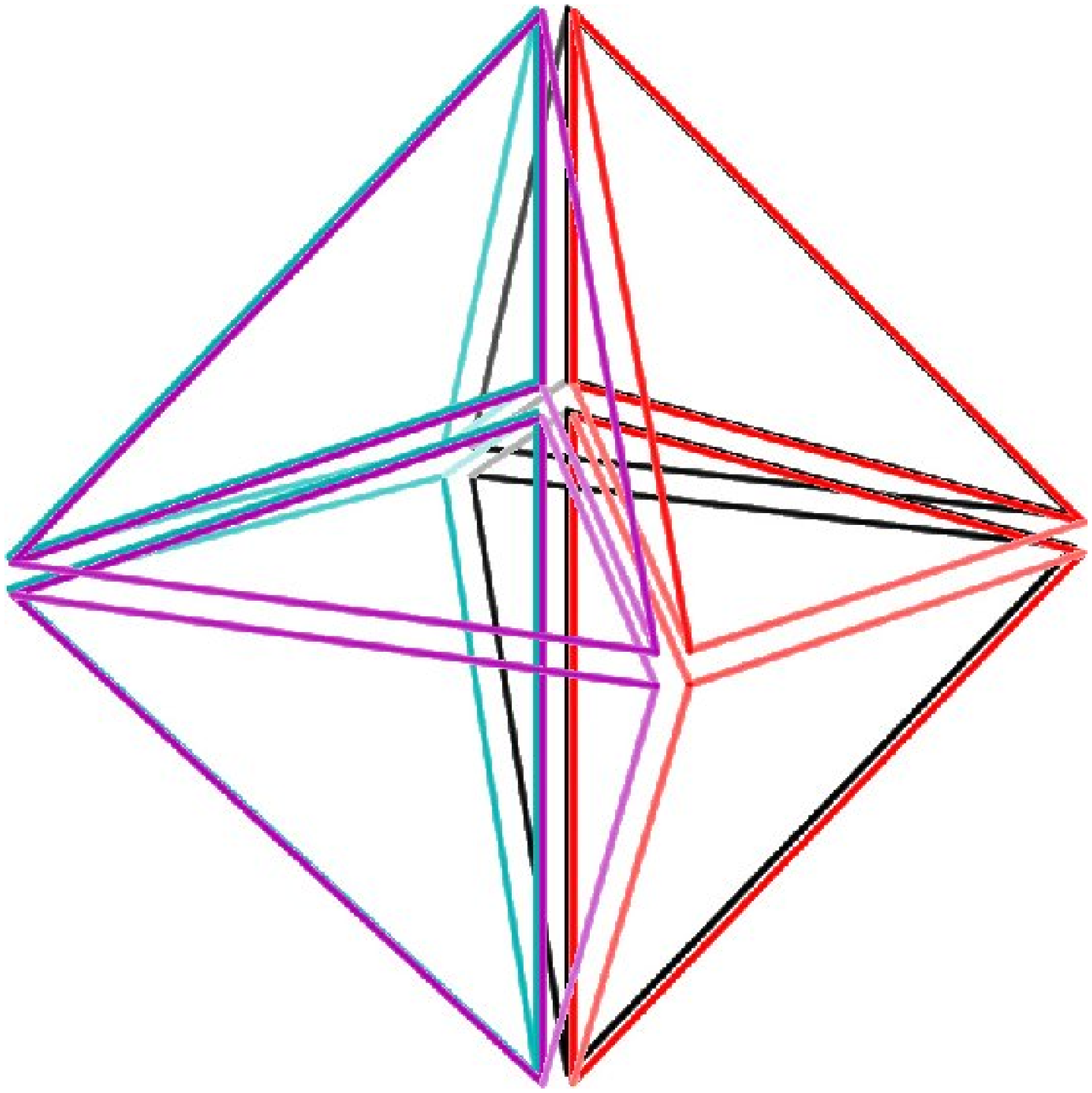}}\\
\caption{Degeneracy during point insertion in 3D. Here
the point that is to be inserted falls onto an edge of the current
tessellation. In this case, we 
need to make a $n$-to-$2n$ flip, where $n$ is the number of tetrahedra that
have the edge in common. In most case this is
either 3 (top left) or 4 (bottom left), like in the
examples shown in this figure, but $n$ can in principle also be larger.
\label{Fign-2n-Flip}}
\ec
\end{figure}

\begin{figure}
\bc
\resizebox{3.3cm}{!}{\includegraphics{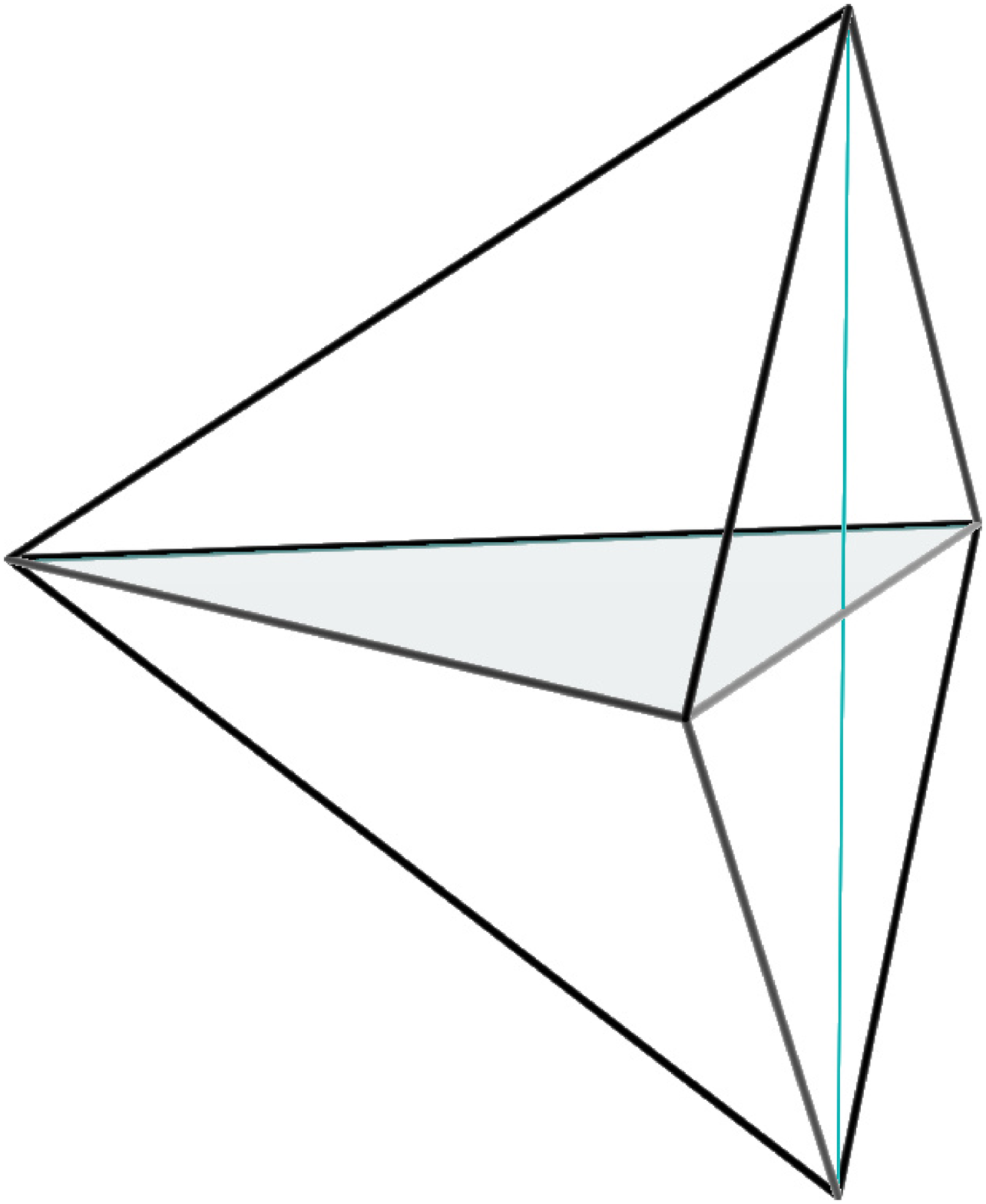}}\vspace*{0.1cm}\\
\resizebox{3.5cm}{!}{\includegraphics{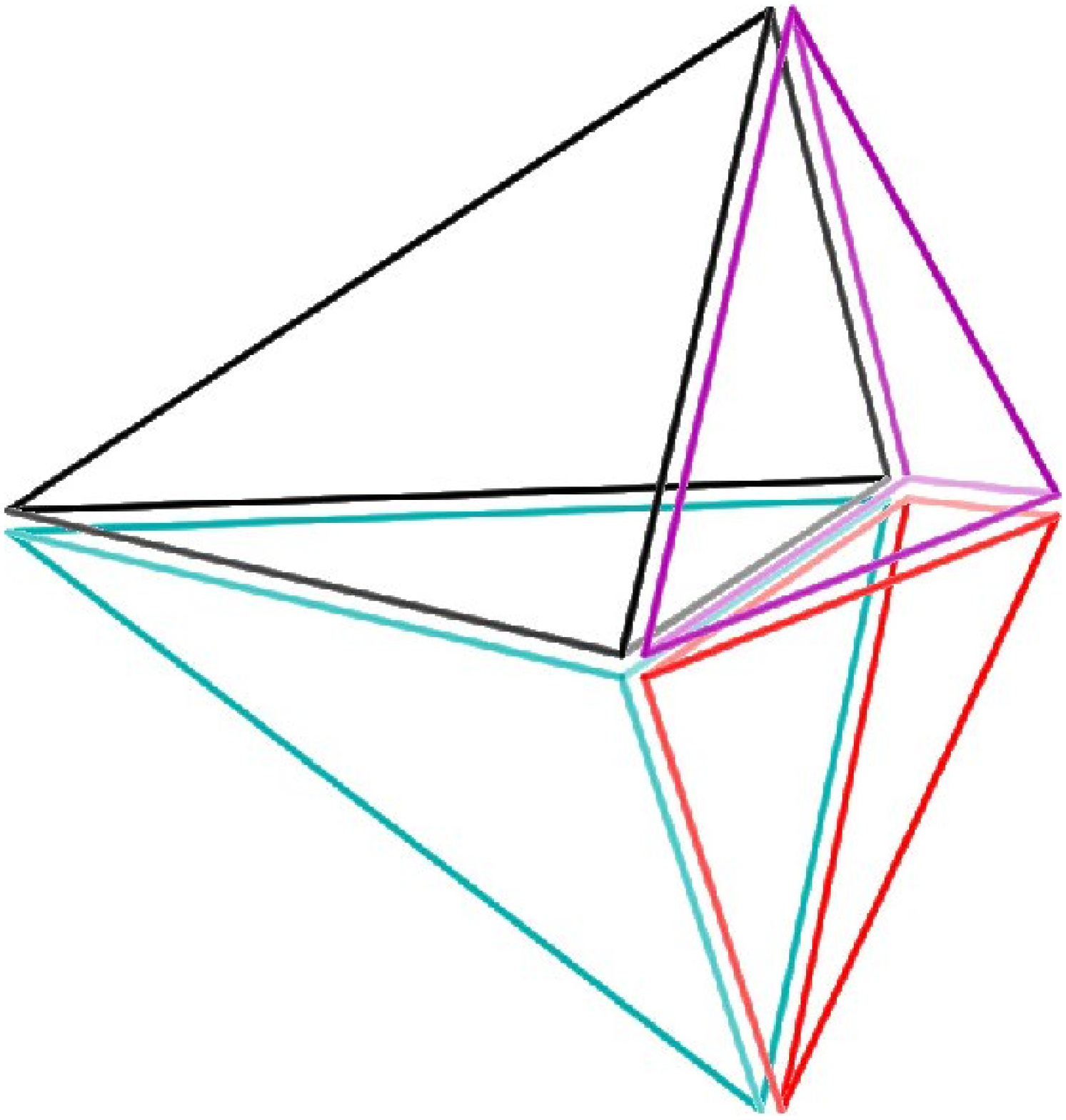}}\ \ %
\resizebox{3.5cm}{!}{\includegraphics{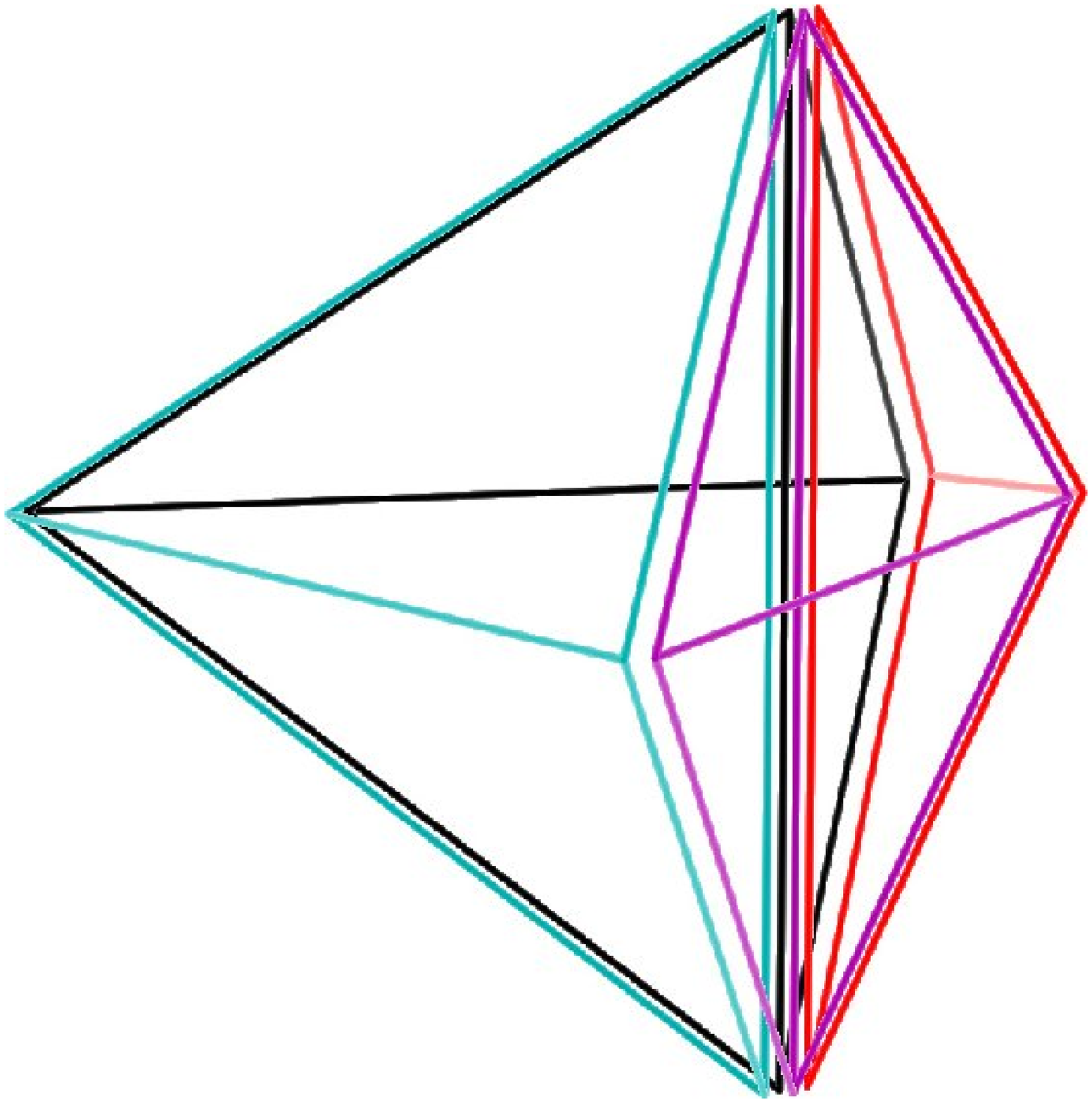}}\\
\caption{Flipping degeneracy in 3D.  If the line that connect the two
  points opposite a common face of two tetrahedra intersects this face
  on one of its edges (like in the sketch on top), the standard 2-to-3
  flip cannot be carried out. Instead, the two tetrahedra may be
  eligible for a 4-to-4 flip. This requires however that the
  intersected edge is the common edge to exactly 4 tetrahedra. In this
  case, four tetrahedra can be replaced by 4 tetrahedra, as shown in
  the bottom of the sketch.
\label{Fig4-to-4-Flip}}
\ec
\end{figure}

A problematic point about incremental insertion is that in this method it can
become hard to deal with degenerate point sets. In particular, the algorithm
described above for constructing the Delaunay triangulation only works
robustly for points in {\em general position}, where in 2D never more than
three points lie on a circle, and never more than 2 points lie on a line. If
we start with a regular point distribution, for example a Cartesian grid, this
condition is evidently strongly violated. But even for random point sets, it
is possible that a degenerate situation approximately occurs, and due to
floating point round-off, we may not be able to correctly decide the outcome
of one of the geometric tests, i.e.~to evaluate the {\em correct sign} of a
nearly degenerate determinant \citep[e.g.][]{Clarkson1992}. However, failure
to do so invariably leads to a breakdown of the mesh construction. In
addition, experience shows that attempts to address this issue with crude
patches, for example in the form of random point perturbations, provides only
unreliable (and inelegant) work-arounds.

One possible approach for solving this issue lies in systematically applying
{\em symbolic perturbations} to the particle coordinates
\citep{Edelsbrunner1990}, which effectively bring the particles into general
position, such that the Delaunay triangulation becomes formally unique and the
algorithm for constructing the tessellation is guaranteed to succeed. There
may then still be triangles/tetrahedra of zero volume attached to the complex
hull of the final tessellation, but their removal represents no major problem.
However, the symbolic perturbation approach still requires robust evaluations
of the correct sign of determinants, which \citet{Muecke1995} proposes to
obtain with long-integer arithmetic.

Another possible method for constructing robust geometric predicates is to
employ {\em exact} floating point arithmetic, implemented through suitable
software packages. This is however very much slower than standard
double-precision arithmetic. An attractive alternative is to use adaptive
precision arithmetic, as proposed and implemented by
\citet{Shewchuk1997}. Here the idea is to monitor the maximum round-off error
that can occur in the evaluation of a geometric test. If there is a risk that
the correct result may be missed with standard floating point arithmetic
(which is carried out in hardware by the CPU), progressively more accurate
approximations to exact floating point arithmetic are employed, until the
correctness of the calculated sign can be guaranteed. Since the exact but slow
floating point arithmetic is only used when it is really needed, this adaptive
precision approach is much faster than using exact floating point arithmetic
throughout.

We use a different method instead which does not need perturbed point
coordinates, but rather relies on modifications of the point insertion
algorithm such that it can directly deal with degeneracies, and we
combine this with a scheme that always guarantees the correct
evaluation of geometric predicates. Let us first discuss the
latter. We here follow the idea of \citet{Shewchuk1997} and estimate
the maximum round-off error in evaluations of in-circle and
orientation tests. When there is a risk of getting the wrong sign with
standard floating point arithmetic, we however evaluate the
determinant with exact long integer arithmetic, which is both simple
and robust. To this end we establish a one-to-one mapping between the
floating point numbers of our point coordinates and the space of
53-bit integers. This is accomplished by mapping our computational
domain to the floating point interval $[1,2[$. All these numbers have
    the same exponent in the standard Institute of Electrical and
    Electronics Engineers (IEEE) representation of double-precision
    numbers, and the 53-bit mantissa effectively provides a linear and
    uniform grid of the possible floating point values in this
    interval. We read out this mantissa and use it to evaluate exact
    geometric predicates using long integer arithmetic with the
    open-source {\small GMP} library, when needed.

Let us now discuss the modifications required in the point insertion algorithm
such that it can deal with degenerate input point sets if correctness of the
geometric tests can be guaranteed. In two dimensions, only one such
modification is required.  We need to detect the case that the point that is
to be inserted lies on an edge of the current tessellation, as illustrated in
Figure~\ref{Fig2Ddegeneracy}. In this case, we can not replace one triangle
with three new ones, but instead need to split both of the triangles that
share the edge into two triangles.

In three dimensions, things are considerably more complicated. Here the point
that is to be inserted may lie on a face of the current tessellation. In this
case, we need to replace the two adjacent tetrahedra with altogether 6 new
tetrahedra. This replacement represents a `2-to-6 flip', as shown in
Figure~\ref{Fig2-6-Flip}. It may also happen that the point falls onto an edge
of the current tessellation. There may be 3, 4, or more tetrahedra present
that share this edge. All of them have to be replaced by two tetrahedra each,
such that we effectively carry out an `$n$-to-$2n$ flip', as illustrated in
Figure~\ref{Fign-2n-Flip}.  Finally, when degeneracies are present, a further
case needs to be considered in the flipping operations that heal the mesh
after a point has been inserted. Recall that the decision whether a 2-to-3 or
3-to-2 flip is carried out when an invalid Delaunay face has been found
depends on the location of the intersection between this triangular face and
the line that connects the tips of the adjacent tetrahedra. Previously, we
discussed the cases where the intersection lies inside the triangle, or
outside. For degenerate cases, it may lie exactly on one of the edges, a case
that requires special treatment. Here a `4-to-4' flip is possible and needs to
be carried out when needed, as illustrated in Figure~\ref{Fig4-to-4-Flip}.

We note that the topology of the resulting Delaunay tessellation is not unique
if degeneracies are present, and the exact outcome (i.e. which of the
different Delaunay tessellations that are possible is realized) depends on the
order in which the points are inserted. However, the corresponding Voronoi
tessellation is still unique, and hence the outcome of our hydrodynamical
calculations is unaffected by the Delaunay non-uniqueness in the presence of
degeneracies, and also does not depend on the order in which the
mesh-generating points are inserted.

A related point concerns the change of the topology of the mesh when the
points are moved. While the Delaunay triangulation changes {\em
  discontinuously} whenever a point is moved into or out of the circumsphere
of another triangle, the corresponding Voronoi tessellation changes {\em
  continuously}. In fact, whenever the Delaunay neighbourhood relations between
two points change, the corresponding Voronoi face shrinks to a vanishing
area. As we will see later on, it is this property that allows the mesh to
deform without running into the mesh-tangling problems that plague other
approaches for moving meshes. Also, note that we can calculate the full motion
of all Voronoi faces based just on the velocity vectors of the mesh generating
points. We will make use of this property in our hydrodynamical schemes, as
discussed in Section~\ref{SecHydro}.

\subsection{Parallelization of the tessellation code} \label{secparallel}

Modern supercomputer platforms feature hundreds to thousands of compute cores,
with a continuing trend to ever larger numbers of cores. Efficient use of this
combined processing power for simulations of dynamically tightly coupled
systems can be quite challenging, especially on distributed memory computers,
which offer the largest and most cost effective performance. Parallelization
of simulation codes for such architectures requires decomposition of a problem
into individual parts, provided we want to avoid complete data duplication,
which is prohibitive if good scalability is desired.

A number of parallel construction algorithms for the Delaunay triangulation
have been proposed, some of them for distributed memory environments
\citep[e.g.][]{Cignoni1998,Lee2001}, others for shared memory machines
\citep{Blandford2006}. However, the approach of \citet{Cignoni1998} replicates
the entire point set on each independent processor, an approach we can not
afford to follow in the interest of scalability.

Rather, we decompose the point set into disjoint spatial domains, each
mapped to a different compute core with its own physical memory. The
idea here is that most of the Voronoi cells of a domain will lie in its
interior and hence only depend on the data local to the processor, while
some cells close to the surface may be affected by data on other
processors, which needs to be dealt with by data communication. Our
strategy to deal with this issue is to construct a locally complete
tessellation by importing {\em ghost points} from neighbouring
processors such that all the Voronoi cells of the points that are local
to the domain are correctly formed. This means that the joint set of all
primary Voronoi cells forms the complete tessellation, but there is no
need to actually ever form it explicitly, i.e.~we do not need to somehow
mesh the tessellations across two neighbouring domains together, which
would be cumbersome. Instead, the ghost points provide the `glue' that
gives the proper connectivity across domains. We will also use ghost
points to implement periodic or reflecting boundary conditions, which
are simply realized through fiducial points that are imported from the
`other side' of the simulation box.  In practice, the use of ghost
points increases the number of cells that need to be considered by a
given processeor typically by $3-10\%$, depending on how many processors
are used. Only if a large number of processors is used for a small
problem, this induces significant overhead and a limit to scalability.

\begin{figure}
\bc
\resizebox{8cm}{!}{\includegraphics{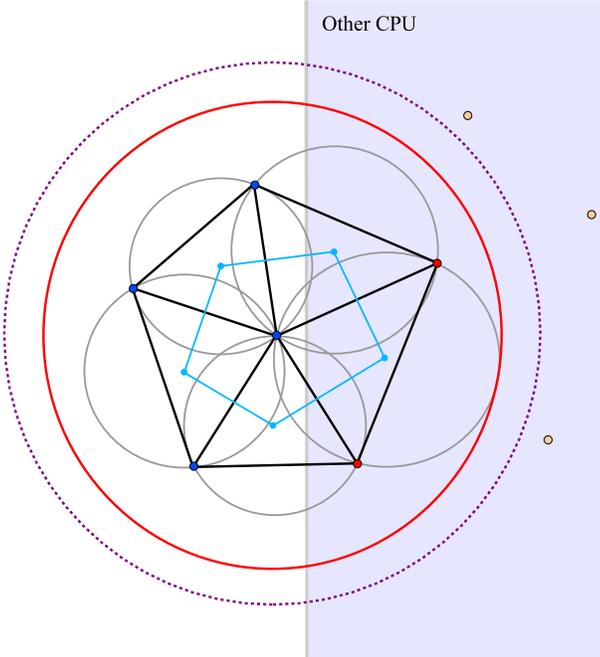}}\\
\caption{Geometry of the test for local completeness of a Voronoi cell. Points
  on the left of the vertical grey line reside on the local domain. The point
  $i$ at the centre of the Voronoi cell (shown in light blue) has a search
  radius $h_i$ equal to the dotted circle; all points on other processors
  inside this circle have been imported as ghost points (red dots) and were
  added to the local tessellation. We now have to decide whether the resulting
  Voronoi cell around point $i$ could still be modified by points not yet
  added to the local tessellation. To this end we consider the circumcircles
  (shown in grey) of all Delaunay triangles that share the central point. The
  smallest circle encompassing all these circumcircles has a radius $s_i$
  equal to twice the maximum circumcircle radius, and is shown in red. If the
  red circle lies inside the dotted circle, i.e.~for $s_i\le h_i$, all the
  Delaunay triangles around the target point $i$ are valid and are part of the
  fiducial global Delaunay tessellation that contains all
  points. Consequently, in this case the Voronoi cell around the target point
  $i$ is complete and unaffected by the presence of the nearby domain
  boundary. For $s_i > h_i$, we increase $h_i$ and add any additional point
  found in the enlarged search region, until the condition $s_i\le h_i$ is
  fulfilled.
  \label{FigSketchPointImport}} \ec
\end{figure}

How can we find the ghost points that need to be imported, and how can
we be sure that our local point set is sufficiently complete? We have
implemented two different algorithms for this task. In the simpler of
the two, we start by first constructing the tessellation for all local
points $\vec{r}_i$ within a domain. With each of the points we associate
a search radius $h_i$, inherited either from the previous timestep, or
initialized with a guess. We then search for all points on {\em other
  processors} that lie within the spheres/circles of radius $h_i$ around
$\vec{r}_i$. The union of the ghost points found in this way is then
added to the local tessellation.  For each local point, we can then
calculate a radius $s_{i}$ equal to twice the maximum radius of all the
circumspheres of the Delaunay tetrahedra that have the point $\vec{r}_i$
as one vertex. The relevant geometry is illustrated for two dimensions
in the sketch of Figure~\ref{FigSketchPointImport}. Here the red circle
has radius $s_{i}$ around the target point $i$, which has an associated
Voronoi cell shown in light blue. This local Voronoi cell around point
$\vec{r}_i$ could only change if there was a further point somewhere
inside the red circle that has not yet been added to the local
tessellation. If we have $h_i>s_i$, then we know that such a point does
not exist, and we are guaranteed that the Voronoi cell around point $i$
is correct.  Otherwise, we need to look whether further points from
other processors need to be added to the local tessellation.  In this
case we increase the radius $h_i$ by some factor and search again for
{\em additional} points on other processors that have not yet been
inserted into the local tessellation. These points are then added to the
local tessellation, and the $s_{i}$ are redetermined. The process is
repeated until the condition $h_i>s_{i}$ holds for all points local to
the processor, at which point the local tessellation is complete,
i.e.~all Voronoi cells of local points are guaranteed to be unaffected
by the presence of the domain boundary.

Note that the `thickness' of the layer of ghost points imported on each
domain is not fixed in this scheme, rather it adjusts to variations of
density along the domain boundaries, as well as to the geometry of the
domains themselves. Once the tessellation is complete, we set $h_i$ to
$s_i$ for use in the next mesh construction; this usually ensures close
to optimum efficiency in finding the minimum required set of ghost
particles.

However, sometimes the above approach may create a ghost layer that is
thicker than really required in situations where the mesh resolution
shows a strong spatial gradient, and a domain boundary lies orthogonal
to this gradient. Since the search region for ghosts is taken to be
spherical, this may lead to the import of a comparatively large number
of ghost points from the side where the mesh resolution becomes
finer. If the mesh resolution changes sufficiently rapidly in space,
this can then incur a substantial overhead that exceeds the usual
$3-10\%$ mentioned above. We have therefore also implemented an
alternative algorithm to determine the ghost region, which is more
efficient in this situation. In this approach we directly search for
possible ghosts in ${\em all}$ circumcircles of those triangles in the
local tessellation that have at least one local particle as one of their
vertices. If a ghost point is found, all triangles modified in the point
insertion step of the ghost will be tested again until no further ghosts
are found. At the end, this method then guarantees that all Delaunay
triangles shared by a local particle are part of the correct global
mesh, and hence the Voronoi cell of this particle is complete. To
prevent that initially very many ghost points are found in the large
triangles present in the first iteration at the surface of the local
domain, we always insert only the closest ghost particle found in a
circumcircle that has not yet been added to the local mesh. Especially
when individual timesteps are used and only parts of the mesh are
constructed for active particles (as discussed later in the context of
our individual time-stepping scheme), this approach is usually more
efficient, despite its larger number of spatial point searches.

The above techniques relies on rapid algorithms to find all particles within a
given sphere of arbitrary radius. To this end we employ a \citet{Barnes1986}
octtree and use the neighbour search algorithms of the parallel SPH code
{\small GADGET-2} \citep{Springel2005}. We also adopted the specific domain
decomposition strategy from the {\small GADGET-2} code, which is based on
subdividing a space-filling Peano-Hilbert curve, an approach that has recently
become popular also in other cosmological simulation codes
\citep[e.g.][]{Shirokov2}. Similar to {\small GADGET-2}, we also use a
`top-level tree' that covers the full simulation volume. This allows us to
quickly decide whether or not a local search region is fully contained in the
local domain, and if not, with which other processors it overlaps. This is
also useful for devising an efficient communication strategy.

The complexity of the tessellation algorithms discussed in this
section might suggest that the resulting computations are quite
expensive and slow, but we want to remark that this is not really the
case.  The geometric tests required to insert a point involve
primarily linear algebra operations that are calculated very
efficiently on modern processors (which often offer combined
multiply-add operations in a single cycle), while the rearrangement of
local triangles or tetrahedra reduces to reorientations of pointers.
As a result, even without significant efforts for speed optimizations,
we reach tessellation speeds on the order of several tens of thousands
of tetrahedra per second. This is comparable to or only slightly more
than the work needed for SPH neighbour search. More importantly, the
computational cost continues to scale just as $N\log N$. There is
hence in principle no obstacle to use the tessellation techniques for
large-scale applications, even if the mesh is reconstructed each
timestep, as in our current approach.

\subsection{Other applications of the tessellation code}

While in the rest of this paper we will focus on applying the Voronoi mesh to
problems of continuum hydrodynamics, we briefly want to mention that the
tessellation methods discussed here are also useful in other contexts.

In particular, Voronoi or Delaunay tessellations are useful for general
density reconstruction tasks.  For example, \citet{Weygaert1994} used Voronoi
tessellation to study cosmic large-scale structure and \citet{Bernardeau1996}
employed them to analyze the statistics of velocity fields.
\citet{Schaap2000} and \citet{Pelupessy2003} proposed to use Delaunay
tessellation as a general estimation tool for linear reconstructions of the
density field, based on the {\em contiguous Delaunay cell} that is formed by
all Delaunay triangles around a given point.

We have applied Voronoi density estimates to calculate the dark matter
annihilation signal expected in high-resolution dark matter simulations
of the formation of a Milky-Way like galactic halo
\citep{Springel2008Nature}. In comparison with SPH, this has the
advantage to provide an unbiased sum of the volumes assigned to each
particle, and to produce less damping of the smallest resolved
structures by smoothing. In our largest simulation, we constructed a
Voronoi mesh for nearly 5 billion particles, composed of about 34
billion tetrahedra in the dual Delaunay tessellation, and with a dynamic
range in point density of more than $10^7$. This may well be one of the
largest Voronoi meshes ever constructed. The mesh construction took 516
seconds on 1024 CPUs of an SGI Altix 4700 (the HLRB-II machine at the
Leibniz-Computing Centre in Garching, Germany).

Outside of astronomy, Voronoi tessellations are widely applied for many
different applications, including point pattern analysis, modelling of spatial
processes, location optimization, and computer graphics, to name just a few. A
comprehensive introduction to these applications can be found in the monograph
of \citet{Okabe2000}.

\section{Finite volume hydrodynamics on a moving Voronoi mesh} \label{SecHydro}

The Euler equations are conservation laws for mass, momentum and energy that
take the form of a system of hyperbolic partial differential equation. They
can be written in compact form by introducing a state vector
\begin{equation} 
\vec{U} = \left(
\begin{array}{c}
\rho\\
\rho \vec{v}\\
\rho e
\end{array} 
\right) = 
 \left(
\begin{array}{c}
\rho\\
\rho \vec{v}\\
\rho u + \frac{1}{2} \rho \vec{v}^2
\end{array} 
\right) 
\end{equation} 
for the fluid, where $\rho$ is the mass density, $\vec{v}$ is the velocity
field, and $e= u + \vec{v}^2 / 2$ is the total energy per unit mass. $u$ gives
the thermal energy per unit mass, which for an ideal gas is fully determined
by the temperature. These fluid quantities are functions of the spatial
coordinates $\vec{x}$ and time $t$, i.e.~$U=U(\vec{x},t)$, but for simplicity
we will often refrain from explicitly stating this dependence in our notation.
Based on $\vec{U}$, we can define a flux function
\begin{equation} 
\vec{F}(\vec{U}) = \left(
\begin{array}{c}
\rho\vec{v}\\
\rho \vec{v}\vec{v}^T + P\\
(\rho e + P)\vec{v}
\end{array}
\right) ,
\end{equation} 
with an equation of state
\begin{equation} 
P= (\gamma-1)\rho u 
\end{equation}  
that gives the pressure of the fluid.  The Euler equations can then be
written in the compact form
\begin{equation}
\frac{\partial\vec{U}}{\partial t} +
\vec{\nabla}\cdot \vec{F} = 0,  \label{EqnEuler}
\end{equation}  
which emphasizes their character as conservation laws for mass, momentum
and energy.

Over the past decades, a large variety of different numerical
approaches to solve this coupled set of partial differential equations
have been developed \citep[see][for comprehensive
  expositions]{Toro1997,LeVeque2002}.  Many modern schemes are
descendants of Godunov's method, which revolutionized the field. By
solving an exact or approximate Riemann problem at cell boundaries,
Godunov's method allows the correct identification of the
eigenstructure of the local solution and of the upwind direction,
which is crucial for numerical stability. While Godunov's original
method offers only first order accuracy and is relatively diffusive,
it can be extended to higher order accuracy relatively simply, and in
many different ways.

We will here employ a so-called {\em finite-volume} strategy, in which
the discretization is carried out in terms of a subdivision of the
system's volume into a finite number of disjoint cells. The fluid's
state is described by the cell-averages of the conserved quantities for
these cells. In particular, integrating the fluid over the volume $V_i$
of cell $i$, we can define the total mass $m_i$, momentum $p_i$ and
energy $E_i$ contained in the cell as follows,
\begin{equation}
\vec{Q}_i = \left(
\begin{array}{c}
m_i\\
\vec{p}_i\\
E_i 
\end{array}
\right)
= \int_{V_i} \vec{U}\,{\rm d} V .
\end{equation} 
With the help of the Euler equations, we can calculate the rate of
change of $\vec{Q}_i$ in time. Using Gauss' theorem to convert the
volume integral over the flux divergence into a surface integral over
the cell results in
\begin{equation}
\frac{{\rm d}\vec{Q}_i}{{\rm d}t}
= -\int_{\partial V_i} \left[ \vec{F}(\vec{U}) - \vec{U}
\vec{w}^T\right] {\rm d}\vec{n} .
\label{EqQevol}
\end{equation} 
Here $\vec{n}$ is an outward normal vector of the cell surface, and $\vec{w}$
is the velocity with which each point of the boundary of the cell moves. In
Eulerian codes, the mesh is taken to be static, so that $\vec{w}=0$, while in
a fully Lagrangian approach, the surface would be allowed to move at every
point with the local flow velocity, i.e.~$\vec{w}=\vec{v}$. In this case, the
right hand side of equation~(\ref{EqQevol}) formally simplifies, because then
the first component of $\vec{Q}_i$, the mass, stays fixed for each
cell. Unfortunately, it is normally not possible to follow the distortions of
the shapes of fluid volumes exactly in multi-dimensional flows for a
reasonably long time, or in other words, one cannot guarantee the condition
$\vec{w}=\vec{v}$ over the entire surface. In this case, one needs to use 
the general formula of equation~(\ref{EqQevol}), as we will do in this
work.

The cells of our finite volume discretization are polyhedra with flat
polygonal faces (or lines in 2D). Let $\vec{A}_{ij}$ describe the
oriented area of the face between cells $i$ and $j$ (pointing from $i$
to $j$).  Then we can define the averaged flux across the face $i$-$j$ as
\begin{equation}
\vec{{F}}_{ij} = \frac{1}{A_{ij}} \int_{A_{ij}} \left[ \vec{F}(\vec{U})
- \vec{U} \vec{w}^T\right] {\rm d}\vec{A}_{ij},
\end{equation}  
and the Euler equations in finite-volume form become
\begin{equation} 
\frac{{\rm d}\vec{Q}_i}{{\rm d}t} = - \sum_j A_{ij} \vec{F}_{ij}.
\end{equation}  
We obtain a manifestly conservative time discretization of this equation
by writing it as
\begin{equation} 
\vec{Q}_i^{(n+1)} = \vec{Q}_i^{(n)} - \Delta t \sum_j A_{ij}
\vec{{\hat F}}_{ij}^{(n+1/2)}, 
\label{eqnupdate}
\end{equation} 
where the  $\vec{{\hat F}}_{ij}$ are now an appropriately time-averaged
approximation to the true flux $\vec{F}_{ij}$ across the cell face.  The
notation $\vec{Q}_i^{(n)}$ is meant to describe the state of the system
at step $n$.  Note that $\vec{{\hat F}}_{ij}$ = $-\vec{{\hat F}}_{ji}$,
i.e.~the discretization is manifestly conservative.

Evidently, a crucial step lies in obtaining a numerical estimate of
the fluxes $\vec{{\hat F}}_{ij}$, and a good fraction of the literature
on computational fluid dynamics is concerned with this problem. This
issue is particularly important since the most straightforward (and
perhaps naive) approach for estimating the fluxes, namely simply
approximating them as the average of the left and right cell-centred
fluxes catastrophically fails and invariably leads to severe numerical
integration instabilities that render such a scheme completely useless in
practice.

One effective cure for the stability problem lies in ``upwind'' schemes that
do not weight the two sides equally, but rather with a bias in the upwind
direction of the flow. This works especially well for simpler equations than
the Euler system, for example the advection equation.  Another, physically
particularly meaningful approach is given by the family of Godunov methods,
which employ analytic solutions of the Riemann problem occurring at each cell
interface, either obtained exactly or approximately.

We will employ Godunov's method in the form of the MUSCL-Hancock scheme
\citep{Leer1984,Toro1997,Leer2006}, which is a well-known and relatively
simple approach for obtaining second-order accuracy in space and time. This
scheme is also popular in astronomy and used in several state-of-the art
Eulerian codes \citep[e.g.][]{Fromang2006,Mignone2007,Cunningham2007}. In its
simplest form, the MUSCL-Hancock scheme involves a slope-limited piece-wise
linear reconstruction step within each cell, a first order prediction step for
the evolution over half a timestep, and finally a Riemann solver to estimate
the time-averaged inter-cell fluxes for the timestep.

\begin{figure}
\bc
\resizebox{8cm}{!}{\includegraphics{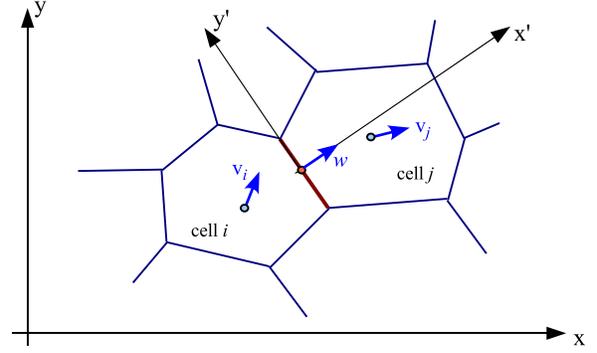}}\\
\caption{Geometry of the flux calculation. We use an unsplit scheme where the
  flux across each face is estimated based on a one-dimensional Riemann
  problem. To this end, the fluid state is expressed in a frame which moves
with the normal velocity $\vec{w}$ of the face, and is aligned with it. Note
that the motion of the face is fully specified by the velocities of the mesh-generating
points of the cells left and right of the face.
\label{FigSketchFluxes}}
\ec
\end{figure}

Figure~\ref{FigSketchFluxes} gives a sketch of the problem of estimating
the flux across the face between two Voronoi cells. Since truly
multidimensional Riemann solvers are not known, we will calculate the
flux for each face separately, treating it as an effectively
one-dimensional problem. Since we do not work with Cartesian meshes, we
cannot use operator splitting \citep{Strang1968} to deal with the
individual dimensions. Rather we use an {\em unsplit} method where all
the fluxes are computed in one step, and are then collectively applied
to calculate the change of the conserved quantities in a cell. For
defining the Riemann problem normal to a cell face, we rotate the fluid
state into a suitable coordinate system with the $x'$-axis normal to the
cell face (see sketch).  This defines the left and right states across
the face, which we pass to an exact Riemann solver. The latter is
implemented following \citet{Toro1997} with an extension to treat vacuum
states, but could easily be substituted with an approximate Riemann
solver for higher performance, if desired.  We have also written an
exact Riemann solver for isothermal gas, similar to the scheme of
\citet{Balsara1994}.  We note that in multi-dimensions the transverse
velocities are also required in the Riemann problem in order to identify
the correct upwind transverse velocity, which is important for an
accurate treatment of shear.  Once the flux has been calculated with the
Riemann solver, we transform it back to the lab frame.

A further important point concerns the treatment of the allowed motion
of cell surfaces in our scheme. In order to obtain stable upwind
behaviour, the Riemann problem needs to be solved {\em in the frame of
  the moving face}. This is important as the one-dimensional Godunov
approach is not Galilean-invariant in the following sense: Suppose left
and right state at an interface are described by $(\rho_{\rm L}, P_{\rm
  L}, v_{\rm L})$ and $(\rho_{\rm R}, P_{\rm R}, v_{\rm R})$, for which
the Riemann solver returns an interface state $(\rho_{\rm F}, P_{\rm F},
v_{\rm F})$ that is the basis for the flux estimate. For example, the
mass flux across the interface is then given by $\rho_{\rm F} v_{\rm
  F}$. Consider now a velocity boost $v$ applied both to the left and
the right side. The new Riemann problem is given by $(\rho_{\rm L},
P_{\rm L}, v_{\rm L}+v)$ and $(\rho_{\rm R}, P_{\rm R}, v_{\rm R}+v)$,
and its solution will be sampled at the fixed coordinate $x=0$ in the
new frame of reference, returning a flux estimate $\rho'_F
v'_F$. However, in general this will yield $\rho'_{\rm F} v'_{\rm F} \ne
\rho_{\rm F}(v_{\rm F}+v)$, which means that the calculated flux
vectors are not Galilean invariant.  This is not necessarily a problem
in practice, but as we will see, it can drastically reduce the accuracy
of Eulerian hydrodynamics in the presence of large bulk velocities. For
this reason, we pay particularly attention to obtain a
Galilean-invariant formulation of our new scheme, which should be
possible if the mesh motion is tied to the fluid motion. 

It is important to note that the Riemann problem itself is an exact
solution of the Euler equations, which is of course
Galilean-invariant. However, the flux vector read off from the Riemann
solution does not transform in a Galilean-invariant way, simply because
the location where the self-similar solution is sampled depends on the
frame of reference.  A timestep in our scheme may also be viewed as a
sequence of reconstruction, evolution, and averaging steps (REA
approach). Both the spatial reconstruction (which is linear in the
primitive variables) as well as the evolution steps (by means of the
Riemann solver) are Galilean-invariant, but the averaging is not; it
depends on the frame of reference. Note that the flux vectors
simultaneously encode the evolution and the averaging, and their
non-invariance ultimately originates in the latter. An immediate and
obvious corollary is that the diffusion error resulting from the
averaging depends on the frame of reference. One may also say in a more
general sense that the truncation error of the Eulerian approach is not
Galilean-invariant. Finally, we would like to stress that this feature
of Galilean non-invariance does of course not mean that the Eulerian
approach necessarily creates incorrect results. It only means that the
errors in the solutions depend on the frame of reference, which is a
highly unwelcome feature. But as higher resolution always helps to
reduce the diffusion error, one should always be able to beat down, at
potentially considerable numerical cost, the additional diffusion error
obtained from some bulk velocity to the point where it lies below a
prescribed tolerance. Nevertheless, it is clearly desirable to have a
numerical scheme where the Galilean-invariance of the Euler equations is
manifestly retained in the discretized forms of the equations, a goal
that is achieved by the method proposed here.

In our new hydrodynamical scheme, each timestep involves the following
basic steps:
\begin{enumerate}
\item Calculate a new Voronoi tessellation based on the current coordinates
  $\vec{r}_i$ of the mesh generating points. This also gives the
  centres-of-mass $\vec{s}_i$ of each cell, their volumes $V_i$, as well as the
  areas $A_{ij}$ and centres $\vec{f}_{ij}$ of all faces between cells.
\item Based on the vector of conserved fluid variables $\vec{Q}_i$ associated
  with each cell, calculate the `primitive' fluid variables
  $\vec{W}_i=(\rho_i, \vec{v}_i, P_i)$ for each cell.
\item Estimate the gradients of the density, of each of the velocity components,
  and of the pressure in each cell, and apply a slope-limiting procedure to
  avoid overshoots and the introduction of new extrema.
\item Assign velocities $\vec{w}_i$ to the mesh generating points.
\item Evaluate the Courant criterion and determine a suitable timestep size
  $\Delta t$.
\item For each Voronoi face, compute the flux $\vec{{\hat F}}_{ij}$ across it
  by first determining the left and right states at the midpoint of the face
  by linear extrapolation from the cell midpoints, and by predicting these
  states forward in time by half a timestep. Solve the Riemann problem in a
  rotated frame that is moving with the speed of the face, and transform the
  result back into the lab-frame.
\item For each cell, update its conserved quantities with the total flux over
  its surface multiplied by the timestep, using equation
  (\ref{eqnupdate}). This yields the new state vectors $\vec{Q}_i^{(n+1)}$ of
  the conserved variables at the end of the timestep.
\item Move the mesh-generating points with their assigned velocities for this
  timestep.
\end{enumerate}
For the sake of definiteness, we will now more explicitly describe the most
important details of these different steps.

\subsection{Gradient estimation and linear reconstruction} \label{SecGradients}

According to the Green-Gauss theorem, the surface integral of a scalar
function over a closed volume is equal to its gradient integrated over
the same volume, i.e.
\begin{equation}
\int_{\partial V} \phi\,  {\rm d} \vec{n} =
\int_{V} \vec{\nabla}\phi \,{\rm d}V.
\end{equation} 
This suggests one possible way to estimate the mean gradient in a Voronoi cell,
in the form
\begin{equation}
\left<\vec{\nabla}\phi\right>_i
\simeq
-\frac{1}{V_i}\sum_j
 \phi(\vec{f}_{ij}) \,\vec{A}_{ij},
\label{Eqnbasicgrad}
\end{equation}
where $\phi(\vec{f}_{ij})$ is the value of $\phi$ at the centroid
$\vec{f}_{ij}$ of the face shared by cells $i$ and $j$, and $\vec{A}_{ij}$ is
a vector normal to the face (from $j$ to $i$), with length equal to the face's
area. Based on the further approximation
\begin{equation}
 \phi(\vec{f}_{ij})\simeq \frac{1}{2}(\phi_i + \phi_j),
\label{eqnmidp}
\end{equation}
this provides an estimate for the local gradient. Note that with the use of
equation (\ref{eqnmidp}), the gradient of cell $i$ only depends on the values
$\phi_j$ of neighbouring cells, but not on $\phi_i$ itself. While the
estimate~(\ref{Eqnbasicgrad}) can be quite generally applied to arbitrary
tessellations, due to the use of only one Gauss point per face it is also
relatively inaccurate and is not exact to linear order in general.

For the special case of Voronoi cells, it is however possible to obtain a
considerably better gradient estimate with little additional effort. The key
is to carry out the surface integral more accurately. Let us assume that in
the vicinity of a point $i$ the scalar function $\phi(\vec{r})$ can be well
approximated linearly as $\phi(\vec{r}) =\phi_i +
\vec{b}\cdot(\vec{r}-\vec{r}_i)$. The vector $\vec{b}$ is the local gradient
that we seek to estimate. We can now write the surface integral as
\begin{equation}
V_i\left<\vec{\nabla}\phi\right>_i = \int_{\partial V_i}\hspace*{-0.2cm} \phi\,  {\rm d} \vec{n} =
\sum_{j\ne i} \int_{A_{ij}}\hspace*{-0.2cm} [\phi_i + \vec{b}\cdot(\vec{r}-\vec{r}_i)]\frac{\vec{r}_{j}-\vec{r}_i}{r_{ij}}
{\rm d}A
\label{eqnaux1}
\end{equation}
where the sum extends over all faces of the Voronoi cell of $i$, and the
integrals extend over each of the faces.  Note that we here already made use
of the fact that the surface normal of each face is parallel to the separation
vector of $i$ and $j$, a property that is in general only fulfilled for
Voronoi tessellations. Following the notation of \citet{Serrano2001}, we now
define $\vec{c}_{ij}$ as the vector from the midpoint between $i$ and $j$ to
the centre-of-mass of the face between $i$ and $j$, i.e.
\begin{equation}
\vec{c}_{ij}\equiv\frac{1}{A_{ij}}\int_{A_{ij}} \left(\vec{r} -
  \frac{\vec{r}_i + \vec{r}_j}{2}\right) {\rm d}A.
\end{equation}
Noting that $\phi_j = \phi_i + \vec{b}\cdot(\vec{r}_j - \vec{r}_i)$,
equation~(\ref{eqnaux1}) can be rewritten as
\begin{equation}
V_i\left<\vec{\nabla}\phi\right>_i =
-\sum_{j\ne i} \left[ \frac{\phi_i+\phi_j}{2}  +
  \vec{b}\cdot\vec{c}_{ij}\right] 
\frac{ \vec{r}_{ij}}{r_{ij}}
A_{ij}.
\label{eqnaux2}
\end{equation}
where $\vec{r}_{ij} = \vec{r}_{i} - \vec{r}_{j}$ is the vector of length
$r_{ij}=|\vec{r}_{ij}|$ connecting the two neighbouring points, 
and $A_{ij}$ is the area of the face.

Next, we can make the replacement $(\vec{b}\cdot\vec{c}_{ij})\, \vec{r}_{ij} =
(\vec{b}\cdot\vec{r}_{ij})\,\vec{c}_{ij} + \vec{b}\times(\vec{r}_{ij} \times
\vec{c}_{ij})$, and set $\vec{b}\cdot\vec{r}_{ij} = \phi_i-\phi_j$.
Then, the term involving the cross products can be rewritten by
reinserting the definition of $\vec{c}_{ij}$:
\begin{equation}
\vec{b}\times \sum_{j\ne i} \frac{\vec{r}_{ij}\times \vec{c}_{ij}}{r_{ij}}
A_{ij}
=\vec{b}\times \sum_{j\ne i} \frac{\vec{r}_{ij}}{r_{ij}}\times 
\int \left(\vec{r} -
  \frac{\vec{r}_i + \vec{r}_j}{2}\right) {\rm d}A
\end{equation}
The term involving $\vec{r}$ is really the surface integral $\int_{\partial V}
\vec{r} \times {\rm d}{\vec{n}}$, which can be cast into a volume integral of
the curl of $\vec{r}$, but $\vec{\nabla}\times \vec{r}=0$ vanishes.  Likewise,
we have $\vec{r}_{ij}\times (\vec{r}_i+\vec{r}_j)/2 = \vec{r}_{ij}\times
\vec{r}_i$, so that the remaining term is proportional to the surface integral
$\int_{\partial V} \vec{r}_i \times {\rm d}{\vec{n}}$, which also vanishes
since $\vec{r}_i$ is a constant vector. As a result, the double cross
product $\vec{b}\times(\vec{r}_{ij} \times \vec{c}_{ij})$ gives a vanishing
contribution to equation~(\ref{eqnaux2}).

We are hence finally left with the following gradient estimate:
\begin{equation}
\left<\vec{\nabla}\phi\right>_i =   \frac{1}{V_i}
   \sum_{j\ne i}  A_{ij} \left( [\phi_j -\phi_i]\,  \frac{\vec{c}_{ij}}{r_{ij}}  
-\frac{\phi_i + \phi_j}{2}\,\frac{\vec{r}_{ij}}{r_{ij}} \right).
\label{eqngrad2}
\end{equation}
Note that this result is exact to linear order, independent of the locations
of the mesh-generating points of the Voronoi tessellation.  Without the term
involving $\vec{c}_{ij}$ this is the same as the simpler Green-Gauss
estimate. However, retaining this extra term leads to significantly better
accuracy, because the gradient estimate becomes exact to linear order for
arbitrary Voronoi meshes. In practice, we shall therefore always use this
gradient estimation in our MUSCL-Hancock scheme for the Euler equations, where
we calculate in this way gradients for the 5 primitive variables $(\rho, v_x,
v_y, v_z, P)$ that characterize each cell.

The result (\ref{eqngrad2}) has also an interesting relation to the formulae
obtained by \citet{Serrano2001} for the partial derivatives of the volume of a
Voronoi cell with respect to the location of one of the points.  As
\citet{Serrano2001} have shown, the derivative of the volume of a Voronoi cell
due to the motion of a surrounding point is given by
\begin{equation}
\frac{\partial V_i}{\partial \vec{r}_j}
= -A_{ij} \left(\frac{\vec{c}_{ij}}{r_{ij}}  
+ \frac{\vec{r}_{ij}}{2r_{ij}} \right) \;\;\; {\rm for}\;\;\; i\ne j .
\label{EqSerr1}
\end{equation}
Furthermore, they show that
\begin{equation}
  \frac{\partial V_i}{\partial \vec{r}_i}
  = - \sum_{j\ne i} \frac{\partial V_j}{\partial \vec{r}_i}.
\label{EqSerr2}
\end{equation}
Using these relations, and noting that according to the Gauss theorem we have
\begin{equation}
 \frac{\phi_i}{V_i}   \sum_{j\ne i}  A_{ij} \frac{\vec{r}_{ij}}{r_{ij}} = 0 ,
\end{equation}
because the summation is just the surface integral of a constant function,
we can also write the estimate for the gradient 
 of $\phi$ at $\vec{r}_i$ more
compactly as
\begin{equation}
\left<\vec{\nabla}\phi\right>_i = -\frac{1}{V_i}\sum_j \frac{\partial V_j}{\partial
  \vec{r}_i} \phi_j .
\end{equation}
An interesting corollary of the above is that
provided $\phi(\vec{r})$ varies only linearly, 
the sum
\begin{equation}
S = \sum_i \phi(\vec{r}_i) V_i
\end{equation}
approximates the integral $\int \phi(\vec{r}) \,{\rm d}V$ {\em exactly},
independent of the positions of the points that generate the Voronoi
tessellation.  This follows because we then have $\partial S/\partial
\vec{r}_i =0$ for all the points $i$.

In our approach, we use the gradients estimated with equation (\ref{eqngrad2})
for a linear reconstruction in each cell around the centre-of-mass. For
example, the density at any point $\vec{r}\in V_i$ of a cell is estimated
as
\begin{equation} \rho(\vec{r}) = \rho_i +
\left<\vec{\nabla}\rho\right>_i \cdot(\vec{r} - \vec{s}_i), 
\end{equation} 
where $\vec{s}_i$ is the centre of mass of the cell. Note that independent of
the magnitude of the gradient and the geometry of the Voronoi cell, this
linear reconstruction is conservative, i.e.~the total mass in the cell $m_i$
is identical to the volume integral over the reconstruction, $m_i =
\int_{V_i}\rho( \vec{r}){\rm d}^3 r$. An alternative choice for the reference
point is to choose the mesh-generating point $\vec{r}_i$ instead of
$\vec{s}_i$. This is the more natural choice if the cell values are known to
sample the values of the underlying field at the location of the
mesh-generating points, then the reconstruction is exact to linear order.
However, our input quantities are cell-averages, which correspond to linear
order to the values of the underlying field sampled at the centre-of-masses of
the cells. For this reason we prefer the centre-of-mass of a cell as reference
point for the reconstruction.

Nevertheless, this highlights that large spatial offsets between the
centre-of-mass of a cell and its mesh-generating point are a source of errors
in the linear reconstruction.  It is therefore desirably to use ``regular''
meshes if possible, where the mesh-generating points lie close to the
centre-of-mass; such meshes minimize the errors in the gradient estimation and
the linear reconstruction. Or in other words, we would like our Voronoi meshes
to be relatively close to so-called {\em centroidal Voronoi meshes}, where the
mesh-generating points lie exactly in the centre of mass of each cell. As we
will discuss in more detail later, we have developed an efficient method for
steering the mesh motion such that this regularity condition can be
approximately maintained at all times (if desired).

In smooth parts of the flow, the above reconstruction is second-order
accurate, with a stencil that consists of the local cell plus all
adjacent cells. However, in order to avoid numerical instabilities the
order of the reconstruction must be reduced near fluid discontinuities,
such that the introduction of new extrema by over- or undershoots in the
extrapolation is avoided.  This is generally achieved by applying slope
limiters that reduce the size of the gradients near local extrema, or by
flux limiters that replace the high-order flux with a lower order
version if there are steep gradients in the upstream region of the
flow. These techniques allow the construction of total variation
diminishing (TVD) schemes, in which spurious oscillations in solutions
can be completely suppressed.

We here generalize the original MUSCL approach to an unstructured grid by
enforcing monotonicity with a slope limiting of the gradients. To this end we
require that the linearly reconstructed quantities on face centroids do not
exceed the maxima or minima among all neighbouring cells \citep{Barth1989}.
Mathematically, we replace the gradient with a slope-limited gradient
\begin{equation}
\left<\vec{\nabla}\phi\right>_i^{'} = \alpha_i
\left<\vec{\nabla}\phi\right>_i ,
\end{equation}
where the slope limiter $0\le \alpha_i\le 1$ for each cell is computed as 
\begin{equation}
\alpha_i = \min(1, \psi_{ij}).
\end{equation}
Here the minimum is taken with respect to all cells $j$ that are
neighbours
of cell $i$, and the quantity $\psi_{ij}$ is defined as
\begin{equation}
\psi_{ij} = \left\{ 
\begin{array}{ccc}
(\phi_i^{\rm max} -  \phi_i)/ \Delta\phi_{ij}  & {\rm for} & \Delta\phi_{ij}>0\\
(\phi_i^{\rm min} -  \phi_i)/ \Delta\phi_{ij}  & {\rm for} & \Delta\phi_{ij}<0\\
1 & {\rm for} & \Delta\phi_{ij}=0\\
\end{array}
\right.
\end{equation}
where
$\Delta\phi_{ij}=\left<\vec{\nabla}\phi\right>_i\cdot(\vec{f}_{ij} -
\vec{s}_i)$ is the estimated change between the centroid
$\vec{f}_{ij}$ of the face and the centre of cell $i$, and
$\phi_i^{\rm max} = \max(\phi_j)$ and $\phi_i^{\rm min} =
\max(\phi_j)$ are the maximum and minimum values occurring for $\phi$
among all neighbouring cells of cell $i$, including $i$ itself.

We note that this slope limiting scheme does not strictly enforce the total
variation diminishing property, which means that (usually reasonably small)
post-shock oscillations are still possible. However, by choosing a slightly
more conservative slope-limiter it is possible to obtain TVD behaviour, at the
price of a more dissipative scheme \citep{Barth2004}.  Finally, we note that
future refinements of the present method could also employ higher-order
polynomial reconstruction schemes, for example based on a larger stencil and
conservative least square reconstruction
\citep[e.g.][]{OllivierGooch1997}. This would be similar in spirit to
higher-order essentially non-oscillatory (ENO) or weighted ENO (WENO) schemes.

\subsection{Setting the velocities of the mesh generators}

A particular strength of the scheme we propose here is that it can be used
both as an Eulerian code, and as a Lagrangian scheme. The difference lies only
in the motion of the mesh-generation points. If the mesh-generating points are
arranged on a Cartesian mesh and zero velocities are adopted for them, our
method is identical to a second-order accurate Eulerian code\footnote{There
  are of course many different variants of 2nd order Eulerian schemes. Our
  method corresponds to the well known MUSCL-Hancock approach.}  on a
structured grid. Of course, one can equally well choose a different layout of
the points, in which case we effectively obtain an Eulerian code on an
unstructured mesh. The real advantage of the new code can be realized when we
allow the mesh to move, with a velocity that is tied to the local fluid
speed. In this case, we obtain a Lagrangian hydrodynamics code, which has some
unique and important advantages relative to an Eulerian treatment.

In fact, our code belongs to the general class of so-called Arbitrary
Lagrangian-Eulerian (ALE) fluid dynamical methods. Unlike other ALE schemes,
the method proposed here however does not rely on remapping techniques to
recover from distortions of the mesh once they become severe, simply because
the Voronoi tessellation produced by the continuous motion of the
mesh-generating points yields a mesh geometry and topology that itself changes
continuously in time, without any mesh-tangling effects. The motion of the
mesh-generating points can be chosen nearly arbitrarily, including cases where
it is prescribed by an external flow field, for example to smoothly
concentrate resolution towards particular regions of a mesh. Also, as we shall
discuss below, we may modify the flow of mesh-generating points such that
certain desired properties of the fluid tessellation are maintained or
achieved, e.g.~a constant mass per cell, or that cell sizes are constrained to
lie within prescribed minimal or maximal bounds.

The most simple and basic approach for specifying the motion of the mesh
generators is to use
\begin{equation}
\vec{w}_i = \vec{v}_i,
\label{eqnvelmesh}
\end{equation}
i.e.~the points are moved with the fluid speed of their cell. This
ansatz is clearly appropriate for pure advection and in smooth parts of
the flow. Whereas it is not strictly Lagrangian because it does not
guarantee that the faces of the cells move with the local velocity and
hence mass exchange can still occur between the cells, it nevertheless
approximates Lagrangian behaviour by minimized the mass flux between
cells. Also, it can be expected that this ansatz will roughly keep the
mass per cell fixed, leading to an adaptive spatial resolution in
situations with strong clustering of matter.

 However, in this scheme there is no mechanism built in that tries to
improve the regularity of the Voronoi mesh in case the mean mass per cell
should develop substantial scatter around a desired mean value, or if a large
number of cells with high aspect ratios occur. If desired, such tendencies of
a growing mesh irregularity can be counteracted by adding corrective velocity
components to the mesh velocities $\vec{w}_i$ given by equation
(\ref{eqnvelmesh}). There are many different possibilities for how exactly to
do this, and we consider this freedom a strength of the formalism. In
Section~\ref{SecMeshRegularity}, we will discuss a few simple regularization
terms that we have explored thus far, and which have proven to be very effective.

\subsection{Flux computation}

An important aspect of our approach is that the specified motion of the
mesh-generating points fully determines the motion of the whole Voronoi mesh,
including, in particular, the velocities of the centroids of cell faces.  This
allows us to calculate the Riemann problem in the rest-frame of each of the
faces.

\begin{figure}
\bc
\resizebox{8cm}{!}{\includegraphics{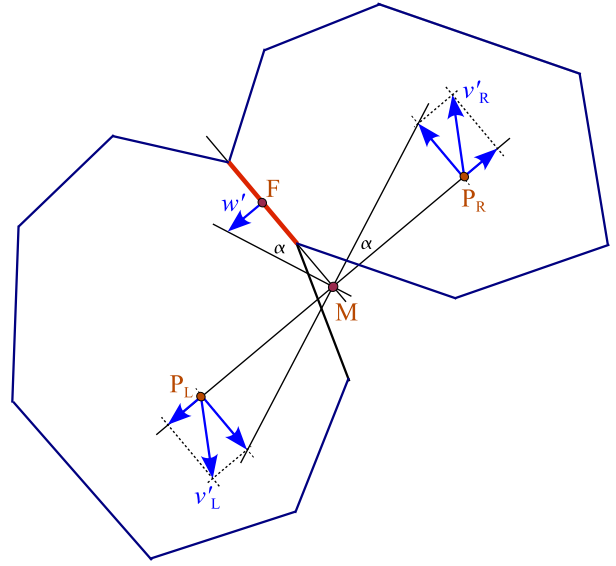}}\\
\caption{Sketch illustrating the calculation of the normal velocity of a
  face based on the motion of its two associated mesh-generating points.
\label{FigSketchVelFace}}
\ec
\end{figure}

Consider one of the faces in the tessellation and call the fluid states in the
two adjacent cells the `left' and `right' states. We first need to determine
the velocity $\vec{w}$ of the face based on the velocities $\vec{w}_{\rm L}$
and $\vec{w}_{\rm R}$ of the two mesh-generating points associated with the
face (they are connected by a Delaunay edge). It is clear that $\vec{w}$ has a
primary contribution from the mean velocity $(\vec{w}_{\rm L}+\vec{w}_{\rm
  R})/2$ of the points, but there is also a secondary contribution $\vec{w}'$
from the residual motion of the two points relative to their centre of
mass. This residual motion is given by $\vec{w}_{\rm R}' = - \vec{w}_{\rm L}'
= (\vec{w}_{\rm R} - \vec{w}_{\rm L})/2$, and we need to determine its impact
on the motion of the face centroid. Figure~\ref{FigSketchVelFace} sketches the
geometry of the situation.  The components of $\vec{w}_{\rm R}'$ and
$\vec{w}_{\rm L}'$ parallel to the line connecting the centroid $\vec{f}$ of
the face with the midpoint $\vec{m}$ of the two mesh-generating points
$\vec{r}_{\rm L}$ and $\vec{r}_{\rm R}$ induce a rotation of the face around
the point $\vec{m}$. We are only interested in the normal velocity component
of this motion at the centroid  of the face. This can be easily computed as
\begin{equation}
\vec{w}' = \frac{(\vec{w}_{\rm L} - \vec{w}_{\rm R})
\cdot
[\vec{f}-(\vec{r}_{\rm R} + \vec{r}_{\rm L})/2]}{|\vec{r}_{\rm
    R}-\vec{r}_{\rm L}|}\,\,
\frac{(\vec{r}_{\rm R}-\vec{r}_{\rm L})}{|\vec{r}_{\rm R}-\vec{r}_{\rm L}|}.
\end{equation}
The full velocity $\vec{w}$ of the face is then given by
\begin{equation}
\vec{w} = \frac{\vec{w}_{\rm R} + \vec{w}_{\rm L}}{2} + \vec{w}'.
\end{equation}
We note that this result can also be used to calculate the rate of 
change of the volume of a cell $i$ due to the motion of its neighbouring
points, viz. 
\begin{equation}
\frac{{\rm d}V_i}{{\rm d}t}
= - \sum_{j\ne i} A_{ij} \left[\frac{\vec{c}_{ij}}{r_{ij}} \cdot(\vec{w}_j -
  \vec{w}_i ) + \frac{\vec{r}_{ij}}{2r_{ij}}\cdot (\vec{w}_j +
  \vec{w}_i)\right],
\label{eqnvolchange}
\end{equation}
where $\vec{r}_{ij} = \vec{r}_i - \vec{r}_j$ is the distance vector between
the two points $i$ and $j$ (with $r_{ij}=|\vec{r}_{ij}|$), $A_{ij}$ is the
area of the common face, and $\vec{c}_{ij} = \vec{f}_{ij} - (\vec{r}_i +
\vec{r}_j)/2$ is a vector pointing to the centre $\vec{f}_{ij}$ of the face
from the midpoint between $i$ and $j$. We note that the same result can also
be obtained with equations (\ref{EqSerr1}) and (\ref{EqSerr2}).

We now calculate the flux across the face using the MUSCL-Hancock approach,
with the important difference that we shall carry out the calculation in the
rest-frame of the face. It is convenient to do this in the primitive variables
$(\rho, \vec{v}, P)$, where we first transform the lab-frame velocities of the
two cells to the rest-frame by subtracting $\vec{w}$,
\begin{equation}
\vec{W}_{\rm L,R}' = \vec{W}_{\rm L,R} - \left(
\begin{array}{c}
0\\
\vec{w}\\
0\\
\end{array}\right) .
\label{eqnrefframe}
\end{equation}
We then linearly predict the states on both side to the centroid of the
face, and also predict them forward in time by half a timestep. This
produces the states 
\begin{equation}
\vec{W}_{\rm L,R}'' = \vec{W}_{\rm L,R}' + \left.\frac{\partial
    \vec{W}'}{\partial \vec{r}}\right|_{\rm L, R} (\vec{f} - \vec{s}_{\rm L,R})
+ \left.\frac{\partial
    \vec{W}'}{\partial t}\right|_{\rm L, R} \frac{\Delta t}{2} .
\end{equation}
The spatial derivatives $\partial \vec{W}'/\partial \vec{r}$ are known, and
given by the (slope-limited) gradients of the primitive variables that are
estimated as described in Section~\ref{SecGradients}. Note that the gradients are
unaffected by the change of rest-frame described by
Eqn.~(\ref{eqnrefframe}). The partial time derivate $\partial \vec{W}/\partial
t$ can be replaced by spatial derivatives as well, based on the Euler equations
in primitive variables, which are given by
\begin{equation}
\frac{\partial \vec{W}}{\partial t}
+ \vec{A}( \vec{W}) \frac{\partial \vec{W}}{\partial \vec{r}} = 0 ,
\end{equation}
where $\vec{A}$ is the matrix
\begin{equation}
\vec{A}( \vec{W}) = \left(
\begin{array}{ccc}
\vec{v} & \rho & 0 \\
 0     & \vec{v} & 1/\rho \\
0 & \gamma P & \vec{v}\\
\end{array}
\right).
\end{equation}
Having finally obtained the states left and right of the interface, we need to
turn them into a coordinate system aligned with the face, such that we can
solve an effectively one-dimensional Riemann problem. The required rotation
matrix $\vec{\Lambda}$ for the states only affects the velocity components,
viz.
\begin{equation}
\vec{W}_{\rm L,R}''' = \vec{\Lambda} \,\vec{W}_{\rm L,R}''=
 \left(
\begin{array}{ccc}
1 &  0 & 0 \\
 0  & \vec{\Lambda}_{\rm 3D} & 0 \\
0 &  0 & 1\\
\end{array}
\right)
\vec{W}_{\rm L,R}'',
\end{equation}
where $\vec{\Lambda}_{\rm 3D}$ is an ordinary rotation of the coordinate
system, such that the new $x$-axis is parallel to the normal vector of the
face, pointing from the left to the right state.

With these final states, we now solve the Riemann problem, and sample
the self-similar solution along $x/t=0$. This can be written as
\begin{equation}
\vec{W}_{\rm F} = R_{\rm iemann}(\vec{W}_{\rm L}''' , \vec{W}_{\rm R}'''),
\end{equation}
where $R_{\rm iemann}$ is a one-dimensional Riemann solver, which returns a
solution for the state of the fluid $\vec{W}_{\rm F}$ on the face in
primitive variables. We now transform this back to the lab-frame, reversing
the steps above,
\begin{equation}
\vec{W}_{\rm lab} = 
\left(
\begin{array}{c}
\rho\\
\vec{v}_{\rm lab}\\
P\\
\end{array}
\right) 
=
\Lambda^{-1}\vec{W}_{\rm F} + \left(
\begin{array}{c}
0\\
\vec{w}\\
0\\
\end{array}
\right) .
\end{equation}
Finally, we can use this state to calculate the fluxes in the conserved
variables across the face. Here we need to take into account that the {\em
  face is moving} with velocity $\vec{w}$, meaning that the appropriate flux
vector in the lab frame is given by
\begin{equation}
\vec{{\hat F}} = \vec{F}(\vec{U}) - \vec{U}\vec{w}^{\rm T}=
\left(\begin{array}{c}
\rho(\vec{v}_{\rm lab}-\vec{w}) \\
\rho\vec{v}_{\rm lab}(\vec{v}_{\rm lab}-\vec{w})^{\rm T} + P\\
\rho e_{\rm lab} (\vec{v}_{\rm lab}-\vec{w})  + P\vec{v}_{\rm lab}
\end{array}
\right),
\label{eqfinalflux}
\end{equation}
where $\vec{U}$ is the state $\vec{W}_{\rm lab}$ expressed in the conserved
variables, and $e_{\rm lab}= \vec{v}^2_{\rm lab}/2 + P_{\rm lab}/[(\gamma-1)
\rho_{\rm lab}]$.  The scalar product of this flux vector with the normal
vector of the face gives the net flux of mass, momentum, and energy that the
two adjacent, moving cells exchange.  It is the flux of equation
(\ref{eqfinalflux}) that can finally be used in the conservative updates of
each cell, as described by equation~(\ref{eqnupdate}).

\subsection{Galilean invariance}  \label{SecGalilei}

In the above scheme, it is clear that the state $\vec{W}_{\rm F}$
sampled from the Riemann solver is invariant under Galilean transformations,
because a special invariant frame for evaluation of the Riemann problem is
adopted, that of the face moving with the flow. As a result, the input left
and right states are invariant under a Galilean boost; any such boost is simply
absorbed into the motion of the face.  This also means that the velocity
difference $\vec{u}\equiv (\vec{v}_{\rm lab}-\vec{w})$ appearing in the flux
of equation (\ref{eqfinalflux}) is invariant as well. It is thus clear that
the mass flux between cells is Galilean-invariant, but this property is much
less evident for the momentum and energy fluxes, as they still have an
additional dependence on the velocity $\vec{v}_{\rm lab}$ in the lab-frame, as
seen in Eqn.~(\ref{eqfinalflux}).

Recall that the desired invariance means that it should not matter whether we
evolve the state of a cell in the current lab-frame, or in a reference frame
that is boosted by a constant velocity relative to it.  In both cases, we
should obtain the same final state when compared again in the same frame.

We can demonstrate this property for the above scheme as follows. Let us assume
for simplicity that there is only one flux vector in or out of a cell. When
calculated in the current lab-frame, the new state after a timestep $\Delta t$
will then be
\begin{equation}
\vec{Q}^{\rm new} = 
\left(\begin{array}{c}
Q_0\\
\vec{Q}_1\\
Q_2\\
\end{array}
\right)
+
\left(\begin{array}{c}
\rho\,\vec{u} \\
\rho\,\vec{v}_{\rm lab}\,\vec{u}^{\rm T} + P\\
\rho\, e_{\rm lab} \vec{u}  + P\,\vec{v}_{\rm lab}
\end{array}
\right) \vec{A}\Delta t ,
\label{eqnstateorig}
\end{equation}
where $\vec{A}$ is the normal vector of the face, and $Q_0$, $\vec{Q}_1$, and
$Q_2$ are mass, momentum and energy of the cell at the beginning.

Let the function $G(\vec{Q},\vec{v})$ return the state vector
 $\vec{Q}'$ of conserved quantities of a cell in a frame that 
is moving with a constant velocity relative to the current frame.
For a Galilean boost with velocity $\vec{v}_0$, the new state is given by
\begin{equation}
\vec{Q}' = G(\vec{Q}, \vec{v}_0) 
\label{eqngal0}
\end{equation}
where
\begin{equation}
G(\vec{Q}, \vec{v}) =
\left(
\begin{array}{c}
Q_0\\
Q_1 + Q_0 \vec{v}\\
Q_2 + \vec{Q}_1\cdot \vec{v} + \frac{1}{2} Q_0 \vec{v}^2\\
\end{array}
\right)
\label{eqngal1}
\end{equation}
defines the boost transformation.
We can now evolve this boosted state over one timestep, which yields
\begin{equation}
\vec{Q}'' = \vec{Q}' + 
\left(
\begin{array}{c}
\rho\,\vec{u} \\
\rho\, (\vec{v}_{\rm lab} + \vec{v}_0) \,\vec{u}^{\rm T} + P\\
\rho\, e_{\rm lab}'\vec{u}  + P\,(\vec{v}_{\rm lab} + \vec{v}_0)
\end{array}
\right) \vec{A}\Delta t ,
\label{eqngal2}
\end{equation}
where
$e_{\rm lab}' = e_{\rm lab}
 + \vec{v}_{\rm lab}\vec{v}_0 + \frac{1}{2}\vec{v}_0^2$.
The flux is here different because $\vec{v}_{\rm lab}$ transform to
$\vec{v}_{\rm lab}+\vec{v}_0$ in the boosted
frame.
Finally, we can take the state $\vec{Q}''$ back to our original frame,
by calculating 
\begin{equation}
\vec{{\tilde Q}}^{\rm new} = G(\vec{Q}'', -\vec{v}_0).
\label{eqngal3}
\end{equation}
Our scheme is Galilean invariant if this state $\vec{{\tilde Q}}^{\rm new}$
agrees with the state (\ref{eqnstateorig}) obtained by evolving the cell in
the original system. Inserting equations (\ref{eqngal0}), (\ref{eqngal1}) and
(\ref{eqngal2}) into (\ref{eqngal3}), and after a bit of algebra, it is seen
that this is indeed the case.  This is an extremely important property not
shared by ordinary Eulerian codes.

As a word of caution, we note that the finite numerical round off errors
always present in ordinary floating point arithmetic will perturb the
exact Galilean invariance of our discretization scheme in practice. In
particular, since the conserved quantities are always stored in the
lab-frame, the effective number of significant bits left for the
internal energy will be reduced for very large bulk velocities, as it is
then defined as the difference of two large numbers. However, with
double precision arithmetic this may only become a problem for really
extremely large Mach numbers, and it could always be solved by the use
of extended floating point precision if needed.

\subsection{Poorly resolved cold flows} \label{seccoldflows}

It is a well known problem in Eulerian finite volume methods that flows that
are dominated by their kinetic energy -- or in other words are very cold and
move supersonically with respect to the calculational frame -- often exhibit
spurious heating in adiabatic parts of the flow. This arises from small
amounts of dissipation occurring in the cold gas, introduced by finite
discretization errors.  Better spatial resolution alleviates the problem, but
if the gas is sufficiently cold, even very small dissipative effects become
readily visible in the evolution of the gas temperature.  Whereas the pressure
forces remain typically negligible as a result of this effect and hence do not
change the gas motion itself, the temperature evolution can be very seriously
in error.  Unfortunately, this problem is ubiquitous in cosmology, where the
early phases of structure formation always involve very cold gas combined with
relatively large velocities that are induced by gravity, resulting in
extremely high Mach numbers. If the spurious dissipation is not prevented, the
temperature of the low-density intergalactic medium can not be trusted and
becomes unusable for quantitative analysis.

Different solutions have been developed in the literature to cope with this
problem \citep[which incidentally is absent in the SPH formalism,
see e.g.][]{Springel2002}.  \citet{Ryu1993} evolve the entropy of the gas as an
additional conserved quantity and define a number of criteria for deciding
whether the energy or the entropy equation should be used. \citet{Bryan1995}
on the other hand propose a `dual energy formalism', where the internal gas
energy is evolved in addition to the total energy. When the gas motion is
highly supersonic, the temperature and pressure of the gas are set based on
the result of the internal energy equation, while otherwise the total energy
is used.

The Galilean-invariance of the moving mesh code suggests that it should in
principle have fewer problems with highly supersonic flow.  However, if the
velocity differences from cell to cell are of the order of the local sound
speed or larger, we find that the moving mesh code can also give rise to
spurious dissipation in the cold gas and as a result can produce an incorrect
temperature evolution. The problem is that in adiabatically evolving gas, any
small difference between the total energy and kinetic energy of a cell
automatically appears in the thermal energy. Even if the discretization errors
from fluid advection are quite small, equal to only a small fraction of the
total energy of a cell, these errors will give rise to spurious changes of the
temperature if the thermal energy is comparably small.

A related problem arises in simulations that are coupled to a collisionless
dark matter or stellar component. The collisionless fluid often dominates the
gravitational field, and is usually treated with the gravitational N-body
approach. The problem is that the resulting gravitational force field is
relatively noisy, and imparts a stochastic driving onto the gas as it flows
through the bumpy potential provided by the N-body system. We have found that
the resulting small-scale velocity fluctuations are readily dissipated away by
the mesh-based hydrodynamics, giving rise to a significant spurious heating of
the gas. Interestingly, SPH is much less susceptible to this effect,
presumably because of its much poorer ability to detect very weak shocks.
Part of the heating effect can be readily understood by the analysis of
\citet{Steinmetz1997}, who showed that two-body encounters between
collisionless particles and fluid elements induce substantial heating in the
gas. In numerical experiments with cold gaseous disks in isolated galaxies
with live N-body dark matter halos, we have found that the effect in the
finite volume hydrodynamics is even stronger than expected based on
\citet{Steinmetz1997}, presumably because the gas also reacts strongly to
slower moving collisionless particles that are not well treated by the impulse
approximation. We also found that the heating can only be efficiently
suppressed if either an extremely large number of N-body particles is used, or
a smooth analytic potential is employed. But such large particle numbers are
impractical in many applications, and also not needed in the SPH approach. We
therefore seek a method that can suppress the spurious heating of the gas
through the N-body component, if needed.

To circumvent the problems described above, we adopt a solution that is
similar to that of \citet{Ryu1993}, but differs in a number of important
aspects. We first define a measure of the total entropy of a cell as
\begin{equation}
S_i = M_i A_i = M_i \frac{P_i}{\rho_i^\gamma },
\end{equation}
where $A_i\equiv P_i / \rho_i^\gamma$ is an entropic function that effectively
labels the entropy per unit mass of the gas, and $\gamma$ is the adiabatic
index.  Note however that the quantity $S$ is not the thermodynamic entropy
itself, but is related to it through a simple monotonic relation. In fact, for
a monoatomic ideal gas the thermodynamic entropy ${S}_{\rm therm}$ per
particle is given by
\begin{equation}
\frac{{S}_{\rm therm}}{N} = \frac{3}{2}k_{\rm B}\left[
\ln \left(\frac{S}{N}\right) + \ln \left(\frac{2\pi m^{5/3}}{h^2}\right)  + \frac{5}{3} \right],
\end{equation}
where $N$ is the number of atoms, $m$ their mass, and $h$ is Planck's
constant. For simplicity, we will call $S$ the total entropy, as it is simpler
to work with than using the thermodynamic entropy directly.

The Euler equations show that outside of shocks, $S$ is a {\em conserved
  quantity}.  We can hence add a further hyperbolic conservation law of the
form
\begin{equation}
\frac{\partial}{\partial t}(\rho A) + \vec{\nabla}\cdot(\rho A \vec{v}) = 0
\label{eqnentrconserv}
\end{equation}
to the set of equations we solve in our finite volume scheme, and treat $S_i$
as a further component in the vector $\vec{Q}_i$ of conserved quantities for
each cell. Furthermore, we may optionally replace the primitive variable $P_i$
with the entropic function $A_i = S_i/M_i$ of a cell. For the vector of
primitive variables $\vec{W}=(\vec{\rho}, \vec{v}, \vec{A}$), the Euler
equations can in this case be written as
\begin{equation}
\frac{\partial \vec{W}}{\partial t}
+ \vec{B}( \vec{W}) \frac{\partial \vec{W}}{\partial \vec{r}} = 0 ,
\end{equation}
where $\vec{B}$ is the matrix
\begin{equation}
\vec{B}( \vec{W}) = \left(
\begin{array}{ccc}
\vec{v} & \rho & 0 \\
 \gamma A \rho^{\gamma-2}     & \vec{v} & \rho^{\gamma-1} \\
0 & 0 & \vec{v}\\
\end{array}
\right).
\end{equation}
This again shows that $A$ stays constant along the flow, making this variable
particularly convenient to characterize adiabatic motion.

We can now apply our usual gradient estimation, spatial reconstruction and
slope limiting procedures to the entropic function $A_i$ (in addition to, or
instead of, the pressure). When the pressure is needed, for example as input
to the Riemann problem, it is calculated as $P= A \rho^\gamma$. Finally, we
compute additional flux components at each cell face, namely the entropy
fluxes corresponding to equation (\ref{eqnentrconserv}), and use them to
update the entropies $S_i$ of all cells, keeping the sum of the total entropy
constant. At each cell face, we take the entropy flux to be $\rho_{\rm F}\,v_{\rm F}\,
A_{\rm U}$, where $\rho_{\rm F}$ and $v_{\rm F}$ are the density and normal
velocity returned by the Riemann solver, while $A_{\rm U}$ is chosen equal to
the entropic function of the {\em upwind side} of the Riemann problem
(i.e.~$A_{\rm U}$ is either equal to $A_{\rm L}$ or $A_{\rm R}$), which
we select based on the sign of $v_{\rm F}$. This hence advects the entropy
assuming that the flow is smooth.

However, normally the result of this entropy advection is {\em discarded} at
the end of each timestep. Instead, we {\em reinitialize} the entropy $S_i$ of
each cell based on the updated values of total energy, total momentum and
mass. This takes care of the fact that in general the entropy will not be
conserved after all. It will tend to increase, either through dissipative
processes in shock fronts as captured by the analytic Riemann solution, or as
a result of the mixing entropy that is generated when the Riemann solution is
averaged over a cell and mapped back to a piece-wise constant state. The
entropy conservation law (\ref{eqnentrconserv}) is therefore essentially
redundant in finite volume methods because the other conservation laws already
fully determine the final averaged state of the cell. This is why in an ordinary
Godunov scheme the entropy is normally not considered explicitly, the scheme
automatically injects exactly the right amount of entropy to satisfy
the conservation laws of total energy, momentum and mass.

However, if the flow is poorly resolved and very cold, or if it is governed by
a noisy external gravitational field, we may give precedence to the entropy
conservation law over that for the total energy, provided the flow is
sufficiently smooth. This can be simply accomplished by keeping the updated
entropy $S_i$ of a cell at the end of a timestep, thereby suppressing local
dissipation.  Instead of reinitializing the entropy with the help of the total
energy equation, the entropy is then used together with the new density to
update the thermal energy, and hence the pressure. In this case there is no
spurious heating of the gas in parts of the flow that are dominated by their
kinetic energy, or are cold and poorly resolved. Instead, the temperature will
evolve adiabatically, as expected for a smooth flow. The catch is that this
procedure temporarily gives up manifest conservation of total energy, as the
thermal energy is now not defined as a difference between total energy and
kinetic energy, but rather based on the value expected for isentropic
evolution of the gas. The resulting errors should normally be negligible
however if the entropy scheme is only applied when the thermal energy is a
negligible part of the total energy, and the pressure forces are unimportant.

The above discussion makes it clear that an important part of this method is
the specific criterion used to decide whether a sufficiently smooth, poorly
resolved cold flow is actually present, and hence a dissipative update of the
entropy via the total energy equation can be delayed. We presently use the
following simple criteria for this purpose.

\begin{figure}
\bc
\resizebox{8cm}{!}{\includegraphics{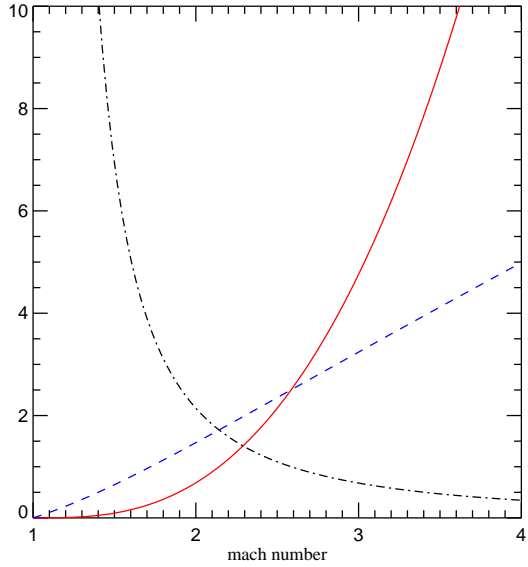}}%
\caption{Dissipative heating rate $f_{\rm diss}({\cal M})$ in a shock as a
  function of Mach number ${\cal M}$ (red  solid line), for a $\gamma=5/3$
  gas. The dashed blue line gives the adiabatic heating rate $f_{\rm adiab}({\cal M})$ 
 of the
  gas as it is compressed at the shock. Finally, the dotted line gives the
  ratio of adiabatic and dissipative heating rates.
  \label{FigShockHeating}}
\ec
\end{figure}

Our primary criterion relies on directly detecting the presence of shocks with
the help of the Riemann problem that we solve for each face. The Riemann
problem yields a contact wave that is sandwiched on both sides either by a
shock wave or a rarefaction fan. The Mach number(s) of the shock(s) present in
the Riemann problem can be easily determined. We hence can find for each cell
the maximum Mach number that occurs in any of the Riemann problems of its
surrounding faces. The idea is to only use the entropy equation whenever this
maximum Mach number is smaller than a prescribed threshold value.

To examine the consequences of such a scheme, we recall the irreversible
thermal dissipation rate of shock as a function of its Mach number. For a
shock propagating with Mach number ${\cal M} = v_1/c_1$ into a medium of
density $\rho_1$, soundspeed $c_1$ and thermal energy per unit mass $u_1$, the
dissipative increase in thermal energy per unit time and unit shock surface
area ${\rm d}F$ can be written as \citep[see also][]{Pfrommer2006}
\begin{equation}
\frac{{\rm d}E_{\rm diss}}{{\rm d}t\,{\rm d}F} 
= \rho_1 v_1  \rho_2^{\gamma-1}  (A_2 - A_1)/({\gamma-1}),
\end{equation}
where $A_1$ and $A_2$ are the pre- and postshock entropic functions, and
$\rho_2$ is the postshock density. The adiabatic heating rate just from  the
reversible compression of the gas is given by
\begin{equation}
\frac{{\rm d}E_{\rm adiab}}{{\rm d}t\,{\rm d}F} 
= \rho_1 v_1  (\rho_2^{\gamma-1} - \rho_1^{\gamma-1}) A_1/({\gamma-1}).
\end{equation}
The jumps in density and entropy can be expressed in terms of Mach number
only:
\begin{equation}
f_{\rho}({\cal M}) \equiv \frac{\rho_2}{\rho_1} = \frac{(\gamma+1){\cal M}^2}
{(\gamma-1){\cal M}^2 +2 },
\end{equation}
\begin{equation}
f_{A}({\cal M}) \equiv \frac{A_2}{A_1} = \frac{2\gamma{\cal M}^2 - (\gamma-1)}{\gamma+1} \left[\frac{(\gamma-1){\cal M}^2 +2 }{(\gamma+1){\cal M}^2}\right]^\gamma
.
\end{equation}
This allows us to express the dissipative heating rate as
\begin{equation}
\frac{{\rm d}E_{\rm diss}}{{\rm d}t\,{\rm d}F} 
= \rho_1 u_1 c_1\, f_{\rm diss}({\cal M}) ,
\end{equation}
with
\begin{equation}
f_{\rm diss}({\cal M}) = {\cal M}[f_A({\cal M}) - 1] f_{\rho}^{\gamma-1}({\cal M}).
\end{equation}
This shows the well-known result that the dissipation rate in a shock depends
very sensitively on Mach number, $f_{\rm diss}({\cal M}) \propto ({\cal M} -
1)^3$. Similarly, we can write the heating rate from the adiabatic shock
compression as
\begin{equation}
\frac{{\rm d}E_{\rm adiab}}{{\rm d}F\,{\rm d}t} 
= \rho_1 u_1 c_1\, f_{\rm adiab}({\cal M}) ,
\end{equation}
with
\begin{equation}
f_{\rm adiab}({\cal M}) = {\cal M}[f_{\rho}^{\gamma-1}({\cal M}) - 1] .
\end{equation}
Note that the adiabatic heating rate increases more slowly with Mach number
than the dissipation, $f_{\rm adiab}({\cal M}) \propto ({\cal M}-1)$. For
very low Mach numbers, the adiabatic heating strongly dominates, and the
dissipative heating becomes comparatively unimportant. 
This is shown in Figure~\ref{FigShockHeating}, where we plot the factors 
$f_{\rm diss}({\cal M})$ and $f_{\rm adiab}({\cal M})$ as a function of Mach
number, as well as their ratio.

This suggests to use a threshold Mach number ${\cal M}_{\rm thresh}$ for
deciding whether the entropy equation may be used to update a cell instead of
the total energy equation. If we pick ${\cal M}_{\rm thresh} \sim 1.1$ and use
the entropy equation only if the maximum Mach number of all shocks in the
Riemann problems surrounding the cell lies below this number, then the entropy
production of very weak shocks that are associated with spurious dissipation
is suppressed. The flow is effectively treated as being smooth and
adiabatic. Note that if real weak shocks of this small strength are present,
they still nevertheless have the correct adiabatic heating rate, which
strongly dominates for these weak shocks (by a factor of more than 100 for
${\cal M} < 1.1$), suggesting that errors in the dynamics should be very
minor. Indeed, we have found that this scheme works especially well for
suppressing artificial heating of the gas from the Poisson noise in the
gravitational field of a N-body system. In this case we have also not been
able to find any detrimental impact on the quality with which the total energy
is conserved (on the contrary actually), which we recall is not manifestly
conserved in self-gravitating systems.

As an alternative to the Mach number switch discussed above, we have also
implemented a scheme that compares the thermal energy of a cell with a
suitably defined kinetic energy in order to determine whether a flow is
cold. This is more similar to the approach of \citet{Bryan1995} and
\citet{Ryu1993}.  In practice, we first determine the expected new thermal
energy $E_{\rm therm} = E_{\rm tot} - M\vec{v}^2/2$ at the end of the
timestep, based on the usual conservation laws.  If this energy is much
smaller than the maximum kinetic energy $E_{\rm kin}^{\rm max}$ among the cell
and all its neighbouring cells,
\begin{equation}
E_{\rm therm} < \alpha_{S}\, E_{\rm kin}^{\rm max},
\label{eqnScrit1}
\end{equation}
then the flow is is considered `cold' and the entropy is kept and not updated
in this step. We typically use $\alpha_{S}\sim 0.01$ for the parameter
$\alpha_{S}$, but the results are not sensitive to this choice provided one
ensures that one switches back to the ``normal'' treatment of the
hydrodynamics once a sufficiently strong dissipative event occurs.

In the Lagrangian mode of the code, we define the kinetic energy $E_{\rm
  kin}^{\rm max}$ of the neighbouring fluid cells relative to the velocity
$\vec{w}_i$ of the current cell. In this way, the criterion (\ref{eqnScrit1})
becomes Galilean invariant and effectively compares the local sound speed with
the size of the velocity changes from cell to cell.  Sometimes this renders
the criterion too restrictive, however, especially in simulations with
self-gravity. We then invoke a further condition,
\begin{equation}
E_{\rm therm} < \beta_{S}\, M_i \, g_i\, R_i,
\label{eqnScrit2}
\end{equation}
which effectively compares the strength of pressure forces to the
gravitational acceleration. Here $R_i$ is the `radius' of the cell (see
below), and $g_i$ is the magnitude of the local gravitational acceleration. If
one of the conditions (\ref{eqnScrit1}) or (\ref{eqnScrit2}) is fulfilled, the
entropy is kept for the current step.  This is based on the idea that if the
pressure forces are negligible compared to the gravitational forces, we are
dealing with an effectively kinematically dominated flow, and it then makes
sense to keep the entropy as this provides for a more accurate temperature
evolution.

Note that the scheme described in this subsection is an optional treatment in
the {\small AREPO} code. Even if enabled, there is no difference to the
ordinary conservative hydrodynamics for sufficiently small values of ${\cal
  M}_{\rm thresh}$, or $\alpha_S$ and $\beta_S$.  Also, if these parameters
are set to unreasonably large values and the entropy production is
artificially suppressed, the dynamics is often still represented surprisingly
accurately.  This is because weak shocks produce only little new entropy. Even
if this entropy production is ignored, the Riemann solver still recovers the
correct jumps in density and velocity and rescues the dynamics.

\subsection{Boundary conditions}

We have implemented two simple boundary conditions thus far, periodic
boundaries and reflective boundaries. In both cases, the computational domain
is restricted to be a rectangular domain of arbitrary aspect ratio. The
implementation of periodic boundaries is realized with the ghost cell
technique discussed earlier. Even if only a single processor is used,
particles close to the edge of the domain will find periodic image particles
`on the other side' of the principal domain, and import those as ghost
particles. While this means that the cells that overlap with the box
boundaries will be duplicated in the mesh construction, the overhead in 
mesh storage this induces is small. But the convenience of this approach lies in
the fact that it does not require a modification of the actual mesh
construction algorithms to make them aware of the periodic boundaries. Also,
this simple technique is readily combined with the approach we adopted to cope
with distributed-memory parallelization.

Reflective boundaries can be realized similarly, except that ghost particles
are now not simply primary particles/cells that are translated by one
box-length. Instead, the spatial location of ghost particles correspond to
{\em mirrored} copies of the primary mesh-generation points. When added to the
primary points, this means that the resulting Voronoi mesh for the principal
domain will always have faces aligned with the box boundaries. It is then
possible to impose different boundary conditions on these faces. For
reflective boundaries, we can simply copy the state of the fluid from the
mirrored point, but with the sign of the normal velocity component
reversed. This will automatically make the mass flux vanish on the surface of
the boundary, and leads to reflective boundary conditions. However, it is also
easily possible with this mirroring technique to realize outflow or inflow
boundary conditions.  Finally, it is possible to arrange for arbitrary
curve-linear boundary conditions by arranging two parallel strings of paired
particles in a suitable way. One of the particles of each pair would
constitute a cell inside the computational domain, the other would be a
fiducial cell outside, and the desired boundary condition can be imprinted at
the face they share. If desired, such a boundary may also be moved in complex
ways. We will discuss an illustrative example of this technique in
subsection~\ref{SecCoffee}.

\section{Mesh regularity} \label{SecMeshRegularity}

As seen in Figure~\ref{FigVoronoiExample}, Voronoi meshes may sometimes look
quite ``irregular'', in the sense that there is a significant spread in sizes
and aspect ratios of the cells, especially for sufficiently disordered point
distributions. While this is not a problem of principle for our approach, it
is clear that the computational efficiency will normally be optimized if
regions of similar gas properties are represented with cells of comparable
size.  Having a mixture of cells of greatly different volumes to represent a
gas of constant density will restrict the size of the timestep unnecessarily
(which is determined by the smallest cells), without giving any giving a
benefit in spatial resolution (which will be limited by the largest cells in
the region).

As we have seen, it is also desirable to have cells where the centre-of-mass
lies close to the mesh-generating point, because this minimizes errors in the
linear reconstruction and limits the rate at which mesh faces turn their
orientation during mesh motion.  Below, we will discuss our approaches for
steering the mesh motion during the dynamical evolution such that, if desired,
mesh regularity in the above sense can be achieved and maintained.

\begin{figure*}
\bc
\resizebox{8cm}{!}{\includegraphics{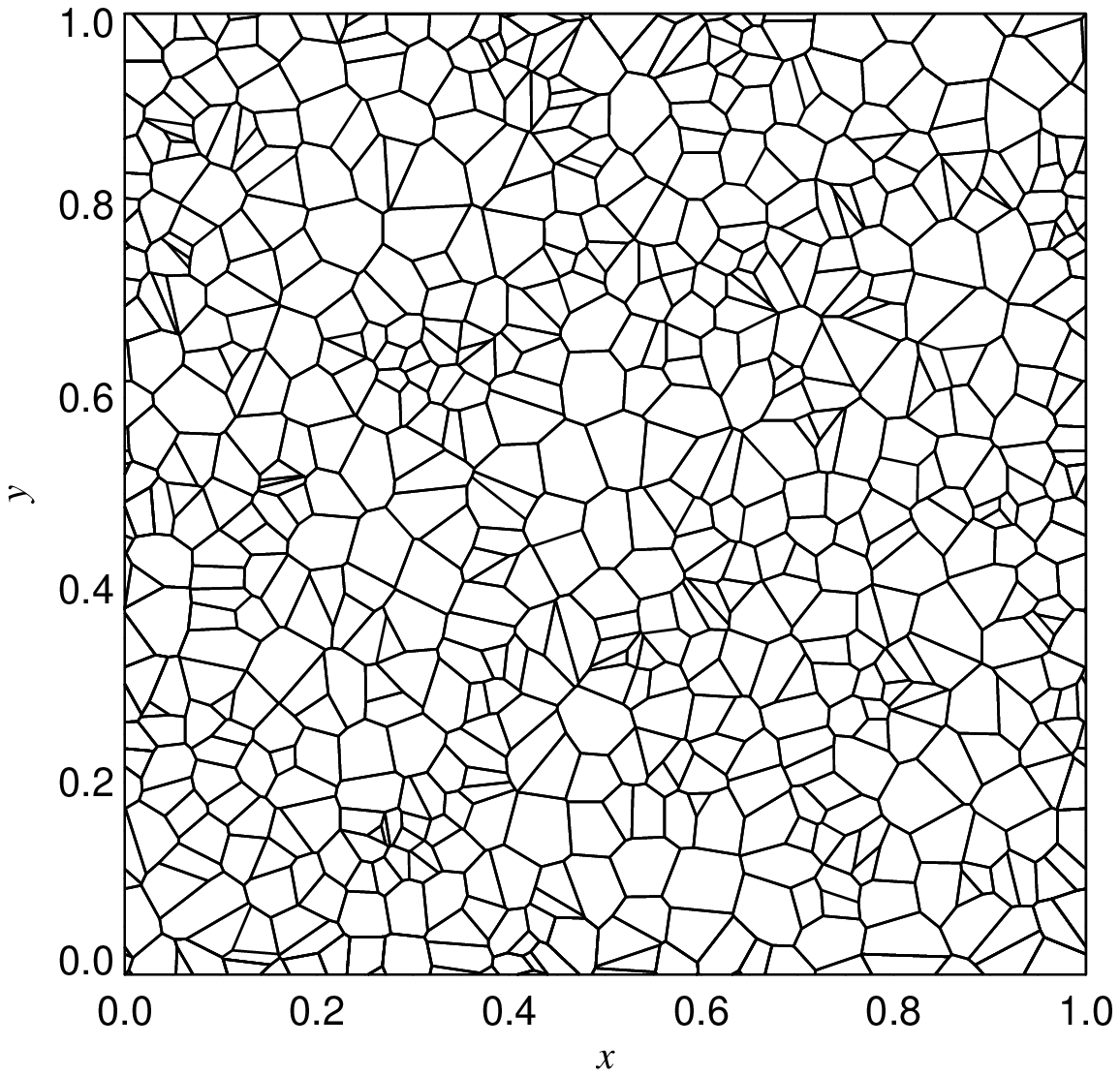}}\ %
\resizebox{8cm}{!}{\includegraphics{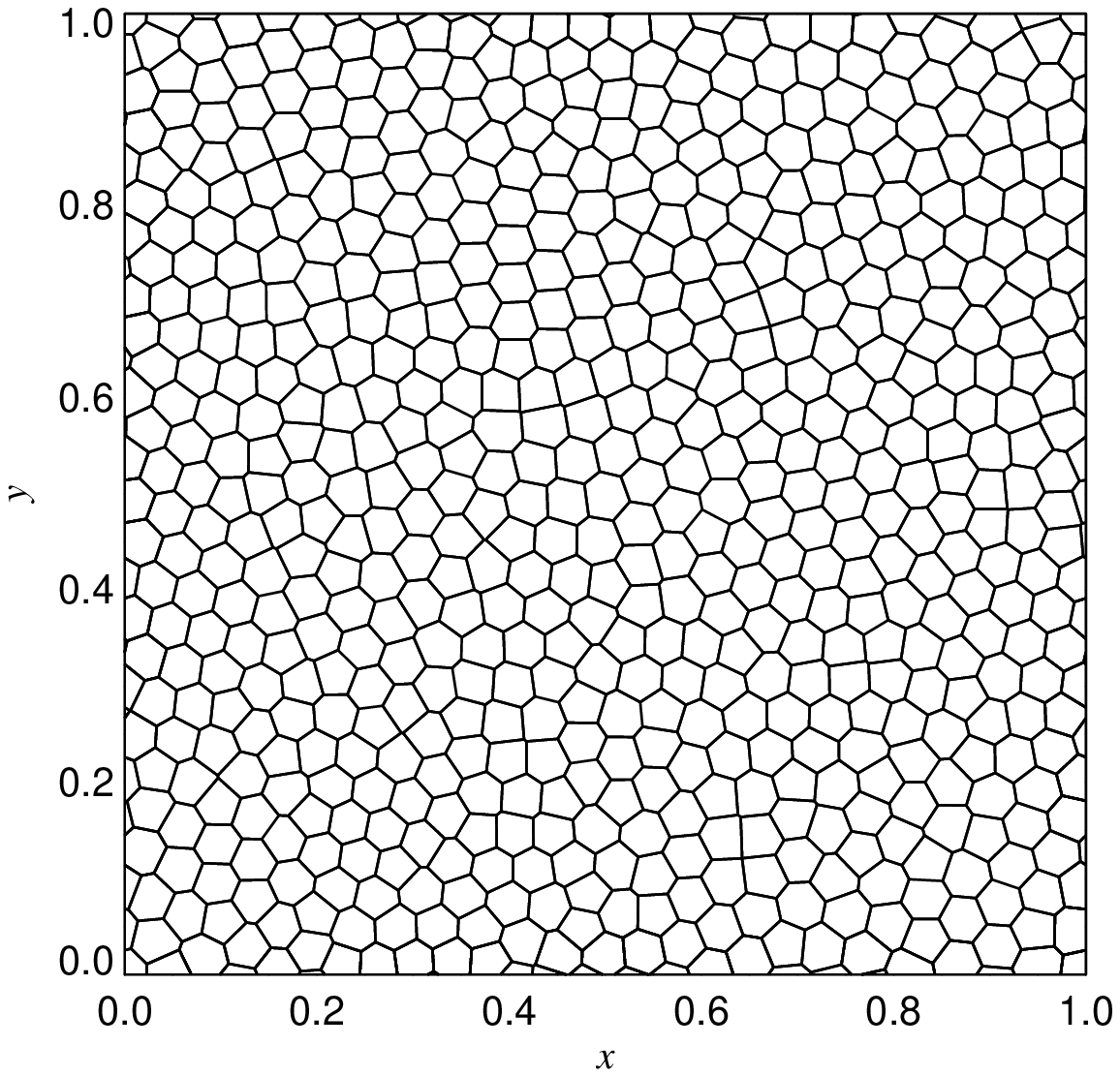}}\ %
\caption{Example for a mesh regularization with Lloyd's algorithm. The
  panel on the left shows the Voronoi mesh of a Poisson sample of 625
  points in the unit square, with periodic boundary conditions. The
  panel on the right hand side is the same mesh after being evolved
  with 50 iterations of Lloyd's algorithm, i.e.~in each step the
  mesh-generating points are moved to the centre-of-mass of their
  cell. The mesh slowly `crystallizes' into a quite regular structure
  with mostly hexahedral cells that are of very similar volume.
\label{FigMeshRegular}}
\ec
\end{figure*}

\begin{figure*}
\bc
\resizebox{8cm}{!}{\includegraphics{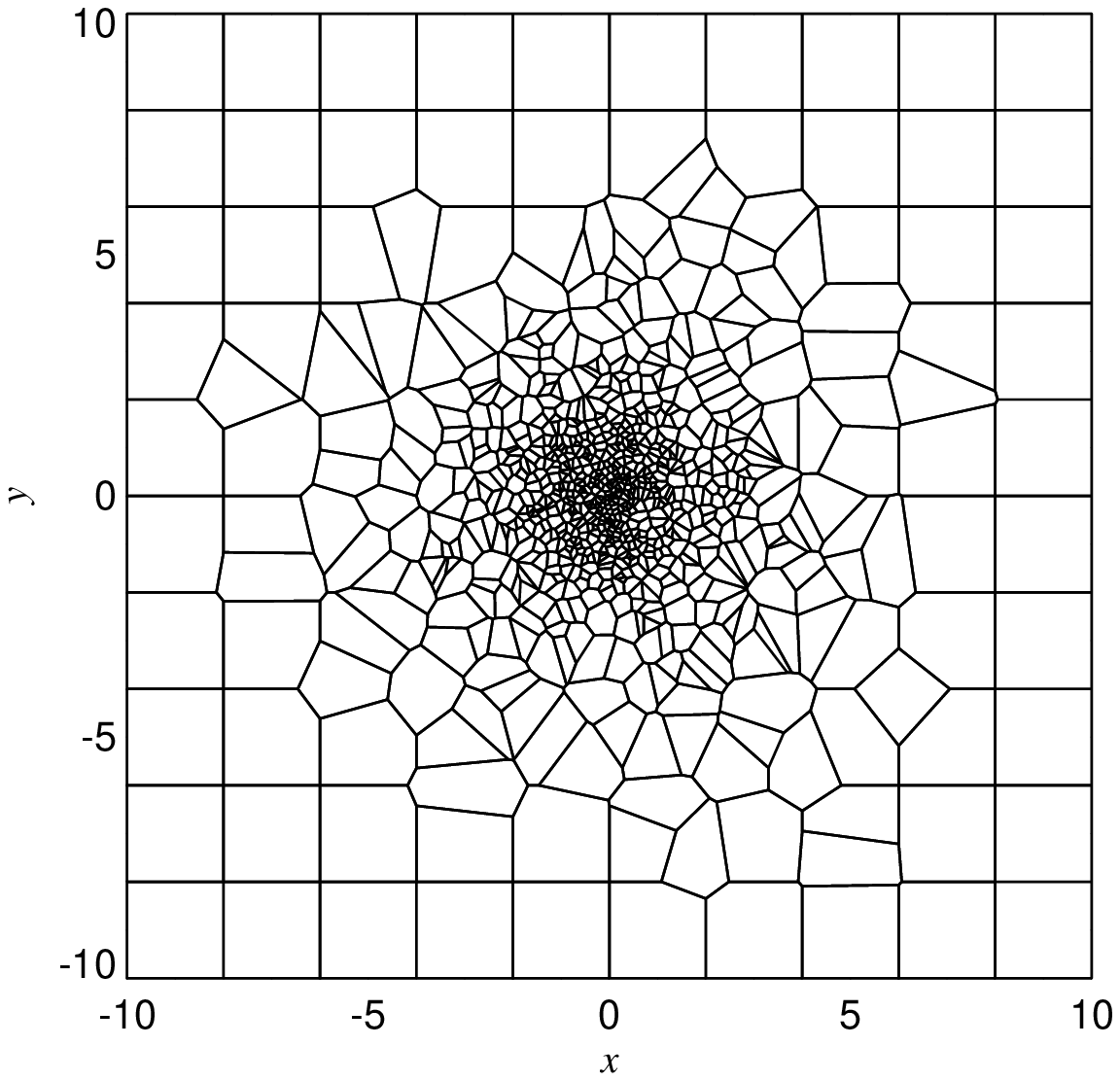}}\ %
\resizebox{8cm}{!}{\includegraphics{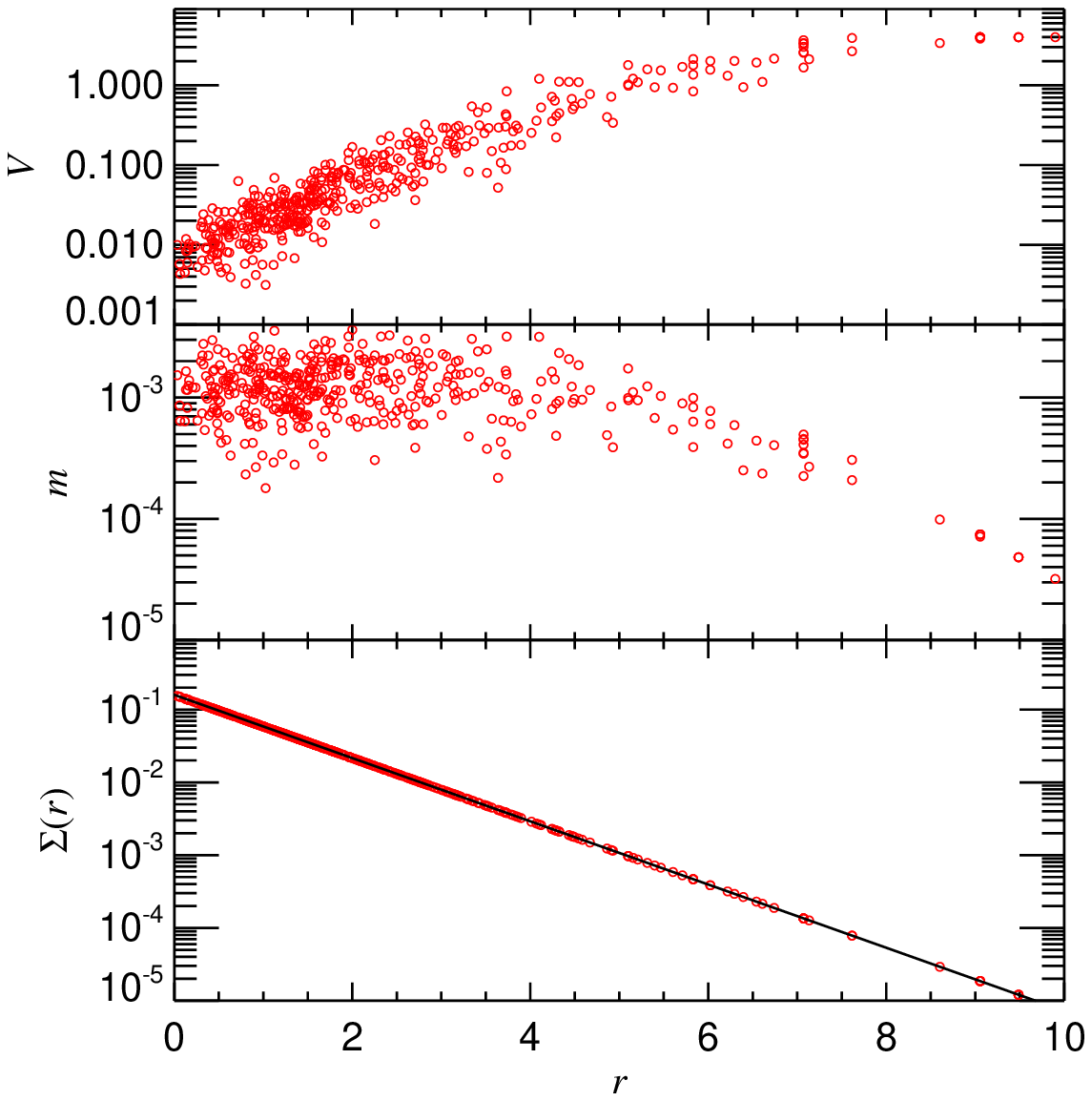}}
\resizebox{8cm}{!}{\includegraphics{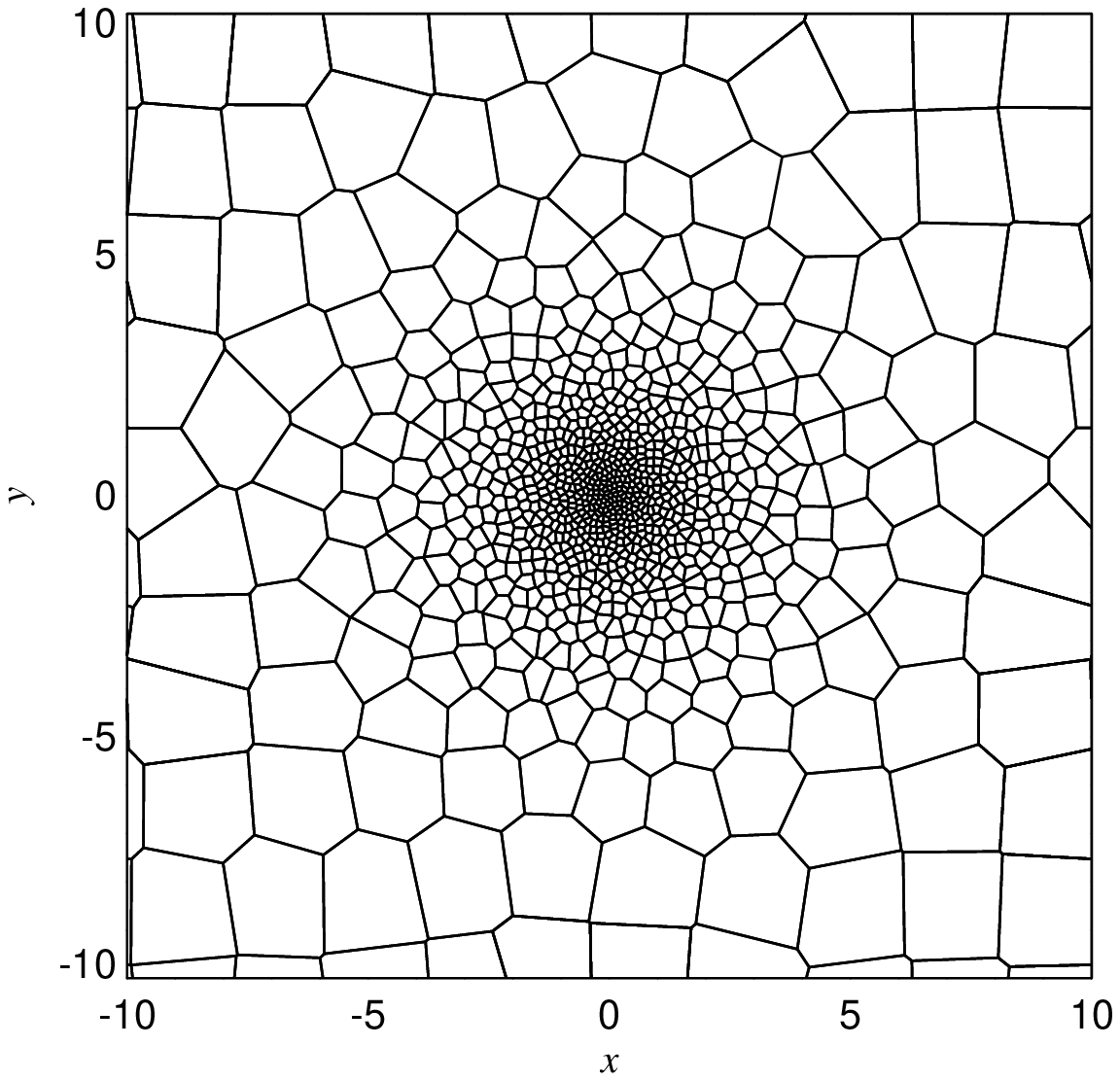}}\ %
\resizebox{8cm}{!}{\includegraphics{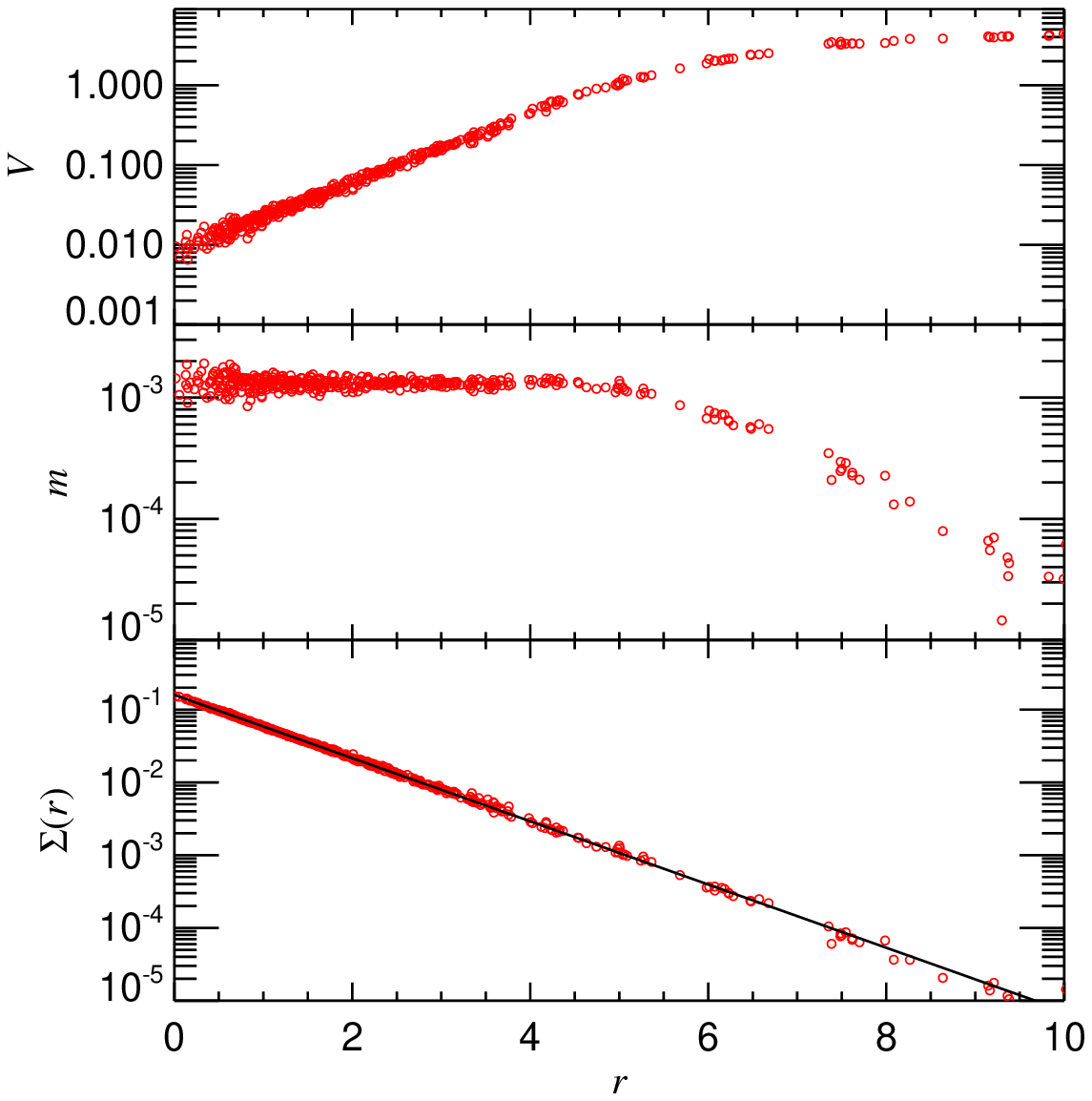}}
\caption{Mesh regularization of the initial conditions of an exponential
disk. We here randomly distributed 750 particles with an exponential surface
profile of scalelength $R_0=1.0$. To represent the `vacuum' outside the disk,
an additional  Cartesian grid of $10\times 10$ particles was placed
into the simulation domain, which is a periodic box of 20 length units on a side. The
resulting Voronoi mesh is shown in the top panel. While it has the right
density field, the mass per cell (and hence the cell volumes) shows
substantial scatter, as seen in the top right panel. However, after the mesh
regularization has been applied, a much better mesh results, as seen in the
bottom panels. While the surface mass density profile has remained the same,
there is now little local scatter in the mass and volume per cell. Inside the
disk, constant mass per cell is reached, while far outside, a constant volume
per cell is obtained, with a smooth transition between the two regimes.
\label{FigMeshRelaxDisk}}
\ec
\end{figure*}

\subsection{Making cells `rounder'} \label{SecRoundCells}

In so-called centroidal Voronoi tessellations, the mesh-generating
points coincide with the centre-of-mass of all cells. There is an
amazingly simple algorithm known as Lloyd's method \citep{Lloyd1982}
to obtain a centroidal Voronoi tessellation starting from an arbitrary
tessellation. One simply moves the mesh-generating points of the
current Voronoi tessellation to the centre-of-masses of their cells,
and then reconstructs the Voronoi tessellation. The process is
repeated iteratively, and with each iteration, the mesh relaxes more
towards a honey-web like configuration in which the Voronoi cells
appear quite `round' and have similar volume -- a centroidal Voronoi
tessellation. This is illustrated in Figure~\ref{FigMeshRegular},
which shows the Voronoi tessellation of a Poisson distribution of 625
points in 2D, and the result of 50 Lloyd iterations applied to it.

Inspired by this algorithm, we (optionally) employ a simple scheme to improve
the local shape of the Voronoi tessellation during the dynamical evolution.
We simply augment equation (\ref{eqnvelmesh}) with an additional velocity
component, which is designed to move a given mesh-generating point towards the
centre-of-mass of its cell.  There are different possibilities to parameterize
such a corrective velocity.  One approach that we found to work very well in
practice is to add a correction velocity whenever the mesh-generating point
is further away from the centre-of-mass of a cell than a given threshold,
irrespective of the actual velocity field of the gas. To this end, we
associate a radius $R_i=(3V_i/4\pi)^{1/3}$ with a cell based on its volume (or
area in 2D).  If the distance $d_i$ between the cell's centre-of-mass
$\vec{s}_i$ and its mesh-generating point $\vec{r}_i$ exceeds some fraction
$\eta$ of the cell radius $R_i$, we add a corrective term proportional to
the local sound speed $c_i$   of the cell
to the velocity of the mesh-generating
point. This effectively applies one Lloyd iteration (or a fraction of it) to
the cell by repositioning the mesh-generating point onto the current
centre-of-mass, ignoring other components of the mesh motion.  In order to
soften the transition between no correction and the full correction, we
parameterize the velocity as
\begin{equation}
\vec{w}_i' = \vec{w}_i + \chi \left\{
\begin{array}{cl}
0 &   \mbox{for}\;d_i/(\eta\,R_i) <  0.9  \\
c_i \frac{\vec{s}_i-\vec{r}_i}{d_i}\frac{d_i-0.9\,\eta R_i}{0.2\,\eta R_i}  
& \mbox{for}\; 0.9 \le d_i/ (\eta\,R_i) <  1.1\\
 c_i\frac{\vec{s}_i-\vec{r}_i}{d_i}  & \mbox{for}\;1.1 \le d_i /(\eta\,R_i)
\end{array}
\right.
\label{EqnShapeCorrVel}
\end{equation}
but the detailed width of this transition is unimportant. In very cold
flows the sound speed may be so low that the correction becomes
ineffective. As an alternative, we therefore also implemented an option
in the code that allows a replacement of $c_s(\vec{s}_i-\vec{r}_i)/d_i$
in equation (\ref{EqnShapeCorrVel}) with
$({\vec{s}_i-\vec{r}_i})/{\Delta t}$. This more aggressive approach to
ensure round cells generally works very well too, but has the
disadvantage to depend on the timestepping.  Our typical choice for the
threshold of the correction is $\eta=0.25$, and we usually set
$\chi=1.0$, i.e.~the correction is, if present, applied in full over the
course of one timestep.  Smaller values of $\eta$ can be used to enforce
round cell shapes more aggressively, if desired. Smaller values of
$\chi$ can be used to apply the corrective velocity more gently in time,
but we have not noticed problems with the choice of $\chi=1.0$ in the
problems we examined thus far.

Because only relatively regular meshes have their centres of mass always close
to their mesh-generating points, the extra velocity component has the tendency
to make the local mesh more regular. Indeed, the above scheme is quite
effective in maintaining low aspect rations for the mesh cells at all times
during the evolution. We therefore found it to be a good default choice for
general simulations with the moving mesh approach. Note that for a reasonably
`roundish' mesh, the correction velocity vanishes and the mesh will be
strictly advected with the fluid in smooth parts of the flow.

\subsection{Maintaining constant mass or volume for the cells} \label{SecInvZel}

In many applications in cosmology, it is desirable to have constant mass
resolution, and to increase/decrease the spatial resolution automatically when
matter clusters or expands.  If the mesh-generating points are moved with the
local fluid velocity, the gas mass in the cells will stay very nearly
constant, thus approximately fulfilling this desired Lagrangian adaptivity
during the course of a simulation. However, some scatter in the mass per cell
will nevertheless occur after a while, and in complicated flows with strong
compressions and shocks, these fluctuations may reach factors of several. This
calls for a method that is automatically able to restore and maintain a
constant mass per cell.

Similarly, we would sometimes like to impose constraints on the volumes of
cells as well, for example by requesting that they should not exceed a maximum
size, or not become smaller than a prescribed scale. A special case of this are
simulations where one would like to have roughly constant volume per cell,
even though large density contrasts develop and at the same time the mesh
should still move with the local flow velocity as far as possible.

In many practical applications, one may in fact request that both the mass and
the volume of cells respect certain regularity conditions. For example, in
situations where a self-gravitating clump of gas (say a galaxy) is embedded in
large regions of essentially empty space, it would be best to have cells of
nearly equal volume in the region that are largely (or completely) devoid of
gas. (In fact, equal mass per cell would be ill-defined in this case as it
basically meant that a single cell would have to represent all of this
volume.) On the other hand, in the regions where the density is large, it
would at the same time be desirable to have equal mass per cell.

We have implemented a scheme to regulate the mesh motion which 
effectively ensures that such prescribed constraints are respected by the
moving mesh. Our method is inspired by the Zeldovich approximation and
requires the solution of a Poisson-like equation. It is very powerful as it
can eliminate even large-scale deviations from the desired distribution of
cells in very few steps, as we discuss next.

Let $n(\vec{x})$ describe the current number density distribution of
mesh-generating points. Let us suppose that this distribution is not quite
ideal yet for the given density field of the gas, according to some suitable
criterion, but that it is not too far away from the ideal distribution
$n_0(\vec{x})$. In the following we will assume that linear order is
sufficient to describe the differences between the current and the ideal
distribution of the mesh-generating points. For each point, let $\vec{q}_i$ be
its ideal coordinate, and $\vec{x}_i$ its current coordinate. They are related
by
\begin{equation}
\vec{x}_i = \vec{q}_i + \epsilon\,\vec{d}_i,
\end{equation}
where $\vec{d}_i$ is the displacement of site $i$ from its ideal coordinate,
and $\epsilon$ is a fiducial dimensionless time variable, with $\epsilon=1$
corresponding to the current situation. Our goal is to estimate $\vec{d}_i$,
such that by applying a coordinate shift $-\vec{d}_i$ to all the points, we
can move the mesh close to the ideal configuration.

We shall now assume that the displacements can be obtained as gradient of a
scalar field $\Psi$,
\begin{equation}
\vec{d}=-\vec{\nabla}\Psi.
\end{equation}
Furthermore, since we only consider linear order, we can write the evolution
of the number density field $n_\epsilon(\vec{x})$ along the particle
trajectories as
\begin{equation}
n_\epsilon(\vec{x}+\epsilon \vec{d}) = \epsilon \,\, n(\vec{x}+\vec{d}) +(1-\epsilon)\,\, n_0(\vec{x}) .
\end{equation}
The `velocities' of each point in this transformation are given by $\vec{v}_i =
{\rm d}\vec{x}_i/{\rm d}\epsilon = \vec{d}_i$. Invoking the Lagrangian
continuity equation for the motion of the points,
\begin{equation}
\frac{{\rm d}n}{{\rm d}\epsilon} + n\vec{\nabla}\cdot\vec{v} = 0,
\end{equation}
and evaluating it at $\epsilon=0$, 
we obtain the Poisson-like equation
\begin{equation}
\nabla^2 \Psi = \frac{n(\vec{x}+\vec{d})}{n_0(\vec{x})} - 1 \simeq \frac{n(\vec{x})}{n_0(\vec{x})} - 1,
\end{equation}
where in the last step we approximated to linear order
$n(\vec{x}+\vec{d})\simeq n(\vec{x})$. What remains to be done is to specify
the desired density of mesh-generating points $n_0$ for the ideal
configuration of the Voronoi cells. Here we use the following ansatz that can
deal with quite general situations, including cases where there is empty
space. We would like that the quantity
\begin{equation}
K_i \equiv \frac{m_i}{\tilde{m}} + \frac{V_i}{\tilde{ V}}
\label{eqnconstantmesh} 
\end{equation}
is equal to a constant value $\tilde{K}$ for all cells,
i.e.~$K_i=\tilde{K}$. Here $\tilde{m}$ is a prescribed constant which
effectively sets the desired (maximum) mass per cell, and $\tilde{V}$ is a
chosen value that determines the desired maximum volume per cell, while $m_i$
and $V_i$ are the actual mass and volume of the cell $i$. For the ideal mesh,
the mass and volume of a cell are given by $m_i = \rho(\vec{q}_i)/n_0(\vec{q}_i)$
and $V_i = 1/n_0(\vec{q}_i)$, respectively. We can hence write
\begin{equation}
n_0(\vec{x}) = \frac{1}{\tilde{K}}\left( \frac{\rho(\vec{x})}{\tilde{m}} + \frac{1}{\tilde{V}}\right).
\end{equation}
Note that the density field itself is assumed to be stationary here; only the
sampling by the mesh points changes.  This leads finally to the following
Poisson-equation to obtain the mesh-displacement vectors
\begin{equation}
\nabla^2\Psi = \frac{\tilde{K}\, n(\vec{x})}{ \rho(\vec{x})/ \tilde{m} +
  1/\tilde{V}} -1.
\label{EqnPoissonMesh}
\end{equation}
This can be solved in the same way as we solve for the gravitational field,
either with particle-mesh (PM) methods in Fourier-space, or in real-space via
a tree, or by a combination of the two (TreePM method). Finally, we estimate
the displacement of a point from its ideal position by evaluation $\vec{d} =
-\vec{\nabla}\Psi$ at its current coordinate instead of the unknown ideal
coordinate, which is again accurate to leading linear order. 
 
We will typically assume periodic boundary conditions for the mesh
regularization. The value of $\tilde{K}$ should then be set such that the
source term on the right hand-side of equation (\ref{EqnPoissonMesh}) 
integrates to zero for the current particle distribution; this is a
prerequisite that the Poisson equation actually has a well-defined solution
for an infinite periodic space.  This means that we should set
\begin{equation}
\tilde{K} = \frac{V_{\rm
    tot}}{\sum_i\left(\rho_i/\tilde{m}+1/\tilde{V}\right)^{-1}},
\end{equation}
where $V_{\rm tot}$ is the total volume of the simulation domain. The $-1$ on
the right hand side of equation (\ref{EqnPoissonMesh}) then eliminates the constant
term in Fourier space. This is similar to the treatment of self-gravity in
periodic spaces, where the mean density needs to be subtracted from the
density field in order to obtain a finite solution of the Poisson equation.

Solving for the displacement field is equivalent to calculating the
gravitational accelerations for a particle distribution with `masses' given by
$\tilde{K}/(\rho_i / \tilde{m} + 1/\tilde{V})$. We use the TreePM formalism
for this, which has the advantage of being free of any restrictions on dynamic
range while at the same time being quite fast. Once we have obtained the
displacement vectors $\vec{d}_i$ for all particles, we add a corrective
velocity to the mesh motion as follows:
\begin{equation}
\vec{w}_i' = \vec{w}_i - \kappa \frac{\vec{d}_i}{\Delta t}.
\end{equation}
We usually set $\kappa=0.5$, such that the estimated displacement from the
ideal position is cut in half in each timestep.  If needed, $\kappa$ is
reduced for the current step such that the maximum displacement of a point
does not exceed half its cell size, which is needed for stability reasons.  As
this scheme is quite effective in maintaining an ideal mesh at all times, the
size of the prefactor $\kappa$ does not really matter much in practice.

We note that the above approach essentially corresponds to an `inverse
Zeldovich approximation', as it is used for example by the {\small
  GADGET-2} code \citep{Springel2005} to produce a gravitational
`glass' \citep{White1996} of constant density, except that we
generalized the approach to allow construction of generalized glasses
for variable density fields that are constrained by the freely
adjustable constants $\tilde{m}$ and $\tilde{V}$.  We note that this
method may also be useful to construct `quiet starts' for SPH
calculations that need to initialize astrophysical objects with a
prescribed density structure, such as stars in simulations of stellar
collisions. The methods most commonly used for this purpose at the
moment rely on settling the particle distribution into equilibrium
with the help of artificial friction or pressure forces
\citep[e.g.][]{Goodman1991}.  We also remark that both schemes for
mesh-regularization discussed above obey the property of
Galilean-invariance of the moving-mesh code. This is because the
primary mesh motion is still given by equation (\ref{eqnvelmesh}); any
Galilean boost would simply be absorbed into it, while the
mesh-correction velocities would remain unaffected.

\subsection{Constructing suitable initial conditions}

The above discussion about mesh regularity also prompts the question of how
suitable initial conditions for a prescribed initial density field can be
constructed.  For many hydrodynamical test problems, constant density fields
are needed that can simply be realized with Cartesian grids. This is also a
possible choice for cosmological initial conditions, where Cartesian grids may
be used for the unperturbed initial conditions. However, sometimes one would
like to start a simulation with a non-trivial density distribution for the
gas, for example in the form of a gaseous disk with a prescribed surface
density profile, or in the form of a spherically symmetric gas cloud that
approximates the gas distribution in a cluster of galaxies.

One popular approach to realize such general density distributions in SPH lies
in randomly sampling the density field, for example with the rejection method
\citep{Press1992}. This effectively produces a Poisson sampling of the
underlying density field. While such a particle distribution can be used as
initial conditions for the mesh-generation points, the quite irregular mesh
this corresponds to represents a significant disadvantage. For example, in the
top left panel of Fig.~\ref{FigMeshRelaxDisk}, we show the Voronoi mesh
resulting from a random realization of a gaseous disk with an exponential gas
surface density profile. In addition to 750 particles used for the primary
disk distribution, a coarse Cartesian grid with $10^2$ points has been used
here to fill the volume with cells that do not exceed a certain maximum
volume. Due to the random sampling, the resulting mesh is characterized by
cells with significant scatter in their volume at any given radius, as seen in
the top panels of Fig.~\ref{FigMeshRelaxDisk}, and since the desired density
profile has been prescribed, this is reflected in an equally large scatter in
the mass per cell.

It is of course nevertheless possible to start a simulation with such a
Poisson distribution and then to let the simulation code improve the mesh with
time. However, if a more quiet start is desired, one can also first relax the
initial mesh with the methods described above, except that the gas
distribution is kept {\em fixed in space} by solving the advection equation
for the moving mesh, instead of the Euler equation. For the advection
equation, we use a simple second-order accurate upwind scheme to determine the
fluxes at all cell faces.

An example for the result of such a relaxation is shown in the two bottom
panels of Fig.~\ref{FigMeshRelaxDisk}. The panel on the left shows the Voronoi
mesh, and the panel on the bottom right the radial profiles of mass and volume
of each cell, as well as the surface density profile. Evidently, the relaxed
mesh is much more regular and features relatively `roundish' cells with low
aspect ratios. Also, at any given radius, there is little scatter in the mean
mass and mean volume of the cells. In fact, their radial variation follows the
imposed constraint of constant ${m_i}/{\tilde{m}} + {V_i}/{\tilde{ V}}$ very
well. This produces a situation where in the inner parts of the disk the mass
per cell is constant, while in the low density outer regions, the volume of
the cells is kept fixed, with a smooth transition in between. This behaviour
is particular useful for structures embedded in nearly or completely empty
space, for example for simulations of isolated or colliding galaxies.

\section{Self-gravity} \label{SecGravity}

Outside of astrophysics, self-gravity of gases plays hardly any role in
computational fluid dynamics. However, gravitational forces are the primary
driver of cosmological structure formation. This fundamental importance of
gravity adds a significant complication to hydrodynamic codes. In fact, in
cosmology there is arguably little value in calculating the hydrodynamics
highly accurately when gravity is not treated with comparable accuracy.

There are some indications that the specific challenges posed by an accurate
treatment of self-gravity have been underestimated when traditional Eulerian
approaches have been employed in cosmology.  This is suggested by recent
comparisons of P$^3$M/Tree/TreePM methods and adaptive mesh refinement (AMR)
codes where both are applied in pure gravity mode to the clustering of
collisionless dark matter.  Both in \citet{Shea2005} and \citet{Heitmann2007}
it was found that state-of-the-art AMR codes like {\small ENZO} and {\small
  FLASH} have significant problems in accurately recovering the low-mass end
of the mass function of dark matter halos. They are only able to match the
results of high-accuracy N-body codes once much finer base meshes and stricter
refinement criteria are used than are normally employed with these codes.
They are then no longer competitive with alternative approaches in
cosmological structure formation in terms of calculational efficiency and
memory consumption \citep{Shea2005}.  While these results have been found in
comparison studies that only tested the AMR gravity solver for a collisionless
fluid, it is clearly worrying that similarly poor behaviour is likely also to
affect calculations of self-gravity in hydrodynamical applications.

The problem seems to be that the adaptive refinement strategy, as presently
applied in the gravity solvers of some of the cosmological AMR codes, does not
work particularly well for the gravitational instability of dark matter, where
structures may grow everywhere from very small seed perturbations. Since the
placement or removal of refinements corresponds to discrete and {\em
  discontinuous} changes in local resolution, the growth of small
perturbations can be delayed if a refinement is placed `too late' onto an
emerging halo. The effectively Lagrangian behaviour of tree codes fares better
in this respect. Here, the spatially homogeneous high force resolution allows
tree codes to be formulated such that they observe the Hamiltonian structure
of the collisionless dynamics of dark matter. This structure is broken each
time the AMR mesh hierarchy is modified, because this changes the effective
gravitational softening associated with the mesh, and modifies the potential
energy stored in the density field. As we will see, in the moving mesh
approach such discontinuous changes in the Hamiltonian structure of
gravitational dynamics can be avoided.

An attractive feature of our new Lagrangian hydrodynamical scheme lies in the
possibility of combining it easily with a particle-based approach to calculate
the gravitational field, in the form of the familiar high-accuracy N-body
solvers for collisionless dynamics, for example Tree or TreePM schemes.  The
simplest approach for this is to treat the mass of each cell as being
concentrated in the centre of the cell, and then to calculate the
gravitational force on a cell as the suitably gravitationally softened N-body
force of the resulting point set.  The hierarchical multipole expansion used
in tree codes, carried out to monopole or quadrupole order, provides an
efficient way to compute these forces.  And thanks to the tree-based approach,
the gravitational resolution then automatically and {\em continuously} adjusts
in a collapsing structure, and the spatial resolution of self-gravity in the
gas is always matched accurately to that of the hydrodynamics \citep[see
also][]{Bate1997}.  We shall employ this approach in this work. The specific
N-body algorithms we adopt for calculating the gravitational forces are those
of an updated version of the code {\small GADGET-2} \citep{Springel2005}.

\subsection{The Euler equations with self-gravity}

If a gravitational field is present, the Euler equations
(\ref{EqnEuler}) are modified by source terms for momentum and energy,
which take the form
\begin{equation}
\frac{\partial\vec{U}}{\partial t} +
\vec{\nabla}\cdot \vec{F} = \left(
\begin{array}{c}
0 \\
-\rho\,\vec{\nabla} \Phi  \\
- \rho \vec{v}\vec{\nabla} \Phi\\
\end{array}\right).
\label{eqnsrc1}
\end{equation}  
The gravitational potential $\Phi$ may be externally specified, or it
describes the self-gravity of the gas as a solution of Poisson's equation,
\begin{equation}
\vec{\nabla}^2\Phi = 4\pi G\, \rho .
\label{eqnsrc2}
\end{equation}  
In the former case, the total energy $E_{\rm tot} = \int (\rho e + \rho \phi)
{\rm d}V$ stays constant if the potential is static. In the more relevant
case of self-gravity, the total energy of the system is given by
\begin{equation}
E_{\rm tot} = \int \left(\rho e + \frac{1}{2} \rho \Phi\right) {\rm d}V  ,
\end{equation} and
is conserved in the dynamics, i.e.
\begin{equation}
\frac{{\rm d}E_{\rm tot}}{{\rm d}t}  = 0.
\end{equation} 

Without gravity, the finite volume formulation for hydrodynamics introduced
earlier ensures conservation of the sum of thermal and kinetic energy to
machine precision. Since the thermal energy in this approach is actually
defined as the difference between the total energy and kinetic energy of a
cell, it is in principle highly desirable to also obtain a discretized
formulation of the dynamics in the self-gravitating case where the conservation
of energy is manifest. Furthermore, it would be convenient if the
gravitational source term could be incorporated into the time integration such
that there is no need to explicitly include gravity in the Riemann solver
\citep[this can be done approximately, however, see for example the PPM scheme
of][]{Colella1984}.

In the following, we first review a standard approach to include self-gravity
in finite volume codes, which however is not explicitly energy-conserving. In
fact, we will show that the resulting errors can be quite substantial for
certain types of problems.  We then briefly discuss an attempt to improve on
this by restoring manifest conservation of the total energy, based on
including the gravitational self-energy in the total energy variable that is
evolved for each cell. Unfortunately, it turns out that this approach is
numerically problematic since it can lead to unphysical changes of the local
thermal energy. We therefore ultimately adopt a different solution that
corrects for the large errors that can appear in the `standard'
approach. While not manifestly conservative, we find that, in practice, the
total energy is conserved quite accurately in this approach.  Since monitoring
the accuracy of total energy conservation can then also serve as a useful
check of the quality of the integration, we consider this as a good
compromise. Finally, we discuss our treatment of locally adaptive,
time-dependent gravitational softening, and how this is accounted for in the
dynamics.

\subsection{A standard approach to include self-gravity} \label{SecEgyStandardApproach}

Arguably the simplest method to include self-gravity lies in an
operator-splitting approach, where one alternatingly evolves the system under
the homogeneous Euler equations and the gravitational source terms.  However,
such fractional step methods are often inadequate for handling the
gravitational source terms, especially in situations with approximate
hydrostatic equilibrium \citep{Mueller1995,LeVeque1998,Zingale2002}. We will
therefore not consider this method here.

Instead, we consider the method suggested by \citet{Mueller1995}, which is
employed in similar form in many current finite volume cosmological codes
\citep[e.g.][]{Truelove1998}.  Since the gravitational energy is nonlocal, an
explicit conservation of total energy in the discretizations of equations
(\ref{eqnsrc1}) and (\ref{eqnsrc2}) cannot easily be obtained with an
extension of the standard flux-based formalism of finite volume methods. We
may therefore give up the property of manifest energy conservation and instead
couple the gravitational field to the Euler equations in a way that resembles
the fractional-step approach, except that gravity is also properly included in
the half-step prediction of the hydrodynamical step.

We begin by noting that, when the Euler equations are expressed in primitive
variable formulation,
\begin{equation}
\frac{\partial \vec{W}}{\partial t}
+ 
\left(
\begin{array}{ccc}
\vec{v} & \rho & 0 \\
 0     & \vec{v} & 1/\rho \\
0 & \gamma P & \vec{v}\\
\end{array}
\right)
\frac{\partial \vec{W}}{\partial \vec{r}} = 
\left(
\begin{array}{c}
0\\
-\vec{\nabla} \Phi\\
0
\end{array}
\right),
\end{equation}
where $\vec{W} = (\rho, \vec{v}, P)$,
the gravitational source term couples only to the momentum equation.  Hence,
to first order in time, the gravitational field does not change the pressure or
density of a fluid, only the velocity is altered. We can therefore account for the
gravitational field in the hydrodynamic flux calculation if we augment
the half-step prediction of the velocities with the gravitational acceleration
according to
\begin{equation}
\tilde\vec{v}_i^{(n+1/2)} = \vec{v}_i^{(n+1/2)} - \frac{\Delta t}{2}
\vec{\nabla}_i\Phi^{(n)} ,
\end{equation}
where the potential $\Phi^{(n)}$ is calculated at the beginning of the
step. Applying the ordinary reconstruction and Riemann solver techniques
discussed earlier, we
can then obtain time-centred flux estimates that solve the homogeneous part of
the Euler equations. To add the gravitational source term into the final
time-advance, we then proceed as follows.
We first use the estimated mass flux to update the mass contained in each cell,
\begin{equation}
m_i^{(n+1)} = m_i^{(n)} - \Delta t \sum_j A_{ij} F_{m}^{ij},
\label{eqnMassgravUpdate}
\end{equation}
which exploits the fact that the gravitational field does not appear in the
mass equation of the conservative form of the Euler equations. With the new
masses in hand, we can calculate the gravitational forces at the end of the
timestep, $\vec{\nabla}_i\Phi^{(n+1)}$. This allows an update of the momentum of
each cell, according to
\begin{eqnarray}
\vec{p}_i^{(n+1)} & = &  \vec{p}_i^{(n)} - \Delta t \sum_j A_{ij}
\vec{F}_{\vec{p}}^{ij}\nonumber \\
& & - \frac{\Delta t}{2} \left[ m_i^{(n)}\vec{\nabla}_i\Phi^{(n)} + m_i^{(n+1)}\vec{\nabla}_i\Phi^{(n+1)}\right],
\label{eqnmomold}
\end{eqnarray}
where $\vec{F}_{\vec{p}}$ is the hydrodynamical momentum flux. Note that this
step also determines the new velocities at the end of the step.  We can use
them to finally obtain a second-order accurate update of the energies of each
cell,
\begin{eqnarray}
\vec{E}_i^{(n+1)} & = & \vec{E}_i^{(n)} - \Delta t \sum_j A_{ij} {F}_{E}^{ij}
\label{eqnEgygravUpdate} \\
& & \hspace*{-1cm} - \frac{\Delta t}{2} \left[
  m_i^{(n)}\vec{v}_i^{(n)}\vec{\nabla}_i\Phi^{(n)} +
  m_i^{(n+1)}\vec{v}_i^{(n+1)}\vec{\nabla}_i\Phi^{(n+1)}\right] .\nonumber
\end{eqnarray}
Here, the term in square brackets is the gravitational work term, while the flux
term involving $F_E$ stems from the homogeneous part of the Euler equations.

The above scheme does not explicitly conserve total energy, but it still
conserves total momentum and mass. Violations of energy conservation can arise
because the gravitational work term, which is estimated effectively with
cell-centred fluxes, may not precisely balance the {\em actual} energy
extracted from the gravitational field, which is determined by the mass fluxes
obtained with the Riemann solver around a cell's boundary. We have found that
this subtle difference can sometimes lead to substantial inaccuracies in total
energy conservation, especially in collapse problems that involve strong
shocks and a conversion of large amounts of gravitational energy into heat
energy. For example, ``Evrard's collapse problem'', to be discussed in
Section~\ref{SecEvrard}, shows large errors of this kind, especially when the
spatial resolution is relatively poor. Note that in this case the violation of
the total energy conservation is first order in time, i.e.~it does not go away
with very fine time-stepping and instead stays constant at a finite (large)
size even in the limit of highly accurate time integration. It is therefore
desirable to obtain a more accurate discretization of the energy equation when
a gravitational field is present.

\subsection{An explicitly conservative formulation to include self-gravity}

One idea for a more accurate discretization of the conservative Euler
equations in the presence of gravity is based on rewriting the standard form
of the energy equation,
\begin{equation}
\frac{\partial }{\partial t} \left(\rho e   \right)
+ \vec{\nabla} \left[ \left( \rho e  + P \right) \vec{v} \right]=
-\rho \vec{v} \vec{\nabla}\Phi ,
\label{eqnEngGrav}
\end{equation}
with the help of the continuity equation as
\begin{eqnarray}
\frac{\partial }{\partial t} \left(\rho e + \frac{1}{2}\rho \Phi \right)
+ \vec{\nabla} \left[ \left( \rho e + \frac{1}{2}\rho \Phi + P \right) \vec{v}
\right]= \nonumber\\
\frac{1}{2}\rho \frac{\partial\Phi}{\partial t} -
\frac{1}{2}\rho\vec{v}\vec{\nabla}\Phi \, .
\label{eqnEngGrav2}
\end{eqnarray}
This suggests redefining the total energy of an individual cell as
\begin{equation} 
E_i =
\int_{V_i} \left(\rho e + \frac{1}{2}\rho\Phi\right)\,{\rm d}V,
\end{equation}
such that the total energy of the system simply becomes the sum of the $E_i$
of all cells.  If we suitably modify the energy flux function in the
hydrodynamical finite volume scheme, the left hand side of equation
(\ref{eqnEngGrav2}), which has the form of a conservation law, can be easily
solved such that the {\em total} energy stays constant. If we can also find an
explicitly conservative discretization of the modified source term on the
right-hand side of equation (\ref{eqnEngGrav2}), we would obtain a scheme that
manifestly conserves the total energy.

This can, in fact, be achieved. We can write the right-hand side of equation
(\ref{eqnEngGrav2}) as
\begin{eqnarray}
\frac{1}{2}\rho \frac{\partial\Phi}{\partial t} -
\frac{1}{2}\rho\vec{v}\vec{\nabla}\Phi
& = &  \\
& & \nonumber \hspace*{-2cm} G \int \rho(\vec{x})\rho(\vec{x}')\frac{ \vec{v}(\vec{x})
   +\vec{v}(\vec{x}')}{2}
\vec{\nabla}_{\vec{x}}\frac{1}{|\vec{x}-\vec{x}'|}{\rm d}^3\vec{x}' .
\end{eqnarray}
If we decompose the $\vec{x}'$-integration into a sum over integrals over all
cells, and integrate the full energy equation over $\vec{x}$ for a cell $i$,
we obtain the discretized form
\begin{equation}
\frac{{\rm d}E_i}{{\rm d}t} 
+ \sum_k \vec{A}_{ik} F_{ik}^{(E)}
= \sum_j \frac{ \vec{v}_i +\vec{v}_j}{2} \vec{f}_{ij} 
\label{eqnEngGrav3}
\end{equation}
for the energy equation, where $\vec{f}_{ij}$ is the gravitational force
between cells $i$ and $j$. We see that the term on the right hand side
effectively symmetrizes the gravitational work term the two cells exert
onto each other. The sum over $j$  extends over all cells, but both relevant terms, the
total force $\sum_j \vec{f}_{ij}$ and the total work term 
 $\sum_j \vec{v}_j \vec{f}_{ij}$
can be accurately and efficiently calculated with a tree algorithm.
The sum over $k$ in equation (\ref{eqnEngGrav3}) accumulates the energy fluxes 
\begin{equation}
F_{ik}^{(E)}
=
\rho_{ik}\left(e_{ik}+\frac{\Phi_{ik}}{2}\right) (\vec{v}_{ik}-\vec{w}_{ik})  
+ P_{ik}\vec{v}_{ik}
\end{equation}
from the neighbouring cells of cell $i$, where the $(\rho_{ik}$, $e_{ik}$,
$\vec{v}_{ik}-\vec{w}_{ik}$, $P_{ik})$ are determined by the Riemann problem
between cells $i$ and $k$, and the potential $\Phi_{ik}$ on the face between
two cells can be defined as the arithmetic mean $\Phi_{ik} =
(\Phi_i+\Phi_k)/2$ of the potentials at the corresponding mesh-generating
points of the cells.  It is not difficult to define a time integration scheme
for equation (\ref{eqnEngGrav3}) that preserves its conservative character in
the discretized form, such that at least formally a finite volume scheme
results that accurately conserves up to machine precision the total energy,
momentum and mass in the presence of a gravitational field.

However, the above approach shows severe short-comings in practice. In
particular, the fact that the temperature of the gas is effectively defined by
subtracting the kinetic energy {\em and} the potential energy from the total
energy associated with a fluid element causes trouble. This can
give rise to spurious local changes in the temperature of the gas due to the
presence of a gravitational field, even though the Euler equations in
primitive variable form show that there should be no first order change in the
temperature due to a gravitational field. We have found that in some cases
this may even drive the temperature to unphysical negative values. Secondly,
in this approach the temperature field couples to discreteness noise present
in the gravitational field, which considerably reduces the accuracy of the
hydrodynamical calculations. In combination, these defects are severe enough that
the `total energy approach' described in this subsection appears not to be a
viable practical solution for implementing self-gravity in the moving-mesh
approach. We therefore refrain from using it in our practical applications.

One exception is the case where gravity is simply described by an external
static gravitational potential $\Phi$. We can then express the conservation of
total energy as
\begin{eqnarray}
\frac{\partial }{\partial t} \left(\rho e + \rho \Phi \right)
+ \vec{\nabla} \left[ \left( \rho e + \rho \Phi + P \right) \vec{v}
\right]= 0,
\end{eqnarray}
which suggests that we include the gravitational energy in the definition of the
total energy of a cell. The thermal energy can then be defined by subtracting
both the kinetic energy and the potential energy from the total energy. We
also need to augment the energy flux term for a cell interface with an
additional gravitational energy flux, with the result that the total energy is
exactly conserved. The inclusion of gravity into the dynamics then proceeds
like in equations (\ref{eqnMassgravUpdate}), (\ref{eqnmomold}) and
(\ref{eqnEgygravUpdate}), except that in equation (\ref{eqnEgygravUpdate}) the
explicit gravitational work term (in square brackets) is omitted as it is
already accounted for by the energy flux.

\subsection{An improved coupling of self-gravity to the Euler equations} \label{SecEgySurface}

In the following, we discuss a method that tries to improve the discretization
of the energy equation used in the `standard approach' discussed above in
Section~\ref{SecEgyStandardApproach}.  Recall that the gravitational work
exerted on a cell over a timestep $\Delta t$ is given by
\begin{equation}
\Delta E^{\rm grav}_{i} = -\int {\rm d} t\int_{V_i} {\rm d} V \rho\,\vec{v}\vec{\nabla}\Phi,
\label{EqnEgw}
\end{equation}
integrated over the moving volume of a cell. 
We may also rewrite this integral as
\begin{equation}
\Delta E^{\rm grav}_{i} = -\int {\rm d} m\int {\rm d}\vec{s} \vec{\nabla}\Phi .
\end{equation}
where ${\rm d}\vec{s}$ is the displacement of each individual mass
element. This highlights that the key to accurate energy conservation in case
of self-gravity is to correctly account for the {\em actual} mass motions that
happen in the system. The problem with equation (\ref{eqnEgygravUpdate}) is
that this is not guaranteed explicitly since it estimates the mass motion with
a cell-centred flux, but the mass fluxes actually used are calculated at the
surfaces of the cells, and may sometimes be quite different.

We suggest another discretization of equation (\ref{EqnEgw}) that improves on
this. First, we introduce the velocity vector $\vec{w}_i$ of the cell's motion,
which splits the integral into two parts, one describing the motion of
the cell itself (with all of its mass), and the other accounting for the
motion of mass elements that are actually exchanged between two adjacent
cells, viz.
\begin{equation}
\Delta E^{\rm grav}_{i} = -  \Delta t\, m_i\vec{w}_i \vec{\nabla}
\Phi_i 
- \Delta t \int \rho_i (\vec{v} - \vec{w}_i) \vec{\nabla}
\Phi_i \,{\rm d}V.
\end{equation}
Instead of approximating the volume integral of the second part with a
cell-centred flux, we transform it into a surface integral. Neglecting
spatial variations in the density, velocity, and force fields for the moment,
we can write $\vec{v} - \vec{w}_i = \vec{\nabla}[(\vec{r} - \vec{r}_i)(\vec{v}
- \vec{w}_i)]$ and apply the Green-Gauss theorem. This yields
\begin{eqnarray}
\Delta E^{\rm grav}_{i} & =& -  \Delta t\, m_i\vec{w}_i \vec{\nabla} \Phi_i\\ 
& & - \frac{\Delta t}{2} \sum_j \rho_{ij} [(\vec{v}_{ij} - \vec{w}_i)\vec{r}_{ij}] [\vec{\nabla}
\Phi_i \vec{r}_{ij}/r_{ij}] A_{ij} , \nonumber
\end{eqnarray}
where the sum is now over all faces of area $A_{ij}$ of a cell, and, as usual,
$\vec{r}_{ij}=\vec{r}_i - \vec{r}_j$ is the displacement vector between the
neighbouring mesh-generating points.  We have also replaced the values of
density and velocity on the surface with those determined by the Riemann
solver.  In fact, the term $\Delta t \rho_{ij} (\vec{v}_{ij} -
\vec{w}_i)\vec{r}_{ij} A_{ij}/r_{ij}$ can be recognized as the integrated
mass flux $\Delta m_{ij}=\Delta t A_{ij}F_m^{ij}$ exchanged between two cells $i$ and $j$, yielding
\begin{equation}
\Delta E^{\rm grav}_{i} = -\Delta t\, m_i  \vec{w}_i \vec{\nabla} \Phi_i 
- \frac{1}{2}\sum_j  \Delta m_{ij} \vec{r}_{ij}\vec{\nabla} \Phi_i .
\label{EqnEgApprox}
\end{equation}
An even more instructive form of this equation is obtained with the
replacement
\begin{equation}
\vec{r}_{ij}\vec{\nabla} \Phi_i\simeq  \Phi_i-\Phi_j  
\end{equation}
which gives
\begin{equation}
\Delta E^{\rm grav}_{i} = -\Delta t \, m_i  \vec{w}_i \vec{\nabla} \Phi_i - \frac{1}{2}\sum_j  \Delta m_{ij} (\Phi_i - \Phi_j).
\label{EqnEgExact}
\end{equation}
If we define the total gravitational energy of the discretized system as
\begin{equation}
E_{\rm pot} = \frac{1}{2} \sum_{ij} G \, m_i \,  m_j \, \phi(r_{ij}) =
\frac{1}{2} \sum_i
m_i \Phi_i
\end{equation}
and the potential as
\begin{equation}
\Phi_i = \sum_{j} G\,  m_j \, \phi(r_{ij}),
\end{equation}
where $\phi_{ij}$ describes the gravitational interaction kernel between two
cells, then it is easy to see that equation (\ref{EqnEgExact}) describes the
gravitational energy change {\em exactly} to linear order in time. This is a
significant improvement compared with schemes based on cell-centred flux for
the gravitational work estimate.  The above can hence replace the energy
update of equation (\ref{eqnEgygravUpdate}) with a more accurate version that
ensures conservation of the total energy in self-gravitating systems. 

In practice, we typically use the version (\ref{EqnEgApprox}) based on the
gravitational forces, instead of equation (\ref{EqnEgExact}) based on the
potential.  In order to render the time integration of the energy equation
second-order accurate, we need to replace the gravitational forces (or
potentials) with averages between the beginning and end of the timesteps,
which can be done as in section~\ref{SecEgyStandardApproach}.  We note that
this method is also applicable in ordinary Eulerian codes, where $\Delta
\vec{w}_i = 0$, not just in the moving-mesh approach developed in this paper.
However, the method may also cause drifts of the temperature of the gas in
certain situations, and is hence not completely free of the problems that were
mentioned earlier.

\subsection{Gravitational softening} \label{secegysoftening}

To calculate gravitational potentials and forces for our unstructured
hydrodynamical mesh, we represent each cell as a mass point with an
appropriate gravitational softening, and employ techniques that are commonly
used in N-body algorithms. In principle, the gravitational field of a single
Voronoi cell could be adopted as the field of a polyhedron of constant
density, with the cell's shape and its total mass. However, this would make an
exact calculation of the field unwieldy and unnecessarily complicated. As we
anyway run out of gravitational resolution on the scale of the mesh cells, the
precise shape of a cell should be unimportant, provided we can ensure that the
generated field is sufficiently smooth and free of anisotropies due to the
mesh geometry. For simplicity, we therefore represent the potential of each
gaseous cell as that of a top-hat sphere of constant density and radius
$h$. In order to improve the smoothness of the potential in light of the
varying geometries of individual cells, we typically choose the volume of this
top-hat sphere to be slightly larger than the cell volume itself. In practice,
we relate $h$ to the volume $V$ of a cell as $h = f_{h} (3 V / 4 \pi)^{1/3}$,
where we choose $f_h\sim 1.0-1.5$. We note that for well-behaved meshes a
softening of the force-law is not strictly necessary because the
mesh-generating points are then always sufficiently distant from each
other. However, a gravitational softening is always required if a
collisionless particle component is present as well, and it allows a
consistent definition of the gravitational binding energy of the gas.

The gravitational potential kernel of a cell of volume $V$ is taken to be
\begin{equation}
\phi(r, h) = -\frac{1}{r}  \left\{\begin{array}{ll}
\frac{r}{2 h}\left[{3} - \left(\frac{r}{h}\right)^2\right]
&\;\mbox{for $r\le h $,}
\\
1 & \;\mbox{for $r> h$,}
\end{array}
\right.
\label{eqngaskernel}
\end{equation}
as a function of distance $r$. We then define the total gravitational
self-energy of the system of Voronoi cells as
\begin{equation}
E_{\rm pot} = \frac{1}{2} \sum_{ij} G\, m_i \,  m_j \, \phi(r_{ij}, h_j).
\label{EqnEpot}
\end{equation}
Ignoring mass exchanges between cells for the
moment, this implies that the gravitational acceleration of a cell is given by
\begin{eqnarray}
m_i\,\vec{a}_i^{\rm grav} & =&  - \frac{\partial E_{\rm
    pot}}{\partial\vec{r}_i} \nonumber \\
& = & - \sum_j G
m_i m_j
\frac{\vec{r}_{ij}}{r_{ij}} \frac{\left[ \phi'(r_{ij},h_i) +
    \phi'(r_{ij},h_j)\right]}{2} \nonumber \\
& & - \frac{1}{2} \sum_{jk} G m_j m_k  \frac{\partial \phi(r_{jk},h_j)}{\partial
h}\,\frac{\partial h_j}{\partial \vec{r}_i} ,
\label{EqnGrAcc}
\end{eqnarray}
where $\phi'(r,h) = \partial\phi / \partial r$ and $\vec{r}_{ij} =
\vec{r}_i - \vec{r}_j$.  The interaction between cells of different
softening lengths is hence symmetrized by averaging the forces, as
opposed to, for example, by averaging the softening lengths. This is
analogues to the formalism employed in \citet{Hernquist1989}. By
defining the potential energy of equation~(\ref{EqnEpot}) slightly
differently, in terms of an interaction potential with a symmetrized
softening length $h_{ij}$, one can however also arrive at a scheme
where the softening lengths are averaged, but then the correction
force derived below is less convenient to calculate.

The last term in equation (\ref{EqnGrAcc}) describes an additional force
component which stems from changes of the gravitational softening lengths. It
has to be included to make the system properly conservative when spatially
adaptive gravitational softening lengths are used \citep{Price2007} that are
allowed to vary in time, which is our default approach to treat self-gravity
in the moving-mesh scheme. Since we tie the gravitational softening length to
the volume of a Voronoi cell, we have
\begin{equation}
\frac{\partial h_j}{\partial \vec{r}_i} = 
\frac{\partial h_j}{\partial V_j} \frac{\partial V_j}{\partial \vec{r}_i} =
\frac{h_j}{3V_j} \frac{\partial V_j}{\partial \vec{r}_i}.
\end{equation}
Defining the quantities
\begin{equation}
\eta_j\equiv  \frac{1}{2} \sum_k  G m_j m_k   \frac{\partial
  \phi(r_{jk},h_j)}{\partial 
h}  \frac{h_j}{3V_j} ,
\label{eqneta}
\end{equation}
we can rewrite the last sum in equation (\ref{EqnGrAcc}) as
\begin{equation}
m_i\,\vec{a}_i^{\rm soft} =  
 -  \sum_{j}   \eta_j  \frac{\partial V_j}{\partial \vec{r}_i} .
\label{eqnsoftfrc}
\end{equation}
Using equations (\ref{EqSerr1}) and (\ref{EqSerr2}) for the partial
derivative of the Voronoi volume \citep[see][]{Serrano2001}, this can be more
explicitly expressed as
\begin{equation}
m_i\,\vec{a}_i^{\rm soft} =  
   \sum_{j\ne i}  ( \eta_j -\eta_i)\, A_{ij} \left(\frac{\vec{c}_{ij}}{r_{ij}}  
- \frac{\vec{r}_{ij}}{2r_{ij}} \right),
\label{eqnsoftc}
\end{equation}
where the sum extends over all the Voronoi neighbours of a cell. Note that
$A_{ij}$, $\vec{c}_{ij}$, and $r_{ij}$ are invariant when $i$ and $j$ are
exchanged, while $\vec{r}_{ij}$ changes sign. The term involving
$\vec{c}_{ij}$ produces therefore an antisymmetric force between $i$ and $j$,
but the same is not obvious for the force from the
$\vec{r}_{ij}$-term. However, according to the Gauss theorem we have
\begin{equation}
\eta_i   \sum_{j\ne i}  A_{ij} \frac{\vec{r}_{ij}}{r_{ij}} = 0,
\label{eqnzero}
\end{equation}
because the summation is just the surface integral of a constant function.
If we subtract equation (\ref{eqnzero}) from equation (\ref{eqnsoftc}), we obtain
\begin{equation}
m_i\,\vec{a}_i^{\rm soft} =  
   \sum_{j\ne i} A_{ij} \left[ ( \eta_j -\eta_i)\,  \frac{\vec{c}_{ij}}{r_{ij}}  
- ( \eta_j +\eta_i)\frac{\vec{r}_{ij}}{2r_{ij}}  \right],
\label{EqnCorrectionForce}
\end{equation}
where now the antisymmetry of the correction force between particles $i$ and
$j$  is manifest.

To properly account for the changes in the gravitational energy when the
softening lengths are varied, we hence need to calculate the quantities
$\eta_i$ given by Equation (\ref{eqneta}), which can be conveniently done
alongside the tree walk used for the gravity calculation. With these values in
hand, we can then calculate the correction force as a surface integral over
the local Voronoi cell.  Finally, the correction force is added to the
ordinary gravitational force, and the resulting total force is used in
equations (\ref{eqnmomold}) and (\ref{eqnEgygravUpdate}), or alternatively in
equations (\ref{EqnEgExact}), replacing $-m\vec{\nabla}\Phi$ where
appropriate.

\begin{figure}
\bc
\resizebox{7.5cm}{!}{\includegraphics{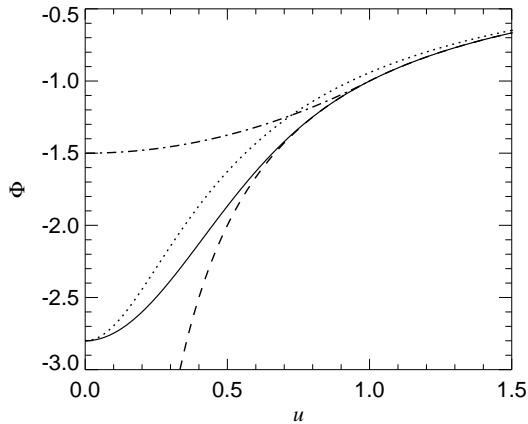}}%
\caption{Softened gravitational potential for cells and particles. The solid
  line shows the spline-based softening we use for collisionless particles,
  while the dot-dashed line is the top-hat like softening of gaseous cells. Both
  potentials become equal to the Newtonian potential for $u\ge 1.0$. For
  comparison, we also show the  Plummer softening with a dotted line. The
  dashed line is the Newtonian potential of a point mass.
  \label{FigPotentialShape}}
\ec
\end{figure}

For the dark matter particles, we employ a softening kernel with a different
shape, the same one as used in the SPH-code {\small GADGET}, which corresponds
to spreading the mass of a particle with the more centrally concentrated SPH
kernel.  This softening kernel is given by
\begin{equation}
\phi_{\rm dm}(r,h)
= - \frac{G}{r} 
\left\{
\begin{array}{ll}
 \frac{14}{5} u - \frac{16}{3}u^3 + \frac{48}{5} u^5 \\
- \frac{32}{5} u^6  ,
&\mbox{$0\le u<\frac{1}{2}$},\\
-\frac{1}{15} +\frac{16}{5} u - \frac{32}{3}u^3  \\
 + 16 u^4 - \frac{48}{5}u^5 +\frac{32}{15}u^6  ,
& \mbox{$\frac{1}{2}\le u <1$}, \\
 1, & \mbox{$u \ge 1$}. \\
\end{array}
\right.
\end{equation}
where $u = r/h$.  Often, we will quote the gravitational softening length for
collisionless dark matter particles in terms of an `equivalent' Plummer
softening length $\epsilon$, defined such that the potential at zero lag is
$m\,\phi(0)=- Gm/\epsilon$. This implies $h=2.8\epsilon$.  We keep the softening
lengths for dark matter particles fixed, which ensures that the collisionless
dynamics is conservative without the need for correction forces like the ones
derived for the gaseous cells.  Figure~\ref{FigPotentialShape} illustrates
the difference in shape between the `particle' and the `cell' kernels, for an
equal choice of $h$. The gravitational softening we have chosen for dark
matter particles results in a slower decline of the force within the softening
length.

In our tree-based gravity calculation, we store for each tree node the maximum
softening length $h_{\rm node}$ of all particles it represents, and we always
open a node if its distance is smaller than $\max(h_i, h_{\rm node})$, where
$h_i$ is the softening of the particle under consideration. As a result,
softened interactions only occur between particles, and not between nodes and
particles.

Finally, a brief comment about our treatment of the gravitational self-energy
of individual resolution elements.  In our tree-based calculation of the
potential, we always sum over all particles, hence the potential at the
location of a particle contains a contribution from the particle itself. In
the case of a dark matter particle of mass $m$ and softening length
$h=2.8\epsilon$, this is $-Gm/\epsilon$. Because the dark matter particle
masses are constant, there is then a finite gravitational binding energy left
even if all particles are spread out to infinity.  While unimportant for the
dynamics itself, we prefer to eliminate this contribution by subtracting the
self-potential $-Gm/\epsilon$ from the calculated potential of a collisionless
particle. For gas particles (which really represent cells of a well-defined
volume), the situation is different. As their mass and volume can change, the
self-energy contribution of a gaseous cell is not constant and hence cannot
simply be subtracted. This is also not necessary in this case.  As the gas
mass is spread out to infinity, its self-energy will automatically tend to
zero, because then the cell volumes and smoothing radii tend to infinity as
well.

\section{Refining or derefining cells} \label{SecRefinement}

In ordinary Eulerian hydrodynamics, adaptive-mesh refinement techniques are
very useful methods for dynamically concentrating the mesh resolution in regions
where it is needed most, while smooth regions or parts of the flow that are
not of interest can be derefined and modelled more coarsely. For applications
with a large dynamic range in density and length scales (which is typical in
cosmology), adaptive mesh refinement is, in fact, often a prerequisite in
Eulerian methods in order to achieve the necessary resolution in the regions
that are of most interest.

\begin{figure}
\bc
\resizebox{5cm}{!}{\includegraphics{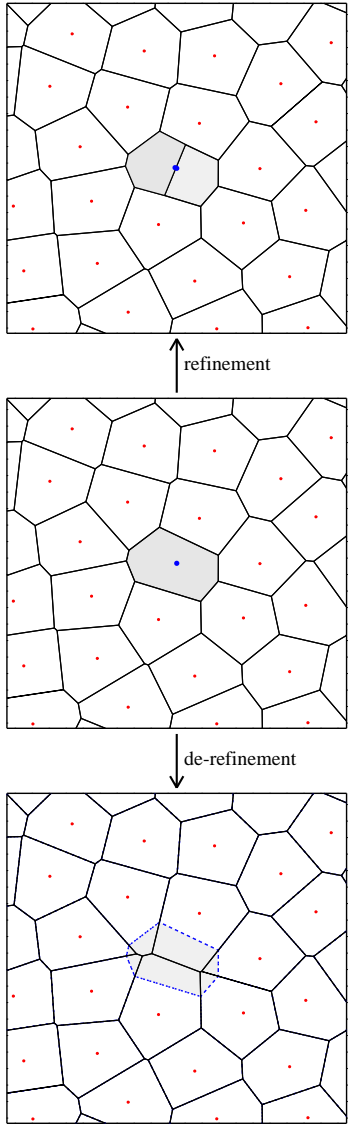}}
\caption{Example for the mesh refinement and derefinement operations.
In the middle panel, a Voronoi cell is marked in grey. In a refinement
operation, the cell is split into two cells, as shown in the top panel. If a
coarsening of the mesh-resolution is desired, the cell may be eliminated from
the mesh in the derefining operation shown in the bottom panel.
\label{FigDerefinementExample}}
\ec
\end{figure}

The Lagrangian moving-mesh methodology introduced in this paper, when combined
with the techniques to steer the mesh-motion, removes much of the need for
adaptive mesh refinement, especially in applications where
quasi-Lagrangian refinement criteria are used, as is typical  in
cosmological structure formation calculations with AMR. In fact, we think that
the Lagrangian moving-mesh approach with its automatic adjustment of
resolution to the local clustering state is ideal for this type of
application, and is arguably more natural than AMR.

One of AMR's particular strengths is however that the refinement criteria can
be nearly arbitrary. This allows resolution to be gained not only where most
of the mass goes, but where resolution is needed or desired most, according to
the problem at hand. For example, one might want to resolve low-density
regions or locate shock fronts particularly well.  Both can be achieved with
AMR by an appropriate choice of the refinement criteria. If the same
flexibility is desired for the unstructured Voronoi code discussed here, one
needs to find ways to refine or derefine the local mesh resolution
dynamically, a topic that we discuss briefly in this section.

In structured AMR, collections of cells can be hierarchically covered with
patches of refined meshes. Since here the geometry of the cells is simple, it
is easy to arrange the daughter meshes to exactly cover a certain set of cells
in the parent mesh. This makes the operations of interpolation and
prolongation straightforward. For our unstructured mesh, the situation is more
complicated. In particular, it is not straightforward to cover a contiguous
set of cells with another set that is better resolved, simply because of the
fact that the cell boundaries are defined as the edges of a Voronoi
tessellation. Finding a new, larger set of points as a replacement for the
points contained in some evacuated region such that the outer convex hull of
the Voronoi cells of all original points remains unchanged is non-trivial in
general. There is one exception, however: If we want to refine just a
single cell, we can split a Voronoi cell into two halfs if we insert a new
mesh-generating point at {\em almost exactly} the same location as the cell's
original point. This will leave all surrounding Voronoi cells {\em unchanged}.

This forms the basic mechanism for mesh-refinement in our code. According to a
criterion of choice, any given cell can be flagged for refinement. It is then
split into two cells by introducing a further mesh-generating point, as
illustrated in Figure~\ref{FigDerefinementExample}. The conserved quantities
of the original cell (mass, energy, momentum) are distributed among the two
halfs in a conservative way, either simply by weighting with the relative
fractions of the volumes occupied by the two new cells, or by using the
estimated linear gradient for a conservative reconstruction combined with a
volume integration. After the new point has been inserted, the
mesh-regularization techniques then dynamically change the local mesh such
that the two nearby points created by the cell split become well separated
from each other over the course of a few timesteps. By introducing the new
point in the direction of fluid gradients, one can furthermore optimize the
direction for which the spatial resolution is gained.

Note that a fundamental difference in this refinement approach compared with
the standard AMR method is that there is no hierarchy of multiple meshes that
cover the same region of interest. Instead, there is always only a single
mesh, albeit with spatially varying resolution. Refinement in our approach
means the dynamic introduction of further cells to locally increase the
resolution.

As we stressed above, the Lagrangian nature of the moving-mesh
approach largely eliminates the need for `mesh derefinements' in many
practical applications.  This is because the mesh follows the flow,
which often means that the resolution automatically stays where it is
needed, and in particular, advection alone does not generate a need
for refinement or derefinement, in contrast to AMR codes.  For
example, if a galaxy that is highly resolved in its centre moves
through space with large velocity, the moving mesh approach
automatically follows the centre well, without any need to introduce
mesh refinements. In Eulerian AMR on the other hand, refinements would
have to be constantly introduced along the path of the galaxy's
centre, and then removed again once it has passed by. Nevertheless, in
certain applications, one may encounter situations also in the
moving-mesh approach where one would like to dynamically reduce the
spatial resolution in special regions of a mesh. However, the geometry
of the Voronoi mesh imposes significant restrictions on a suitable
mesh coarsening operation. One possibility is to basically try to
reverse the refinement operation discussed above. To this end one can
move two mesh-generating points close together over the course of a
couple of timesteps, until they have essentially identical
position. Once this is achieved, the Voronoi cells corresponding to
the two points can simply be merged by replacing the two points with a
single mesh-generating point at the same location.  The new cell then
inherits the sum of the conserved fluid variables of the two merged
cells.

However, there are several technical difficulties in this approach that make
its application problematic in practice. For example, the decision for
derefinement is not made for a single cell, but for two neighbouring cells
simultaneously. In addition, several timesteps are needed to bring two
mesh-generating points close together in a smooth fashion, either during the
actual time evolution (over which the conditions for derefinement may well
change) or during a pseudo-evolution where the density field is kept static
and only the advection equation for the deforming mesh is solved.

Because of these difficulties, we have implemented an alternative derefinement
strategy where a cell is dissolved instantly by simply removing its
mesh-generating point from the tessellation. This means that the volume of the
removed cell will be claimed by the surrounding Voronoi cells, as illustrated
in Figure~\ref{FigDerefinementExample}. It is then also natural to distribute
the conserved fluid quantities (mass, energy, momentum) of the dissolved cell
among these neighbours, in proportion to the claimed volume fractions. Working
out the corresponding geometrical factors requires the construction of the
Voronoi diagram of the neighbouring cells with and without the point that is
removed.

A small complication in this approach is that the removal of a cell changes
the geometry of all the neighbouring cells. This in turn may well change the
outcome of the derefinement criterion for these neighbouring cells. For
example, if all cells below a certain size are supposed to be derefined and
two neighbouring cells are candidates for the derefinement, then the removal
of one of them will make the other larger, so that it may no longer fulfill the
derefinement criterion. To make the order of derefinement a well-defined
procedure, we construct the list of cells that are derefined in a given
timestep in the following way. First, we restrict ourselves to derefinement
criteria that allow a priority ordering of some kind, i.e.~we need to be able
to unambiguously identify the cell that should  `most urgently' be
derefined. Starting with this cell, we then flag cells for derefinement in the
order of this urgency parameter. However, we skip all cells that already have
a neighbouring cell that is flagged for derefinement in the same timestep. In
this way we always have a well-defined set of cells that can be derefined
in a given timestep, and only cells whose derefinement criteria are
independent from each other are derefined in the same step. This also means
that two neighbouring cells are never derefined in the same timestep. In
Section~\ref{SecHydroTests}, we will discuss a test problem (the Noh problem)
where we apply both the refinement and derefinement strategy described here.

\section{Time integration}  \label{SecTimeintegration}

In this section, we discuss issues of time integration. In particular, we
introduce an individual timestep scheme that can be used for our finite volume
discretization on an unstructured mesh. We will also address how we combine
the hydrodynamics with the integration of a collisionless N-body system that
represents dark matter or stars in galaxies. Finally, we detail how we
implemented cosmological integrations in an expanding universe, and we explain
the general structure of our new simulation code {\small AREPO} that
implements the methods discussed in this paper.

\subsection{Timestep criterion}

For hydrodynamics with a global timestep, we employ a simplified CFL
timestep criterion in the form
\begin{equation}
  \Delta t_i = C_{\rm CFL} \frac{ R_i}{c_i + |\vec{v}'_i|}  \label{EqnTiStep}
\end{equation}
to determine the maximum allowed timestep for a cell $i$. Here $R_i$ is the
effective radius of the cell, calculated as $R_i=(3V_i/4\pi)^{1/3}$ from the
volume of a cell (or as $R_i=(V_i/\pi)^{1/2}$ from the area in 2D), under the
simplifying assumption that the cell is spherical. The latter is normally a
good approximation, because we steer the mesh motion such that the
cell-generating point lies close to the centre-of-mass of the cell, which
gives it a ``roundish'' polyhedral shape.  $C_{\rm CFL} < 1$ is the
Courant-Friedrichs-Levy coefficient (usually we choose $C_{\rm CFL}\simeq
0.4-0.8$), $c_i = \sqrt{\gamma P/\rho}$ is the sound speed in the cell, and $
|\vec{v}'_i| = |\vec{v}_i-\vec{w}_i|$ is the velocity of the gas {\em relative
  to the motion of the grid}. In the Lagrangian mode of the code, the velocity
$|\vec{v}'_i|$ is close to zero and usually negligible against the sound speed,
which means that larger timesteps than in an Eulerian treatment are possible,
especially if there are large bulk velocities in the system.

If the code is operated with a global timestep, we determine the next system
timestep as the minimum
\begin{equation}
\Delta t = \min_i \Delta t_i
\end{equation}
of the timestep limits of all particles. In simulations with gravity, we also
impose a second kinematic timestep criterion for each particle, as described
in \citet{Springel2005}, and we restrict the maximum allowed timestep to a
suitable value. However, we have also implemented an individual timestep
scheme, where the different timestep conditions of different cells are treated
in a more flexible and computationally efficient fashion. This is discussed
next.

\begin{figure*}
\bc
\resizebox{16.5cm}{!}{\includegraphics{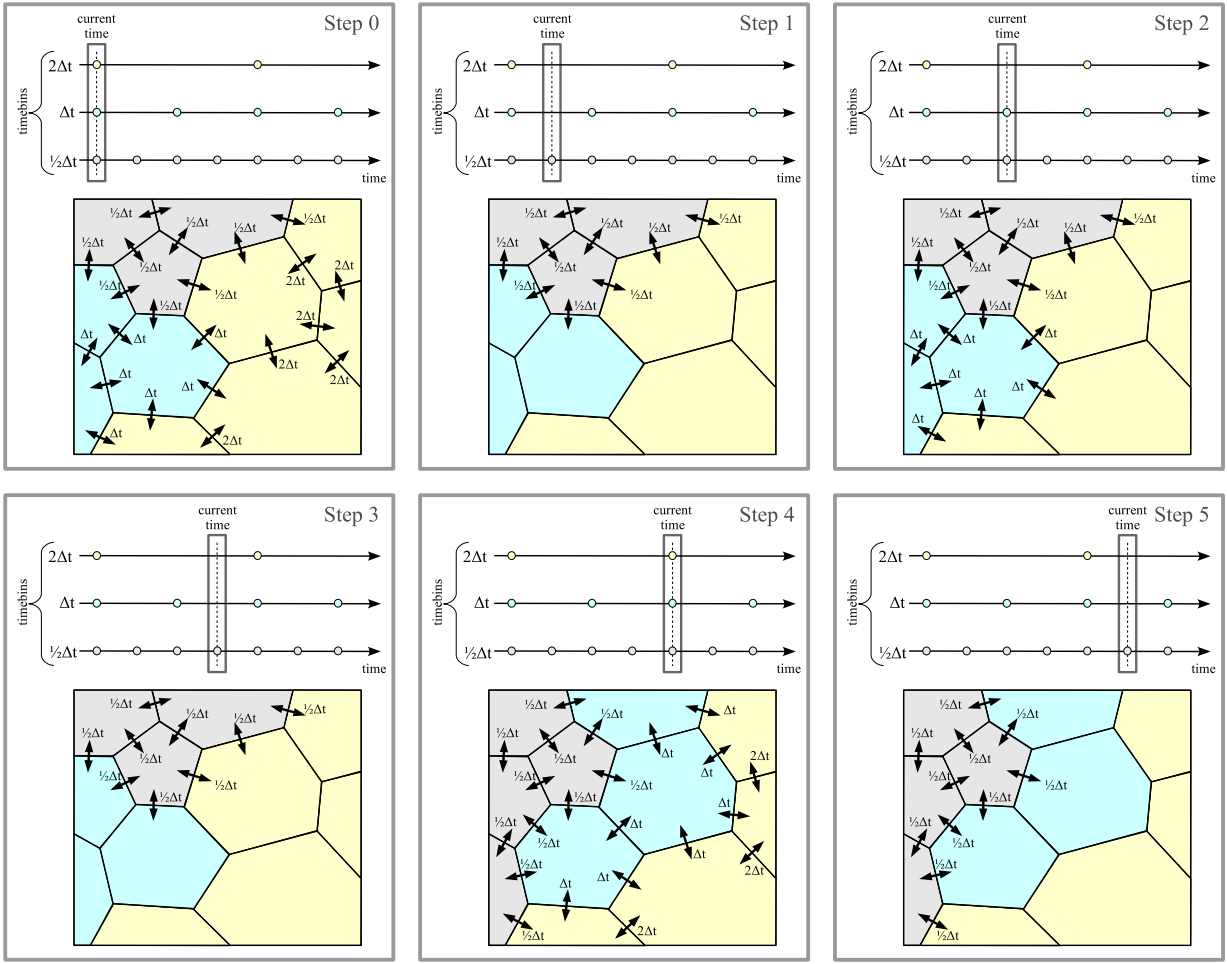}} %
\caption{Sketch of the individual time integration scheme used in {\small
    AREPO} for the hydrodynamics. The mesh cells in this example occupy three different
  timebins, and are coloured accordingly in the sketch. In each
  timestep, fluxes are calculated for all faces where at least one of the
  cells is active on the current timestep. The cells then exchange conserved
  quantities on the smaller of the two timestep sizes of the two neighbouring
  cells of each active face. A
  cell has the possibility to reduce its timesteps at the end of each step, 
  but can only move to a timestep twice larger every second step in
  order to maintain a nested synchronization of the timestep hierarchy.
  \label{FigSketchTiIntegration}}
\ec
\end{figure*}

\subsection{Individual timesteps}

In typical cosmological simulations, a large dynamic range in densities
quickly occurs as a result of gravitational clustering. Accordingly, local
dynamical times can vary by orders of magnitude. It has hence long been common
practice to use individual timesteps for the collisionless N-body problem, a
technique that has also been extended to hydrodynamical SPH simulations
\citep[e.g.][]{Katz1996,Springel2001gadget}.  However, the use of individual
timesteps in mesh-based finite volume codes is more problematic and appears to
be rarely used, except in the context of AMR simulations. In the latter,
individual refined grid patches are typically subcycled in time (frequently by
a factor of 2 if the refinement factor is 2) relative to their parent
grid. Refluxing techniques are then used to assure that a fully conservative
solution is obtained on the coarser parent grid as well. Note that in this
approach the same volume is effectively covered multiple times.

\begin{figure*}
\bc
\resizebox{16.5cm}{!}{\includegraphics{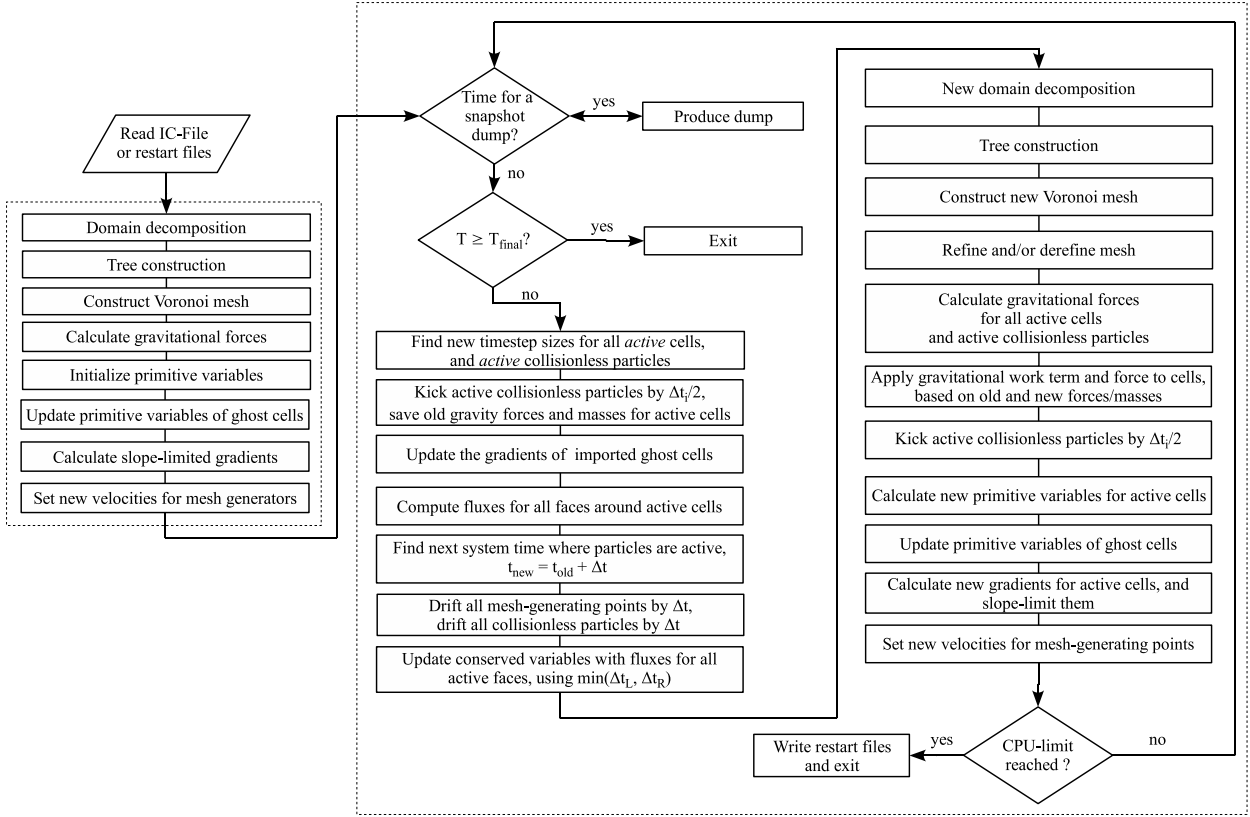}}
\caption{Flow chart of the simulation code. After a number of initialization
  steps (marked by the dashed box on the left), the code enters a main
  loop. In each iteration of the main loop, the system advances by a time
  interval $\Delta t$ that corresponds to the smallest occupied timebin. ${\rm
    Active}$ cells or collisionless particles are always those with a timestep
  that is in sync with the current time of the system. In the first phase,
  operations correspond to active particles beginning their timestep, in the
  second phase to those that end their timestep.
\label{FigFlowChart}}
\ec
\end{figure*}

We aim for another solution, because in the moving mesh approach the
cell size may vary greatly without an associated nested grid structure. In
order to save computational time in simulations with a large dynamic range, we
would like to be able to evolve only certain parts of the mesh with a small
timestep, and other parts with a larger step size. At the same time we
want to retain the conservative character and the stability of the
finite-volume approach that is obtained for a global timestep.

Our approach to address these requirements is based on a discretization of the
allowed timestep sizes into a power-of-two hierarchy, similar to the approach
frequently adopted in cosmological SPH codes \citep[e.g.][]{Katz1996}. This
means that the actual timestep $\Delta t_i$ of a cell $i$ is determined by
taking the largest power of 2 subdivision that is smaller than the timestep
criterion of equation~(\ref{EqnTiStep}). This effectively puts the cells into
a set of timestep bins that form a nested hierarchy of possible timesteps,
providing for a partial synchronization of the timesteps of different
cells.

Our individual timestep integration of the unstructured mesh is based on the
principle that, if two adjacent cells have different timesteps, their common
face is evolved with the smaller of the two steps. This leads to a
time-integration scheme that is graphically explained in the sketch of
Figure~\ref{FigSketchTiIntegration}.  In this example, in `Step 0', the
current system time is synchronized with the start of all three timestep sizes
that are present. As a result, the Voronoi mesh needs to be generated for all
cells present in the system, and fluxes are estimated for all of the
faces. However, the flux estimate is done for different timestep sizes,
depending on the timesteps of the involved cells, as indicated in the
sketch. For each face, always the smaller timestep of the two neighbouring
cells is used as actual timestep, and the time-integrated fluxes estimated for
the faces are used to update the conservative quantities of the two adjacent
cells.

Once the step is completed, the system time is advanced to the next
beginning/end of occupied timestep bins, and step 1 in
Fig.~\ref{FigSketchTiIntegration} begins. For cells that have completed their
timestep (these are the ones marked with timestep size `$1/2 \Delta t$'), new
primitive fluid variables and gradients are estimated, but the other cells
continue to use their old primitive variables and gradients for half-step
predictions and flux estimates. Also, their mesh-generating points continue to
move with the velocities assigned in their last active step.  In step 1, only
a much smaller set of the faces is active during the step, and only those
cells of the Voronoi mesh need to be constructed that have at least one active
face. We achieve this by inserting only the mesh-generating points of such
active cells into the mesh, and by ensuring the completeness of their Voronoi
cells with the search radius technique discussed in
subsection~\ref{secparallel}.  Fluxes are estimated only for the active faces,
and used to update the conservative quantities of the involved cells. This
process repeats again in step~2 and step~3. Whenever a cell has completed its
timestep, its primitive variables are updated based on the accumulated changes
of its conserved quantities. Also, the cell may then change its timestep
size. The timestep can always become smaller after a step has been completed,
but it can only increase if the higher timestep level is synchronized with the
current time, i.e.~if the target timebin starts one of its steps at the
current time. This means that a cell may increase its timestep only every
second step. An example for timestep changes is seen in step 4 of the sketch
in Fig.~\ref{FigSketchTiIntegration}: After having completed step 3, a few
cells reduce their timestep, and others increase it. With this change, the
system is then integrated forward in time through steps 4 and 5.

By construction, the above scheme is conservative as it only involves pairwise
exchanges of conserved fluid quantities. We have also found it to perform
accurately in practice, in the sense that we obtained comparable accuracy in
simulations where a global timestep or the more efficient individual timestep
scheme were used. A specific test of this will be discussed in
Section~\ref{SecHydroTests}.

A crucial point lies in the determination of suitable individual
timesteps; this obviously can have a large impact on the accuracy of the
individual timestep scheme, as well as on the efficiency gain that can be
realized with it. The timestep criterion of equation (\ref{EqnTiStep}) is
purely local, and is only appropriate for hydrodynamical waves that travel
with the local sound speed. If a supersonic shock wave is approaching, the
local gas element would be ignorant of it until the shock has arrived, and may
therefore be put on an inadequately large timestep just before the shock
strikes. We hence need to determine adequate timesteps by somehow taking
information about distant regions into account. The idea is that any given
cell should estimate the earliest time when it could become affected by the
gas present in some other cell, and this would then provide a suitable maximum
individual timestep.  To make this concept more explicit, we define a `signal
speed' \citep{Whitehurst1995,Monaghan1997} between two cells $i$ and $j$,
\begin{equation}
v^{\rm sig}_{ij} = c_i + c_j  - \vec{v}_{ij}\cdot \vec{r}_{ij} / r_{ij},
\end{equation}
where the velocity difference $\vec{v}_{ij}$ of the two cells is
projected onto their separation vector.  We then require that the
timestep of cell $i$ should be smaller than the travel time of this
signal over the distance $r_{ij}$ of cells $i$ and $j$. This means we
replace the timestep criterion of equation (\ref{EqnTiStep}) with
\begin{equation}
\Delta t_i = C_{\rm
  CFL}\min\left( \tau_i,\; \frac{R_i}{c_i+|\vec{v}_i'|} \label{EqnTiStepIndividual}
\right)
,
\end{equation}
where 
\begin{equation}
 \tau_i =  \min_{j\ne i}\left( \frac{r_{ij}}{c_i+c_j -
    \vec{v}_{ij}\cdot \vec{r}_{ij} / r_{ij}} \right).
\end{equation}
The timestep of equation (\ref{EqnTiStepIndividual}) is the maximum
allowed timestep for cell $i$, and can be used in our individual
timestep approach.

A brute-force calculation of this timestep criterion would be a prohibitive
$N^2$ process. However, we can use a hierarchical tree-based grouping of the
particles for a much more rapid evaluation of the timestep criterion, with a
cost of order ${\cal O}(\log N)$ per particle.  For this purpose we use the
same oct-tree that we anyway employ for neighbour search (as needed in the
parallelized mesh construction, see Fig.~\ref{FigSketchPointImport}) and the
gravity calculation. For each tree node, we store the maximum sound speed
$c^{\rm max}$, and the maximum velocity magnitude $v^{\rm max}$ for all the
cells with their mesh-generating points contained in the node. The calculation
of the maximum allowed timestep is then done with a special tree walk. We
start the walk with a first guess for the timestep, equal to $\Delta t_{\rm
  current} = R_i/(c_i+|\vec{v}_i'|)$. If a single particle $j\ne i$ is encountered, its value
of $r_{ij}/v_{ij}^{\rm sig}$ is computed and used to update $\Delta t_{\rm
  current}$ if it is smaller. If a tree node is encountered, we calculate a
special tree opening criterion, of the form
\begin{equation}
d_{\rm min} < \Delta t_{\rm current} \,(c_i +  c^{\rm max} + |\vec{v}_i|+ v^{\rm max} ),
\end{equation}
where $d_{\rm min}$ is the smallest distance of the point $\vec{r}_i$
to the boundaries of the node under consideration. If this condition
is fulfilled, the tree node is opened and its daughter nodes are
considered in turn, otherwise the tree walk along this branch of the
tree can be discontinued because there cannot be a particle inside the
node that would require a smaller timestep than the current one. When
the tree walk finishes, the timestep of cell $i$ is finally given by
$\Delta t_i = C_{\rm CFL} \Delta t_{\rm current}$.

We note that this scheme is more general and
flexible than the suggestion by \citet{Saitoh2008} to restrict the
timestep choices of a particle (cell) by the timesteps of its
immediate neighbours. Our approach can choose optimum timesteps even
under extreme conditions. For example, one can imagine a high-speed
collision of two self-gravitating cold blobs of gas. While the blobs
are still separate, our scheme would assign large timesteps to them,
allowing them to efficiently propagate through space, but right
before the physical collision starts, the timesteps would be
reduced appropriately.  We also note that the above scheme produces
Galilean-invariant timestep choices.

\begin{figure}
\bc
\resizebox{7.5cm}{!}{\includegraphics{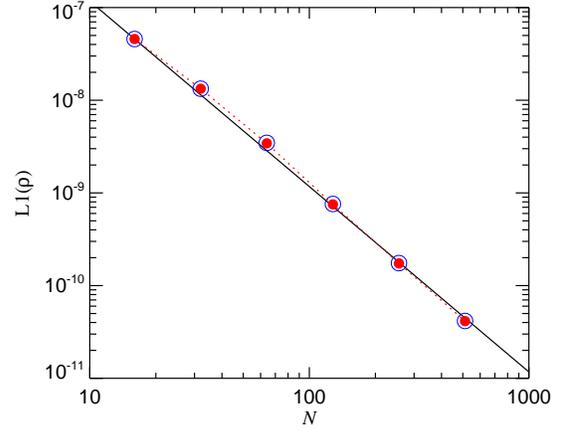}}
\ec
\caption{L1 error norm for an acoustic wave in one dimension, calculated with
  the moving-mesh code (red circles), or with a fixed mesh (blue open
  circles). The two schemes produce nearly identical errors (the moving mesh
  code lies only $\sim1-2\%$ lower), and the solution converges as $L1\propto
  N^{-2}$, i.e.~with second order accuracy.
  \label{FigL1Accoustic}}
\end{figure}

\begin{figure}
\bc
\resizebox{7.5cm}{!}{\includegraphics{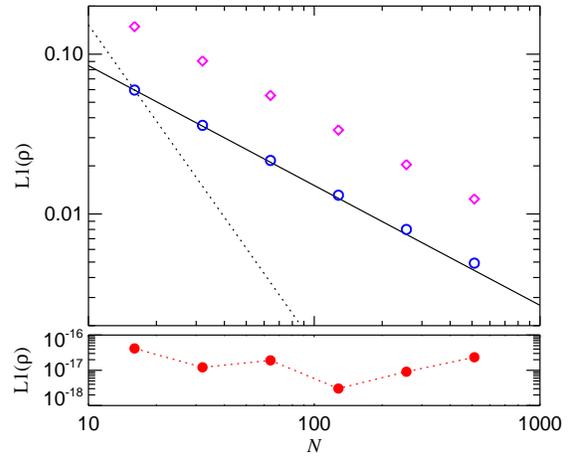}}
\ec
\caption{L1 error norm for a contact discontinuity (a density step from
  $\rho=1$ to $\rho=2$ that moves once through a box. The blue circles show
  the error as a function of resolution when we use our code with a Cartesian
  mesh with a fixed grid. In the presence of the discontinuity and the need to
  advect it the global rate of convergence is reduced to only about $L1\propto
  N^{-3/4}$. However, for the moving-mesh code, shown with red circles, the
  error is {\em consistent with zero} within floating point rounding
  errors. For comparison, we also show with diamonds the L1 error of the same
  test carried out with {\small ATHENA} (using 2nd order reconstruction and
  the Roe solver). The dotted line illustrates a second-order scaling.
  \label{FigL1Contact}}
\end{figure}

\subsection{Cosmological integration}

In an expanding Friedman-Lemaitre cosmology, the Euler equations need
to be modified by source terms that describe the decay of velocities
and energies due to the expansion of space. It is convenient to
describe the fluid positions in terms of comoving coordinates $\vec{x}
= a\vec{r}$, where $a$ is the cosmological scale factor $a=1/(1+z)$
and $z$ is the redshift. We also define a comoving density $\rho_c
\equiv a^3 \rho$, and a `comoving pressure' $P_c \equiv
(\gamma-1)\rho_c u$. The Euler equations in an expanding universe can
then be written as
\begin{equation}
\frac{\partial \rho_c}{\partial t} + \frac{1}{a}\vec{\nabla}_c (\rho_c
\vec{v}) = 0,
\end{equation}
\begin{equation}
\frac{\partial( \rho_c\vec{v})}{\partial t} + \frac{1}{a}\vec{\nabla}_c [(
\rho_c \vec{v}\vec{v}^T + P_c)\vec{v}] = - H(a)\, \rho_c \vec{v} - \frac{\rho_c}{a^2} \vec{\nabla}_c\Phi_c,
\end{equation}
\begin{equation}
\frac{\partial( \rho_c e)}{\partial t} + \frac{1}{a}\vec{\nabla}_c [(
\rho_c e + P_c)\vec{v}] = - 2H(a)\, \rho_c e - \frac{\rho_c\vec{v}}{a^2}\vec{\nabla}_c\Phi_c  .
\end{equation}
Here $\vec{v} = a\, \dot\vec{x}$ is the peculiar velocity.  The specific
energy is defined in terms of the peculiar velocity as $e = u + \vec{v}^2/2$.
The gradient operator $\vec{\nabla}_c$ acts on the comoving coordinates
$\vec{x}$, and $H(a)= \dot a /a$ is the Hubble expansion rate.  $\Phi_c$ is
the comoving peculiar gravitational potential, which is the solution of
\begin{equation}
\nabla_c^2\, \Phi_c = 4\pi G\, [\rho_c(\vec{x}) - \overline{\rho_c}],
\end{equation}
where $\overline{\rho_c}$ is the mean comoving density of the universe.

Integrating these equations over the comoving volume of a Voronoi
cell, it is easy to see that suitable `conservative' variables are
still given by mass, momentum and energy of a cell, except that the
total momentum and total energy of the simulated system are not
strictly conserved any more due to the presence of the terms involving
$H(a)$. However, surface integrals over cells simply yield the
ordinary fluxes of the Euler equations, evaluated with the physical
fluid quantities and the physical areas of the cell interfaces. Hence
we can continue to use the Godunov approach for determining the
fluxes, and they themselves are still fulfilling a detailed balance
between cells.

To incorporate the loss terms due to cosmological expansion, we
proceed similarly as for the gravitational source terms. To calculate
time-centred fluxes, we incorporate the decay terms in the half-step
prediction of the primitive variables and then obtain a second-order
accurate update for the full step by evaluating the loss terms at the
beginning and end of the step. For example, for the energy contained
in a cell, this takes the form
\begin{equation}
E_{n+1} = E_{n} + \Delta E_{\rm flux} - [H(a_n)E_n
+ H(a_{n+1})E_{n+1}]\,{\Delta t},
\label{Eqsinkupdate} 
\end{equation}
where $E_{n+1}$ is the energy at the end of the step, and $\Delta
E_{\rm flux}$ denotes the accumulated energy flux into the cell across
its surface.  Note that equation~(\ref{Eqsinkupdate}) can be easily
solved for the new energy $E_{n+1}$ at the end of the step.

\begin{figure*}
\bc
\resizebox{8cm}{!}{\includegraphics{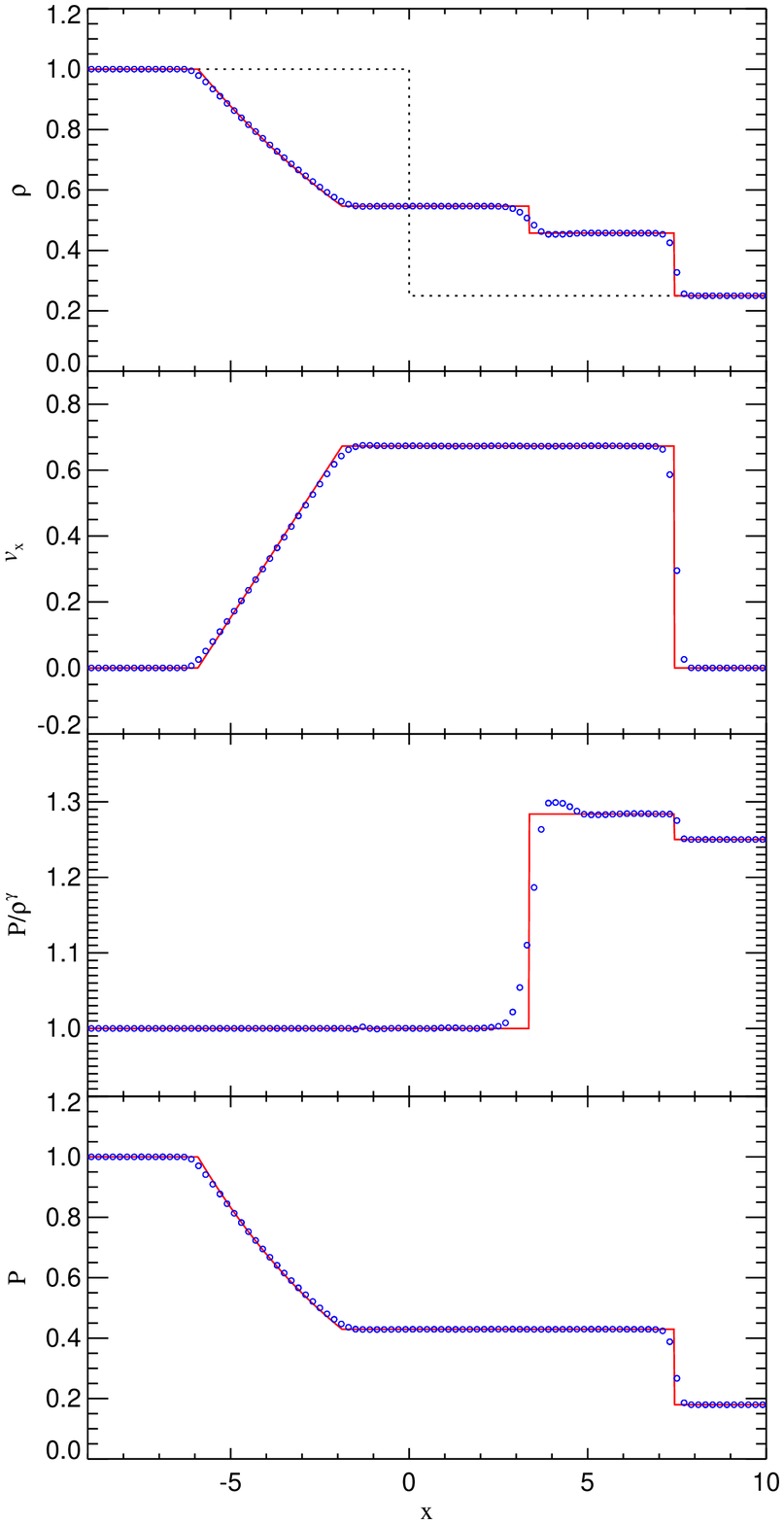}} %
\resizebox{8cm}{!}{\includegraphics{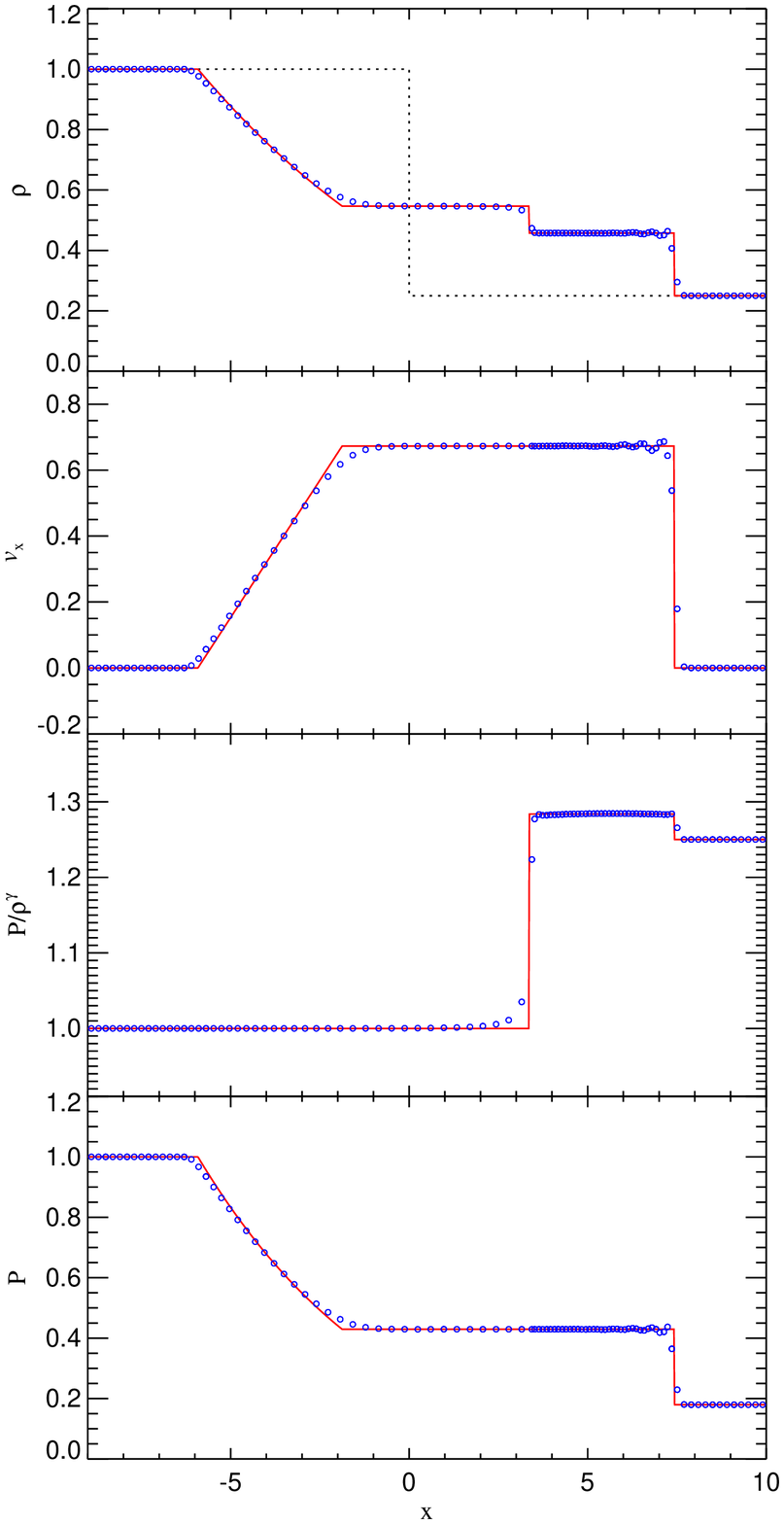}}\\
\caption{A shock tube test with initial conditions frequently used in
  previous code tests
  \citep{Hernquist1989,Rasio91,Wadsley2004,Springel2005}. The panel on
  the left shows the result (symbols) of a 2D test with equal volume per
  cell and a fixed mesh, the right when the mesh is allowed to move. The
  solid lines show the analytic solution, and the dotted lines in the
  top row mark the initial conditions.
  \label{FigSod1}}
\ec
\end{figure*}

\subsection{Structure of the AREPO code}

In Figure~\ref{FigFlowChart}, we show a basic flow-chart of the new
cosmological hydrodynamical code {\small AREPO} that implements the methods
described in this paper, and which was used to calculate all the test problems
discussed in the next sections.  This code is parallelized for distributed
memory computers, and is written in ANSI-C. Its input and output files largely
match those of the TreePM/SPH code {\small GADGET-2}, such that a comparison
of moving-mesh calculations with corresponding ones done with SPH is
straightforward.

The {\small AREPO} code allows a variety of different types of simulations,
both in 2D and 3D. Self-gravity of the gas can be included, and is either
computed with a pure Tree or a TreePM approach. A collisionless dark matter or
stellar fluid can be optionally included as well. Simulations both in
Newtonian space, or in an expanding universe are possible. Also, fully
adaptive, individual timesteps are supported both for the gas and the dark
matter particles. The flow chart of Figure~\ref{FigFlowChart} shows how the
code arranges the different calculational steps. We have also implement
additional physics modules into our new code, such as radiative cooling, star
formation, and energy feedback processes, following the treatment in the most
recent version of the {\small GADGET} code.  Details of these implementations
will be described elsewhere.

\begin{figure*}
\bc
\resizebox{8.0cm}{!}{\includegraphics{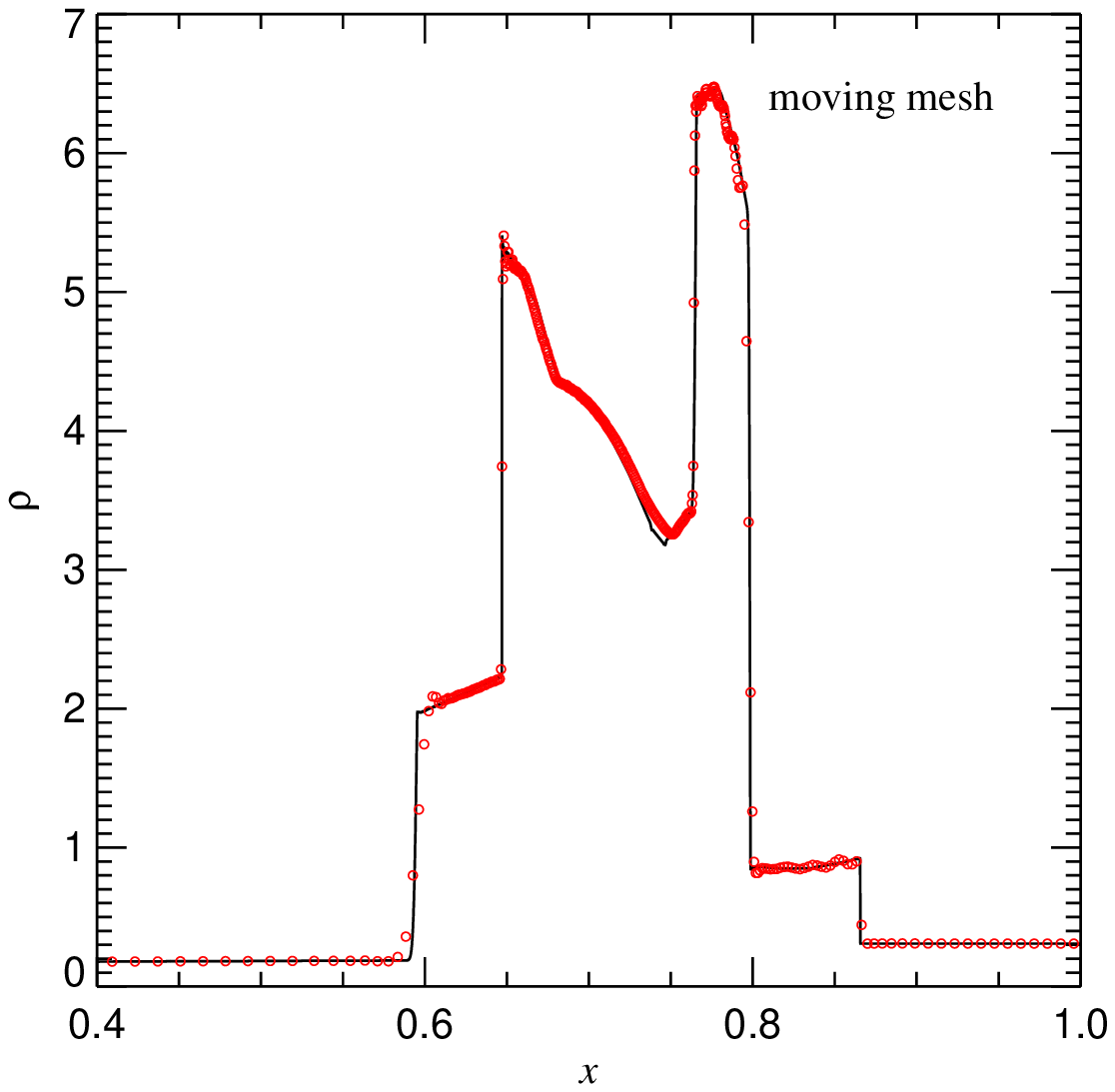}} %
\resizebox{8.0cm}{!}{\includegraphics{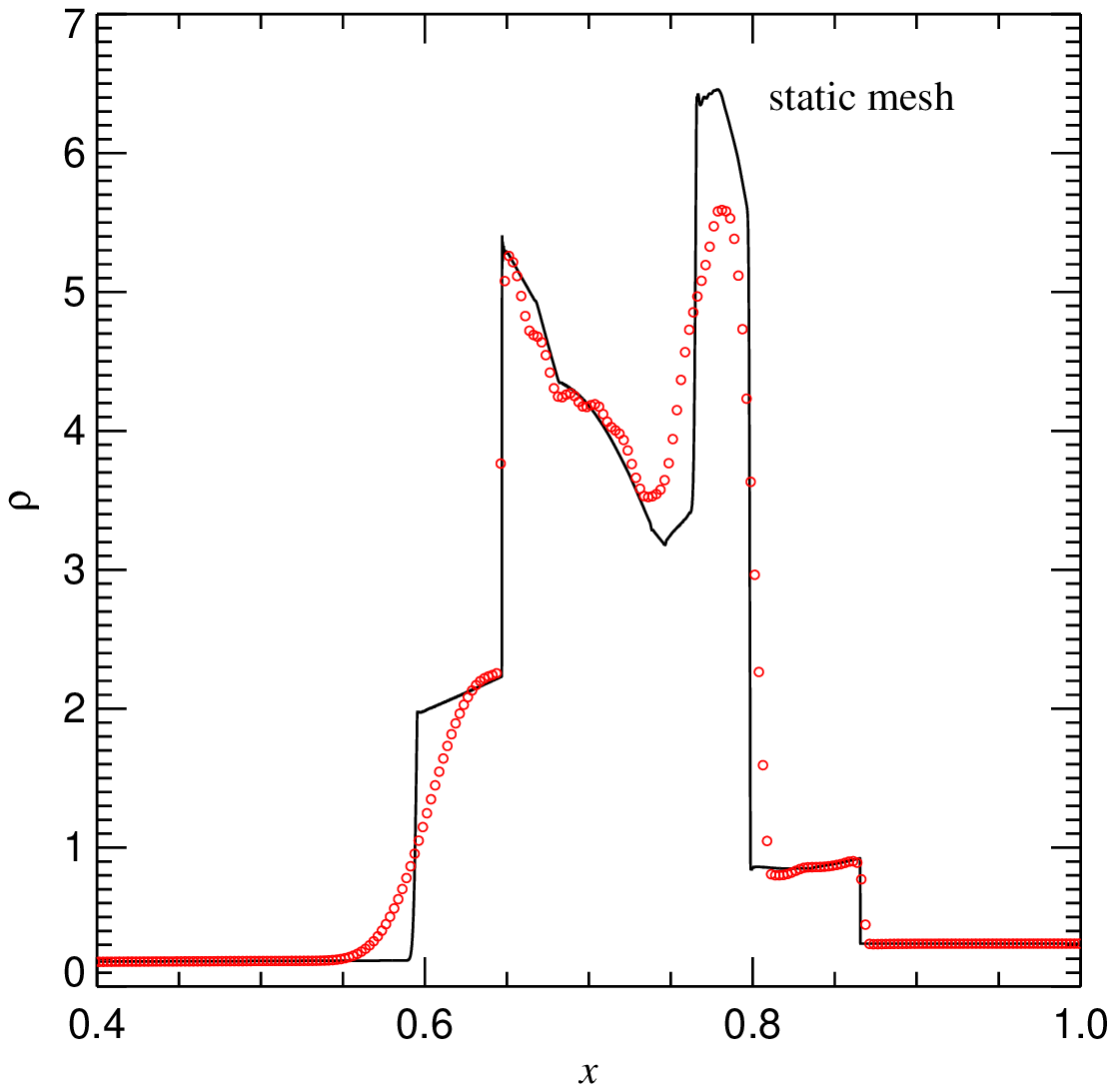}}\\
\caption{Density profile at time $t=0.038$ of the interacting double blast wave
  problem of \citet{Woodward1984}, for a resolution of 400 cells in the domain
  $[0,1]$. The panel on the left shows our result for
  the moving mesh-code, while the panel on the right is based on the same
  initial conditions but using a fixed mesh. The red circles mark density  values for
  individual cells of the 400-cell calculation, while the solid line is a
  high-resolution calculation with 20000 cells and a fixed mesh, for comparison.
  \label{FigDoubleBlast}}
\ec
\end{figure*}

\section{Hydrodynamical test problems} \label{SecHydroTests}
 
In this section, we consider a number of hydrodynamical test problems in order
to assess the accuracy and robustness of our new moving-mesh code. We
will frequently compare the results with calculations that start from
identical initial conditions but do not allow for a motion of the
mesh-generating points. In this case, our code should behave equivalently to a
standard Eulerian scheme with second-order accuracy in space and time. For a
few of the test problems we investigate this aspect explicitly by also
comparing with the publicly available, high-accuracy MHD code {\small ATHENA}
\citep{Stone2008}. A number of our test problems also allow comparisons with
results published in the literature for other codes. Note that tests involving
self-gravity are discussed separately in Section~\ref{SecGravityTests}.

\subsection{One-dimensional waves}

We begin with arguably one of the most elementary hydrodynamical test
problems, the treatment of simple waves in one dimension. This, in particular,
can serve as a sensitive test of the convergence rate of the code \citep[see,
e.g., the discussion in][]{Stone2008}. We first consider simple acoustic
waves. Following \citet{Stone2008}, we initialize a traveling sound wave of
very small amplitude $\Delta\rho/\rho=10^{-6}$ (to avoid any wave steepening)
and with unit wavelength in a periodic domain of unit length and unit
density. We use a one-dimensional version of the code in this test, where the
Voronoi faces can be easily constructed at the mid-points of the
mesh-generating points. However, we have checked that the same results are
also obtained with the two-dimensional version of the code.  The pressure is
set to $P=3/5$, such that the adiabatic sound speed is $c_s=1$ for a gas with
$\gamma=5/3$. The mesh-generating points are moved with the local velocity of
each cell, without terms for mesh regularization.

We let the wave travel once through the box, and compare the
final result with the initial conditions in terms of an L1 error norm. We
define the latter as
\begin{equation}
{\rm L1} = \frac{1}{N} \sum_i |\rho_i - \rho(x_i)|,
\end{equation}
where $N$ is the number of cells, $\rho_i$ the numerical solution for cell
$i$, and $\rho(x_i)$ is the expected analytic solution for the problem (which
is equal to the initial conditions in this first test).

In Figure~\ref{FigL1Accoustic}, we show results for the error norm for the
acoustic sound wave test as a function of the number of cells, both for a
fixed-mesh, and for the moving mesh. Reassuringly, the results demonstrate
global second order convergence of the code, as expected for a smooth problem
like this one. This is true both for the moving-mesh approach, as well as when
we keep the mesh fixed, with almost identical errors. Furthermore, we note
that the absolute size of the errors are very similar to what
\citet{Stone2008} achieved with {\small ATHENA}.

Next, we consider a more demanding test, the advection of a contact
discontinuity once through the box. To this end, the left half $x<0.5$ of the
box is set to density $\rho=1$, and the right half to density $\rho=2$, with
pressure $P=3/5$ everywhere. We now let this contact discontinuity move once
through the box with velocity $v_x=1.0$ everywhere.  In
Figure~\ref{FigL1Contact} we show the L1 error as a function of resolution if
a fixed mesh is used. Now the convergence is in fact only $L1\sim
N^{-0.75}$. This is simply reflecting the numerical diffusivity of the
Eulerian approach for contact discontinuities. We have checked that
{\small ATHENA}  also shows the same scaling of the error if second-order
reconstruction is used. On the other hand, for our moving mesh code, the error
is {\em consistent with zero to machine precision}, $L1 \la 10^{-17}$. This is
of course the expected result for a Galilean-invariant scheme, as for $v_x=0$
the fixed mesh recovers the analytic result. This illustrates in a first
practical application the accuracy gain offered by a moving-mesh: pure
advection errors are reduced or eliminated. On the other hand, the error for
the Eulerian result is primarily set by the distance over which the
discontinuity needs to be advected, largely independent of the velocity of the
flow. It is hence a strong function of the reference frame picked for the
calculation.

\subsection{Shock-tube test}

We continue our investigation of basic hydrodynamical test problems with a
one-dimensional Sod shock tube.  For definiteness, we pick a left state
($x<0$) described by $P_1=1$, $\rho_1=1$, and $v_1=0$, and a right state
($x\ge 0$) given by $P_2=0.1795$, $\rho_2=0.25$, and $v_2=0$, in a gas with
adiabatic index $\gamma=1.4$.  Of course, a large number of other simple
Riemann problems are equally well possible. We have adopted these parameters
because they were previously used in a number of other code tests
\citep[][among others]{Hernquist1989,Rasio91,Wadsley2004,Springel2005}.

We sample the problem with points of spacing $\Delta x =0.2$ along the x-axis,
and examine the solution at time $t=5.0$. Note that in our moving
mesh-calculation this means that the cells start out with unequal masses. We
however refrain here from trying to adjust the mesh motion such that the
masses per cell become equal. Rather, the mesh motion is simply taken to be
given by the local flow velocity in the moving-mesh case.  We have set-up this
problem in a two-dimensional domain to also test the 2D mesh generation, even
though this problem could of course be more efficiently calculated with the 1D
version of our code.

In Figure~\ref{FigSod1}, we compare the shock-tube results, both for our
moving-mesh code and the fixed-mesh case, with the analytic result expected for
this Riemann problem. Both calculations produce a sharply resolved shock (of
Mach number ${\cal M}=1.48$), but there is a trace of small post-shock
oscillations in the Lagrangian calculation. Presumably, this could be avoided
with a more sophisticated wave-by-wave flux limiting procedure that would give
our scheme the total-variation-diminishing (TVD) property, which it presently
does not have due to the simpler MUSCL-Hancock approach. Note that the contact
discontinuity is smoothed out quite noticeably in the Eulerian calculation,
which also gives rise to a corresponding error in the entropy profile across
the contact discontinuity. As we have also seen above, this is a generic
feature in Eulerian methods and results from advection errors in evolving the
moving contact discontinuity. In contrast, the contact discontinuity is very
sharp in the moving-mesh calculation, and it {\em stays sharp} as a
function of time. The conclusion that one may draw from this test is hence
that the moving-mesh approach can resolve shocks just as well as an Eulerian
method on a fixed mesh, but it is able to produce more accurate results for
contact discontinuities.

\begin{figure*}
\bc
\resizebox{5.5cm}{!}{\includegraphics{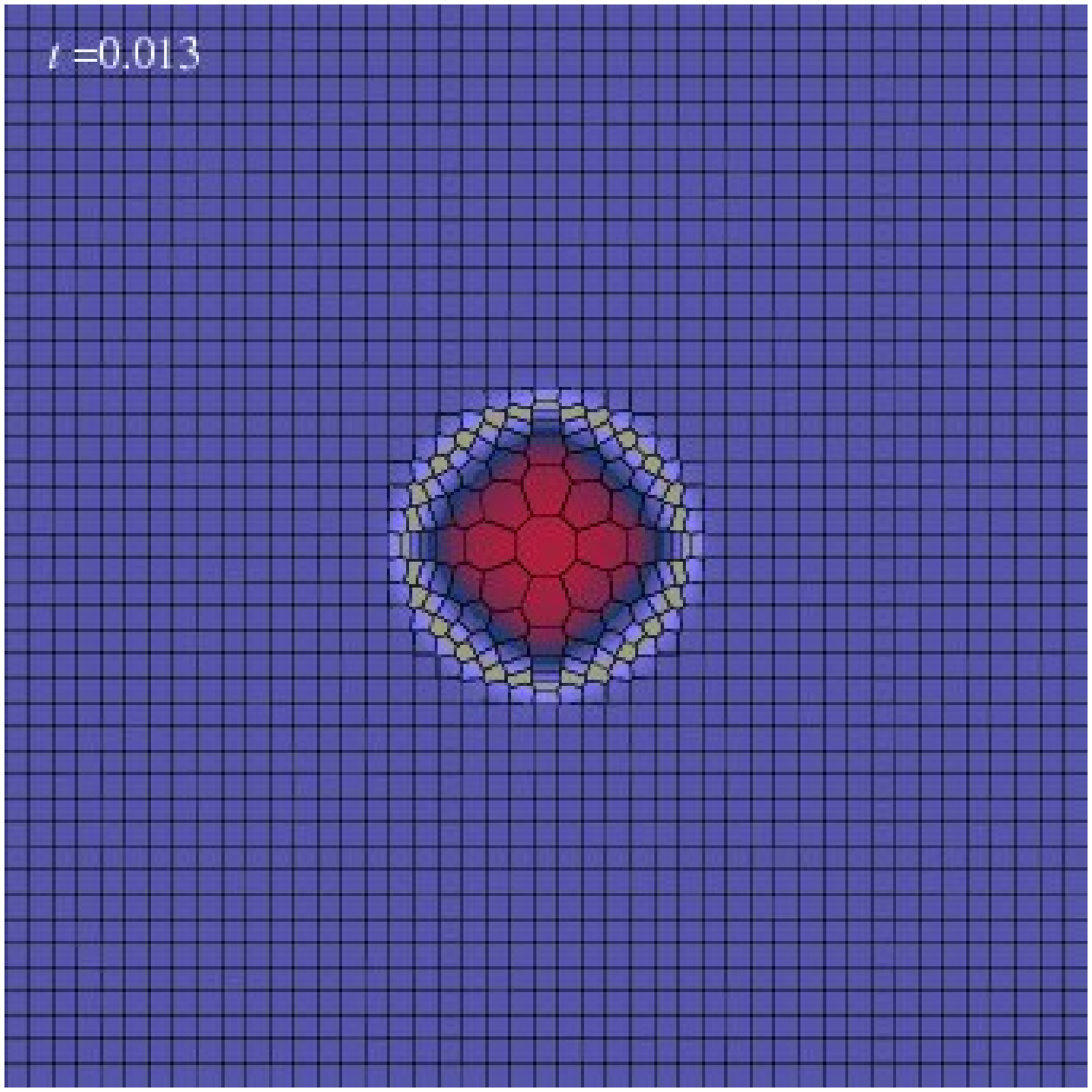}}\hspace*{0.1cm}%
\resizebox{5.5cm}{!}{\includegraphics{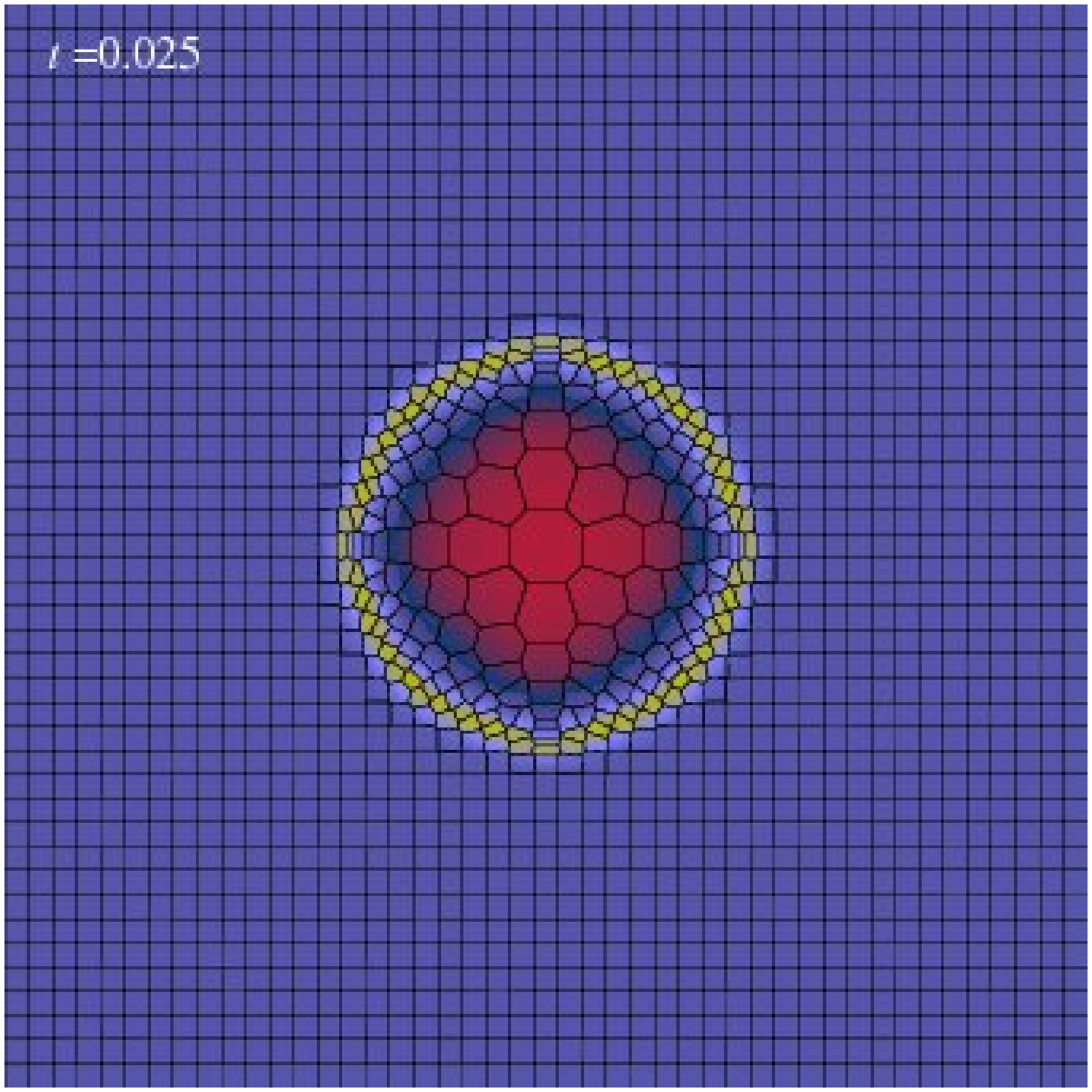}}\hspace*{0.1cm}%
\resizebox{5.5cm}{!}{\includegraphics{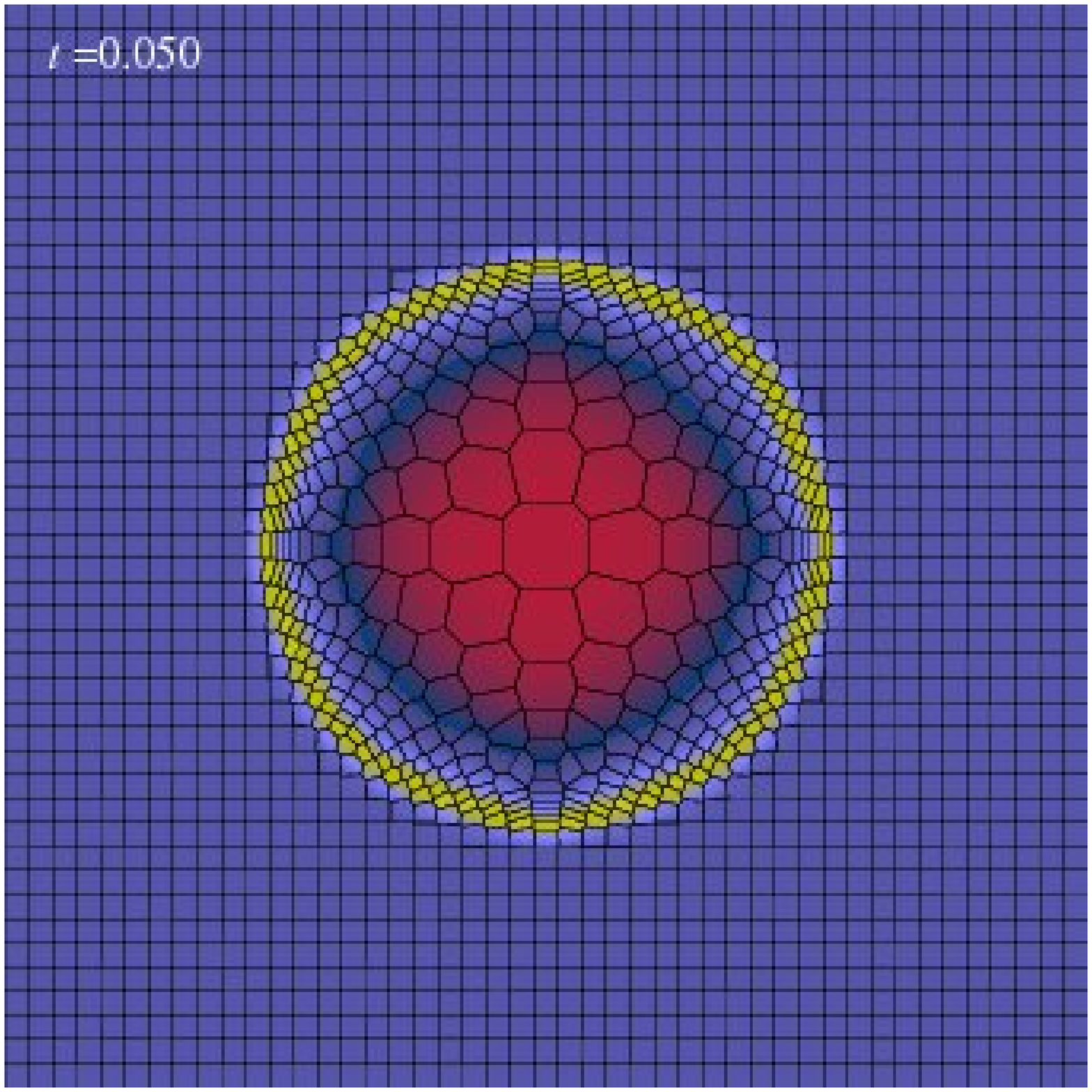}}\vspace*{0.1cm}\\
\resizebox{5.5cm}{!}{\includegraphics{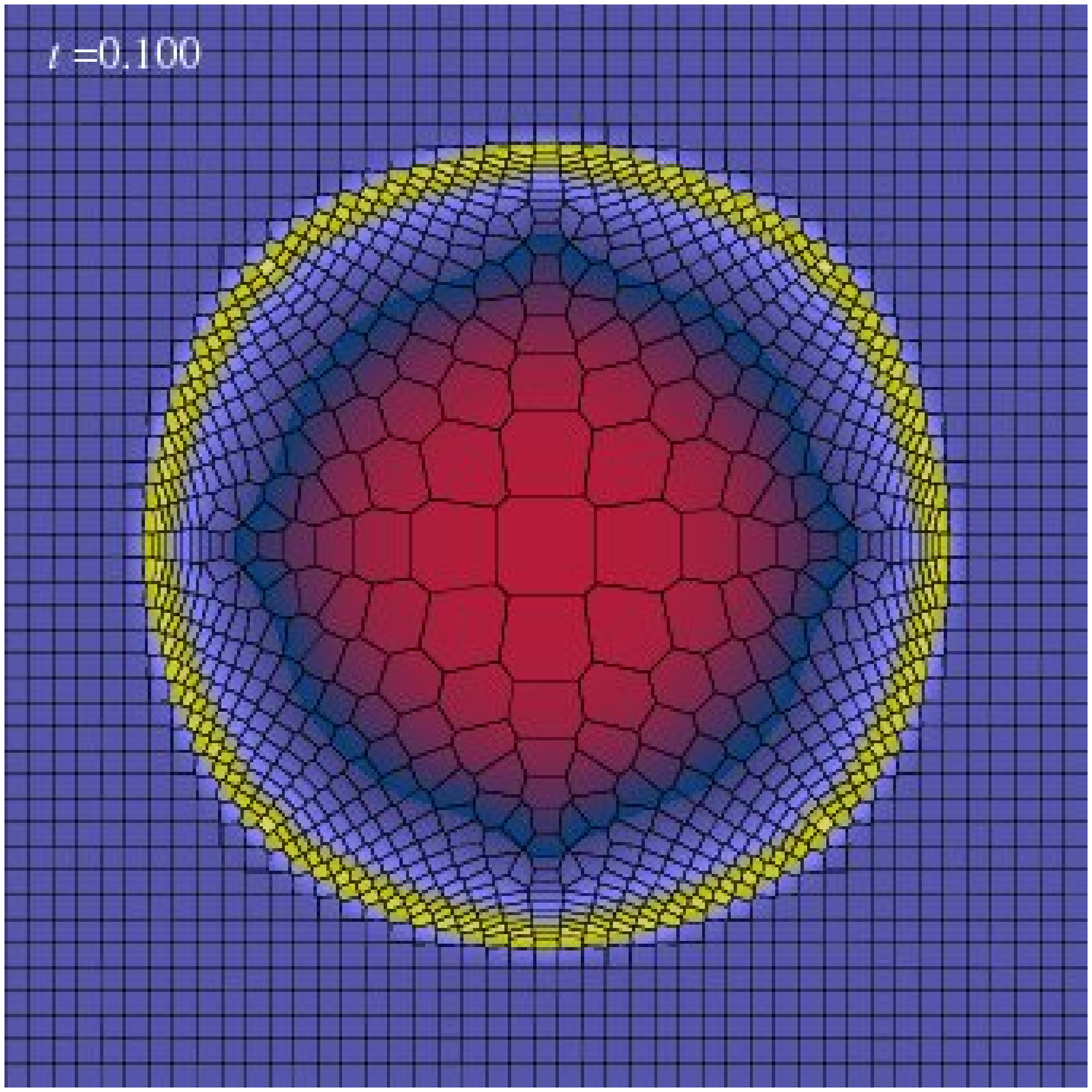}}\hspace*{0.1cm}%
\resizebox{5.5cm}{!}{\includegraphics{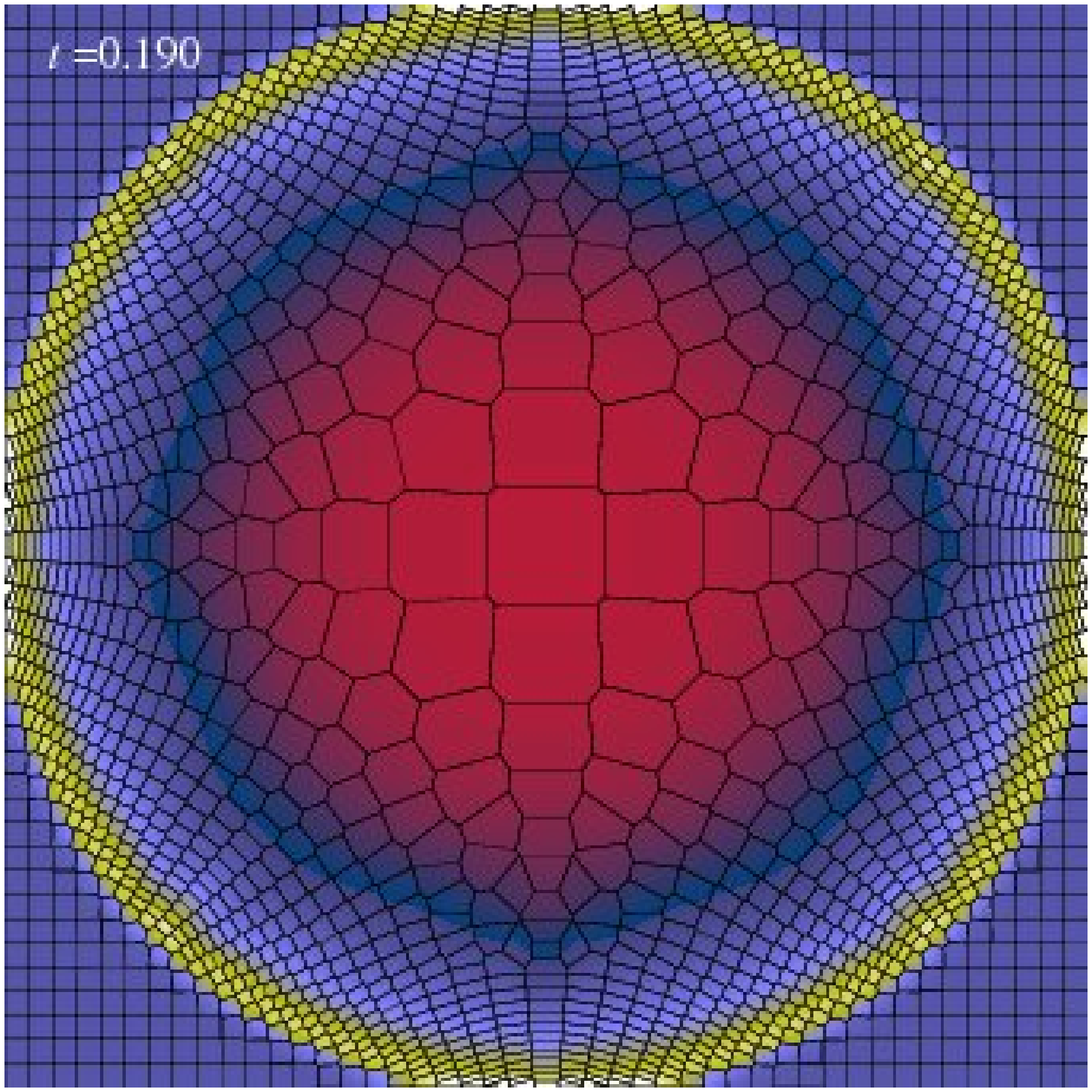}}\hspace*{0.1cm}%
\resizebox{5.5cm}{!}{\includegraphics{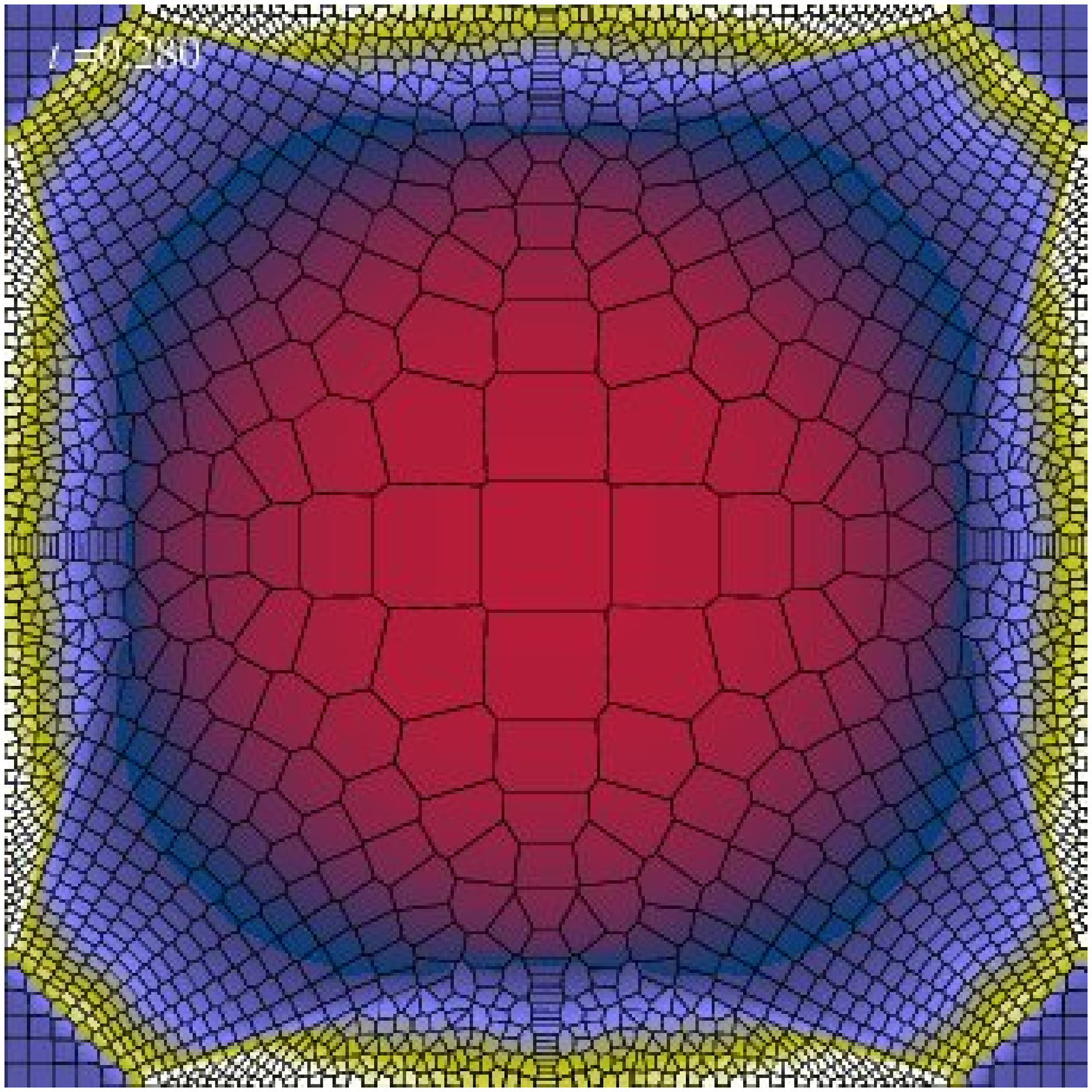}}\vspace*{0.1cm}\\
\resizebox{5.5cm}{!}{\includegraphics{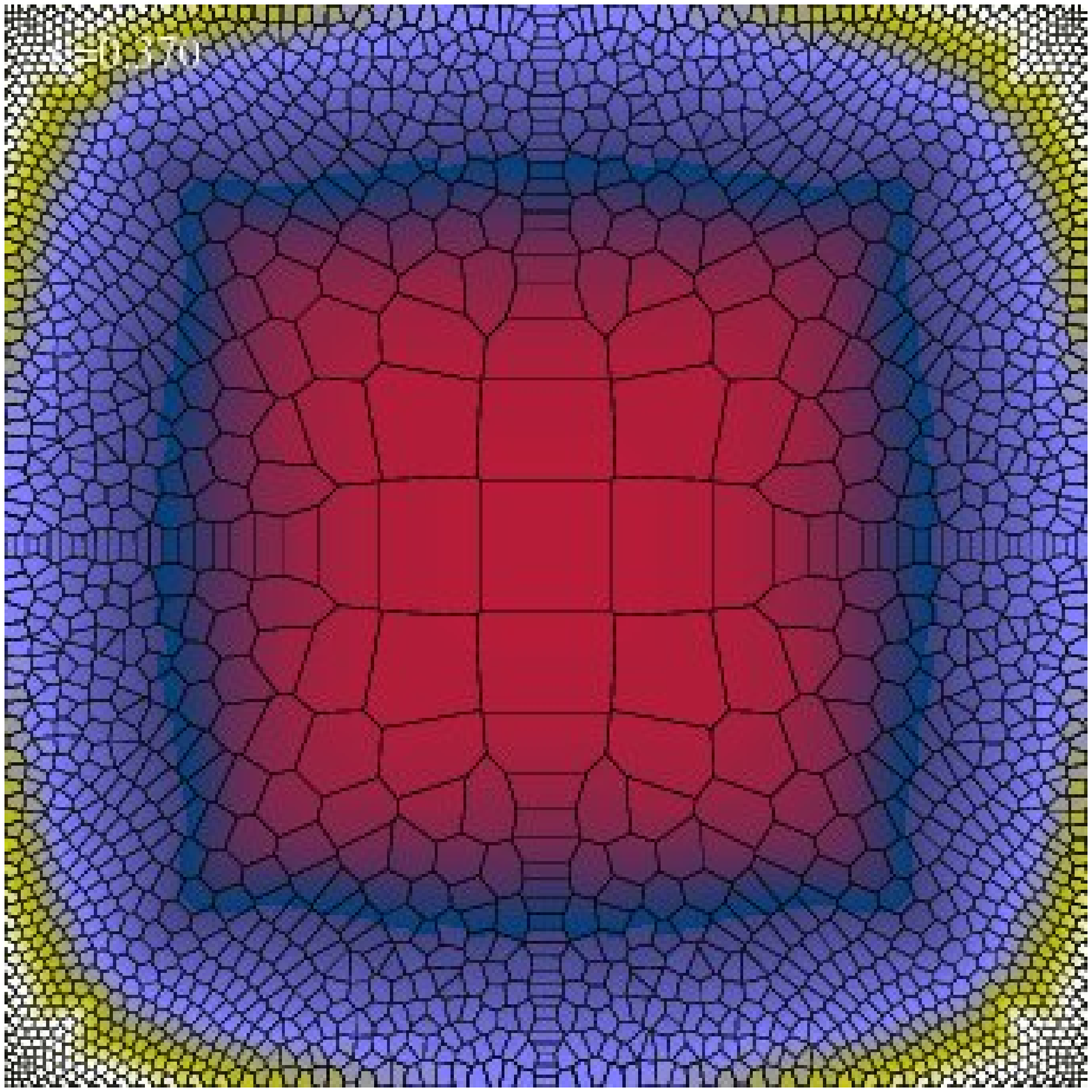}}\hspace*{0.1cm}%
\resizebox{5.5cm}{!}{\includegraphics{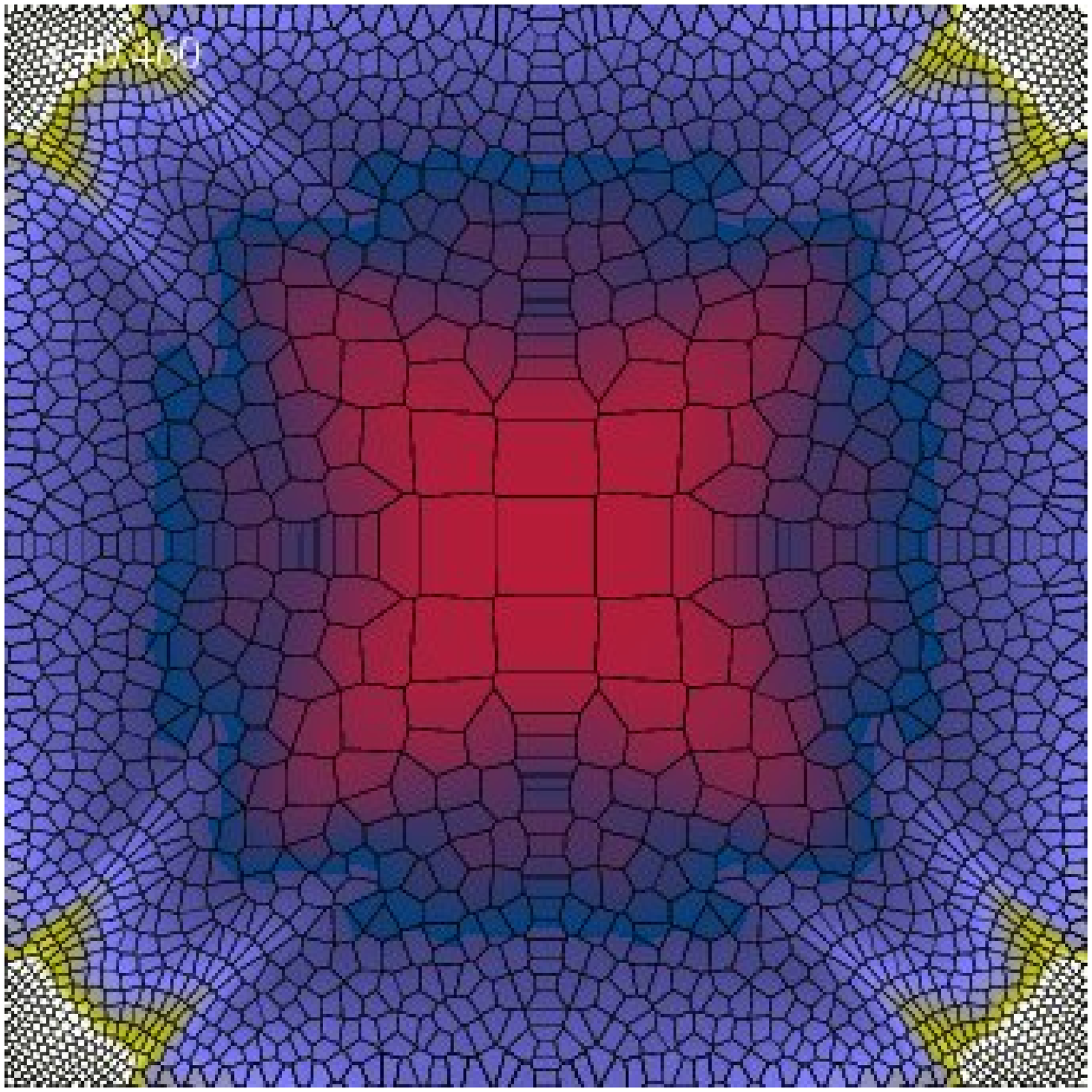}}\hspace*{0.1cm}%
\resizebox{5.5cm}{!}{\includegraphics{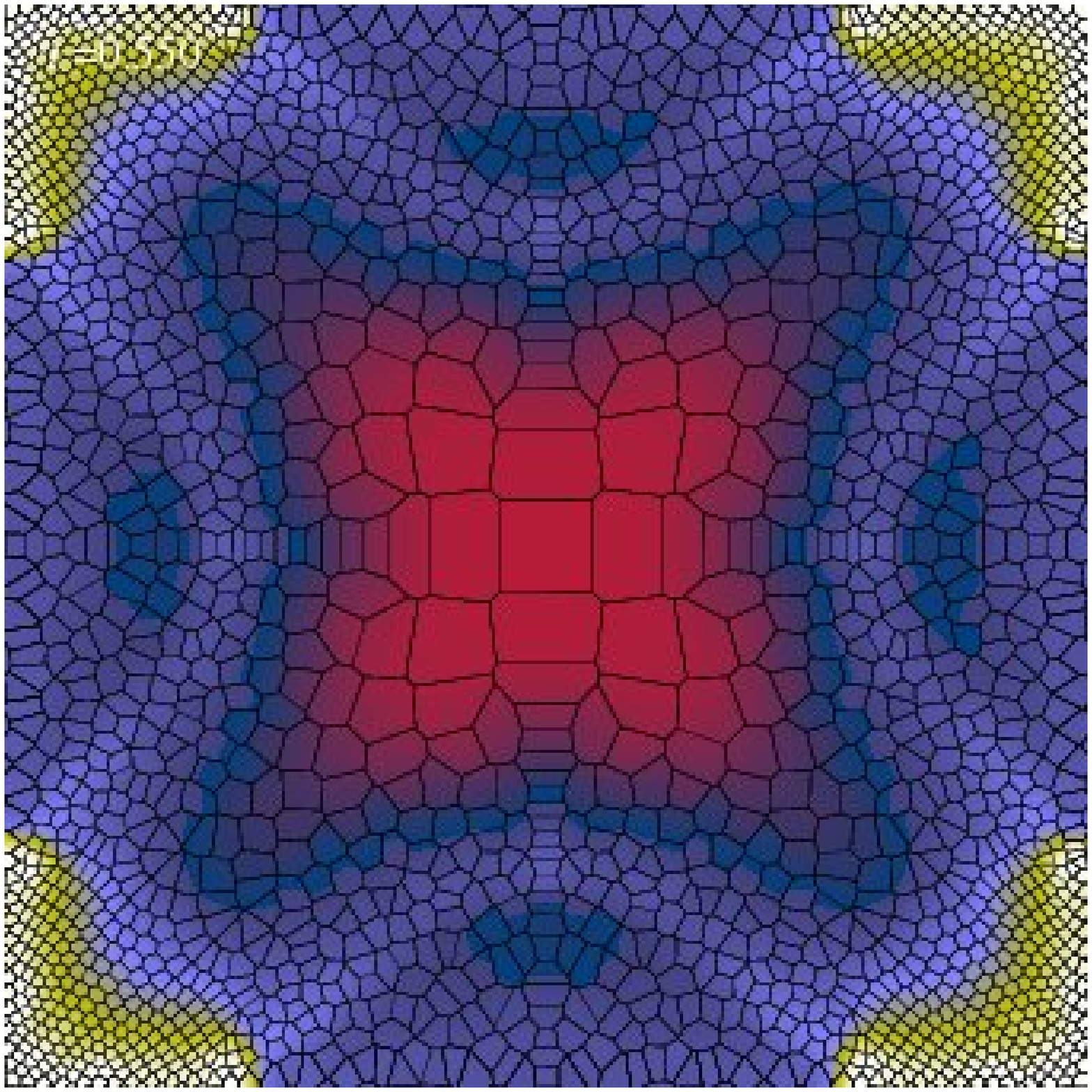}}\\
\caption{Time evolution of the density field in a 2D Taylor-Sedov blast wave 
 calculation with the moving-mesh code. The time of each snapshot is
 indicated in the panels. The evolving Voronoi mesh is overplotted,
 and has a resolution of $45\times 45$ cells. Roughly at time
 $t=0.19$, the shock reaches the periodic boundaries of the domain of unit
 side length $L=1$,  and effectively collides with the blast wave of the periodic grid of
 explosions described by this set-up. This compresses much of the
 matter into the corners of the domain, a process that is well
 followed by the moving mesh.
 \label{FigSedovEvol}}
\ec
\end{figure*}

\begin{figure*}
\bc
\resizebox{8.0cm}{!}{\includegraphics{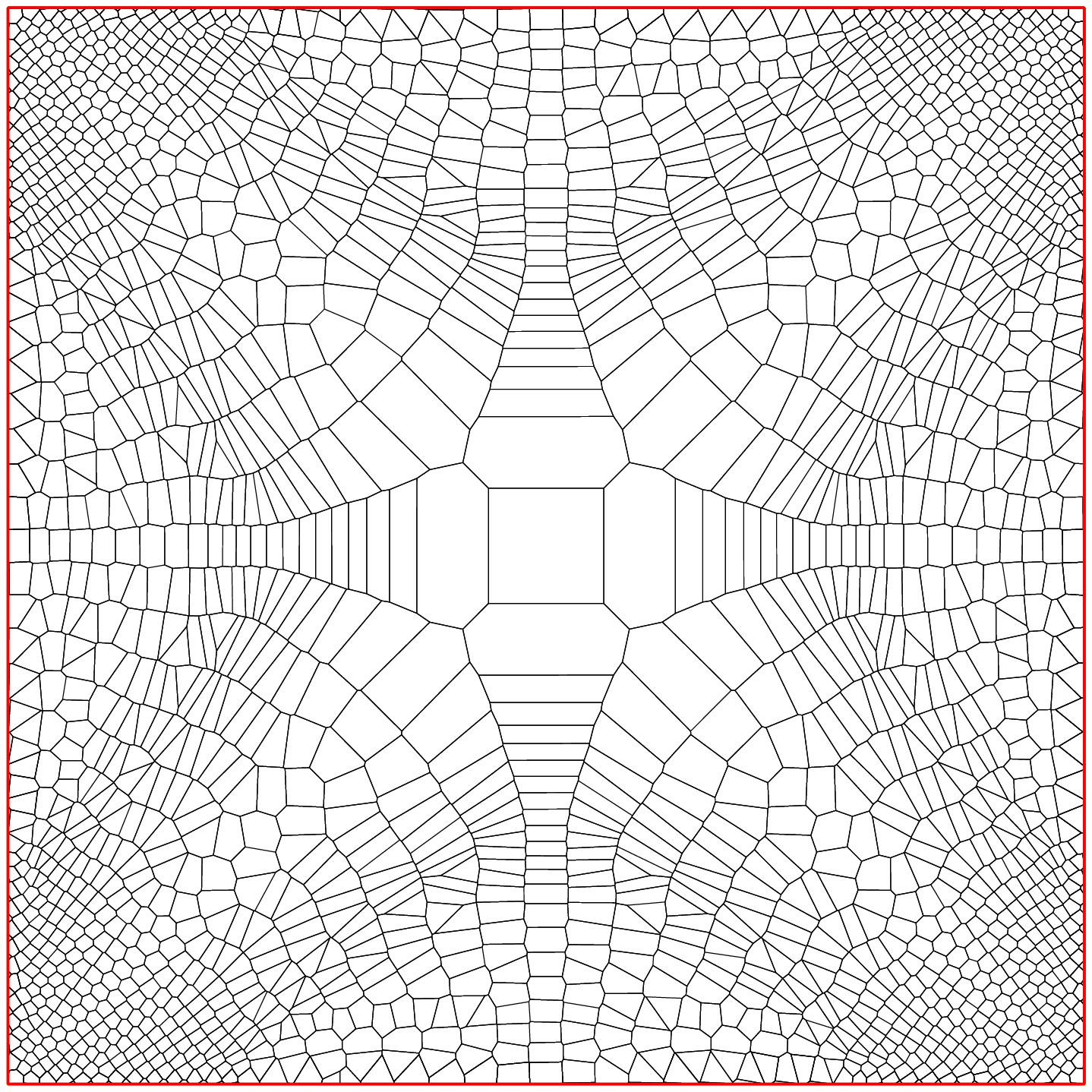}}
\resizebox{8.0cm}{!}{\includegraphics{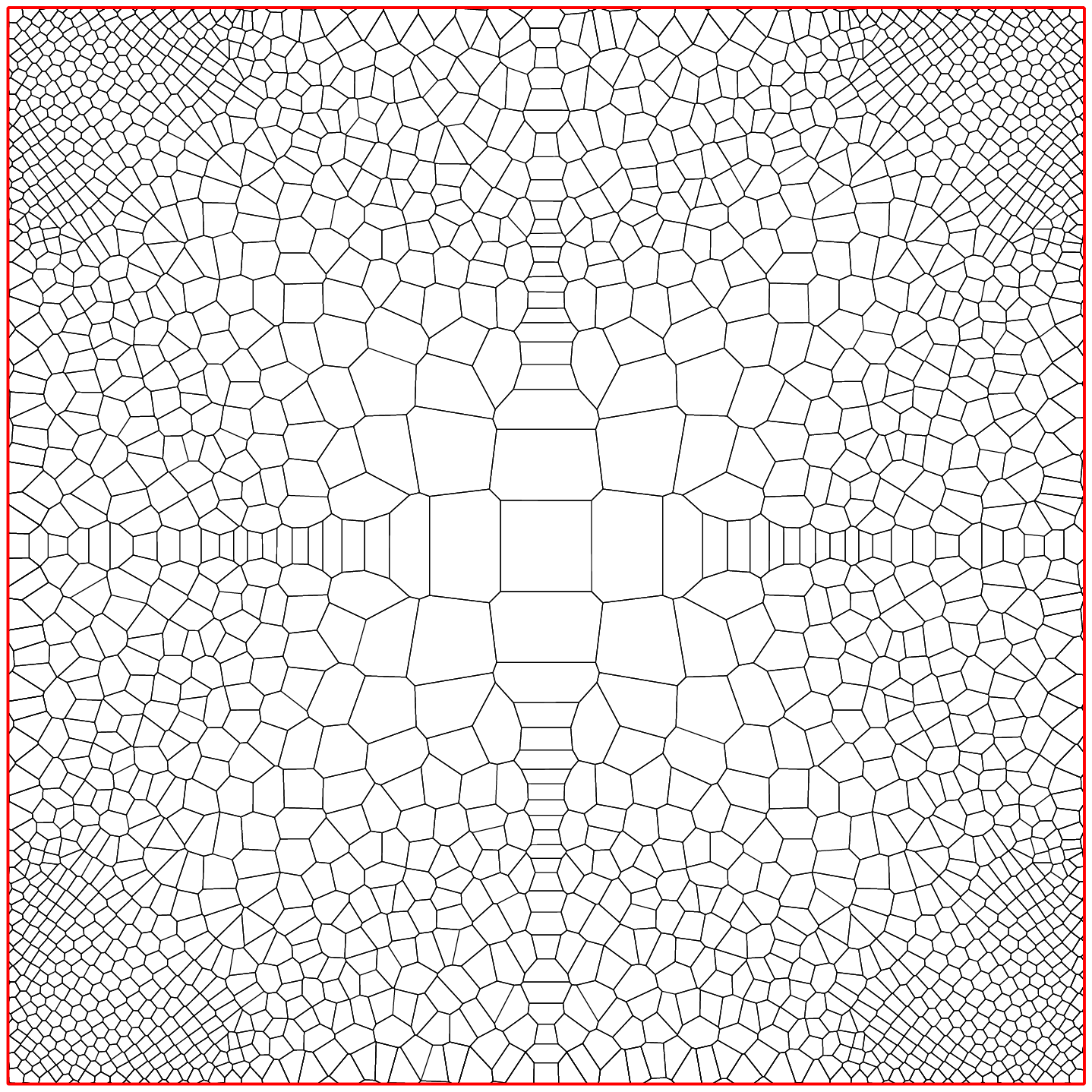}}
\caption{Effect of mesh regularization on the geometry of the Voronoi
  mesh. The panel on the left shows the Voronoi mesh obtained at
  $t=0.55$ for the Sedov-Taylor blast wave test of
  Figure~\ref{FigSedovEvol} when the mesh-generating points are only
  moved with the local gas velocity. While this produces a mesh well
  adjusted to the particular flow properties and symmetries of this
  problem, the high aspect ratio of some cells may be unfavourable in
  more general flows. The panel on the right shows the Voronoi mesh if
  we apply our standard mesh regularization procedure during the mesh
  motion. This tends to make the cells 'rounder' and more isotropic.
Note that the predicted density distributions of both simulations are
very similar.
\label{FigSedovMeshRegularity}}
\ec
\end{figure*}

\begin{figure*}
\bc
\resizebox{7cm}{!}{\includegraphics{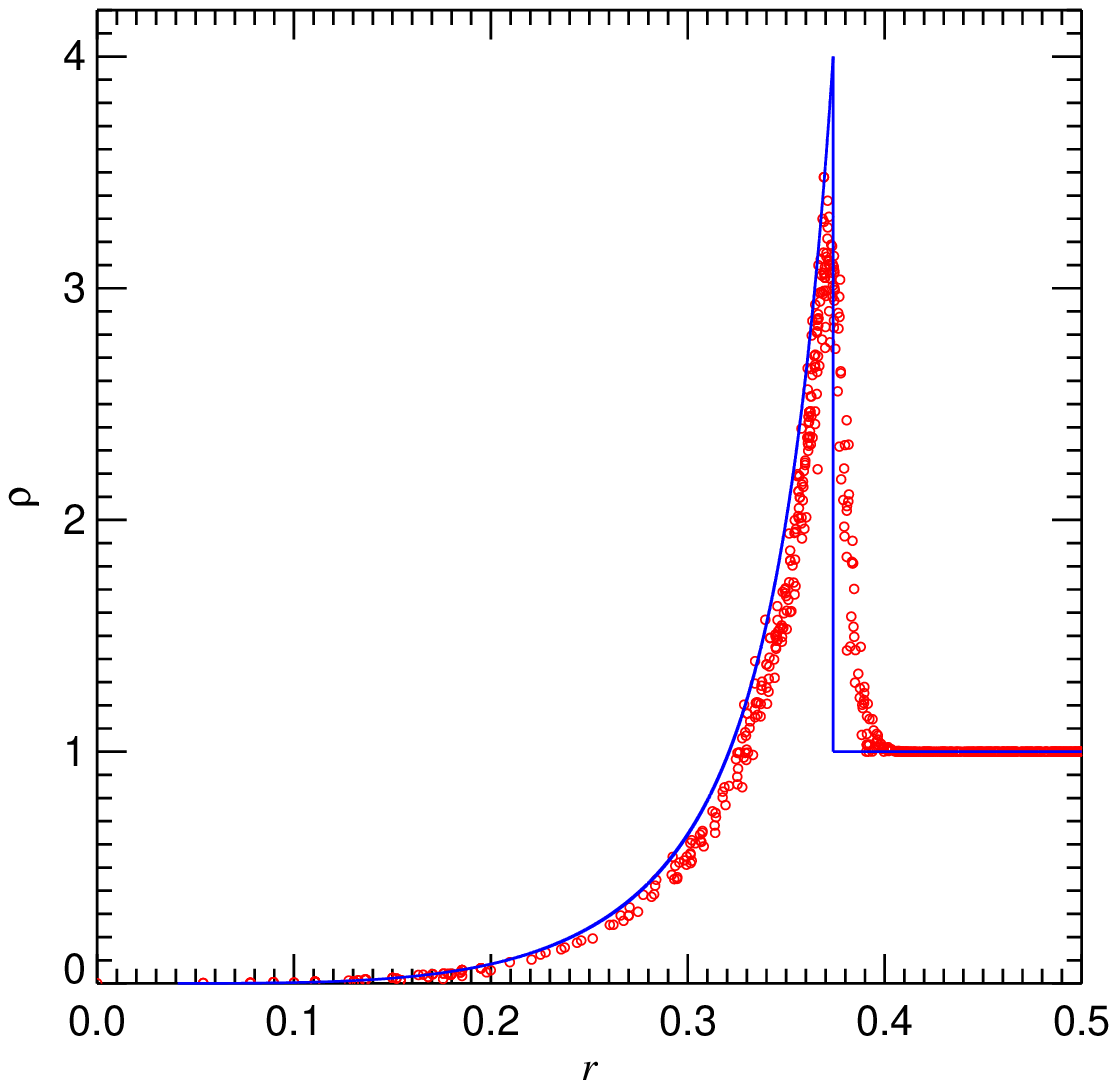}} %
\resizebox{7cm}{!}{\includegraphics{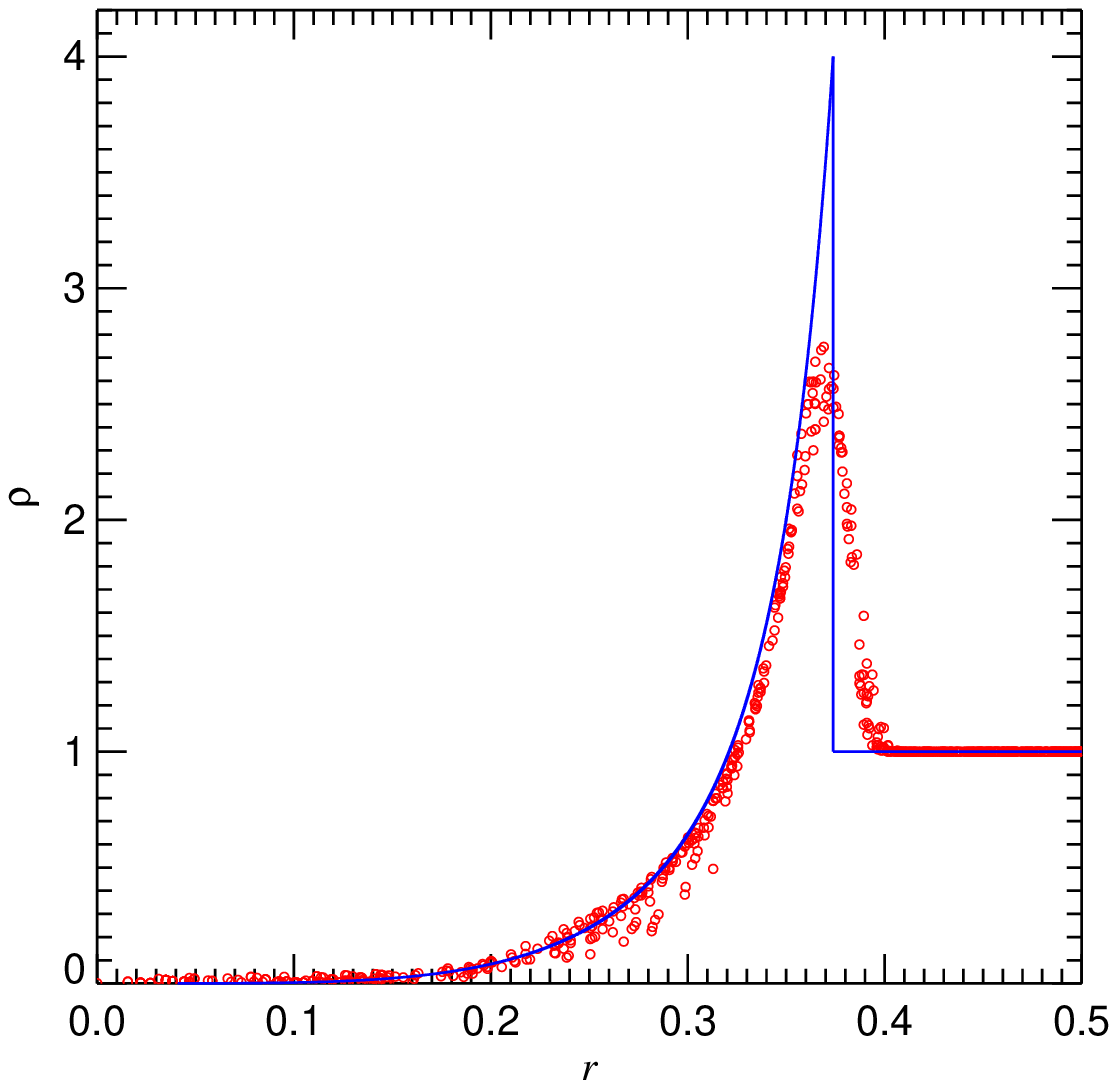}}
\caption{Density profile for a 3D Taylor-Sedov blast wave calculation
at time $t=0.06$. The initial resolution was $64^3$ cells, with all
the explosion energy injected into a single cell. We compare results
for our moving-mesh code (left panel) with the result obtained for a fixed
Cartesian mesh (right panel). The circles give the densities of individual cells,
which have been randomly subsampled by about $1/200$ to avoid too
strong crowding. 
\label{FigSedovProfile}}
\ec
\end{figure*}

\begin{figure*}
\bc
\resizebox{5.5cm}{!}{\includegraphics{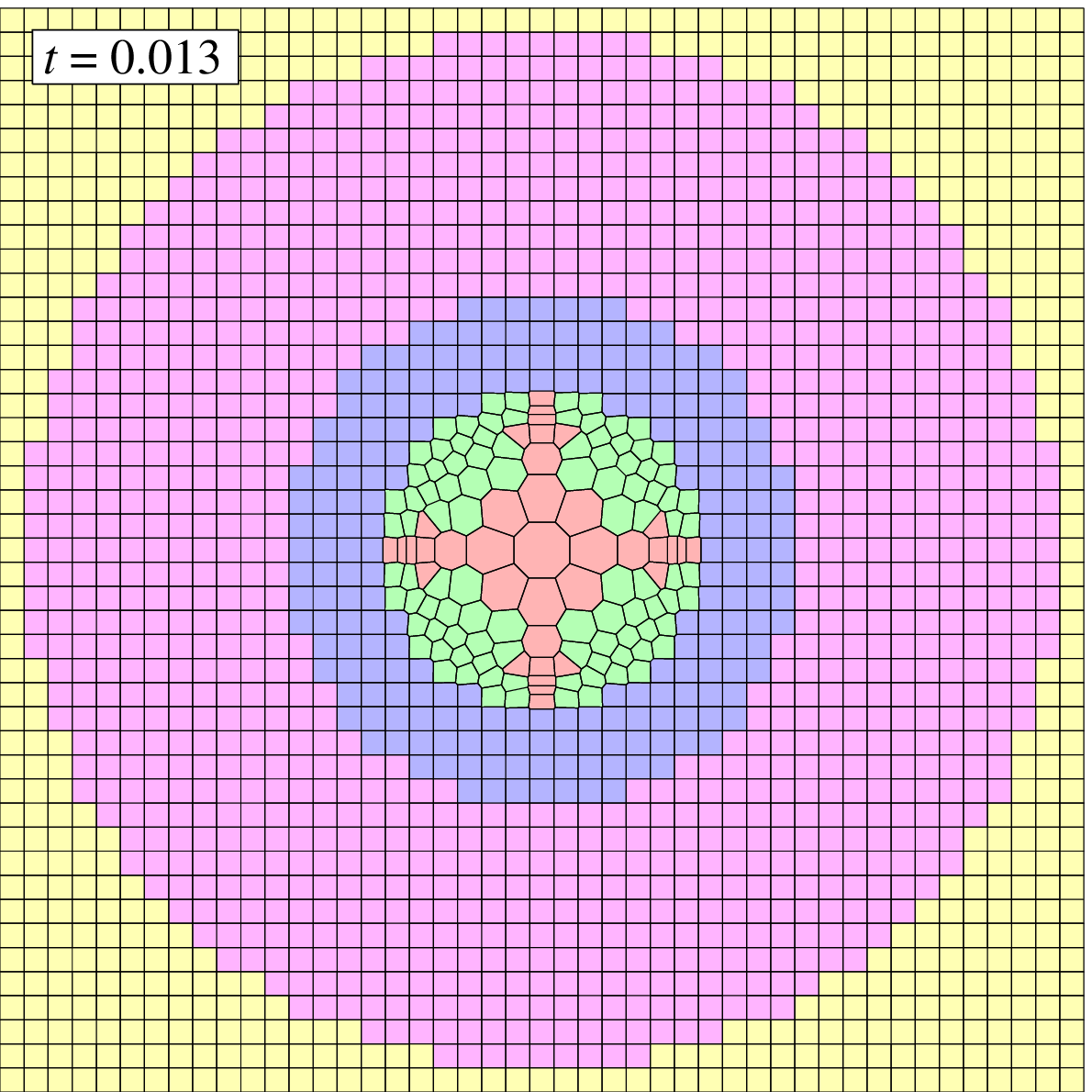}}\hspace*{0.3cm}%
\resizebox{5.5cm}{!}{\includegraphics{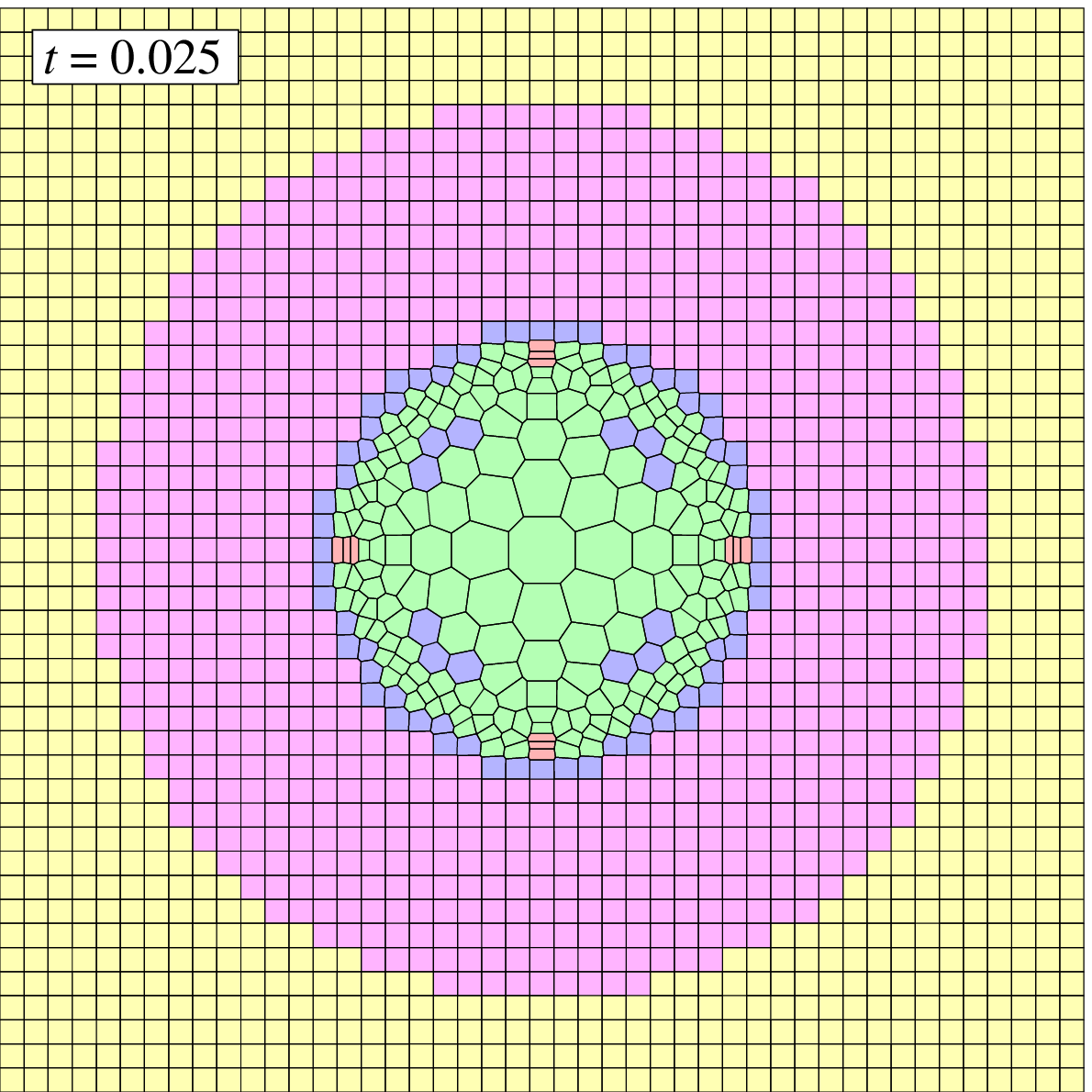}}\hspace*{0.3cm}%
\resizebox{5.5cm}{!}{\includegraphics{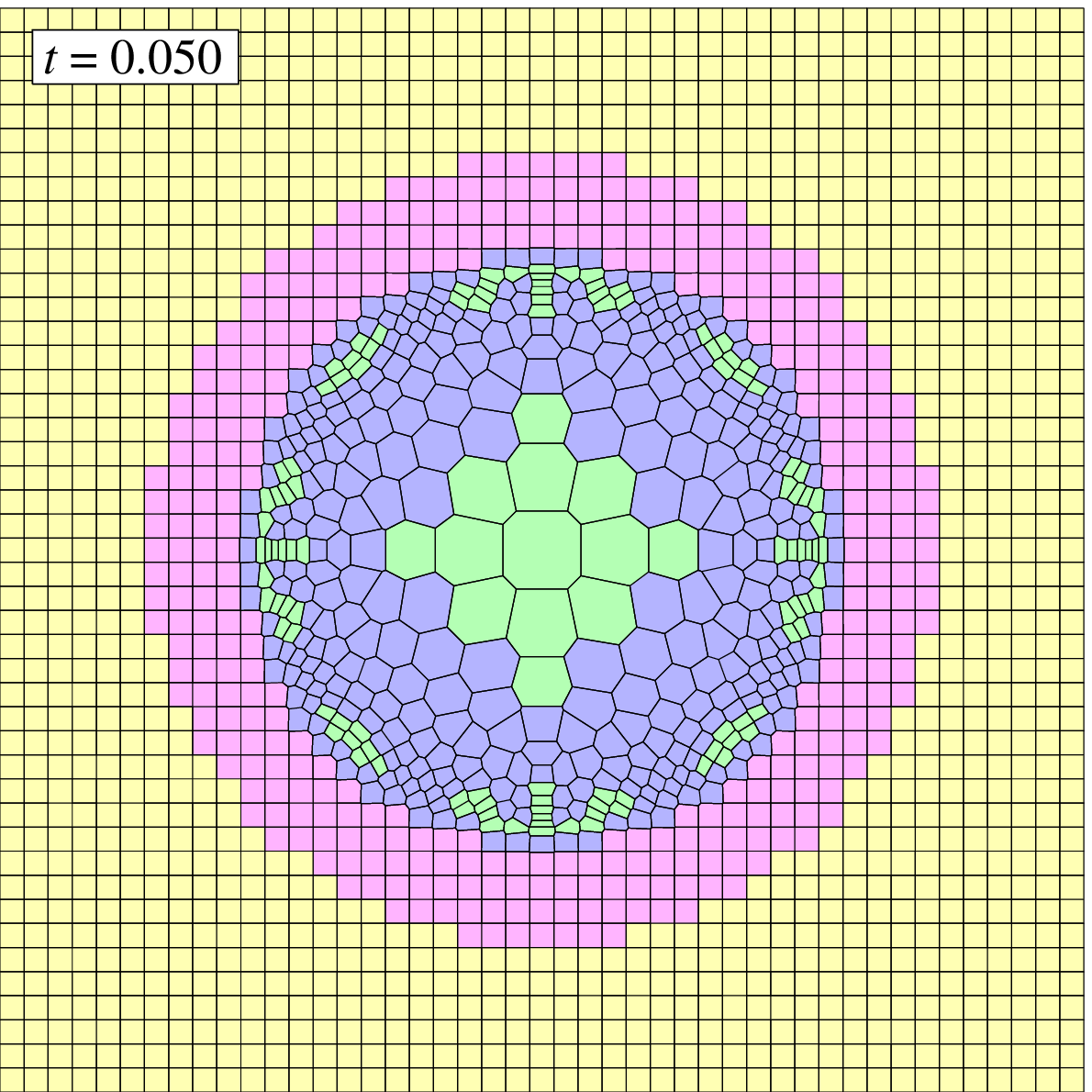}}\vspace*{0.3cm}\\
\vspace*{-0.2cm}\resizebox{4.0cm}{!}{\includegraphics{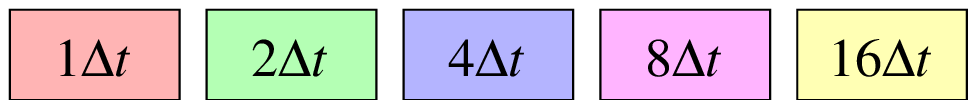}}
\caption{Spatial distribution of the sizes of individual timesteps in
  an integration of the two-dimensional Taylor-Sedov blast
  problem. The cells shaded in different colours are evolved with
  different timesteps in a power-of-two timestep hierarchy, as
  labeled. Fluxes between cells are always calculated on the smallest
  timestep of the two adjacent cells. Our tree-based approach to
  calculate the first possible arrival time of a signal from any other
  cell puts cells well ahead of the blast wave on sufficiently small
  timesteps.
\label{FigSedovTiSteps}}
\ec
\end{figure*}

\subsection{Interacting blast waves}

Another classic one-dimensional test problem is the interaction of two strong
blast waves, as introduced by \citet{Woodward1984}. Here a gas of density
$\rho=1$ with adiabatic index $\gamma=1.4$ in the domain $x\in [0,1]$ is
initially at rest. The pressure is set to $P=1000$ for $x<0.1$, to $P=100$ for
$x>0.9$ and to $P=0.01$ elsewhere. The boundary conditions are reflective on
both sides. The time evolution of this problem features multiple interactions
of strong shocks and rarefactions, which provides for a sensitive test of a
hydrodynamical code.

We follow \citet{Stone2008} and study a low resolution calculation of the
problem with 400 equally spaced points in the domain of width $L=1$. We
consider both a calculation with a fixed mesh, and one with a moving mesh; in
the latter case, the mesh-generating points are moved with the local flow
velocity, so that the calculation is effectively Lagrangian, and
mesh-regularization is carried out with $\eta=0.1$ and $\chi=1.0$. We use the
1D version of the code.  For comparison purposes, we also compute a
high-resolution result with a fixed mesh of 20000 cells, which serves as a
proxy for a nearly exact solution.

In Figure~\ref{FigDoubleBlast}, we show the density profile at time $t=0.038$,
at which point our results can also be compared with those of
\citet{Stone2008} and \citet{Woodward1984}. Our `Eulerian' fixed-mesh solution
is similar in quality to that obtained with {\small ATHENA} by
\citet{Stone2008}, except that it shows slightly more diffusion in the contact
discontinuities. This presumably reflects the benefits of the third-order
reconstruction that \citet{Stone2008} had used for this problem, while we have
only employed our standard second-order scheme. Nevertheless, both Eulerian
results show quite sizable smoothing of the contact discontinuities,
especially for the one at $x\simeq 0.6$. On the other hand, the moving-mesh
solution does much better in this respect. The deviations to the
high-resolution result are much smaller everywhere, and in particular, the
density maximum at $x\sim 0.8$ is recovered quite well and the discontinuities are
resolved sharply. For the same number of cells, the moving-mesh code therefore
clearly produces a more accurate solution. Similar to the simple shock tube
problem, we see that it is again the contact discontinuities that are improved
most.

\subsection{Point explosion}

In this test, we inject an energy $E$ into a point-like region in an initially
homogeneous cold gas of density $\rho$. This results in a spherical
Taylor-Sedov blast-wave, which has a well-known analytic self-similarity
solution \citep[e.g.][]{Landau1966}.  After a time $t$, the blast wave
propagates to a distance $r(t)= \beta ({E t^2}/{\rho})^{1/5}$, where the
constant $\beta$ depends on the adiabatic index $\gamma$ ($\beta \simeq 1.15$ for
$\gamma=5/3$ in 3D), $E$ is the explosion energy, and $\rho$ describes the
initial density of the ambient gas.  Directly at the spherical shock front,
the gas density jumps to a maximum compression of $\rho'/ \rho =
(\gamma+1)/(\gamma-1)$, with most of the mass inside the sphere being swept up
into a thin radial shell.  Behind the shock, the density rapidly declines and
ultimately vanishes towards the explosion centre.

We first consider the 2D case, which allows us to illustrate the mesh motion
in an easy way.  In Figure~\ref{FigSedovEvol}, we show the time evolution of
the density field with the mesh overlaid for a low-resolution calculation of
the blast wave problem. Initially, the mesh-generating points for a gas of
unit density are arranged in a $45\times 45$ Cartesian mesh, and an energy of
$E=1$ is injected into the central cell. The mesh is allowed to move with the
local flow velocity. We see that the propagation of the blast wave is
reflected in an evolving mesh geometry, with the smallest cells occurring where
the mass piles up behind the shock. Periodic boundary conditions are used in
this problem, so that the shock eventually collides with its mirror copies at
the boundaries of the box. This compresses most of the mass temporarily into
the corners of the box. However, the moving mesh algorithm can deal with this
gracefully and robustly.

Actually, for the simulation displayed in Figure~\ref{FigSedovEvol}, the mesh
was not moved just with the local flow velocity of each cell, but in addition
the correction scheme of equation (\ref{EqnShapeCorrVel}) was applied (using
$\eta=0.3$ and $\chi=1.0$), which tries to keep mesh cells round. The effect
of this can be seen in Figure~\ref{FigSedovMeshRegularity}, where the mesh
geometry at time $t=0.55$ is compared with (right) and without (left) any mesh
regularization. Clearly, when the cells are moved with the local flow
velocity alone, the mesh acquires some cells of quite high aspect ratio. Actually,
the shape of the cells tends to adapt to the local flow features, for example,
the cells become elongated parallel to the blast wave. This improves the
spatial resolution in the direction of propagation of the shock front, which
can be desirable in principle.  In fact, this automatic resolution adjustment
mimics attempts to make SPH more adaptive to local resolution requirements
with the help of anisotropic kernels \citep{Owen1998}.  However, for general
flow problems, we argue that it is safer and more robust to avoid high aspect
ratios, as one cannot rely on local symmetries for long, and the
next shock wave may strike from another random direction. Also, as we
discussed earlier, `roundish' cells offer the best accuracy for spatial
reconstruction and the treatment of self-gravity.

\begin{figure*}
\bc
\resizebox{5.5cm}{!}{\includegraphics{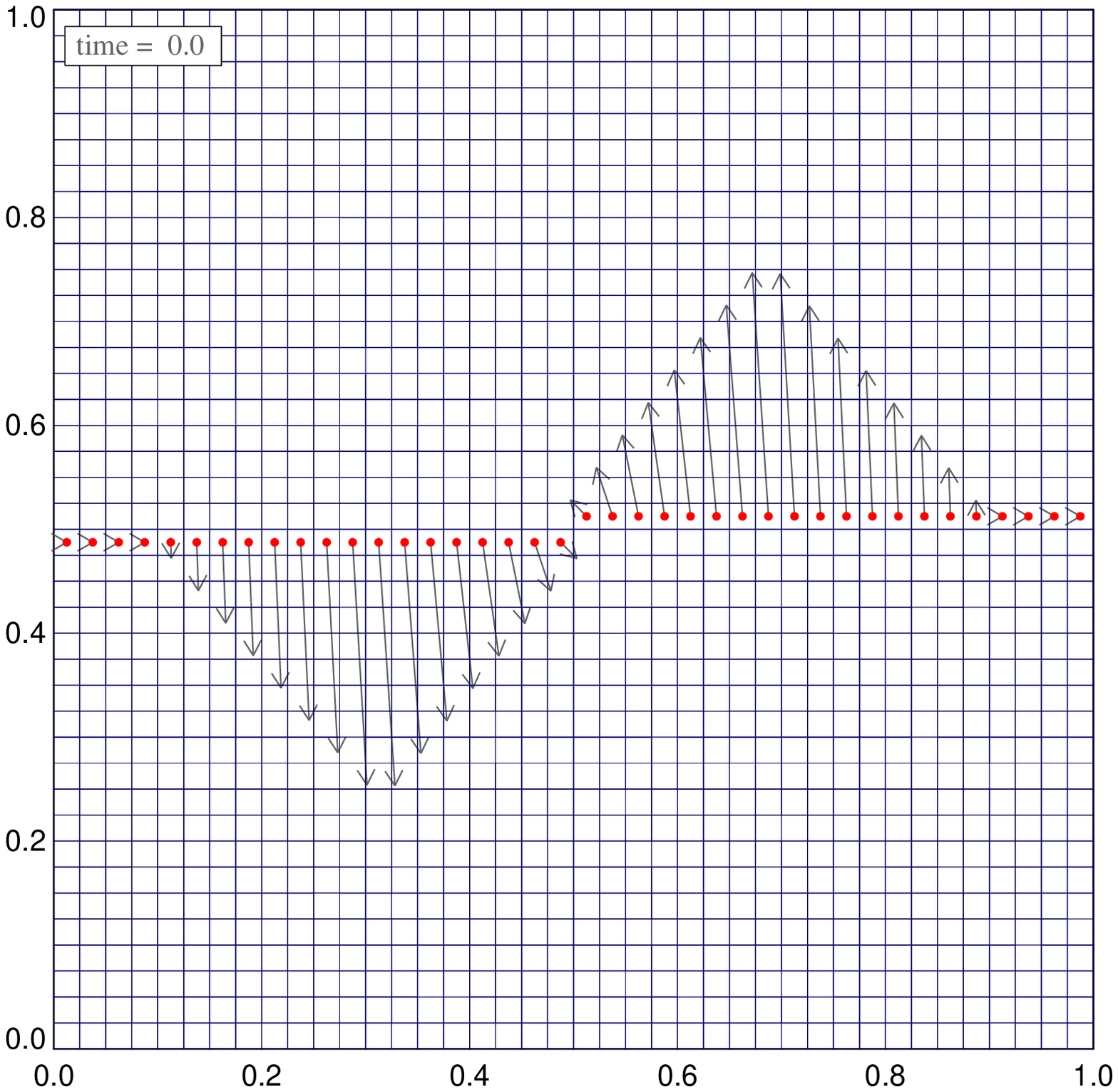}} %
\resizebox{5.5cm}{!}{\includegraphics{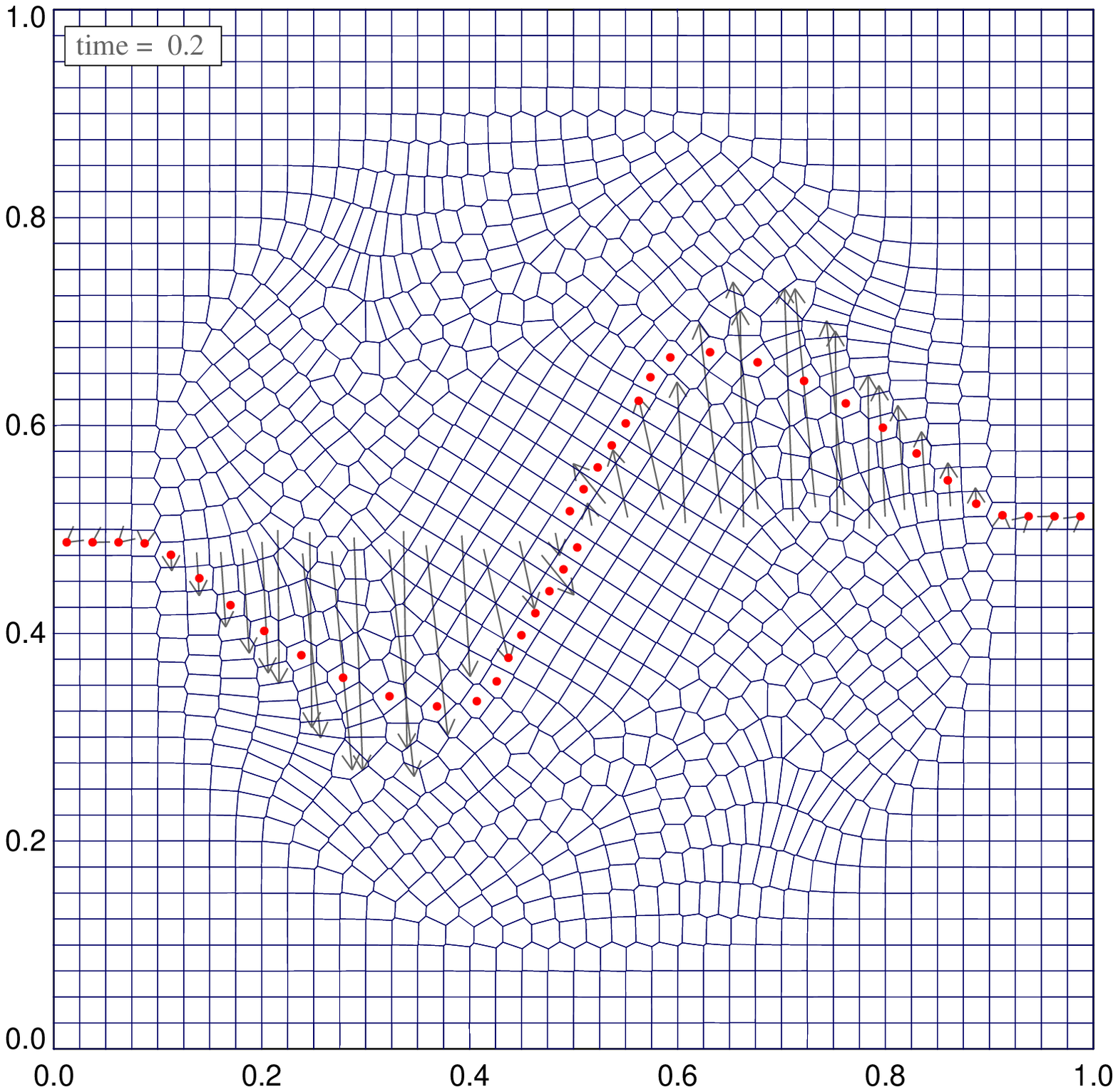}} %
\resizebox{5.5cm}{!}{\includegraphics{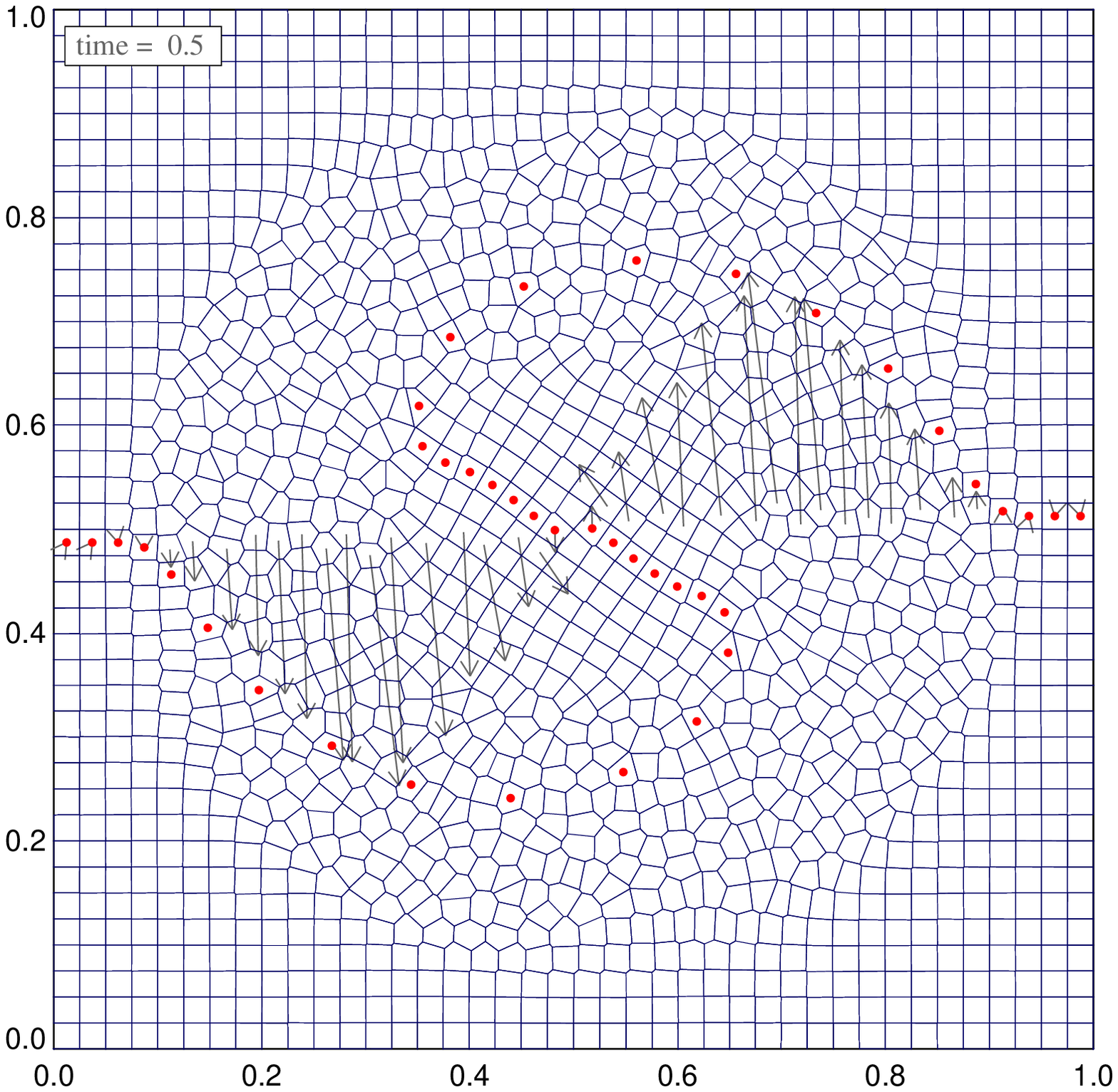}}\\
\resizebox{5.5cm}{!}{\includegraphics{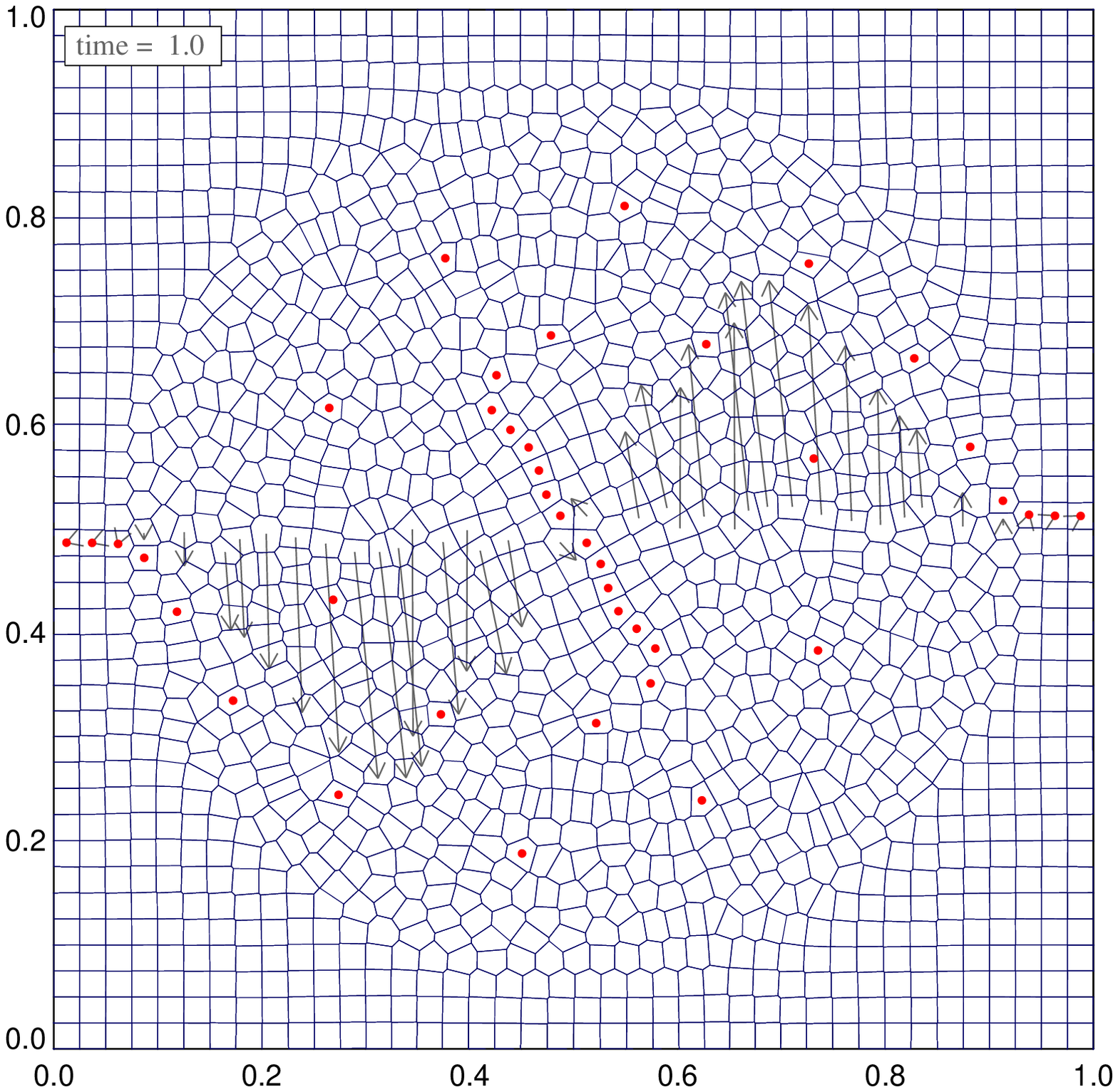}} %
\resizebox{5.5cm}{!}{\includegraphics{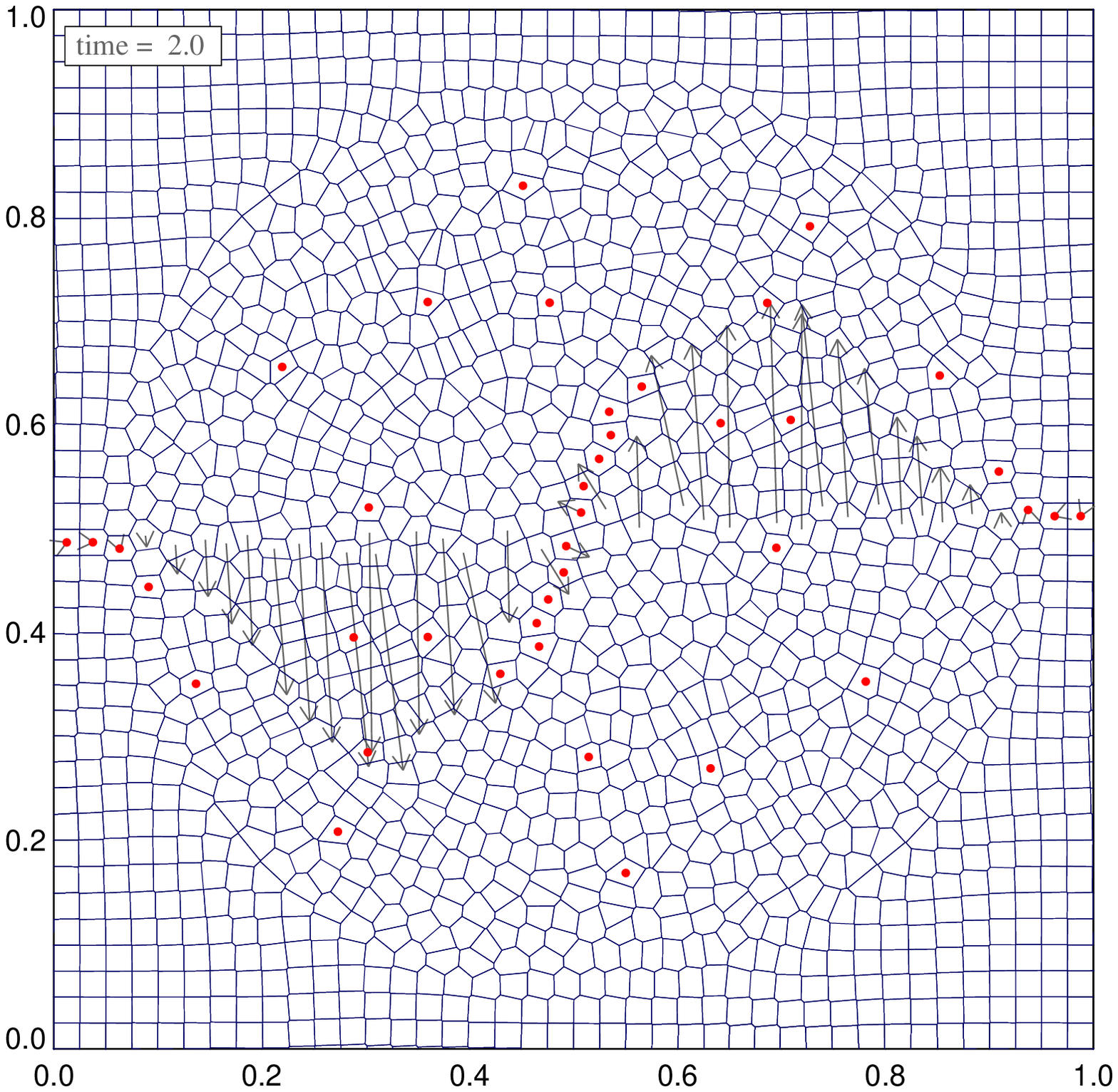}} %
\resizebox{5.5cm}{!}{\includegraphics{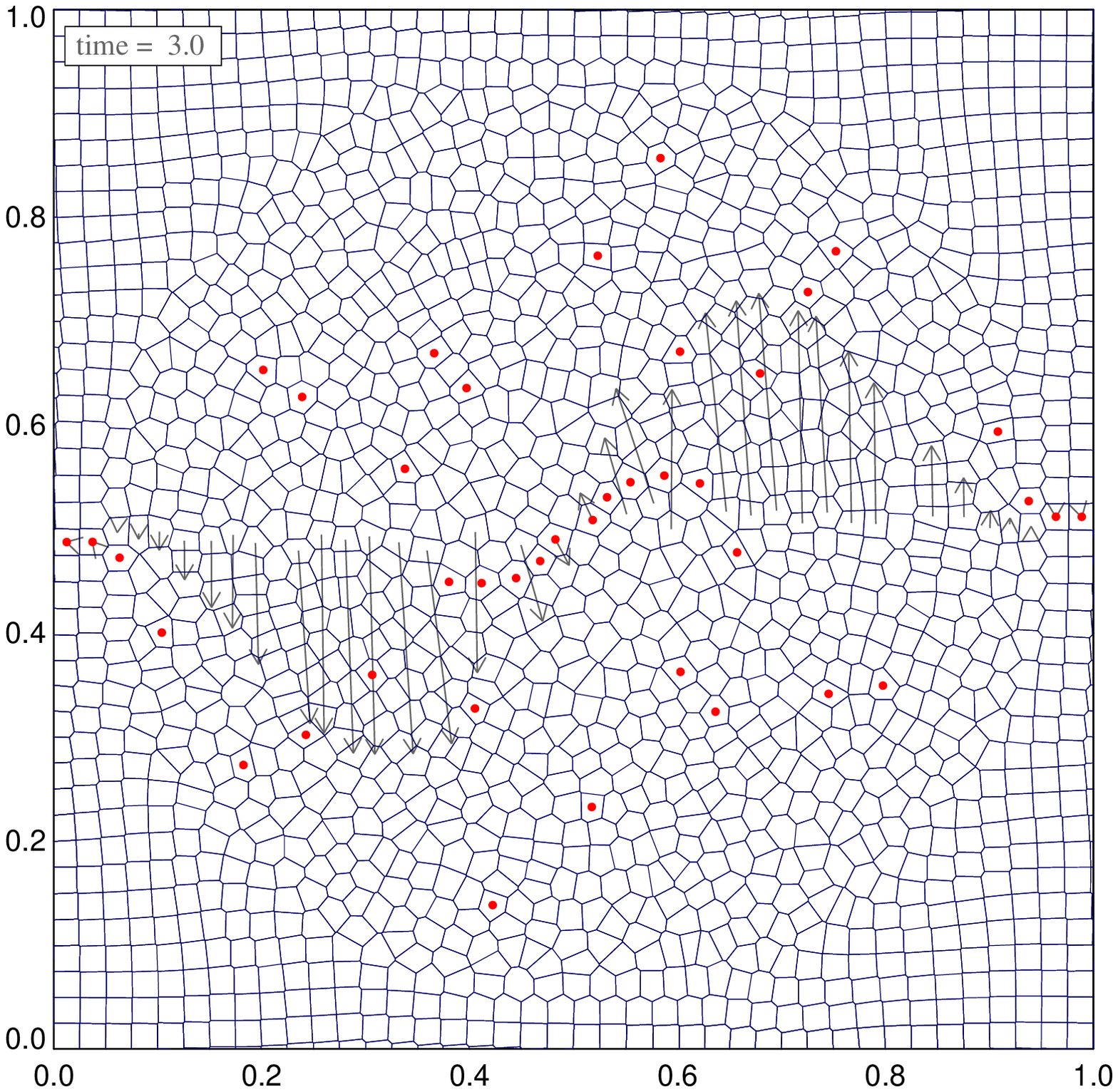}}\\
\caption{Mesh-motion in the `triangular' vortex problem of Gresho, calculated
with a $40\times 40$ grid of  particles. One horizontal row of particles in
the initial distribution is marked with circles, and the same cells are
labeled with circles in all subsequent frames. 
\label{FigGreshoMeshMotion}}
\ec
\end{figure*}

We now consider the accuracy of the shock front by comparing with the
analytic solution. In Figure~\ref{FigSedovProfile}, we compare the
densities of individual cells as a function of radial distance to the
explosion centre, both for the moving-mesh approach, and for the code
run with a fixed Cartesian mesh. The comparison is made at time
$t=0.06$, for an initial grid of $64^3$ cells, now in 3D. Clearly, the
moving-mesh approach resolves the sharp density spike of the blast wave
better, due to its improved spatial resolution in regions of high
density. It also shows slightly weaker deviations from spherical
symmetry at $r\sim 0.25$ compared with the Cartesian grid.  There is a
small phase error in the sense that the numerical simulation appears
slightly more evolved than the analytical solution; the origin of this
lies in the poorly resolved early phase of the point explosion. At later
times, or for better resolution (which is essentially the same for this
self-similar problem), this error becomes ever smaller.  We note that
\citet{Feng2004} give $256^3$ results for their WENO solver, which
curiously look somewhat worse than our results here despite their higher
mesh resolution.

Finally, in Figure~\ref{FigSedovTiSteps} we illustrate the behaviour
of our individual timestep integration scheme for the 2D Taylor-Sedov
blast wave problem. We show the mesh at three different times
(corresponding to the first three panels shown in
Fig.~\ref{FigSedovEvol}), with each cell shaded according to its
assigned timestep. Far away from the explosion site, the allowed
timesteps are significantly larger than close to the shock wave and in
the heated central bubble. The timesteps are restricted in a sequence
of spherical shells even ahead of the shock, such that the arriving
shock wave is guaranteed to be integrated accurately in time, even
though the cold gas far away can be integrated on timesteps that can
in principle be orders of magnitude larger. This choice of timesteps
is made possible by our tree-based scheme to estimate the earliest
possible arrival time for every cell of a signal from any other cell.

We note that the results of the individual timestep scheme are
essentially indistinguishable from a fixed timestep integration, but
require significantly less computational effort.  Compared to the
equivalent calculation with a global timestep (set equal to the
minimum of the local timestep constraint of all cells), 4.3 times
fewer flux computations and Riemann problems have to be calculated
over the course of a calculation from $t=0$ to $t=0.1$. For higher
resolution or in 3D, the saving would be still larger. In fact, many
physical applications in cosmic structure formation feature such a
large dynamic range in timescales that individual timesteps are
mandatory to make large simulations still tractable.

\begin{figure*}
\bc
\resizebox{5.5cm}{!}{\includegraphics{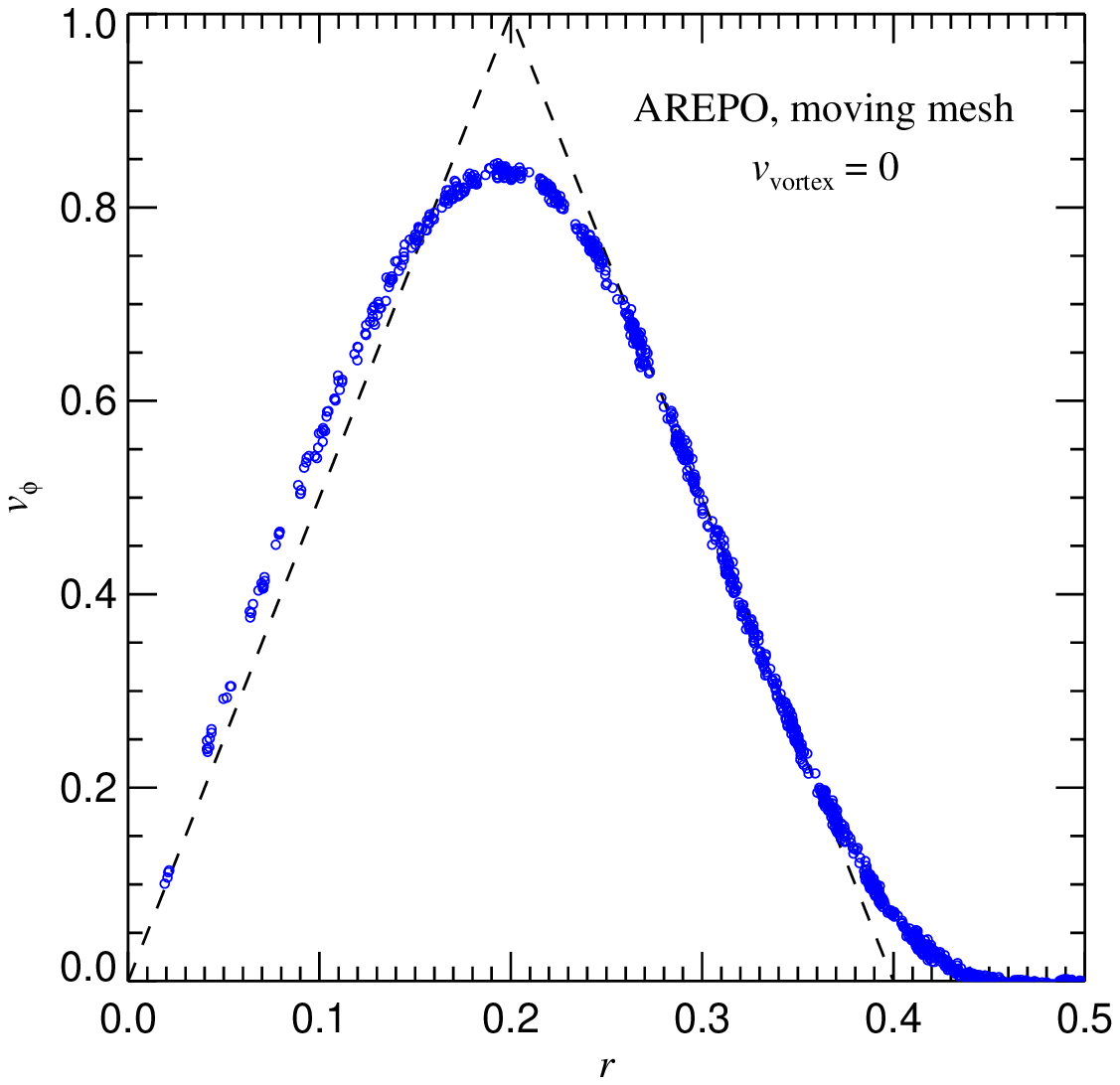}} %
\resizebox{5.5cm}{!}{\includegraphics{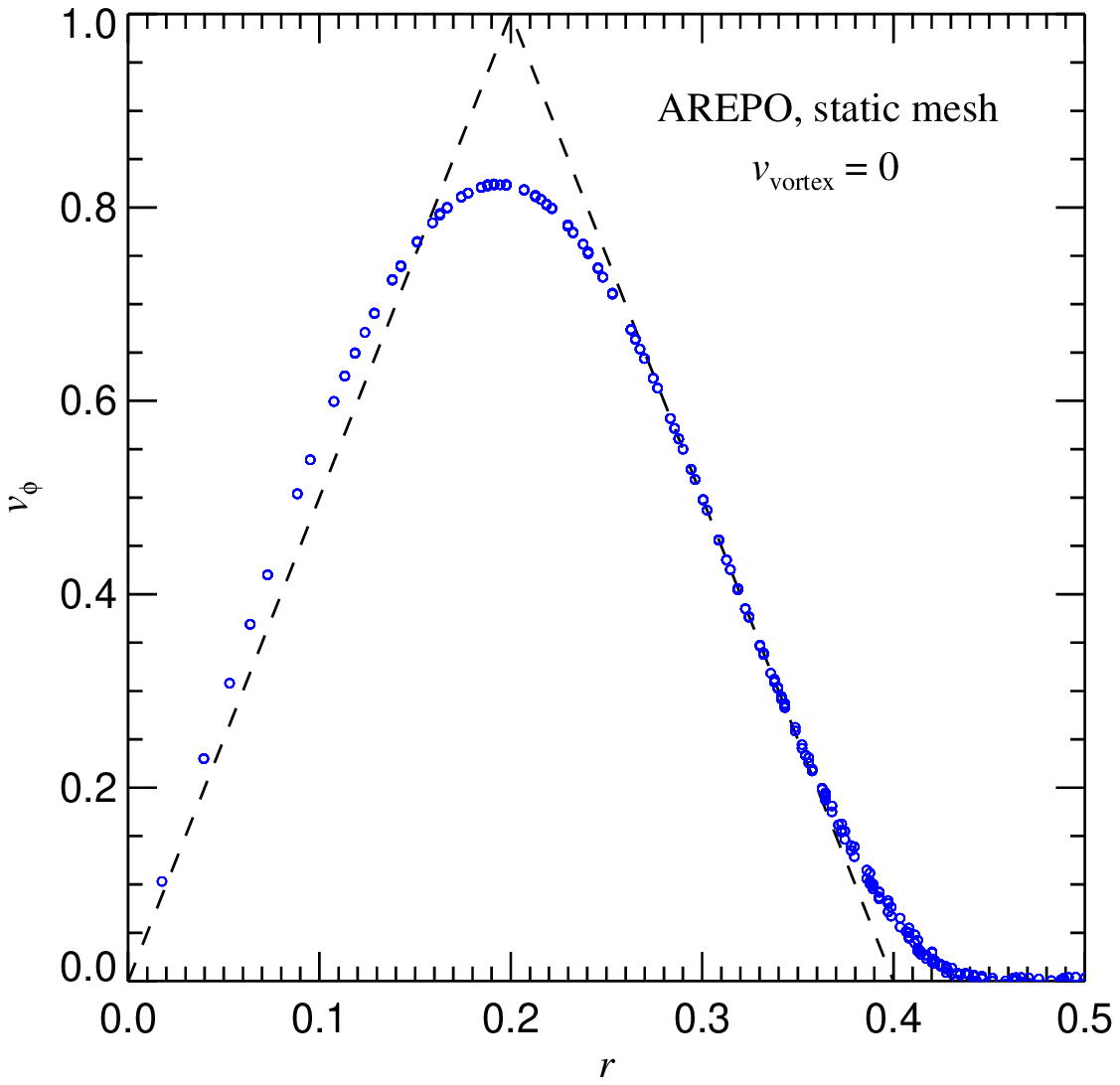}} %
\resizebox{5.5cm}{!}{\includegraphics{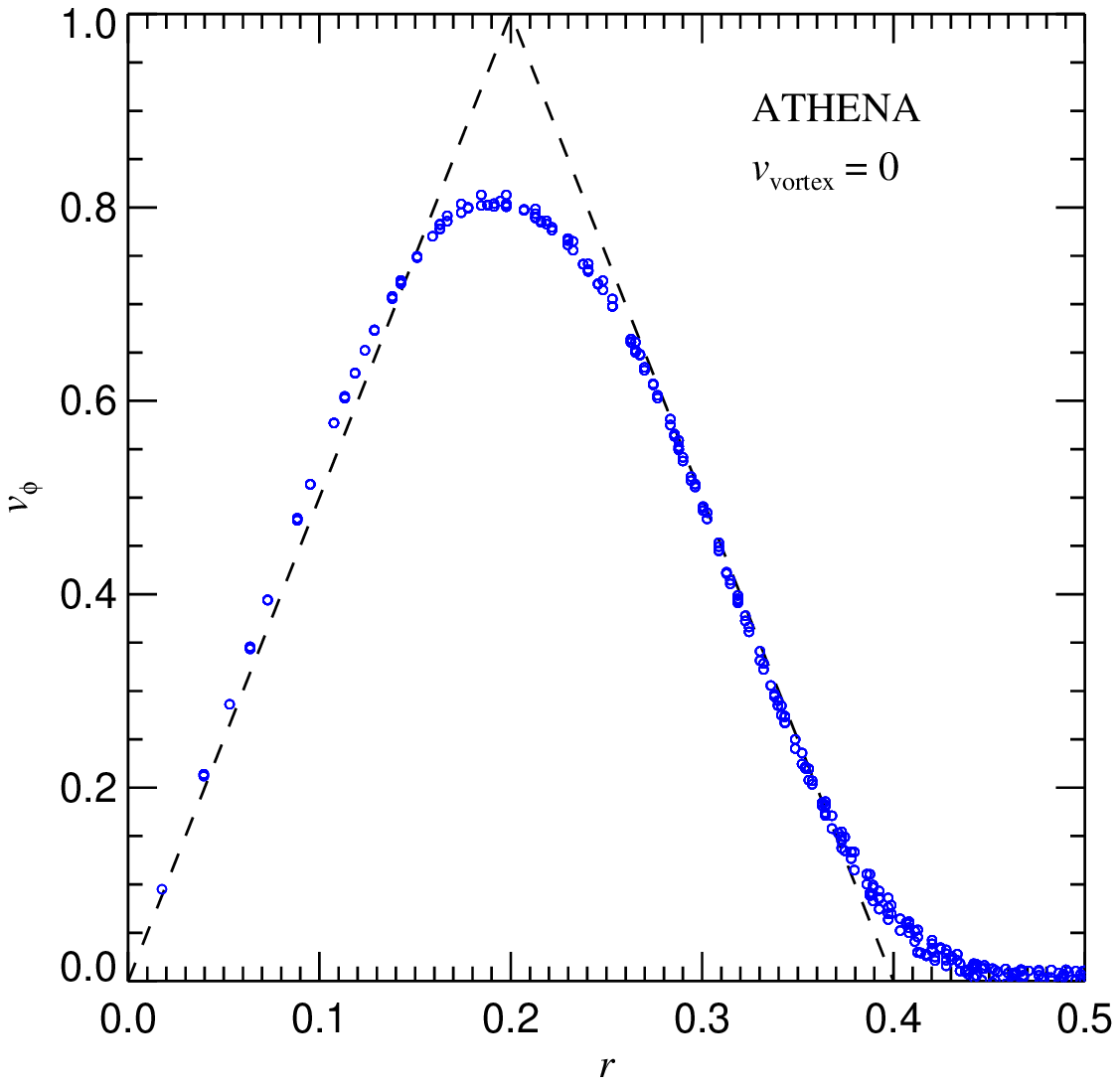}}\\
\resizebox{5.5cm}{!}{\includegraphics{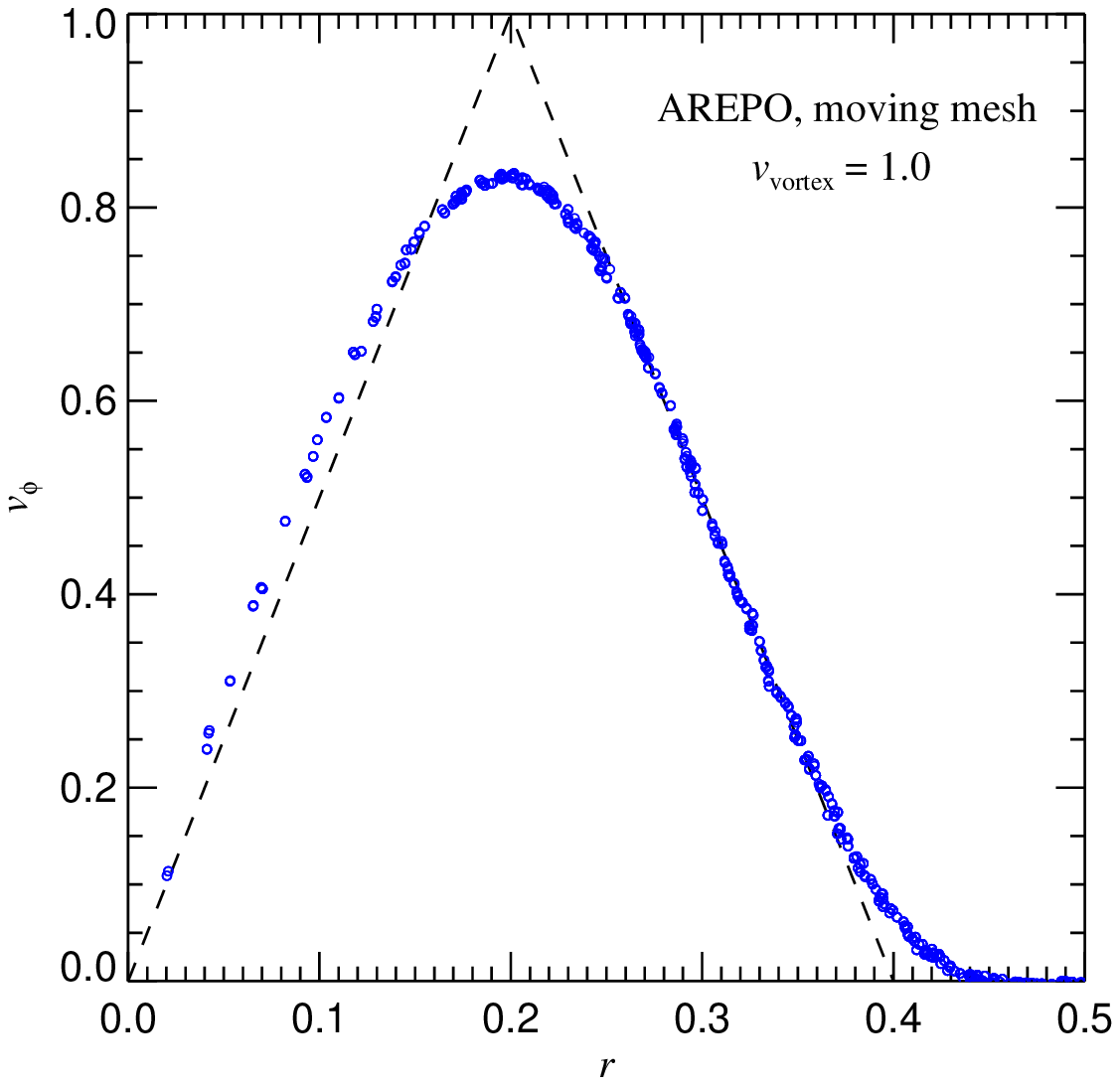}} %
\resizebox{5.5cm}{!}{\includegraphics{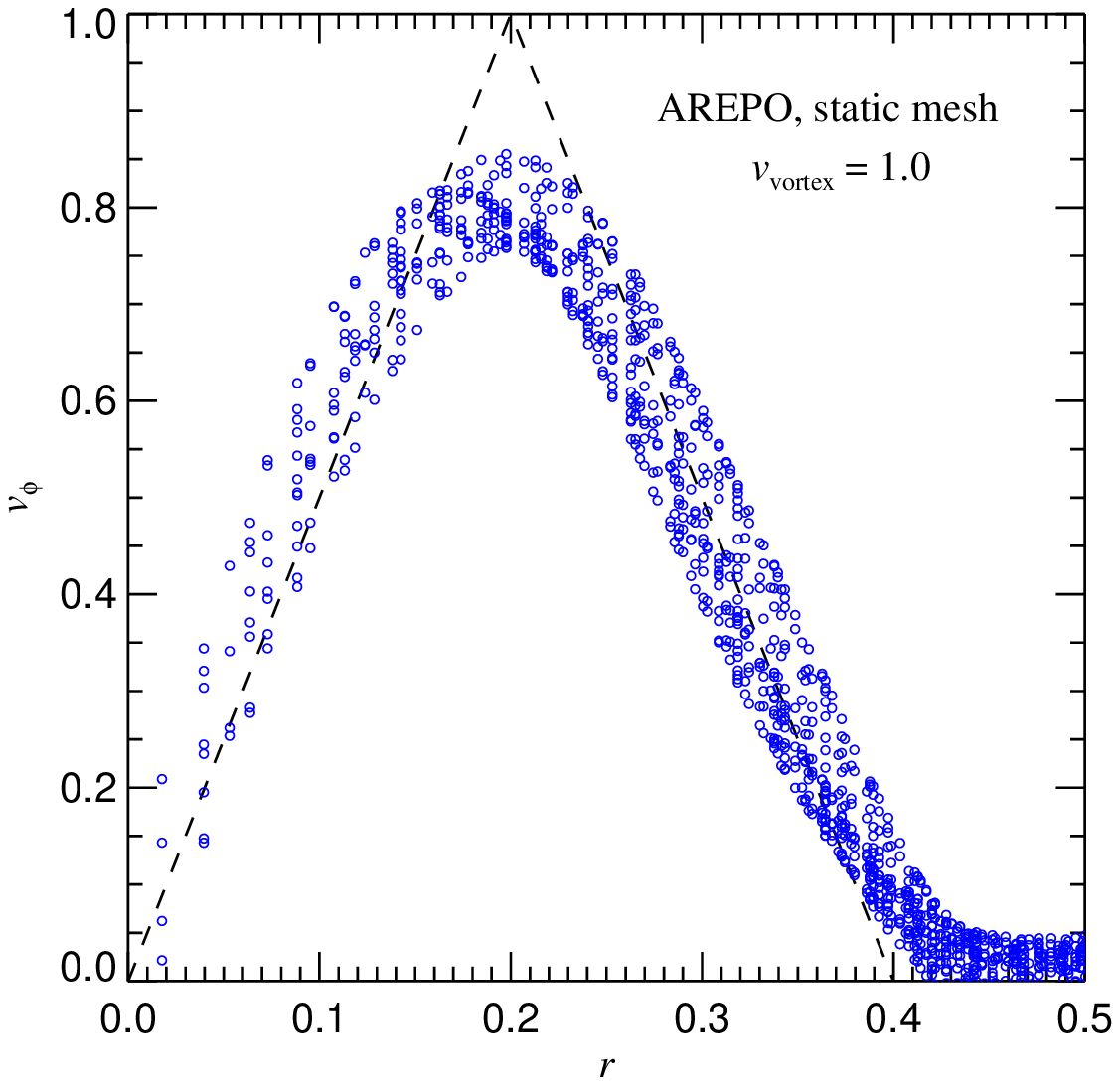}} %
\resizebox{5.5cm}{!}{\includegraphics{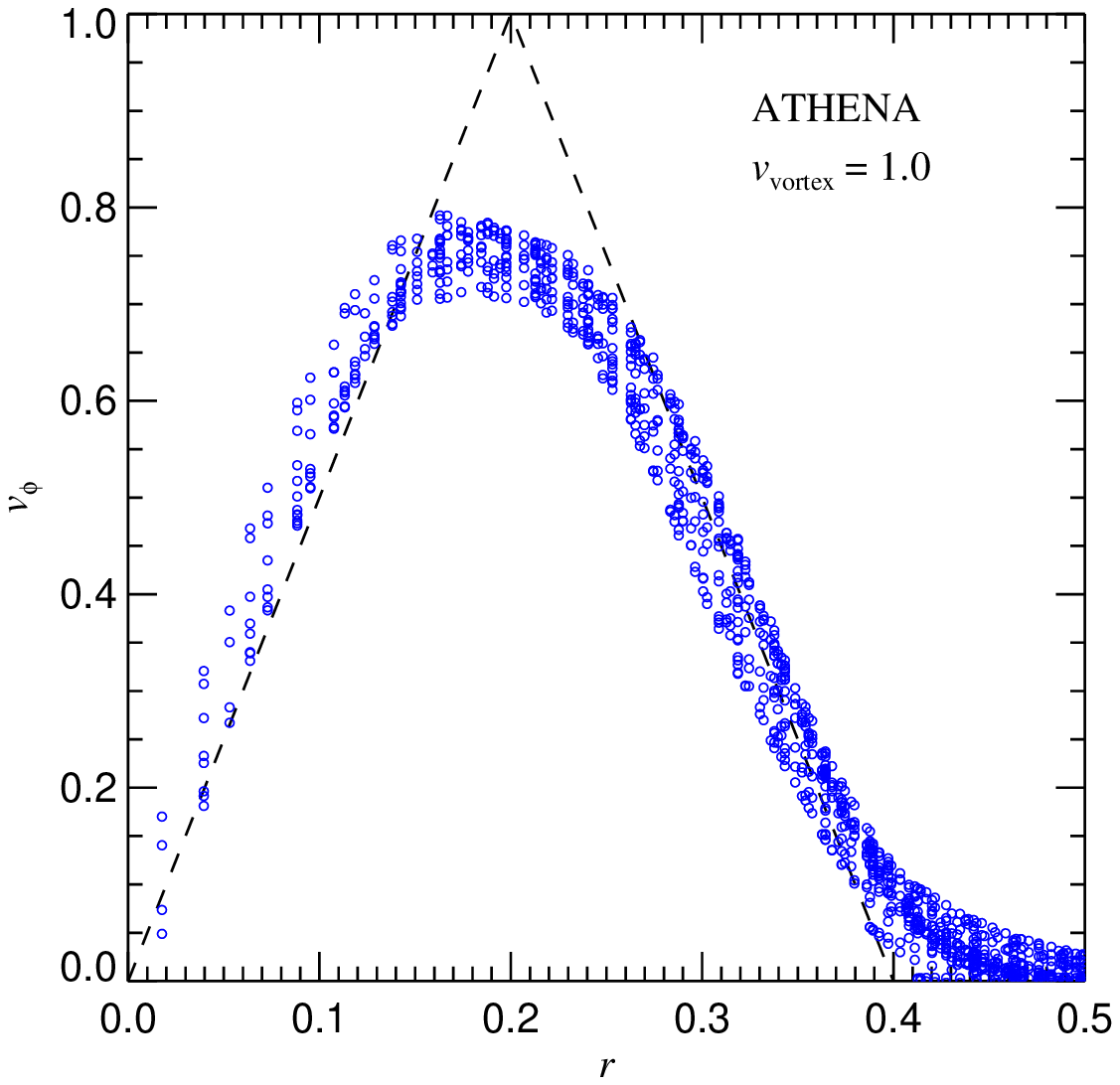}}\\
\resizebox{5.5cm}{!}{\includegraphics{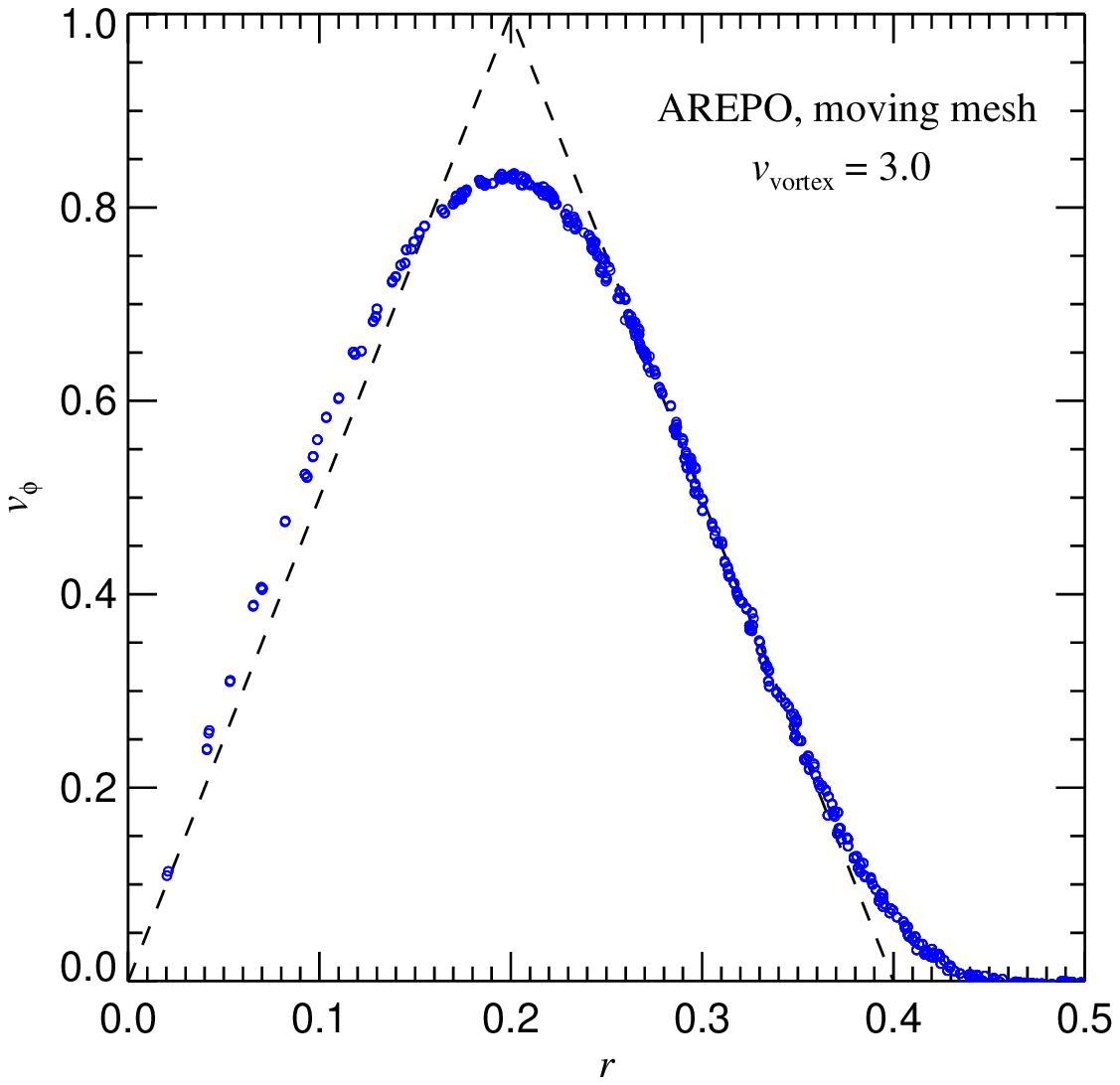}} %
\resizebox{5.5cm}{!}{\includegraphics{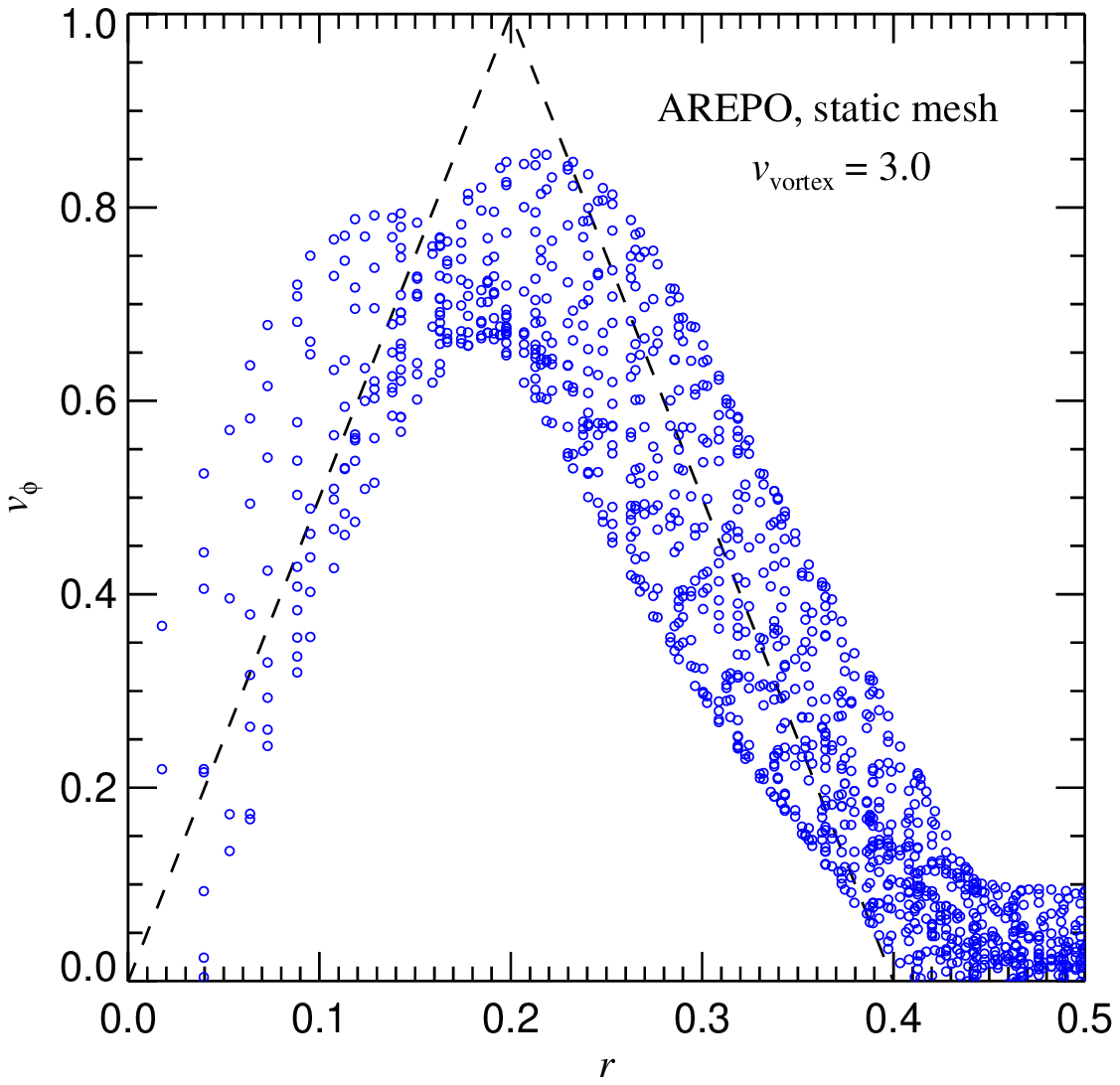}} %
\resizebox{5.5cm}{!}{\includegraphics{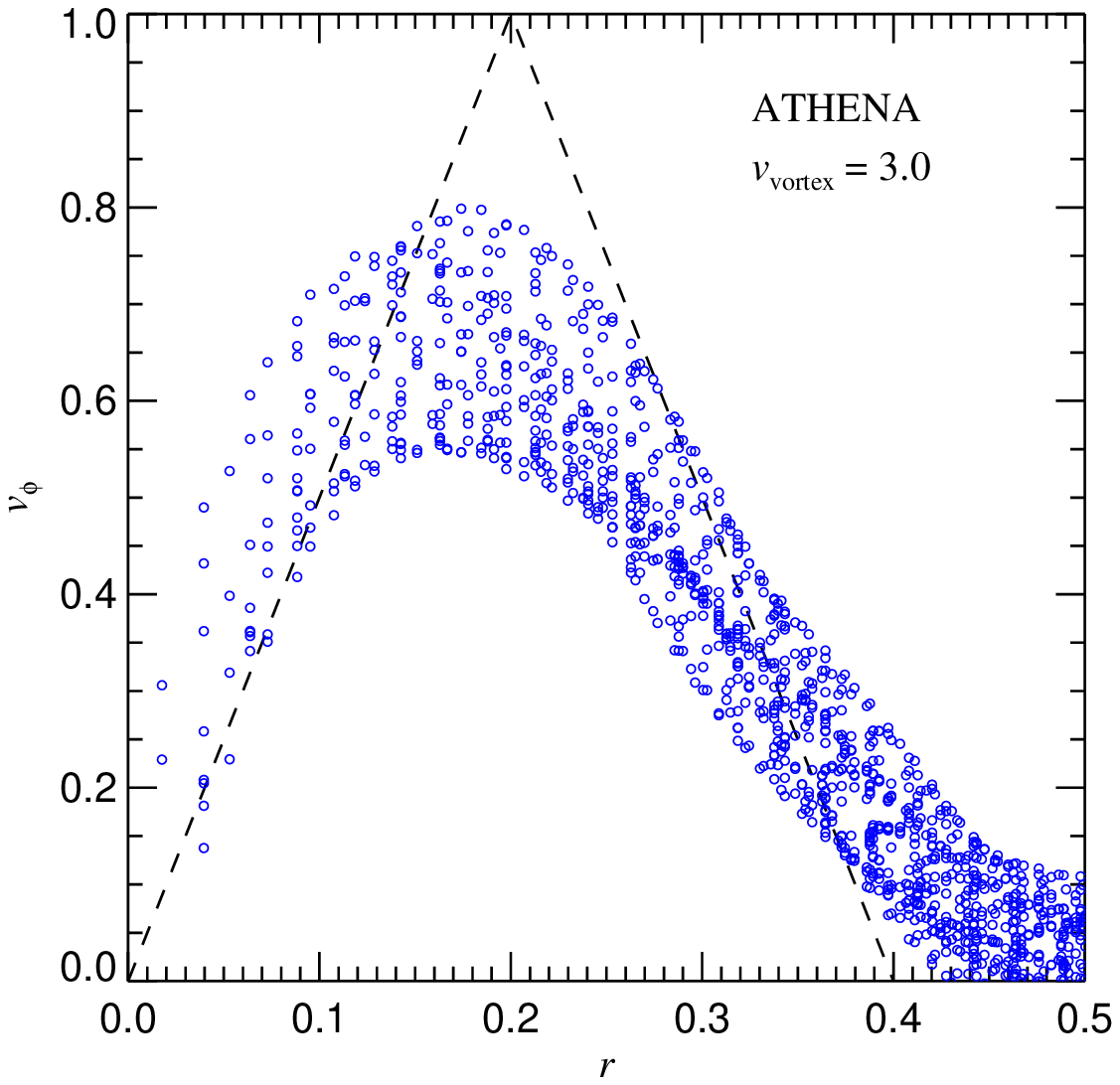}}\\
\caption{Azimuthal velocity profiles at time $t=3.0$ for Gresho's `triangular'
  vortex problem when calculated with different codes and for different bulk
  velocities of the vortex. In the panels of the top row, the vortex is
  stationary, while in the middle row it moves with a speed $v_x=1.0$, and in the
  bottom row with $v_x=3.0$, which is comparable to the speed-of-sound of the
  gas. We compare results calculated with our {\small AREPO} code, both for a
  moving mesh and a fixed Cartesian mesh, with those obtained using the {\small ATHENA} code
  \citep{Stone2008}.  For $v_x=0.0$, all three methods produce results of comparable
  quality. However, if the vortex is non-stationary, the Eulerian approaches
  develop significant asymmetries and show elevated diffusivity and angular
  momentum transport due to the increase in advection errors. 
  In contrast, the Lagrangian moving-mesh result is invariant when the vortex is set in motion.
  \label{FigGreshoProfiles}}
\ec
\end{figure*}

\subsection{The Gresho vortex problem}

An interesting test for the conservation of vorticity and angular momentum is
provided by the `triangle vortex' problem of \cite{Gresho1990}, which we apply
here to the Euler equations in 2D, following \citet{Liska2003}. The 
vortex is described by an azimuthal velocity profile
\begin{equation}
v_\phi(r) = \left\{
\begin{array}{ll}
5r & {\rm for}\;\;\; 0\le r <0.2 \\
2-5r & {\rm for}\;\;\; 0.2\le r <0.4 \\
0 & {\rm for}\;\;\; r \ge 0.4 \\
\end{array}
\right.
\end{equation}
in a gas of constant density equal to $\rho=1$.
Thanks to a suitable pressure profile \citep{Liska2003} of the form
\begin{equation}
P(r) = \left\{
\begin{array}{ll}
5 + 25/2 r^2 & {\rm for}\;\;\; 0\le r <0.2 \\
9 + 25/2 r^2 - \\\;\;\;\;\;\;20 r + 4 \ln (r/0.2) & {\rm for}\;\;\; 0.2\le r <0.4 \\
3+4\ln 2 & {\rm for}\;\;\; r \ge 0.4 \\
\end{array}
\right.
\end{equation}
the centrifugal force is balanced by the pressure gradient and the vortex becomes
independent of time. Note that in principle an arbitrary constant pressure
could be added to the pressure profile.

We shall consider three different variants of this test. In the first, the
vortex is at rest in the calculational frame, which we describe by $40\times
40$ cells in the unit domain at our default resolution (arranged initially as
a Cartesian mesh). In the second and third variants, we follow
\citet{Liska2003} and let the vortex move with a constant velocity $v_{\rm
  vortex}$ along the positive $x$-direction, i.e.~all the gas gets an
additional velocity component of $\Delta v_x= v_{\rm vortex}$. We consider the
choices $v_{\rm vortex}=1$ and $v_{\rm vortex}=3$, and use periodic boundary
conditions, such that the vortex moves 3 and 9 times, respectively, through
the box over the simulated time span of $t=3$ time units. The additional gas
motion in the $v_{\rm vortex}>0$ case makes the problem more difficult for the
Eulerian approach because it becomes more demanding to advect the gas
accurately over the grid.  In all cases, we run the problem for $t=3$ time
units, and then compare the azimuthal velocity profiles of the final with the
initial state.

Before we discuss these results, we first illustrate in
Figure~\ref{FigGreshoMeshMotion} the time evolution of the mesh geometry
produced by our moving-mesh code in the stationary vortex case. In order to
guide the eye and to show the motion of individual mesh cells, a horizontal
row of cells has been marked with circles in the initial conditions, and the
same cells are then labeled again in all subsequent time frames. Also, for a
vertical strip of cells, velocity vectors are added in the plots at each
output time. It is nicely seen how the central region of the mesh accurately
follows a solid body rotation in the early evolution, and how it is surrounded
by an outer shell that exhibits strong shear. However, there is no
pathological mesh twisting or tangling due to this shear. Rather, the Voronoi
mesh transforms its geometry continuously, and changes the local neighbourhood
relations between cells in a smooth fashion. As time goes by, the initial
symmetry in the mesh geometry slightly deteriorates, but the mesh motion stays
nicely regular.

In Figure~\ref{FigGreshoProfiles}, we compare the results for the azimuthal
velocity profiles at the final time of all of our runs. In the top row of
panels, we show calculations where the vortex was stationary relative to the
computational frame. We give results obtained with our new code both with a
fixed mesh and with a moving mesh, based on identical initial conditions.  To
compare our results with another high-accuracy Eulerian code, we have also
computed this problem with the publicly available code {\small ATHENA} by
\citet{Stone2008}. For the latter, we used second-order spatial reconstruction
and the Roe solver.  After 3 time units, all three codes show some significant
smoothing of the initial velocity profile, but they do not differ strongly in
the quality of the results.

\begin{figure}
\bc
\resizebox{7.8cm}{!}{\includegraphics{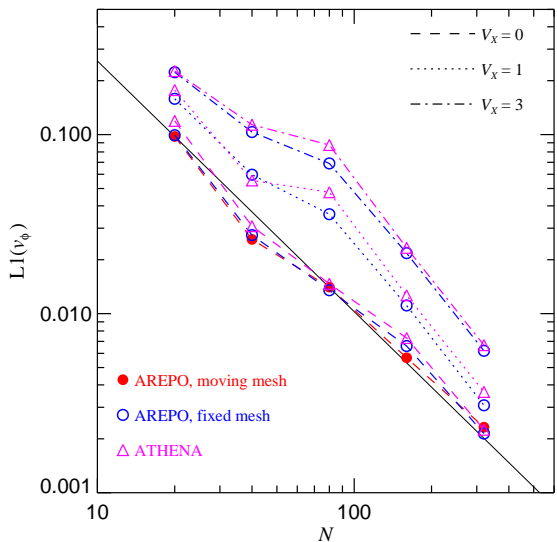}}
\caption{L1 error norm of the azimuthal velocity profile
of the Greho vortex problem at time $t=3.0$, as a function of initial mesh
resolution ($N\times N$). We show results for three
different bulk velocities of the vortex, $v=0$, $v=1$ and $v=3$. The
results for
our moving-mesh code are shown as  red circles -- they are independent of the
bulk velocity to  machine precision. If we use a fixed Cartesian mesh instead,
we obtain the results shown by the open blue circles. The three sets of
results correspond to the different bulk velocities. In contrast
to the moving-mesh
code, the errors grow with increasing bulk velocity. Finally, for comparison
with an independent Eulerian hydrodynamics code, we show the
results obtained with {\small ATHENA} as triangles.
\label{FigGreshoL1Error}}
\ec
\end{figure}

It is now interesting to consider the changes in these results when the vortex
is set in motion. We show the corresponding results in the middle and bottom
row of Figure~\ref{FigGreshoProfiles}, for the cases $v_{\rm vortex}=1$ and
$v_{\rm vortex}=3$. Of course, in principle nothing should change, as the
physical problem only differs by a Galilean transformation from the original
setup of a stationary vortex.  Indeed, our moving-mesh code produces the same
result as before, and proves completely insensitive to this velocity boost, as
expected for a Galilean-invariant formulation. Quite in contrast, both our own
code when used with a fixed mesh as well as {\small ATHENA} show substantially
degraded results in the moving vortex case. In particular, the symmetry of the
vortex motion is partially lost \citep[consistently with the results
of][]{Liska2003}, and there is a larger degree of smoothing of the azimuthal
velocity profile. Clearly, this is the result of additional numerical
diffusivity and advection errors that now occur in the Eulerian treatment. A
particularly troubling aspect of these errors is that they are a strong
function of the velocity with which the vortex moves, as this determines the
distance over which the system has to be advected during the simulated
timespan.  This becomes clear by comparing the results for different vortex
velocities; increasing $v_{\rm vortex}$ and keeping the simulated timespan
fixed, the error in the Eulerian calculations can be increased nearly
arbitrarily. In contrast, the moving mesh code retains its original solution
independent of the bulk motion of the vortex, which is physically a much more
meaningful behaviour.

We now examine more quantitatively the convergence rate for this vortex
problem. To this end we measure the L1 error for the azimuthal velocity
profile at time $t=3.0$, as a function of the mesh resolution. We compare the
results obtained for the moving-mesh approach with those of the fixed mesh,
again for our three different bulk velocities, $v_{\rm vortex}=0$, $v_{\rm
  vortex}=1$ and $v_{\rm vortex}=3$.  Our results are summarized in
Figure~\ref{FigGreshoL1Error}, where we also include results obtained with
{\small ATHENA}, for comparison. Clearly, for zero bulk velocity, the errors
of our code {\small AREPO} are quite similar between the moving-mesh and the
fixed-mesh, and also very close to the independent code {\small ATHENA}. Note
that the results converge only approximately as ${\rm L1}\propto N^{-1.4}$, as
indicated by the solid line in the plot. This is presumably a consequence of
the discontinuities in the vorticity profile present in this problem, at
$r=0.2$ and $r=0.4$. If a non-vanishing bulk velocity is included, we 
see significant accuracy differences between the moving-mesh and the
fixed mesh. Whereas the moving-mesh results {\em do not change at all}, the
error increases for the Eulerian approach with growing bulk velocity. The
magnitude of this deterioration is consistent between {\small AREPO} and
{\small ATHENA}. We argue that this highlights an important 
shortcoming of traditional Eulerian approaches.

In passing, we want to note that we also tried SPH on this problem, with the
implementation of SPH in the {\small GADGET3} code. It turns out that this is
a hard problem for SPH, and a direct comparison with the results presented
above shows SPH to be substantially less accurate when the same number of
particles is used. In fact, for the lower resolutions, the vortex typically
does not survive until $t=3.0$; the angular momentum is transported to the
boundaries of the domain before this time, where it is then effectively
canceled by encounters with oppositely moving gas from the adjacent periodic
image domains. We defer a more detailed comparison of SPH with the moving-mesh
code to a future study.

\subsection{The Noh problem}

\begin{figure*}
\bc
\hspace*{0.2cm}\resizebox{9.0cm}{!}{\includegraphics{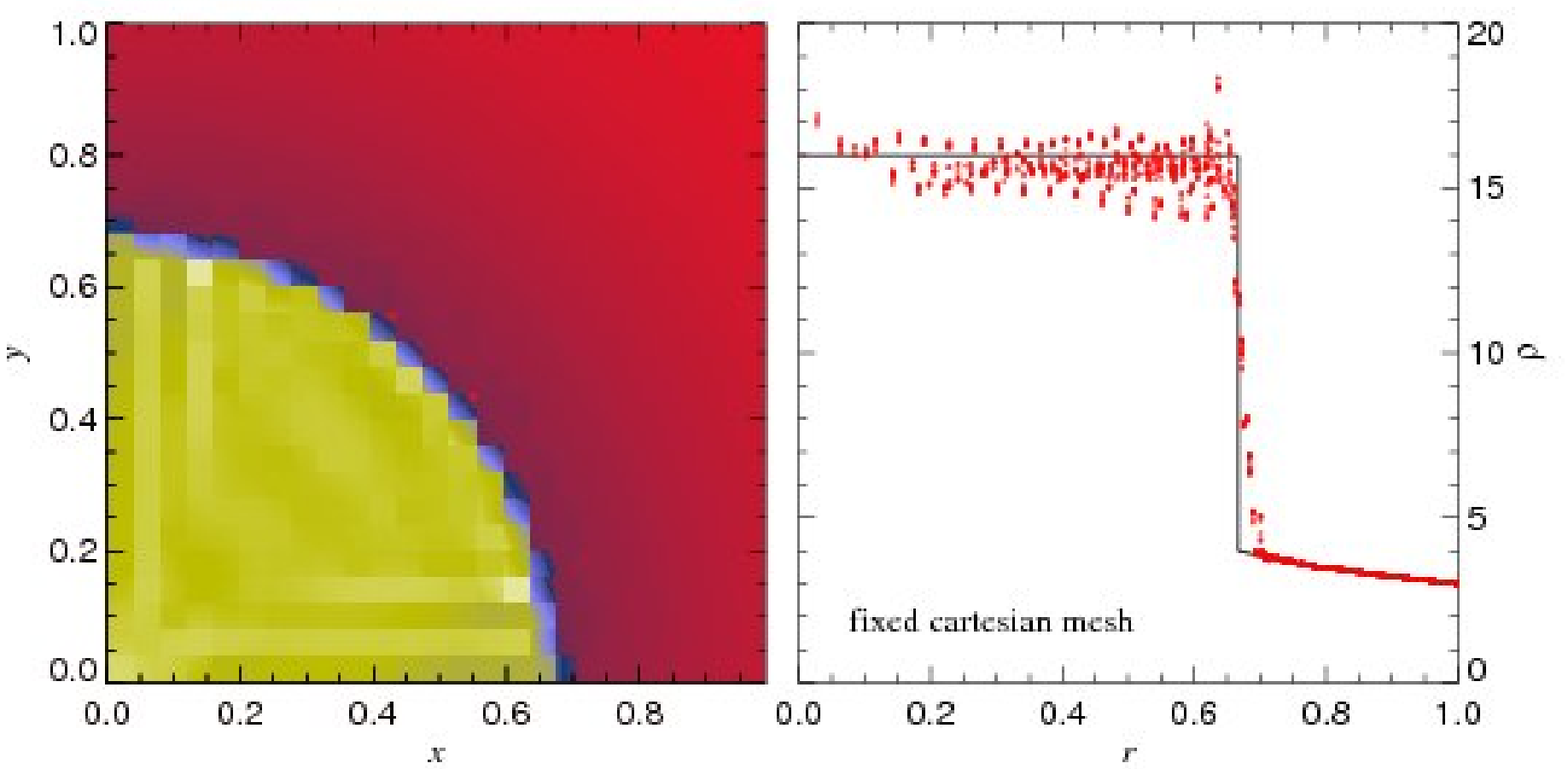}}\hspace*{-0.2cm}%
\resizebox{9.0cm}{!}{\includegraphics{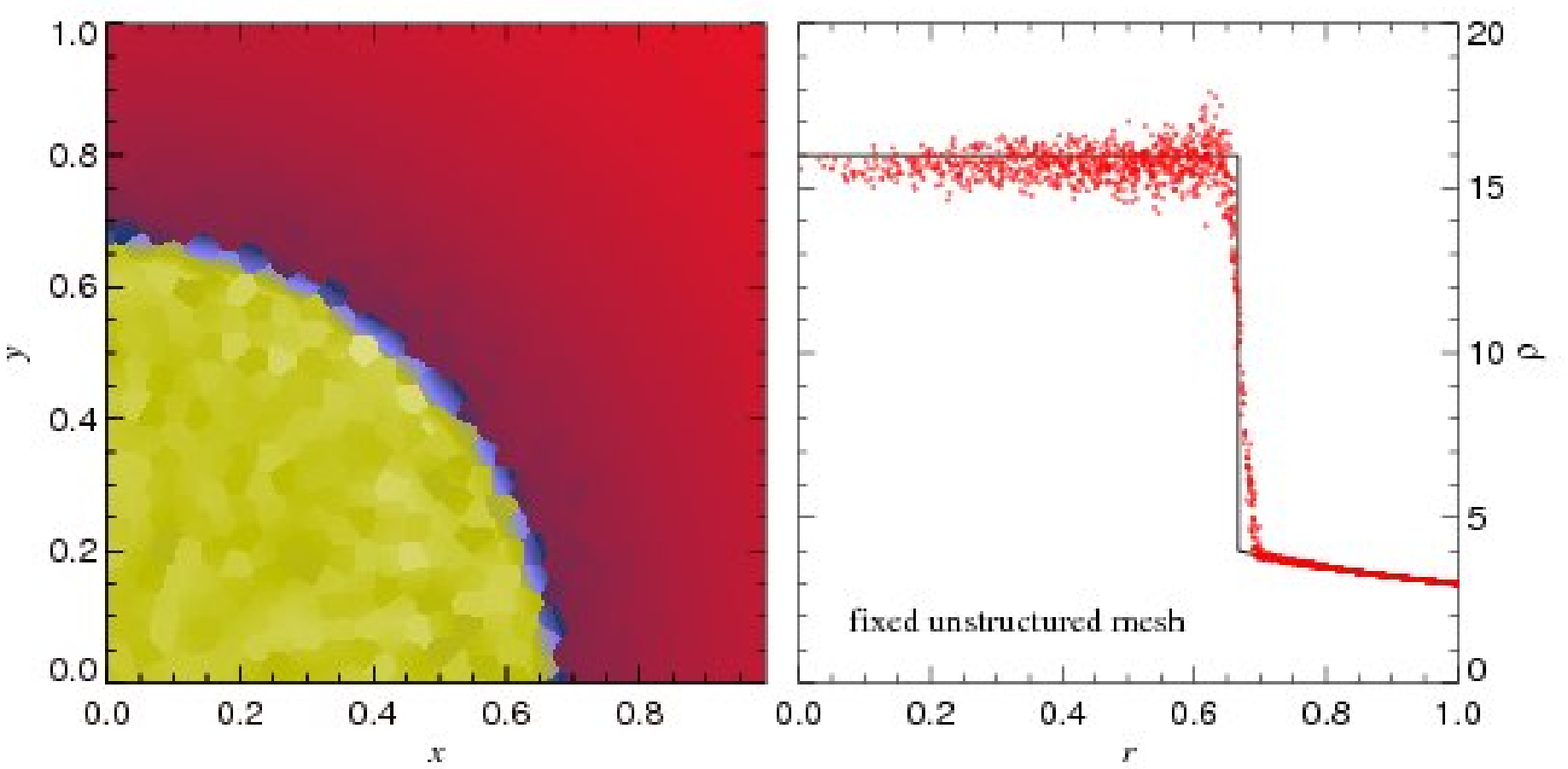}}\\
\hspace*{0.2cm}\resizebox{9.0cm}{!}{\includegraphics{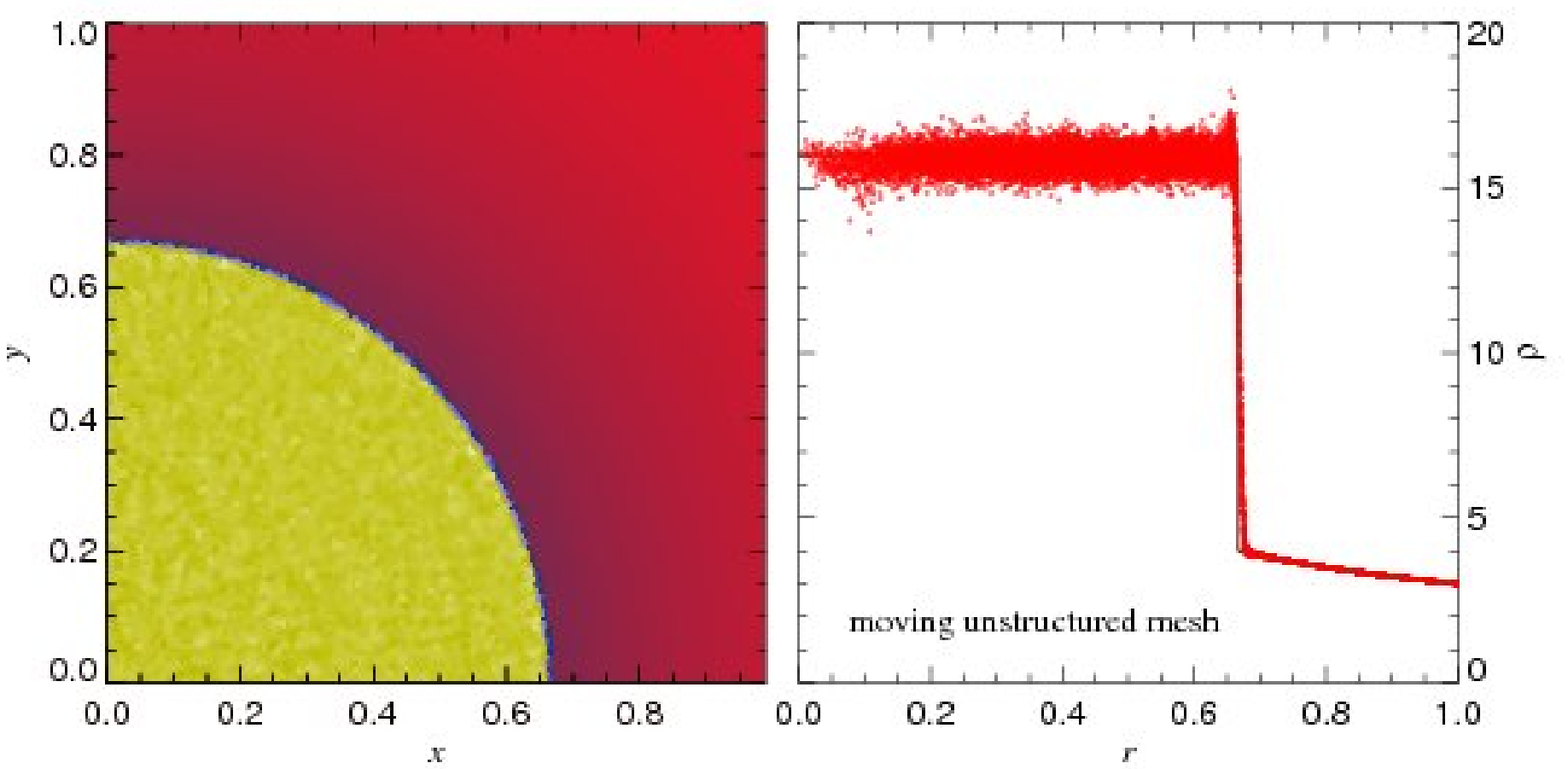}}\hspace*{-0.2cm}%
\resizebox{9.0cm}{!}{\includegraphics{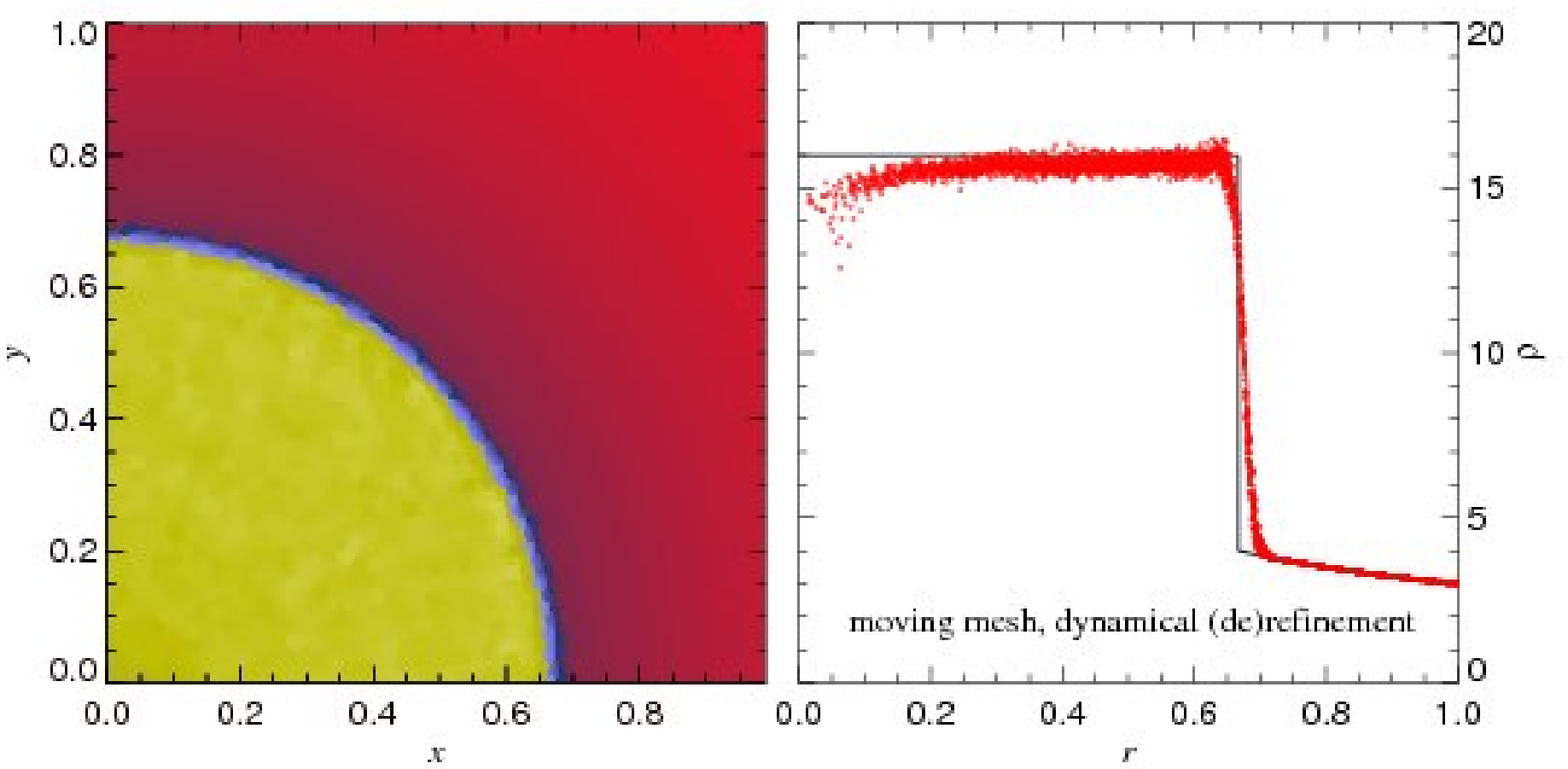}}\\
\resizebox{5.8cm}{!}{\includegraphics{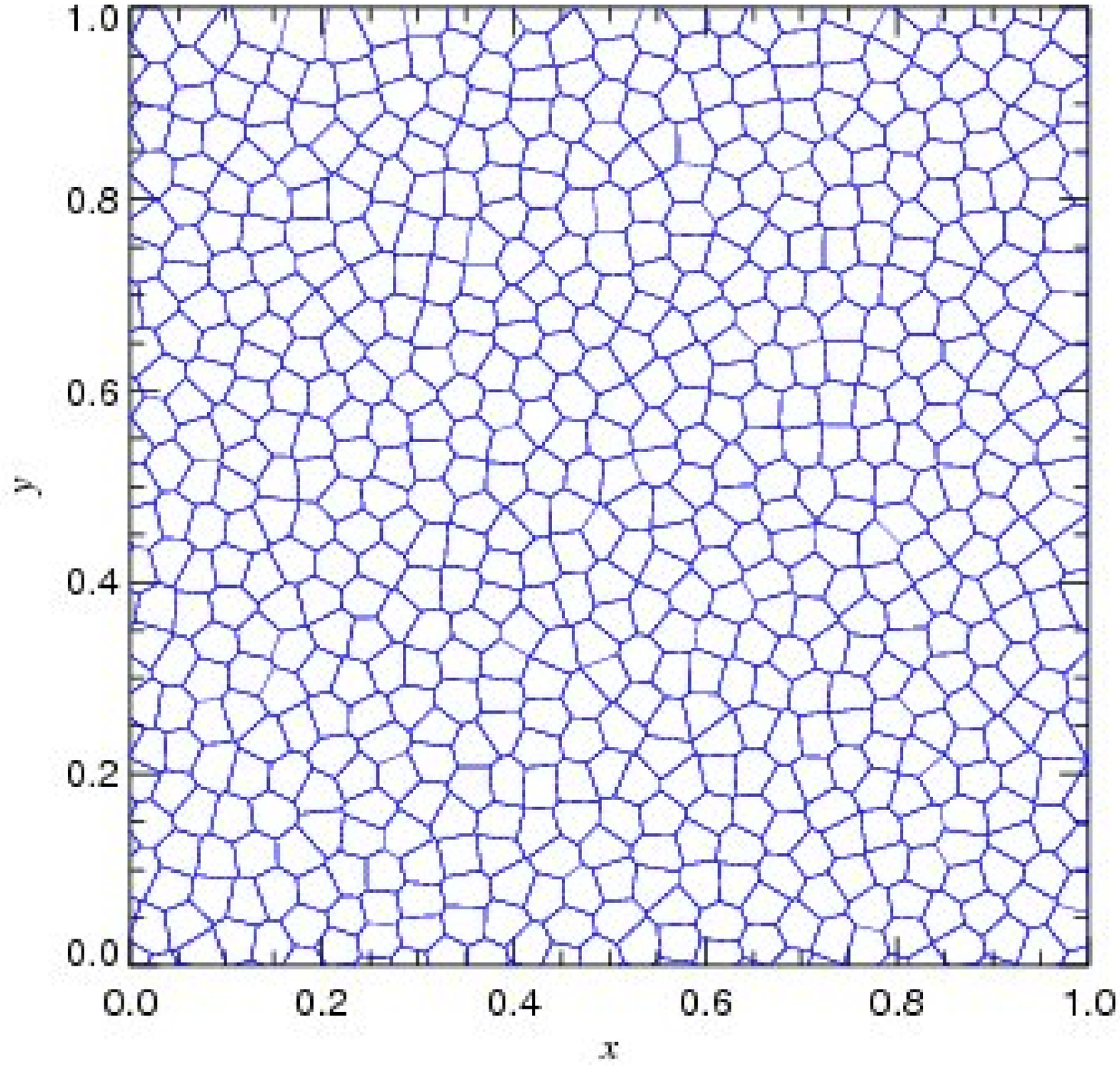}}%
\resizebox{5.8cm}{!}{\includegraphics{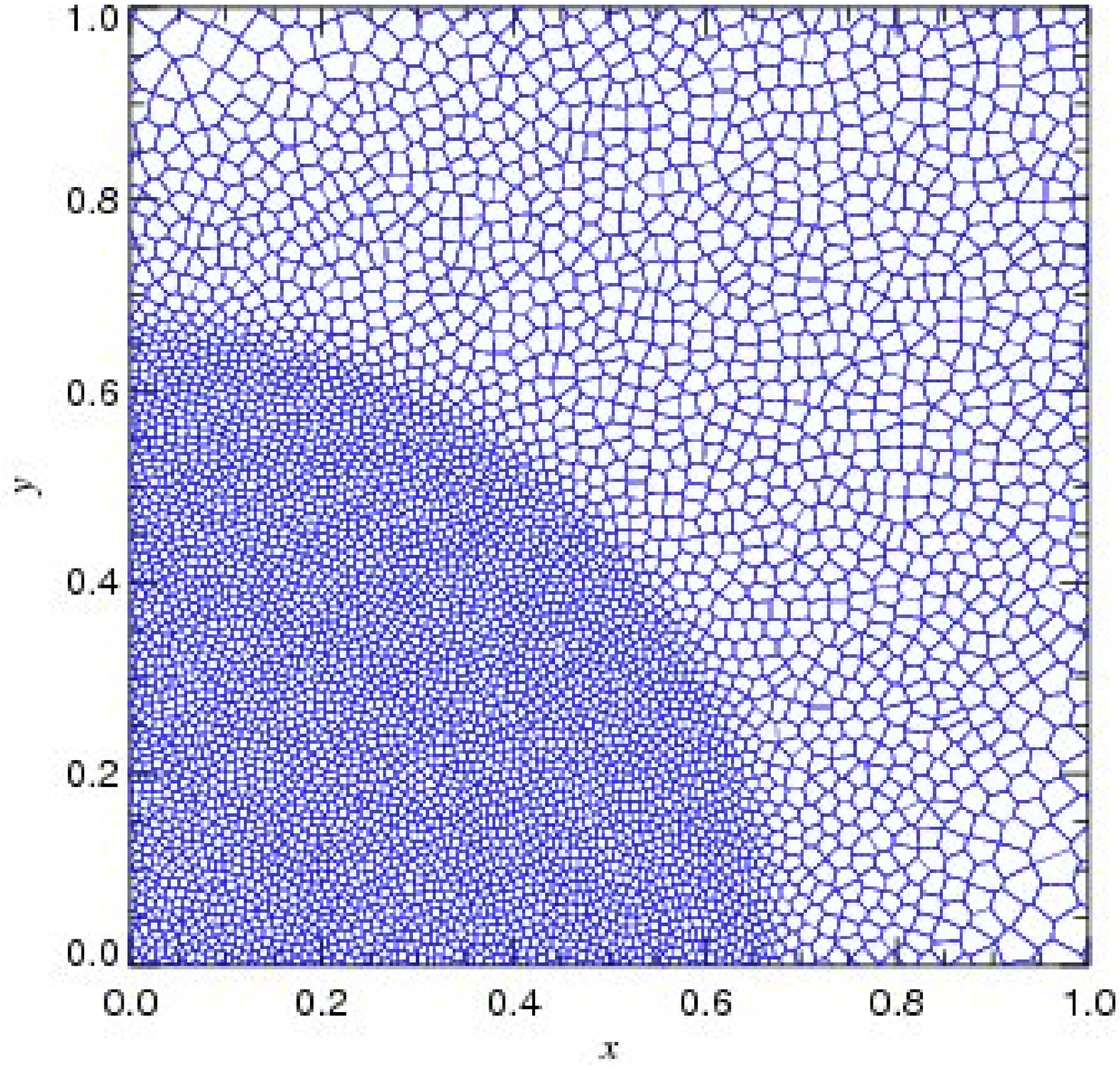}}%
\resizebox{5.8cm}{!}{\includegraphics{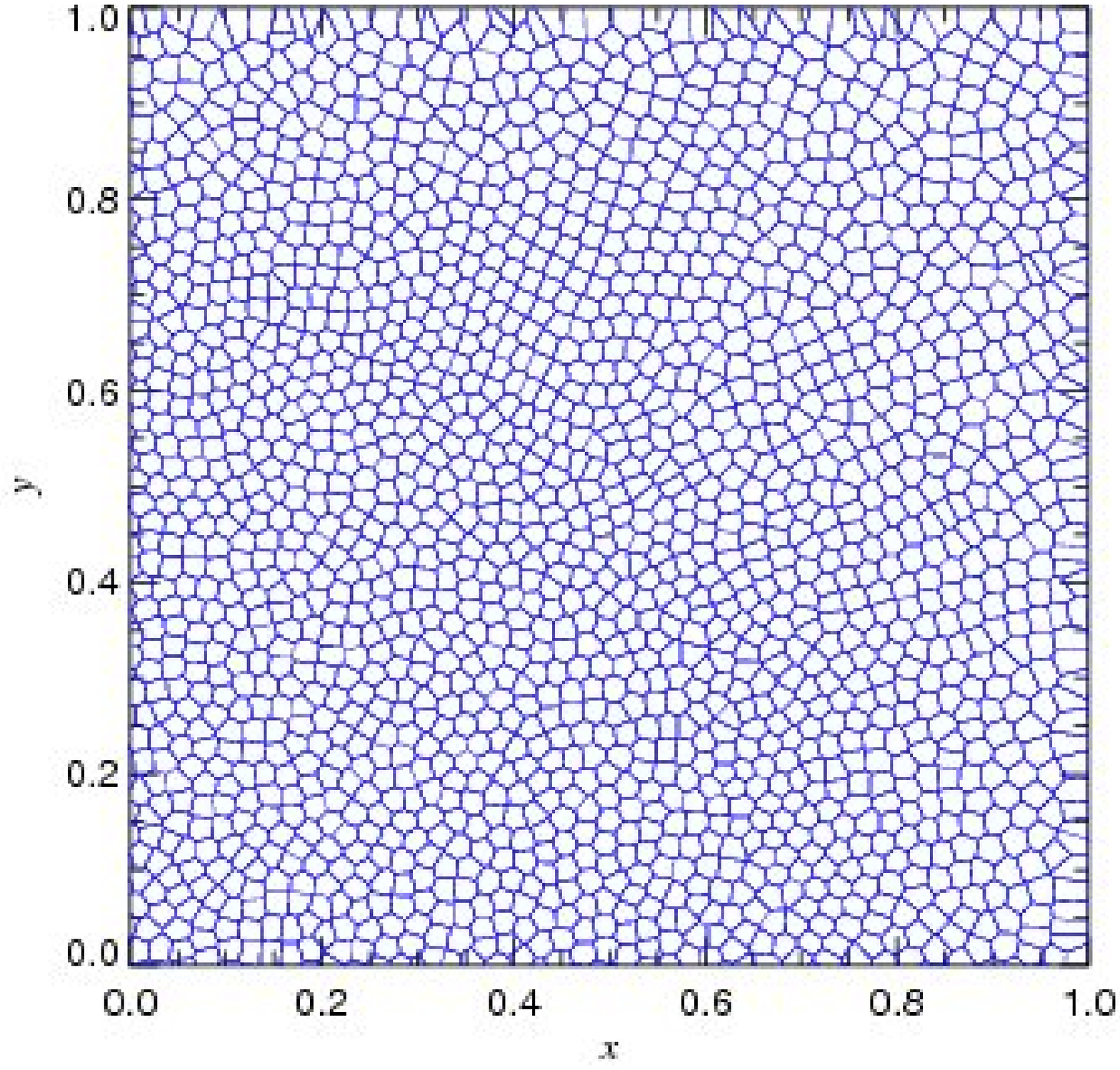}}\\
\caption{The Noh-problem in 2D at low resolution, calculated with four different
  strategies for the treatment of the mesh. The top four pairs of panels
  illustrate the result of the calculation at time $t=2.0$ for each of these
  schemes. In each case, we show a projected density field in one quadrant of
  the implosion, and we give the radial density profile where we compare the
  densities of all mesh cells with the analytic solution. In the top left, the
  result for a stationary Cartesian mesh with $25^2$ cells in the unit
  quadrant is shown. For comparison, the top right gives the result for an unstructured
  stationary mesh with the same number of cells. The middle left pair of
  panels shows the case of an unstructured moving mesh with constant mass
  resolution of $m_i\simeq 1/25^2$. Finally, the fourth case (middle right
  panels) is for a dynamically refined/derefined mesh, where cells are split
  if $m_i> 1.5 /25^2$ or eliminated if their volume falls below
  $V_i<0.25/25^2$.
The bottom three panels show the mesh geometries at the final time 
of the three schemes where an
unstructured mesh is used. From left to right: the stationary unstructured
case, the moving-mesh case with constant mass resolution, and finally the
dynamically refined/derefined mesh.
\label{FigNoh2D}}
\ec
\end{figure*}

We now consider the strong shock test proposed by \citet{Noh1987}, which
has an analytic solution. This is generally considered a very difficult
problem, and quite frequently, numerical methods have problems running
this test without crashing. In fact, in the test suite of
\citet{Liska2003}, only four out of the studied eight schemes managed to
run this problem at all. The set-up consists of a $\gamma=5/3$ gas that
has initially uniform density equal to $\rho_0=1$, vanishingly small
pressure, and everywhere a radial inflow velocity towards the origin of
$v = -1$. As a result of the inflow, a strong spherical shock wave of
formally infinite Mach number develops and travels outwards with a speed
$v_s=1/3$. Inside of the shock front, the density is constant; it has a value of
4 in the 1D case, 16 in the 2D case, and 64 in the 3D case.  Outside of
the shock, the density profile is given by
\begin{equation} 
\rho(r,t) = \rho_0 \, (1+t/r)^n ,
\end{equation} 
with $n=2$ in the 3D case, $n=1$ for the 2D case, and $n=0$ for the 1D case.

The problem has been considered in 1D, 2D, and in 3D in the literature, but we
restrict ourselves to two- and three-dimensional tests. As in previous studies
of this problem, we calculate only one quadrant  when a
Cartesian mesh is used, and apply reflective boundary conditions at the inner
boundaries. However, when an unstructured mesh is used, we calculate all four
quadrants in order to avoid imposing mirror symmetry along the coordinate
axes.  The outer boundaries are modelled with a special inflow boundary that
makes use of the analytic solution known for the problem.

We begin by considering the 2D problem carried out with different strategies
for treating the mesh. The simplest approach is a fixed Cartesian mesh of
resolution $25\times 25$ cells in one quadrant. Our second calculation was
done with an unstructured mesh that has the same total number of cells as in
the Cartesian case, but is also kept stationary. Comparison of these two
schemes allows an assessment of how well the unstructured mesh performs relative
to a Cartesian mesh of equal spatial resolution. Our third calculation uses a
moving unstructured mesh where the mesh cells are moved with the local
velocity of the gas, such that the mass per cell stays constant to good
approximation. We here use an initial mesh that has been extended to
$[-3,3]\times[-3,3]$ in order to provide enough mesh area for the implosion
problem. Finally, in our last variant of this problem we also use a moving
mesh but exercise our schemes for dynamically refining and derefining the
mesh, as described in Section~\ref{SecRefinement}. Specifically, we split
cells into two  when their mass has gone up above 1.5 times the initial
average mass of a cell, $\overline{m}=1/25^2$. This automatically generates
new cells in the inflow region, and maintains a constant mass resolution
there. In this case we  do not have to extend the mesh beyond the
$[-1,1]^2$ domain; rather, new mesh cells appear dynamically as needed. In
addition, we also derefine cells (i.e.~delete them) if their volume falls
below $0.25$ times the initial volume $\overline{V}=1/25^2$ of the cells. This
prevents the spatial resolution in the high-density region from getting better
than a certain limit. In fact, it limits the maximum effective resolution per
dimension to $N_{\rm eff}= 1/(0.25\times \overline{V})^{1/2} = 50$.

\begin{figure*}
\bc
\resizebox{8.0cm}{!}{\includegraphics{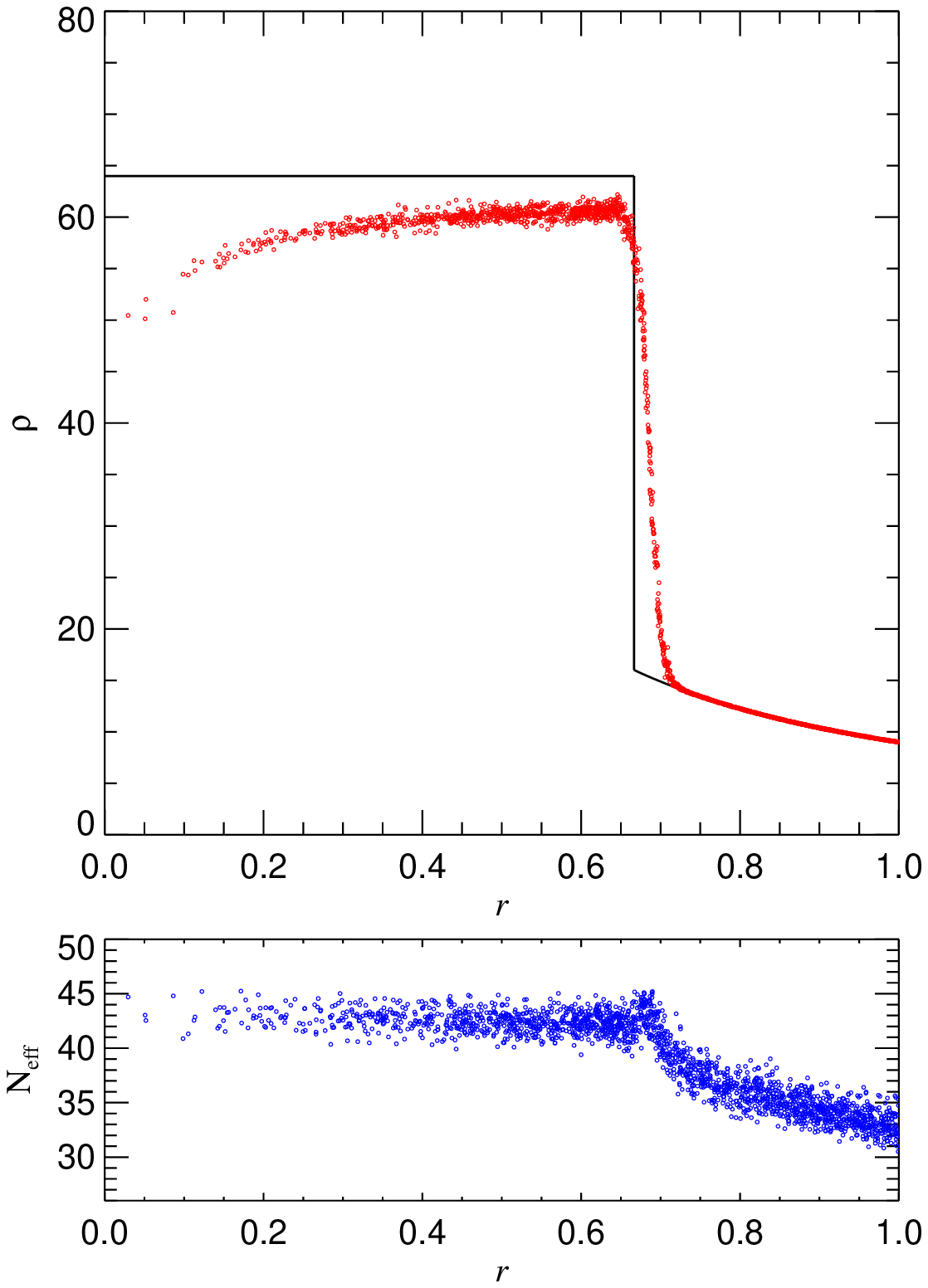}}\hspace*{0.5cm}%
\resizebox{8.0cm}{!}{\includegraphics{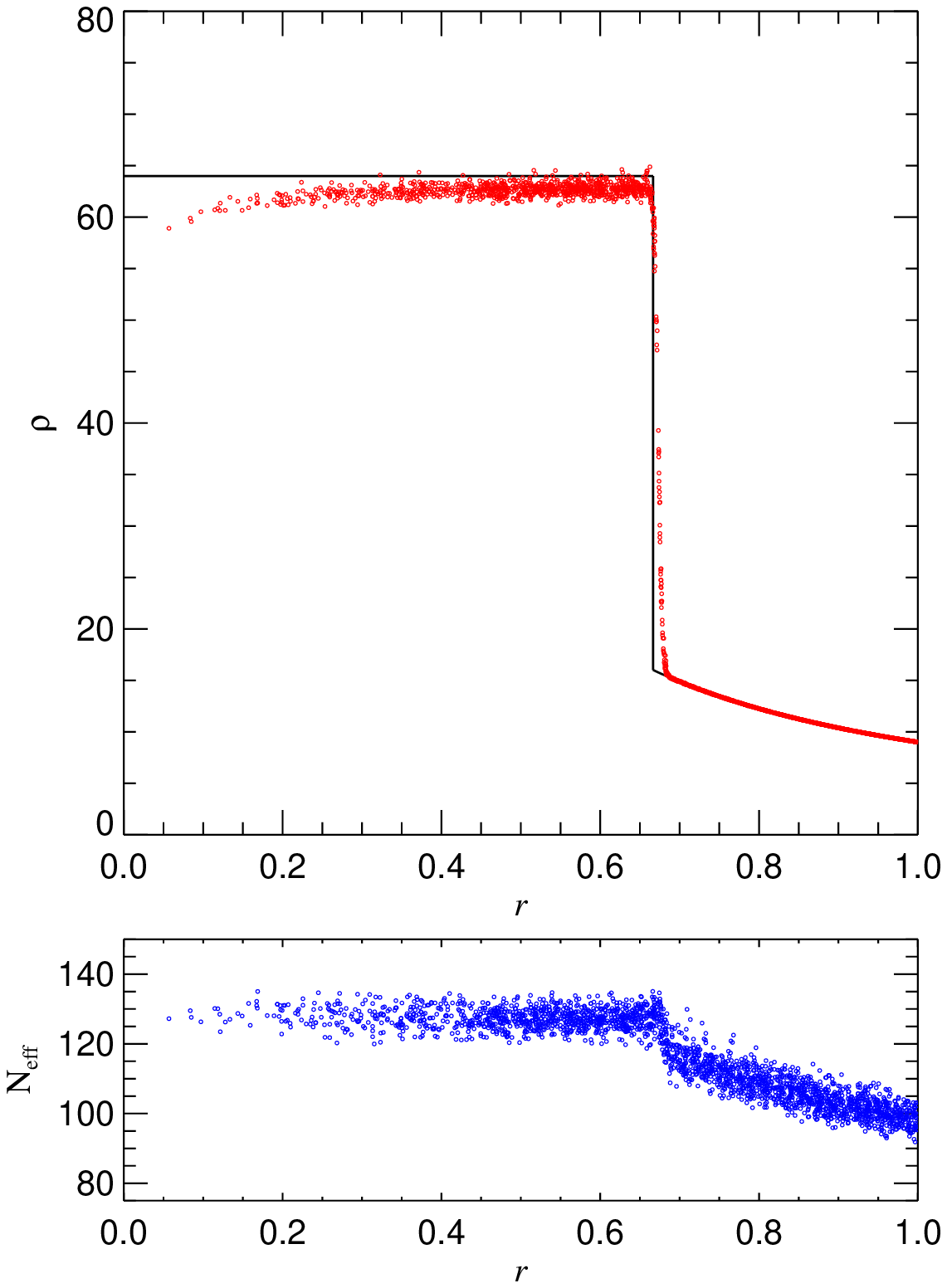}}
\caption{The Noh-problem in 3D at two different effective resolutions, 
calculated with dynamical mesh creation and mesh derefinement. The panels on
the left are for a low-resolution calculation, where cells are created at the
edge of the box such that the mass resolution does not drop below
$1.5\times \overline{m}$, with $\overline{m}=2.16\times 10^{-3}$, and cells in
the high-density region are eliminated if their volume drops below $V_i<0.05
\overline{V}$, with $\overline{V}=2.16\times 10^{-3}$. The high-resolution
calculation on the right hand side has 27 times better mass and volume resolution.
In both cases, the bottom panels give the effective resolution
per dimension for the cells as a function of radius, 
defined as $N^i_{\rm eff}= 1/V_i^{1/3}$, where $V_i$ is the volume of a cell.
Only a random subset of 2000 of the cells is shown in each panel.
\label{FigNoh3D}}
\ec
\end{figure*}

In Figure~\ref{FigNoh2D}, we compare the results of these four different mesh
strategies. In each case we show a projected density field at time $t=2.0$,
and we compare the densities of individual mesh cells with the expected
analytic solution for the radial density profile.  In all four cases, the
postshock flow shows some substantial density scatter. This is presumably a
combination of weak post-shock oscillations present in our scheme, and
geometrically induced asphericities. The oscillatory behaviour is particularly
noticeable and coherent in the Cartesian case, where also the so-called
`carbuncle' phenomenon is at work, which can produce artefacts for very strong
grid-aligned shocks. \citet{Stone2008} invoke a special cure for this problem
and achieve a quiet post-shock flow in the Noh problem with the help of
judiciously introduced extra dissipation. We expect that the unstructured
fixed mesh should be less susceptible to such grid alignment effects. Indeed,
the result for this case (top right panel of Figure~\ref{FigNoh2D}) shows no
coherent oscillations and preferred directions of the kind seen for the
Cartesian grid, but it nevertheless exhibits a similar degree of noise.

Our calculation with a {\em moving} unstructured mesh and constant mass
resolution (middle left in Figure~\ref{FigNoh2D}) benefits from an
automatically higher spatial resolution in high density regions. As a
result, the position of the shock front is recovered more accurately,
and the shock is nicely round. Nevertheless, this solution suffers from
a similar degree of oscillatory behaviour in the post-shock
region. Finally, we consider our calculation with a dynamically
refined/derefined moving mesh, shown in the middle right of
Figure~\ref{FigNoh2D}. Here the shock is also recovered well, with an
accuracy that is slightly better than for the fixed mesh results, thanks
to slightly smaller cell sizes at the shock front. In the postshock
region, the oscillations are noticeably reduced. This is a result of the
derefinement procedure that is active in this region, which tends to
smooth out high frequency noise. It is in any case reassuring that our
dynamical mesh refinement and derefinement schemes work robustly in this
difficult hydrodynamic problem without introducing any artefacts. The
suppression of the density at the origin is perhaps caused by so-called
`wall heating' \citep{Ryder2000}, which is commonly seen at a similar
level in other calculations of the Noh problem
\citep[e.g.][]{Liska2003,Stone2008}.

The bottom three panels of Figure~\ref{FigNoh2D} show the mesh geometry at the
final time of the three calculations that use an unstructured mesh. It is
nicely seen how the constant mass-resolution case (bottom middle panel)
produces a mesh that varies strongly in spatial resolution. On the other hand,
the dynamically created and derefined mesh (bottom right panel) shows almost
no trace of the spherical shock front, due to the particular derefinement
criterion used. Note that the latter is arbitrary, and if desired, one could
for example refine the mesh only in the region of the shock, and derefine it
elsewhere.

\begin{figure*}
\bc
\resizebox{7cm}{!}{\includegraphics{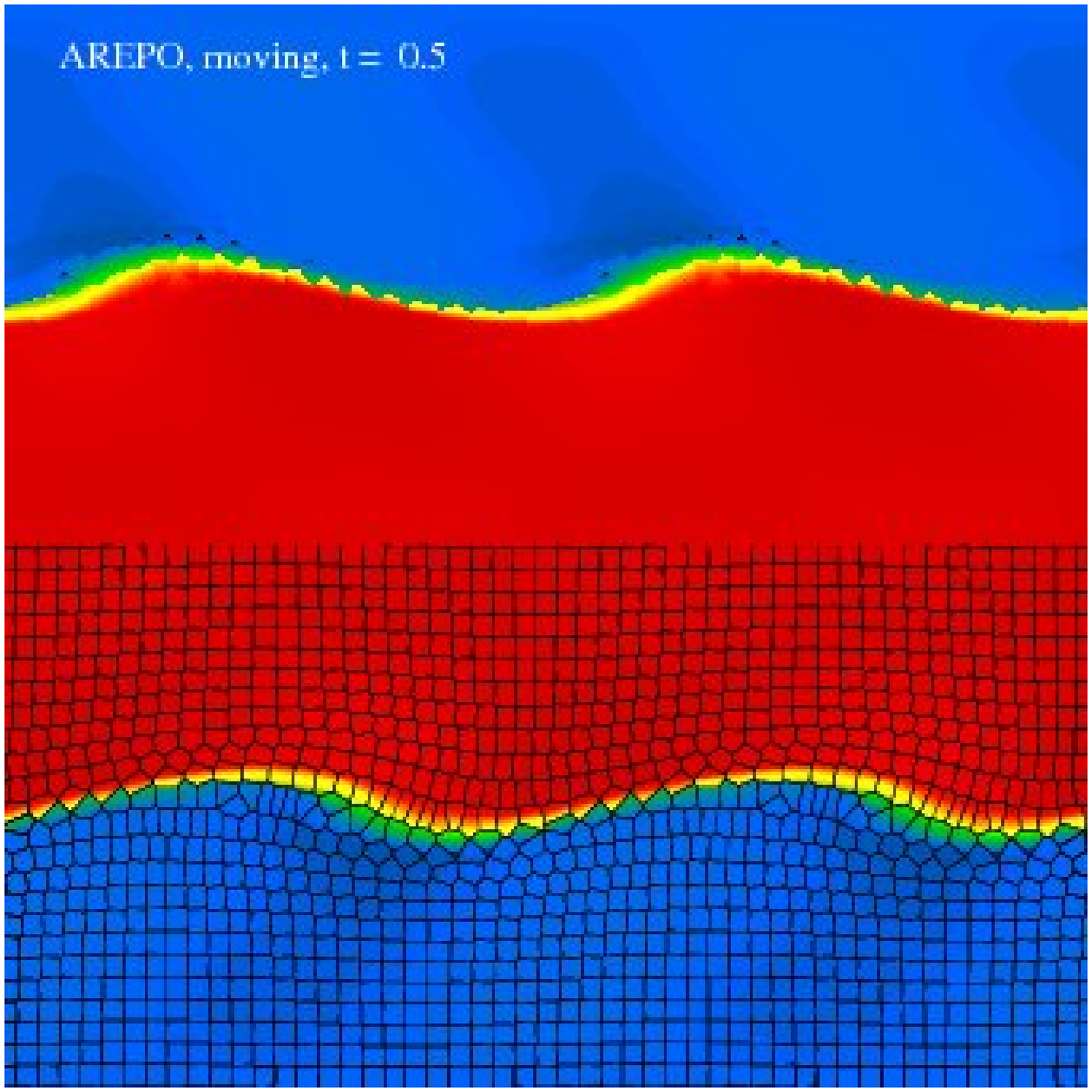}}\ %
\resizebox{7cm}{!}{\includegraphics{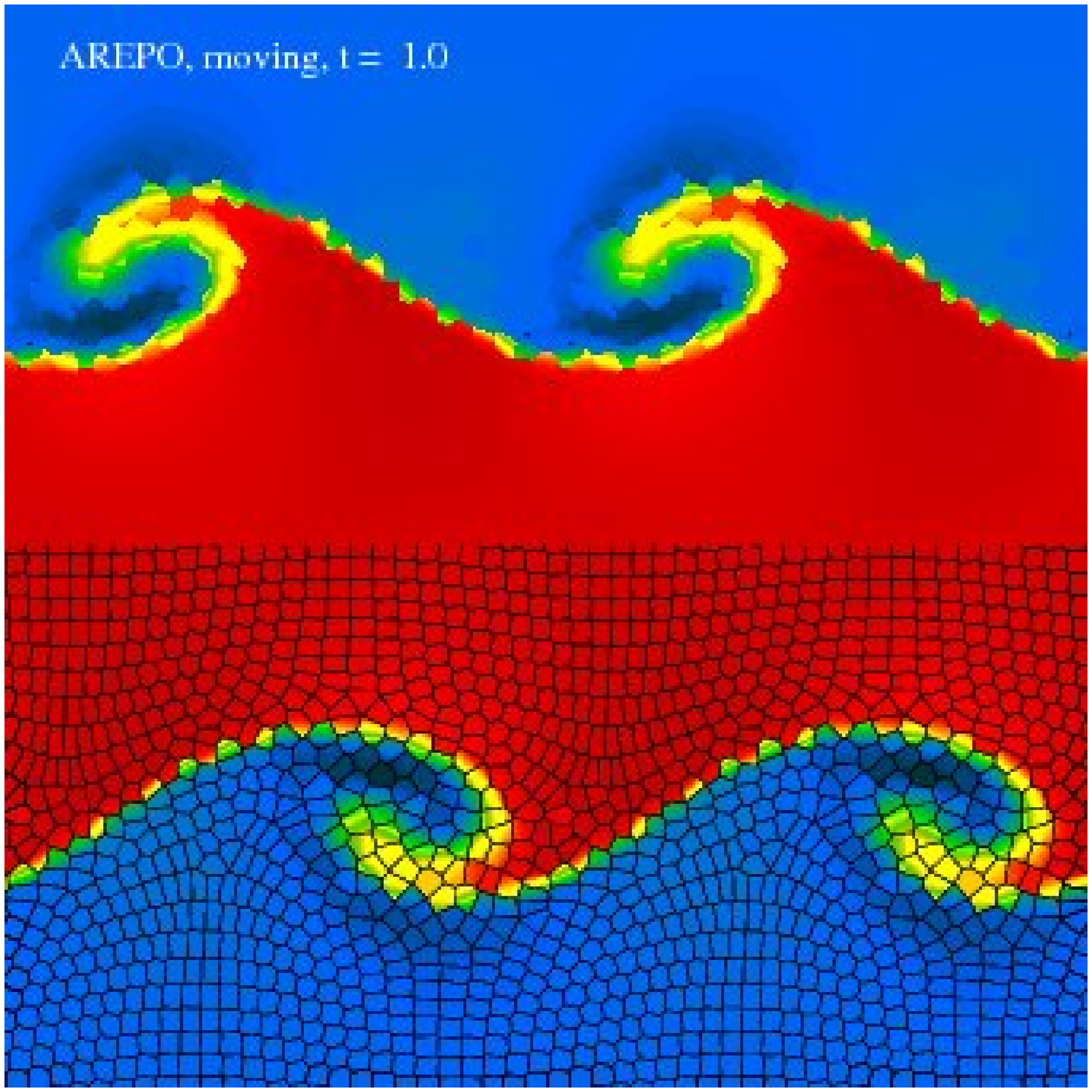}}\\
\resizebox{7cm}{!}{\includegraphics{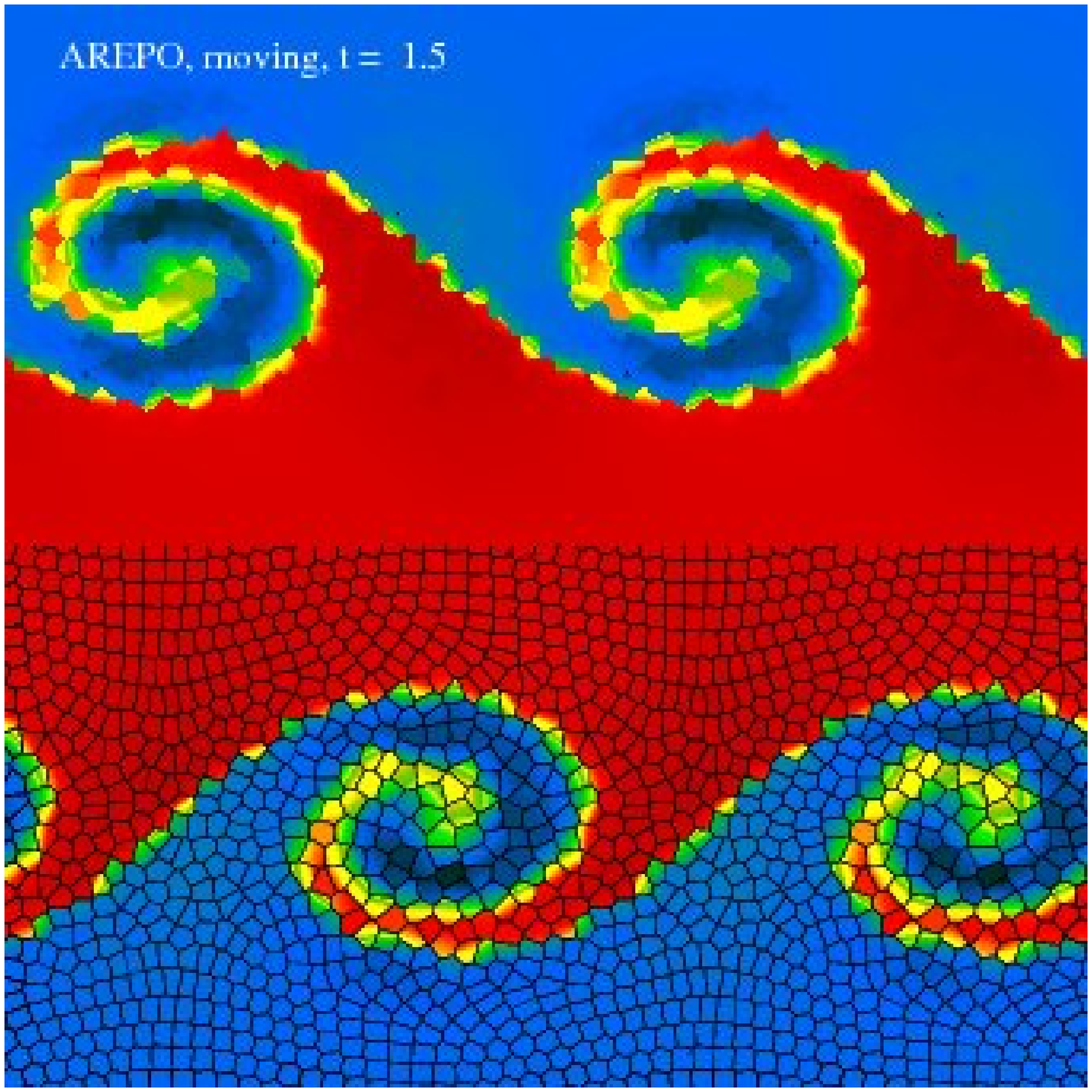}}\ %
\resizebox{7cm}{!}{\includegraphics{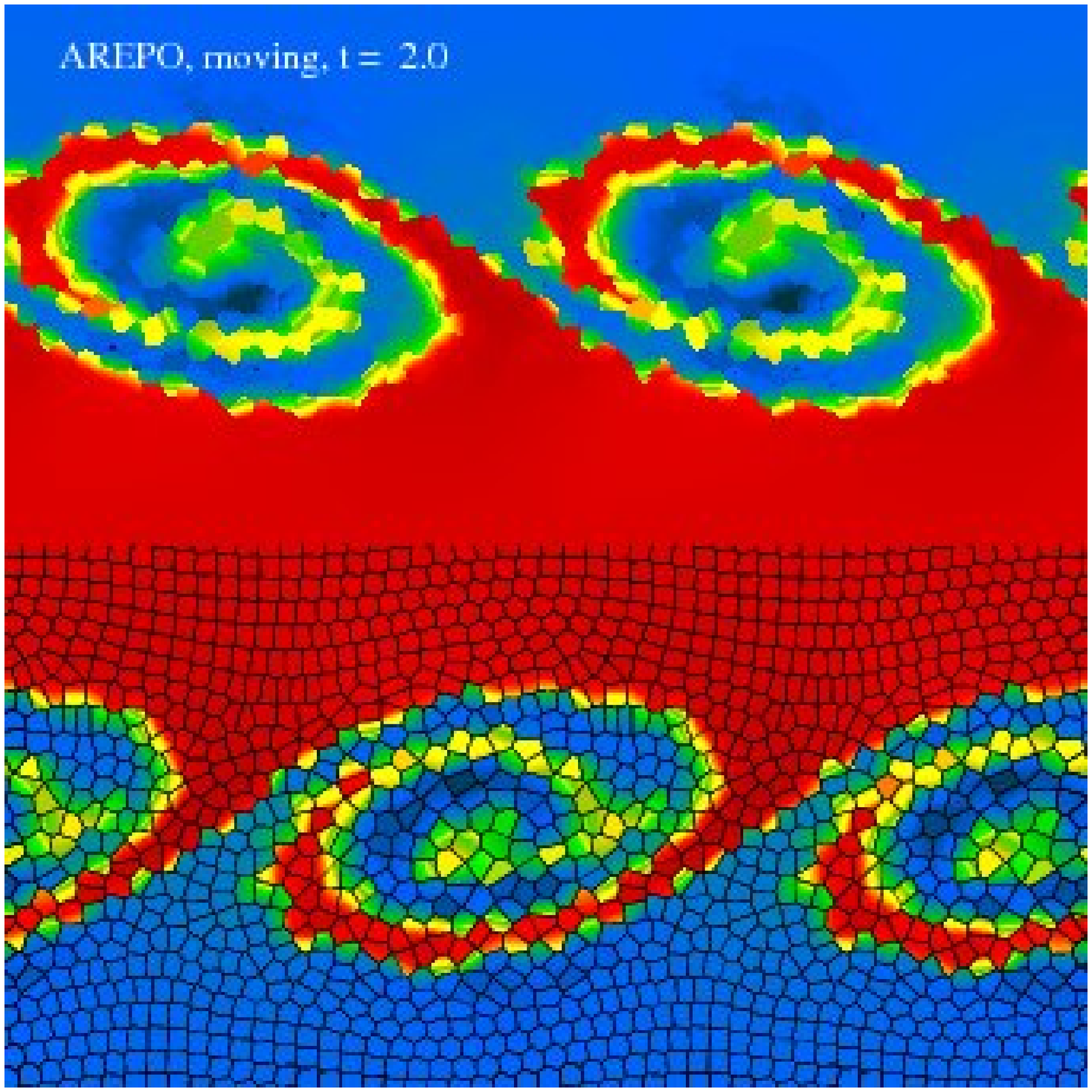}}\\
\ \\
\resizebox{7cm}{!}{\includegraphics{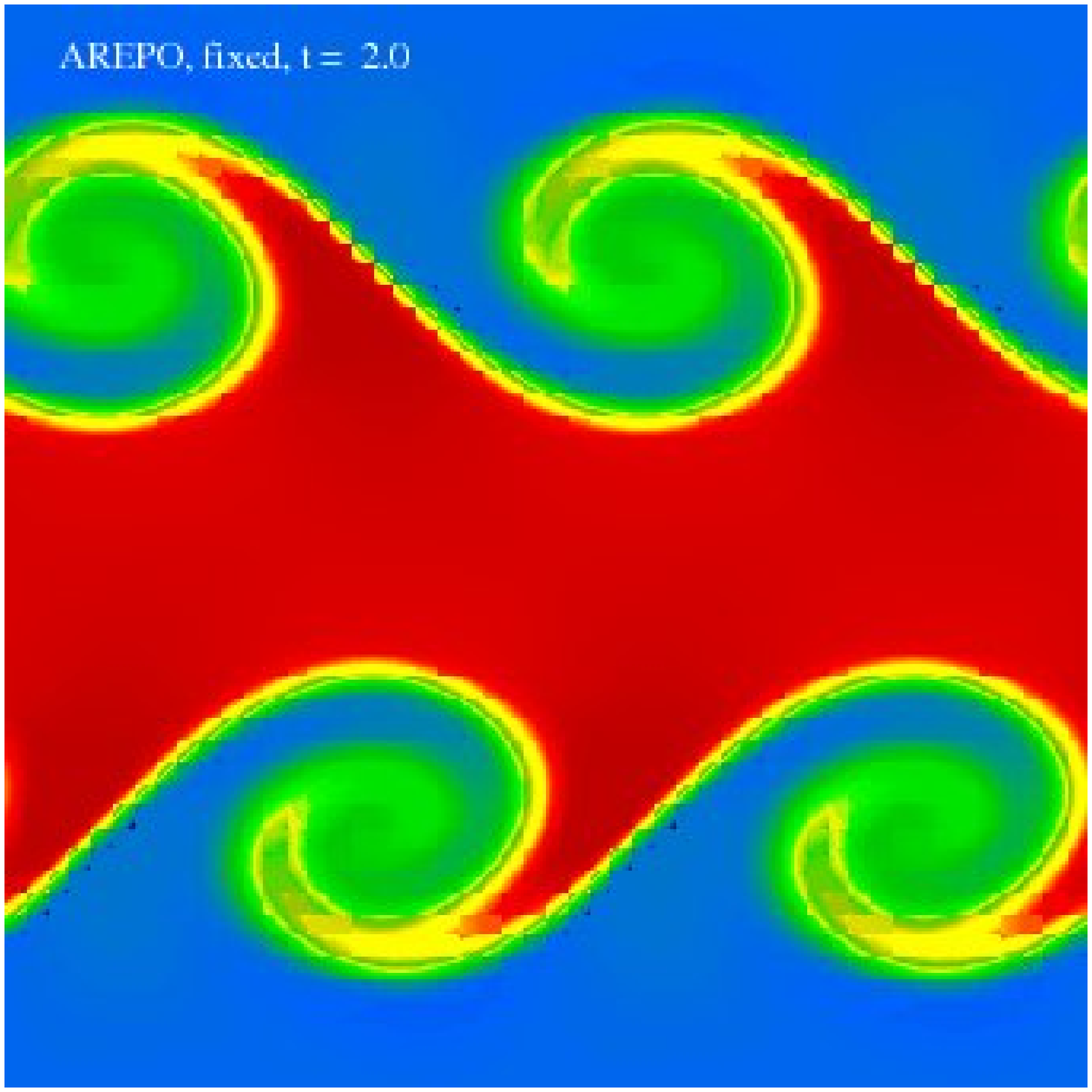}}\ %
\resizebox{7cm}{!}{\includegraphics{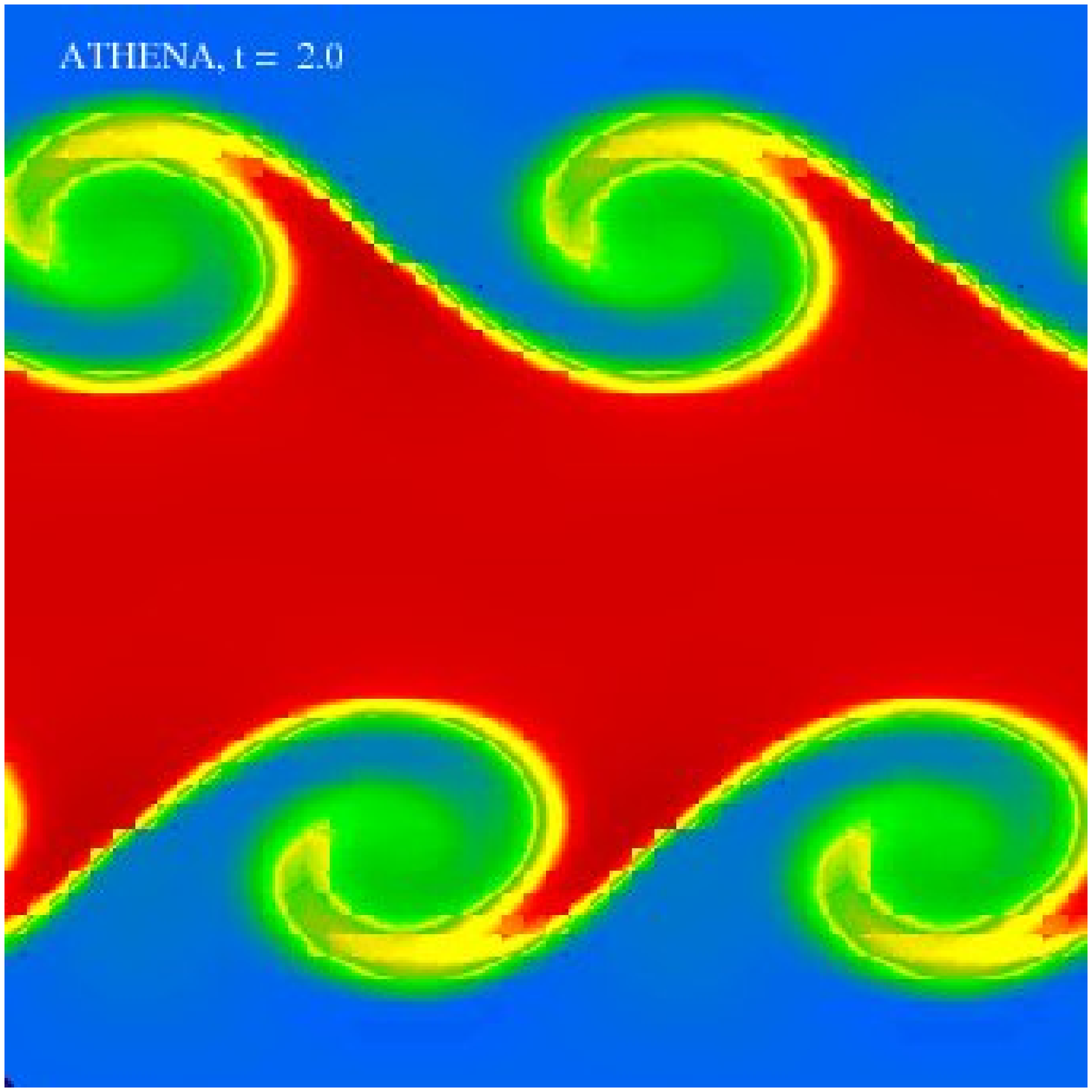}}
\caption{The top four panels show the time evolution of the Kelvin Helmholtz
  instability in a low resolution ($50\times 50$) test calculation with the
  moving-mesh method. Each panel gives the density field (at times $t=0.5$,
  $1.0$, $1.5$ and $2.0$), with the Voronoi mesh overlaid in black in the lower
  half of the box. For
  comparison,  the lower two panels show the results for the same initial conditions, but
  this time computed keeping the initial Cartesian mesh fixed. The panel on
  the bottom left shows the result
  at time $t=2.0$  obtained with our code {\small AREPO} for a fixed mesh, 
  while the bottom right gives the result of {\small ATHENA} (with second
  order reconstruction and the Roe solver). The latter two results are nearly
  identical. Note however that in the non-linear regime the KH instability appears to evolve 
  somewhat faster for the moving-mesh code compared with the fixed grid.
  \label{FigKHTimeEvolv}}
\ec
\end{figure*}

\begin{figure*}
\bc
\resizebox{5.5cm}{!}{\includegraphics{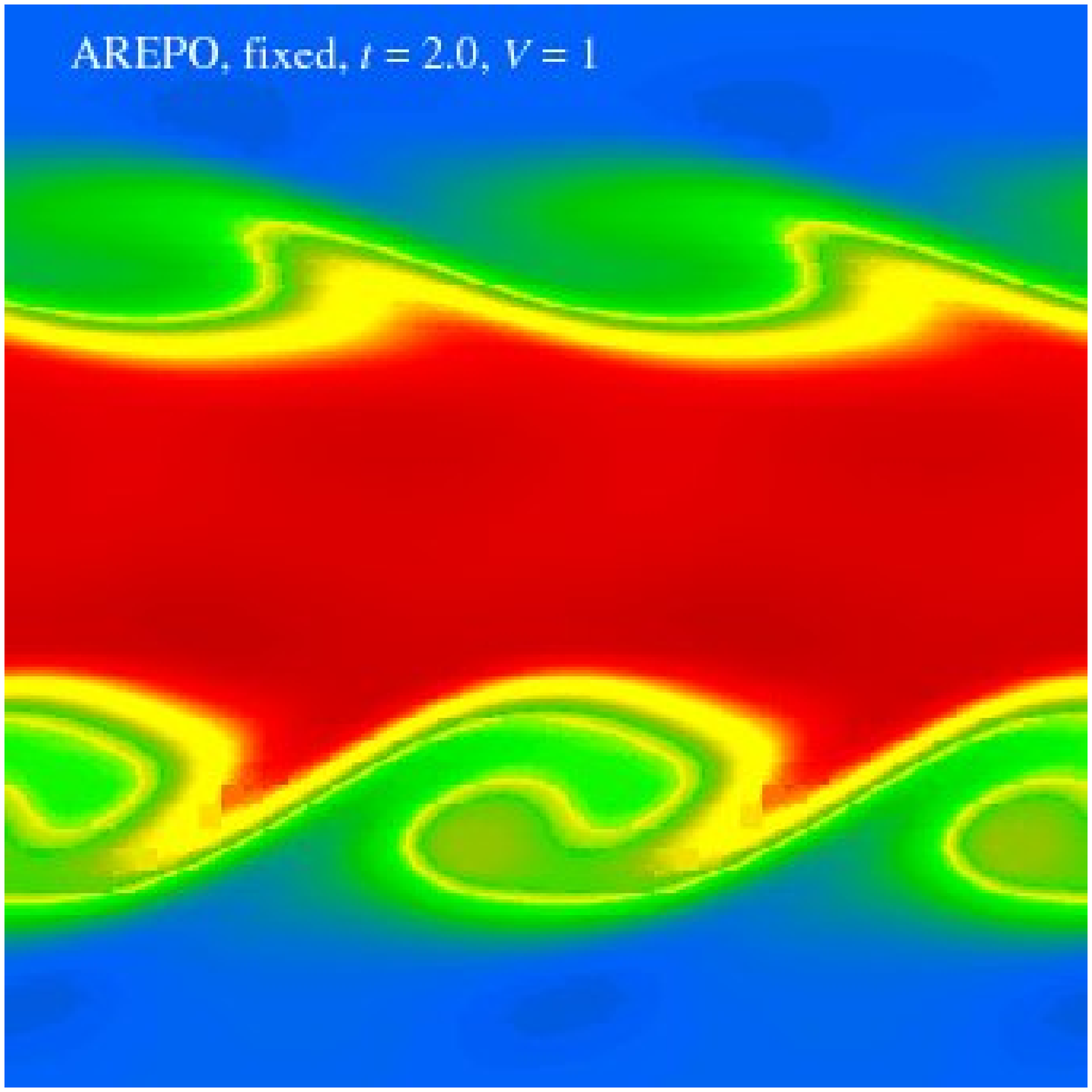}}\
\resizebox{5.5cm}{!}{\includegraphics{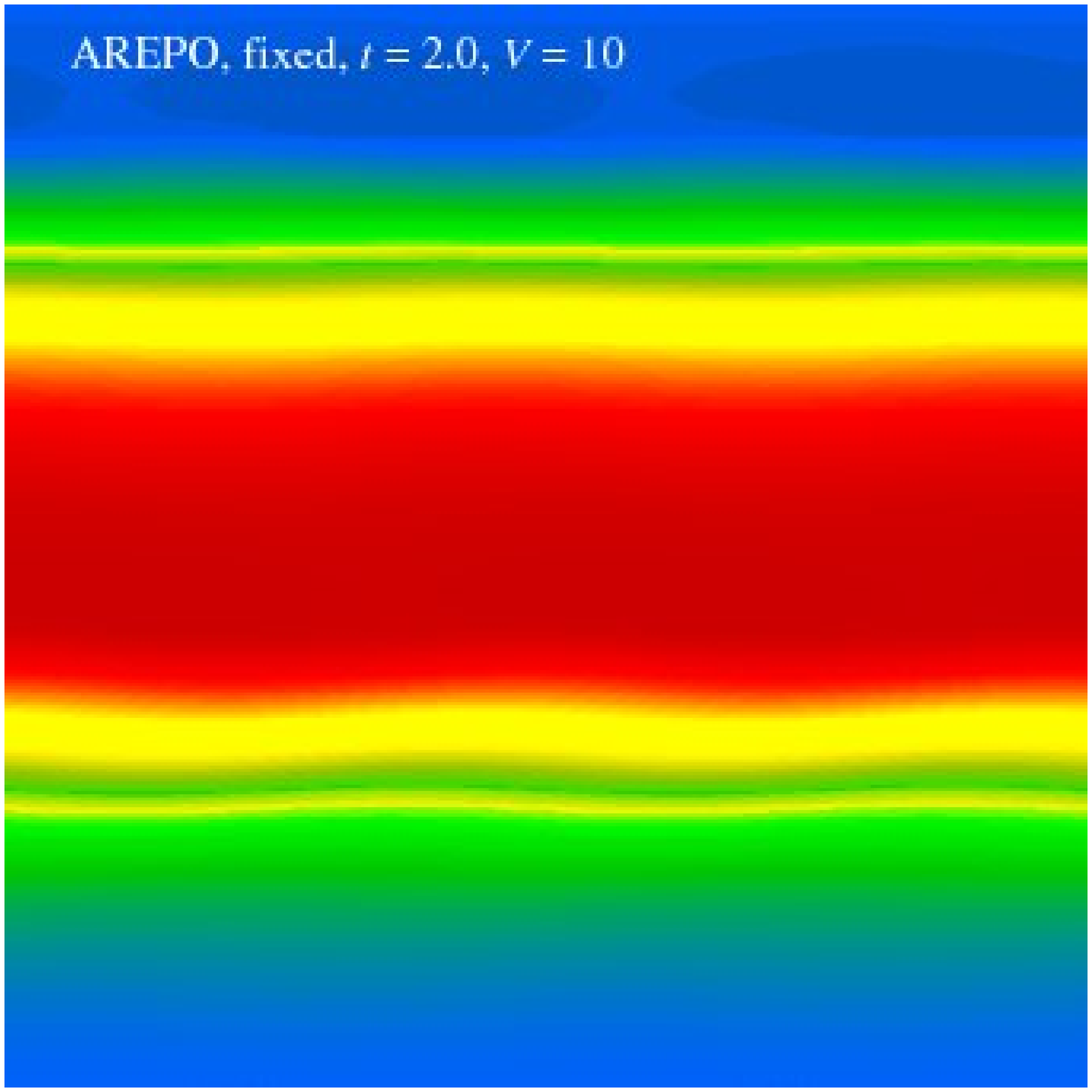}}\
\resizebox{5.5cm}{!}{\includegraphics{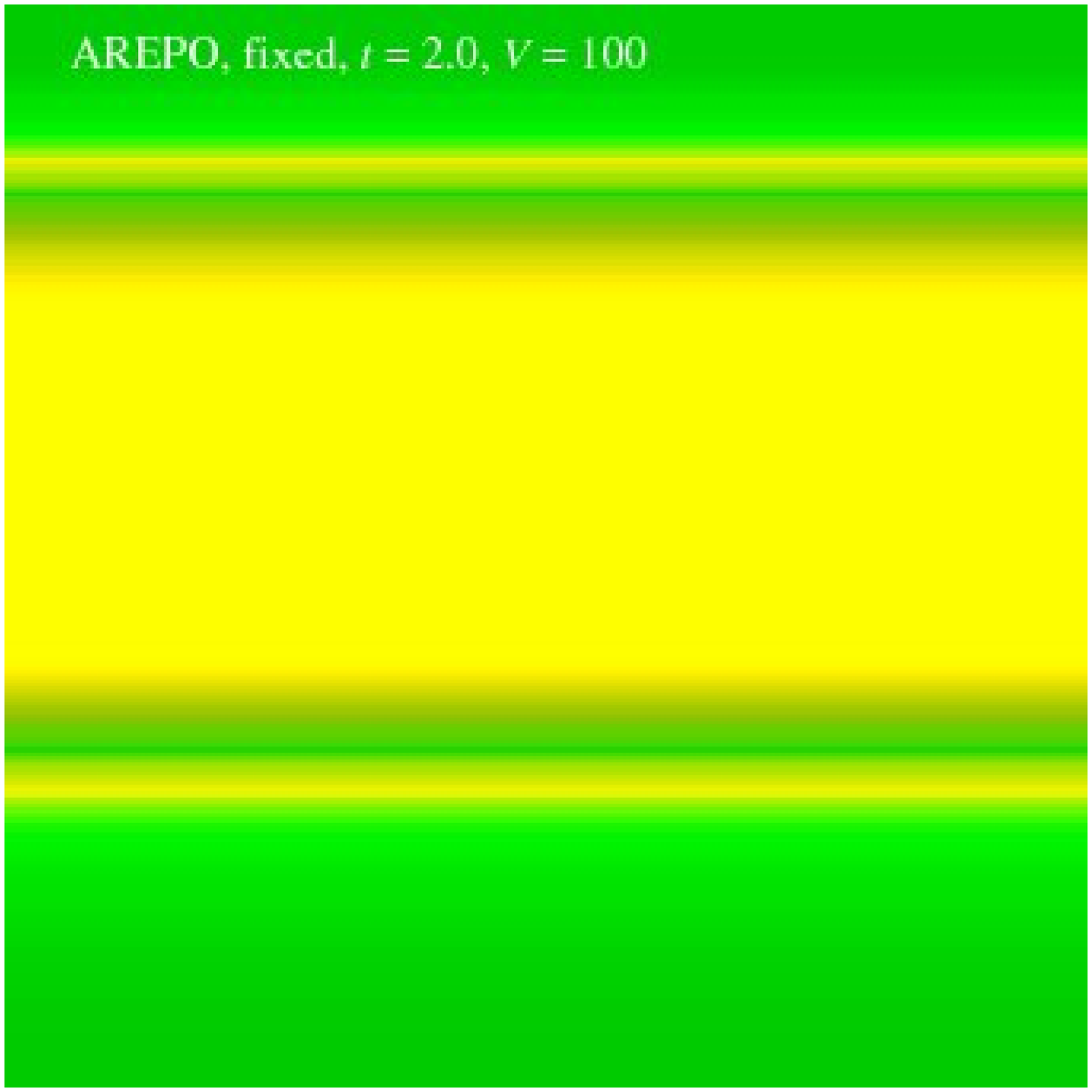}}
\caption{Kelvin Helmholtz instability test at time $t=2.0$, computed with
  {\small AREPO} with a fixed mesh. In the three cases, different boost
  velocities along both the $x$- and $y$- directions have been applied. The
  fact that the results do not agree (and in particular not with the $V=0$
  result shown in the bottom of Figure~\ref{FigKHTimeEvolv}) is direct
  evidence for a violation of Galilean invariance of the Eulerian approach. We
  note that we have obtained nearly identical results for this test when it is
  carried out with {\small ATHENA} instead of our code {\small AREPO}.
  \label{FigKHCompFixedDifferentVx}}
\ec
\end{figure*}

Finally, we now consider 3D calculations of the Noh problem. This represents a
still more demanding test than the 2D problem due to the larger density
contrast reached in the 3D case. To test many of the new features of our
moving-mesh code, we carry out this test with dynamic generation of mesh
cells, and dynamic derefinement in the high-density region. Specifically, a
cell is created if its mass content lies above $m_i> 1.5\,\overline{m}$, and
it is dissolved if its volume has dropped below $V_i<0.05\,\overline{V}$,
where $\overline{m}$ and $\overline{V}$ are the initial average mass and
volume per cell.  In Figure~\ref{FigNoh3D}, we give results for two different
initial resolutions, corresponding to $\sim 16.7^3$ ($\overline{m}=
\overline{V} = 2.16\times 10^{-3}$) and $\sim 50^3$ cells in the unit
octant. With the above refinement/derefinement criteria, by $t=2.0$ the
effective resolution in the lower resolution calculation has grown to $N_{\rm
  eff}\sim 34$ at $r=1$, and in the central high-density region, it is limited
to a maximum of $\sim 45.2$. For the higher resolution calculation, these
numbers are three times as large.  We note that in the lower resolution
calculation, about 812620 cells have been created, and 225209 were destroyed
during the course of the calculation. For the higher resolution calculation,
these numbers are a factor $\sim 27$ higher. In Figure~\ref{FigNoh3D}, we see
that the moving-mesh code with dynamic mesh refinement/derefinement is able to
integrate this problem robustly, with satisfactory accuracy given the
effective resolutions employed here, nevertheless some limited postshock
oscillations are present. Also, some `wall heating' is clearly present at the
centre \citep{Ryder2000, Liska2003}.

\subsection{Kelvin-Helmholtz instability}

Fluid instabilities are among the most interesting phenomena of hydrodynamics,
and they play a crucial role in mixing processes and the production of
turbulence.  Their importance in cosmological gas dynamics is potentially very
large. For example, fluid instabilities are thought to be important for an
accurate treatment of stripping of gas from satellite galaxies, and for
calculating the correct level of turbulence and entropy expected in the
intracluster gas of clusters of galaxies. Recently, numerical inaccuracies of
SPH in the treatment of fluid instabilities across contact discontinuities
with large density jumps have caused concern about the scheme's ability to
adequately treat such problems \citep{Agertz2007}. Fixes have been proposed
for this issue \citep{Price2007KH,Wadsley2008}, but it is not clear yet
whether they can be applied successfully in general calculations without
introducing inaccuracies in other regimes.

\begin{figure*}
\bc
\resizebox{8.0cm}{!}{\includegraphics{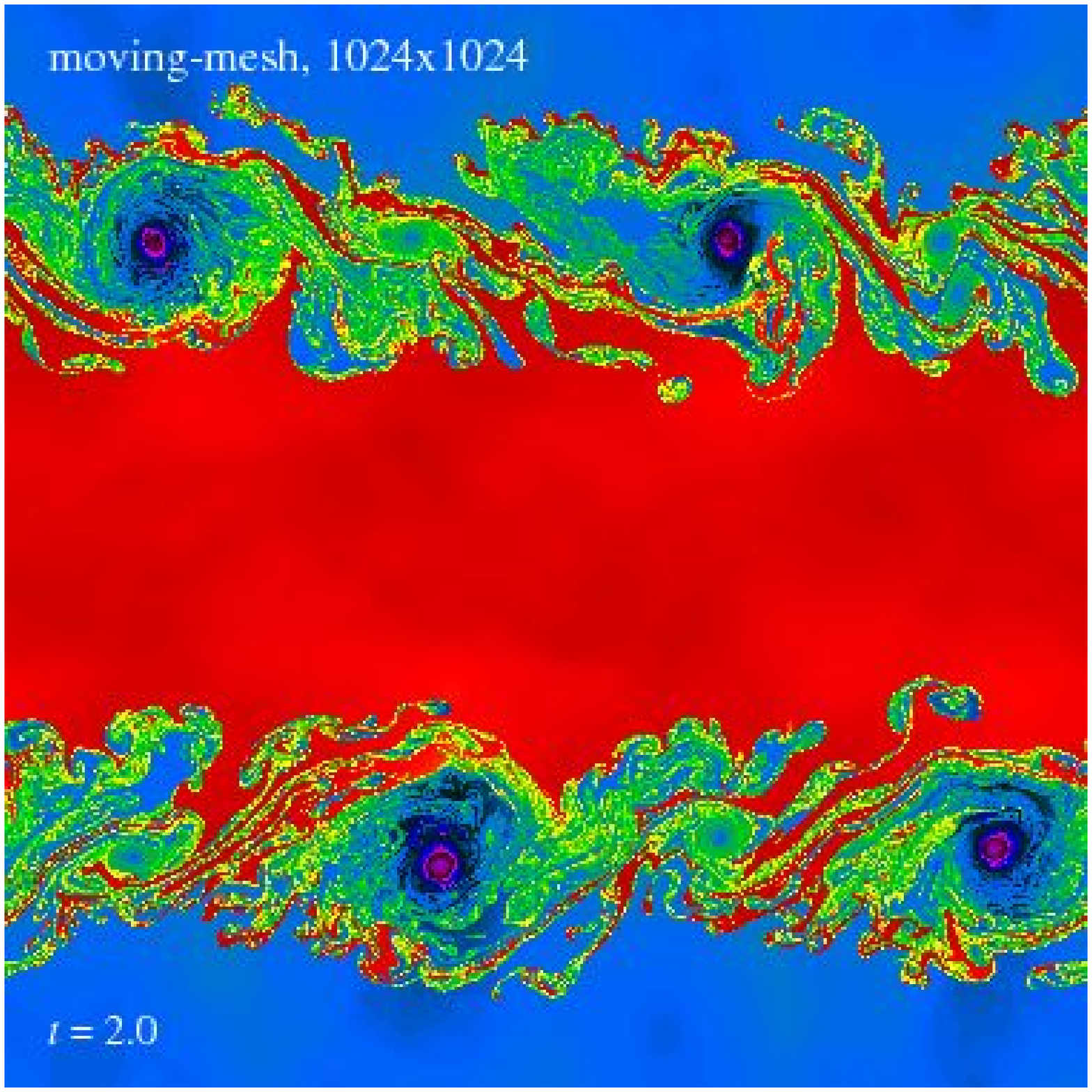}}\ %
\ %
\resizebox{8.0cm}{!}{\includegraphics{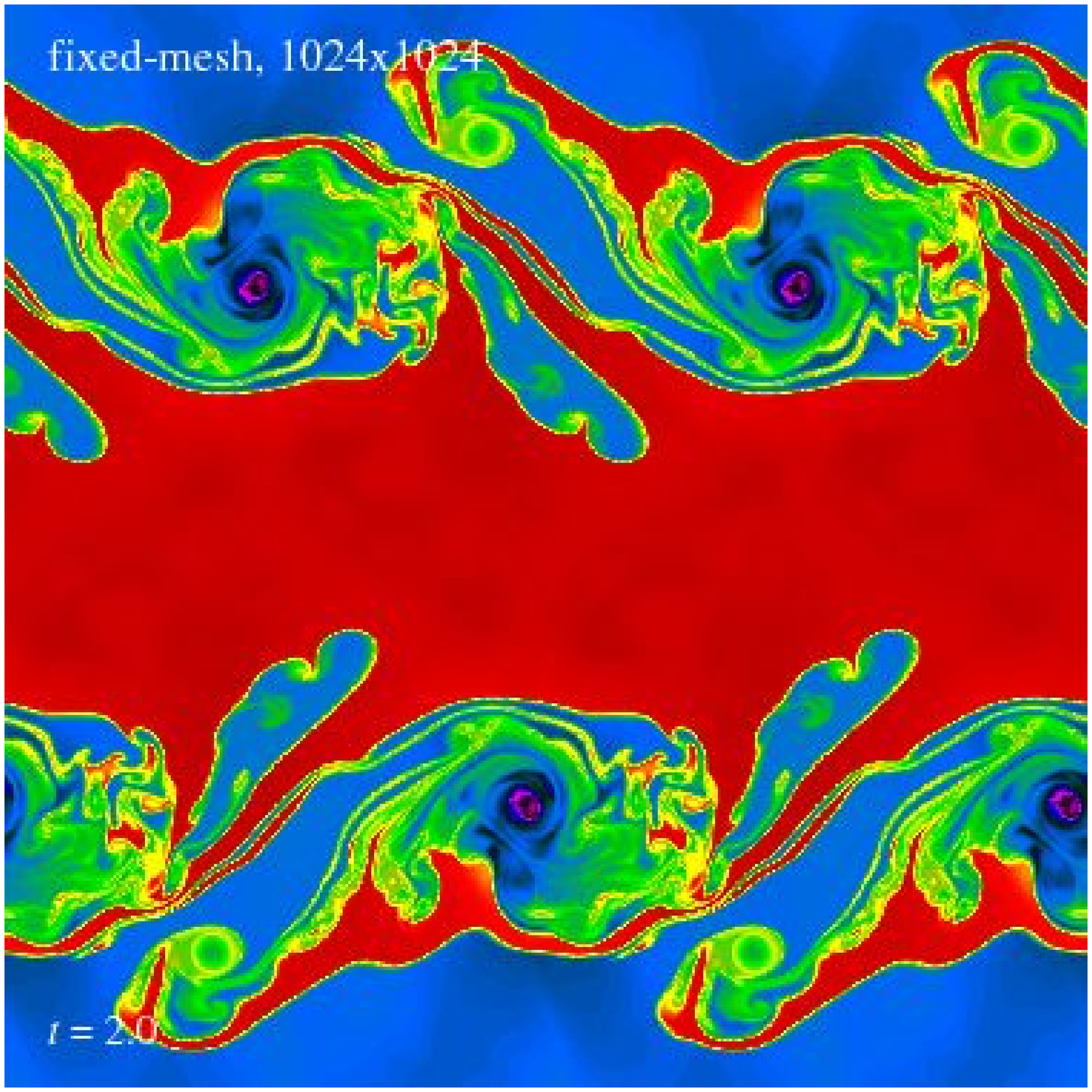}}\\
\caption{Kelvin Helmholtz instability test at high resolution, using a
  $1024\times1024$ initial mesh. We compare results obtained with the
  moving-mesh approach (left panel) to
  those on a fixed grid (right panel), at time $t=2.0$.
Quite strikingly,  small-scale features of the flow are preserved in the moving mesh code
with much less mixing, albeit at the  price of earlier generation of asymmetries in the flow.
\label{FigKHDiffRes}}
\ec
\end{figure*}

On the other hand, it is not obvious that Eulerian methods provide superior
accuracy for fluid instabilities in all regimes, even though this is often
assumed by advocates of these schemes. One can certainly expect that problems
due to the Galilean non-invariance of Eulerian codes could be a source of
concern here. In this subsection we will examine these issues with the
important example of the Kelvin-Helmholtz (KH) instability. This occurs across
contact discontinuities in the presence of a tangential shear flow.

For simplicity, we first consider a simple shear-flow in two dimensions, where
we strongly excite a single mode by imposing a suitable velocity perturbation.
In a periodic domain of unit length on a side, with principal coordinate range
$[0,1]^2$, we set-up gas with density $\rho=2$ in the central horizontal strip
described by $|y-0.5|<0.25$, and give it velocity $v_x=0.5$ to the right,
whereas the rest of the box is filled with gas of density $\rho=1$ that moves
to the left with speed $v_x=-0.5$. The pressure is set to $P=2.5$ everywhere,
with $\gamma=5/3$. There are hence two contact discontinuities along which KH
instability can develop.

To make sure that initially a single mode will dominate the linear growth of
the instability, we excite a single mode with a wave-length equal to half the
box size by perturbing the $v_y$ velocity field according to
\begin{eqnarray}
v_y(x,y) & = & w_0  \sin(4 \pi x) \times\\
& &  \left\{
      \exp\left[ - \frac{(y-0.25)^2}{2 \sigma^2}\right] + 
\exp\left[- \frac{(y-0.75)^2}{2 \sigma^2}\right]\right\} \nonumber
\end{eqnarray}
with $w_0 = 0.1$ and $\sigma = 0.05/\sqrt{2}$.  The two exponential damping
factors restrict the perturbation to the region close to the two
interfaces. The details of how this perturbation is imparted are relatively
unimportant for the test.

We first carry out a test at low resolution, using $50\times 50$ cells that
are initially arranged as a Cartesian mesh. This allows us to visualize the
motion of the mesh as a function of time when our moving-mesh approach is
used. This is illustrated in the time-sequence shown in the top four panels of
Figure~\ref{FigKHTimeEvolv}\footnote{A video of this simulation as well as
  other videos of our example calculations may be found at
  http://www.mpa-garching.mpg.de/$\sim$volker/arepo}, which includes an
overlay of the Voronoi mesh. We see that the moving-mesh approach develops
well-defined KH-billows and is able to maintain a relatively sharply defined
boundary between the two fluids, with only a small amount of mixing between
them.  Also, the moving mesh approach has no problem coping with the strong
shear present in this simulation. This has traditionally been a significant
challenge for Lagrangian hydrodynamics codes.

\begin{figure*}
\bc
\resizebox{14cm}{!}{\includegraphics{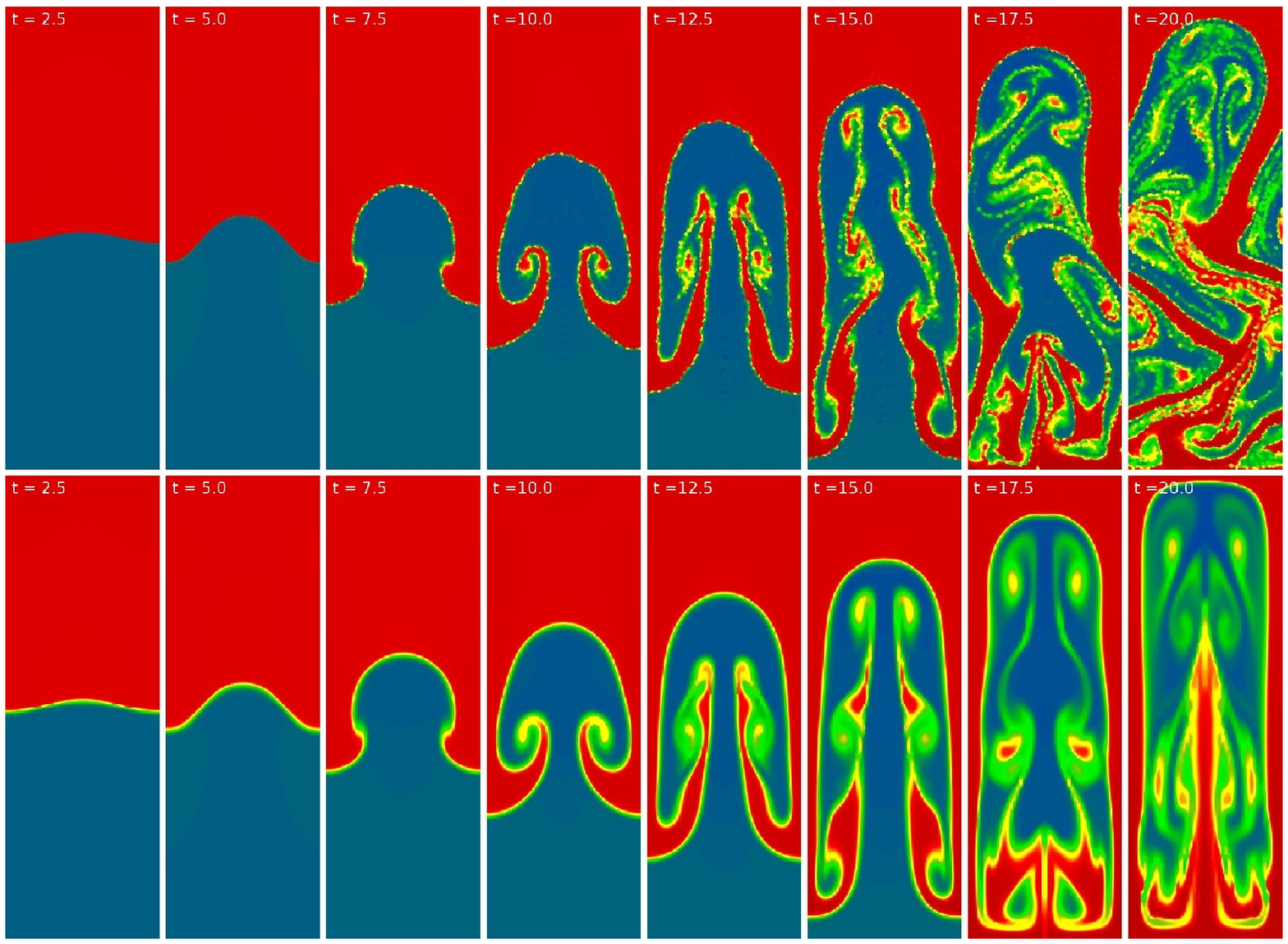}}
\caption{Time evolution of a Rayleigh-Taylor instability in simulations
  with a moving (top) and a static (bottom) mesh. The resolution is
  quite low, only $48\times 144$ cells have been used for this test.
\label{FigRTevolv}}
\ec
\end{figure*}

In the bottom two panels of Figure~\ref{FigKHTimeEvolv} we show the final
result at $t=2.0$ obtained for the same initial conditions when the mesh is
kept fixed instead, in one case calculated with our own code, in the other
with {\small ATHENA}. Reassuringly, the two codes give nearly indistinguishable
solutions when a fixed mesh is used.  This confirms that our code {\small
  AREPO} is comparable in accuracy to state-of-the-art second order accurate
Eulerian codes when run with a fixed mesh. The two fixed-mesh calculations of
Figure~\ref{FigKHTimeEvolv} can also be compared to the moving-mesh result
shown in the top four panels.  Clearly, the results are qualitatively similar,
but there is substantially more mixing in the fixed-mesh calculations, which
wash out the KH-billows to nearly constant density at this resolution. While
this effect is expected to become smaller with increasing resolution, the
effective numerical diffusivity of the Eulerian calculation is clearly much
higher than that of the moving mesh approach, a finding that is also expected
based on the advection tests of a contact discontinuity carried out at the
beginning of this section. There is also a further important difference
between the moving-mesh and the fixed-mesh results. In the non-linear regime,
the KH instability appears to evolve somewhat faster for the moving-mesh
approach than for the fixed mesh. In fact, the fixed-mesh result at $t=2.0$ is
more similar to the $t=1.5$ moving-mesh output than to its $t=2.0$ output.

We next consider the ability of the schemes to cope with additional bulk fluid
motion, or in other words, with a transformation into a boosted frame of
reference. To this end we simply add a constant velocity vector to all cells
of the initial conditions, and we shall again compare the results at time
$t=2.0$.  The physics does not change due to such a Galilean boost, and we
should therefore get the same results. We have already seen that this is not
in general the case for Eulerian codes. Here we test how strong the resulting
effects are in practice.  In Figure~\ref{FigKHCompFixedDifferentVx}, we show
what {\small AREPO} returns for velocities equal to $v=1$, $10$, or $100$,
imposed both in the $x$- and $y$-directions, {\em if the mesh is kept fixed}.
Thanks to the periodic boundary conditions, the system will have returned at
time $t=2.0$ again to the its original position, after the code had to advect
it one or several times through the box. In principal, all three results
should therefore be identical. But the actual results are very different; in
this Eulerian mode, the code's result and hence the error in the calculation
is a strong function of the magnitude of the bulk velocity relative to the
rest-frame of the calculation. Especially when these velocities become
supersonic, the calculated solution can become qualitatively and
quantitatively inaccurate. For the somewhat brutal test of $v=100$, the
calculated solution is completely dominated by advection errors, with the
density field in the box becoming almost homogeneous.  We have also evolved
the same initial conditions with the {\small ATHENA} code, finding very
similar results.  We expect this behaviour to be generic for Eulerian
codes. The accuracy with which fluid instabilities are calculated across
contact discontinuities quickly deteriorates if the contact discontinuity
moves. In contrast, when we run the same initial conditions with the {\em
  moving mesh} of {\small AREPO}, we recover the same results in all three
cases, and they are identical to the $v=0$ result shown in
Figure~\ref{FigKHTimeEvolv}.

Is this a serious problem for Eulerian codes? This very much depends on the
problem that is studied. In many applications, an individual system is studied
and one can freely choose a convenient reference frame for the
calculation. One will then pick one in which velocities relative to the
calculational frame are small. The issue of Galilean non-invariance may then
not be of great concern. However, we argue that this is not the case for
cosmological simulations, where multiple objects are simulated at the same
time, many of them moving with large velocities compared to their sound
speed. In this case, the accuracy of the calculation correlates strongly with
the bulk velocity of the galaxies, a rather worrying effect.

However, Galilean non-invariance may not be the only problem that troubles the
Eulerian approach when the Kelvin-Helmholtz instability is considered.  We
have found that in an Eulerian calculation of this problem at high resolution,
multiple secondary KH billows are spawned in the early evolution due to grid
irregularities. On the other hand, the moving-mesh method appears less
susceptible to this problem, thanks to its ability to advect the mesh with the
contact discontinuity. Some of these grid-induced features can even affect the
long-term evolution of the instability.  In Figure~\ref{FigKHDiffRes}, we
compare fixed and moving-mesh versions of the same KH instability test at a
resolution of $1024^2$. The moving-mesh preserves much more fine detail in the
flow. This is because contact discontinuities between different phases can be
advected with large speeds without being necessarily mixed. We think this is a
very interesting difference, which makes the moving-mesh code particularly
attractive for the study of multi-phase media.

\begin{figure}
\bc
\resizebox{7cm}{!}{\includegraphics{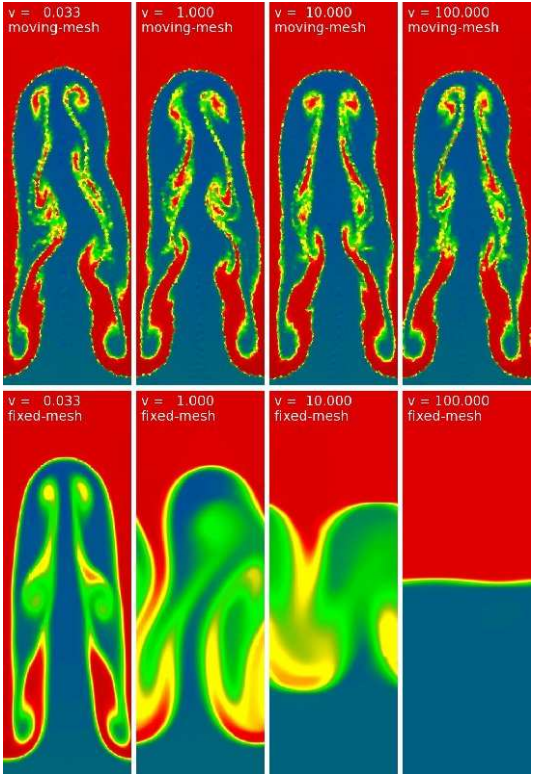}}
\caption{Rayleigh-Taylor instability calculated with different
  Galilean boosts $v_x$ in the horizontal direction (the simulation
  domain is periodic in the $x$ direction). 
  The correct result should in principal be independent of
  $v$. The top row shows the result at time $t=15.0$ computed with
  our moving-mesh approach, while the bottom row of panels gives the
  corresponding results for a fixed-mesh calculation with {\small AREPO}.
  \label{FigRTDifferentVx}}
\ec
\end{figure}

\begin{figure}
\bc
\resizebox{8cm}{!}{\includegraphics{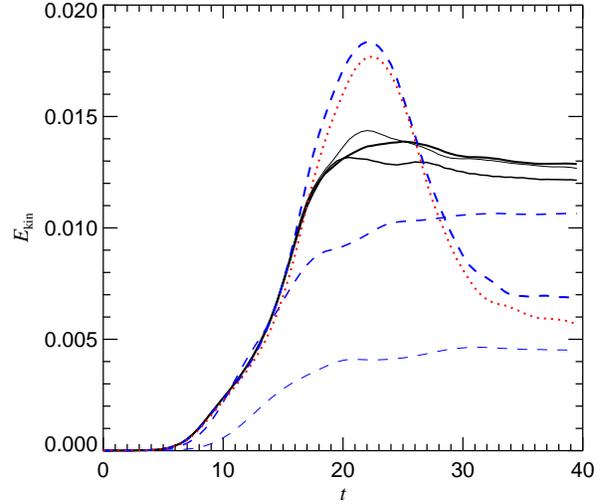}}
\caption{Kinetic energy in our Rayleigh-Taylor test calculations as a
  function of time. For the boosted simulations, the kinetic energy
  was defined by first subtracting the horizontal boost from the $v_x$
  velocity component. The black solid lines show the moving-mesh
  calculations, for velocity boosts $v_x=0$, $v_x=1$ and $v_x=10$ (the
  thickness of the lines decreases in this sequence). The dashed lines
  give the results for fixed-mesh calculation with our code, for the
  same three velocities, from top to bottom.  Finally, for comparison,
  the dotted line gives the result obtained with {\small ATHENA} for
  $v_x=0$.
\label{FigRTKineticEnergy}}
\ec
\end{figure}

\subsection{Rayleigh-Taylor instability}

Another important type of fluid instability arises in stratified
atmospheres in approximate hydrostatic equilibrium if a denser fluid
lies above a lighter phase.  In such a Rayleigh-Taylor (RT) unstable
state, energy can be gained if the lighter fluid rises in the
gravitational field, triggering buoyancy-driven fluid motions.

We consider a simple test in 2D where we excite a single Rayleigh-Taylor 
mode for clarity. Our setup is a small variation of a similar test considered
in \citet{Liska2003}, and also similar to a test described in Jim Stone's test
suite of {\small ATHENA}. The computational domain is chosen as $x\in
[0,\;0.5]$ and $y\in[0,1.5]$, with periodic boundary conditions at the
$x$-boundaries, and reflecting walls at the top and bottom of the domain.  The
density is $\rho=2$ in the top half of the domain, and $\rho=1$ in the bottom
half. The pressure in the vertical midplane is $P_0=2.5$ (with $\gamma=1.4$)
and varies vertically as $P(y) = P_0 + g\,(y-0.75)\rho$, where $g=-0.1$ is an
imposed external gravitational field. This ensures an initial hydrostatic
equilibrium. The initial velocities are zero everywhere, except for a small
perturbation that is designed to excite a single mode for the Rayleigh-Taylor
instability. We adopt for this perturbation
\begin{equation}
v_y(x,y) = w_0\, [1 - \cos(4\pi x)][1 -  \cos(4\pi y/3)],
\end{equation}
where $w_0= 0.0025$.  For the tests discussed in the following, we
deliberately use a comparatively low resolution of $48\times144$ cells.

In Figure~\ref{FigRTevolv}, we first compare the time evolution of the
system between a calculation carried out with a static Cartesian mesh,
and one with our new moving-mesh approach. Clearly, the evolution is
rather similar during the early linear growth of the
perturbation. However, the moving-mesh approach is able to maintain a
sharper contact discontinuity, as expected. Eventually, the evolution
starts to differ markedly once the dynamics becomes very non-linear.
The moving-mesh solution loses vertical symmetry and starts to develop
turbulence. In contrast, the fixed mesh calculation is able to maintain
perfect symmetry for a longer time, but it shows much stronger mixing
than the moving-mesh calculation, which also damps the fluid motion. In
both cases, the loss of symmetry is caused by small round-off errors,
but in the moving mesh approach their growth in the transverse direction
is faster and less benign. This is due to a kind of bending instability
in the mesh. When the mesh is compressed strongly in one direction, it
automatically means that the cells develop a large aspect ratio, which
can only be relaxed (in the sense that the cells become rounder again)
through some transverse motions. It turns out that the mesh likes to
respond to transverse perturbations in this situation; they tend to grow
quickly, and numerical round-off is sufficient to trigger this. This
also means that the moving mesh can itself be a source of unwanted
perturbations, but they only really become relevant in poorly resolved
flows, where the shape of an individual cell directly matters. This is
similar to the Kelvin-Helmholtz problem on a Cartesian mesh, where at
high resolution small-wavelength secondary billows are triggered by
perturbations originating at the mesh corners. If the Rayleigh-Taylor
problem is simulated with higher resolution and with resolved
(i.e.~softened) phase boundaries, symmetry is maintained much longer in
the moving-mesh approach.

We have also used this Rayleigh-Taylor test to investigate once more the
question of Galilean invariance. To this end we have added a constant velocity
$v_x$ to the initial state. As the system is periodic in the $x$-direction,
the evolution should in principle not change if viewed in the rest-frame of
the moving fluid. In Figure~\ref{FigRTDifferentVx}, we show a comparison of
the resulting fluid state when either a moving mesh or a fixed mesh is used in
our code {\small AREPO}.  We carry out this comparison for horizontal flow
velocities of $v_x=0.033$, $v_x=1$, $v_x=10$, and $v_x=100$. It is seen that
the Eulerian result changes significantly along this sequence. The horizontal
motion leads to a damping of the growth rate of the instability, additional
mixing, and also to a loss of symmetry. In fact, for a sufficiently large
velocity boost, the instability is suppressed entirely. We note that in this
case the Eulerian calculations also become much more expensive, as their
Courant condition becomes limited by the bulk velocity. The moving-mesh
calculation does not suffer from this problem, and it produces a
Galilean-invariant solution as expected. However, the way the symmetry is lost
in the calculation tends to vary, as this is determined by small numerical
noise in the mesh motion.

In Figure~\ref{FigRTKineticEnergy} we show the kinetic energy in the
simulations as a function of time.  The thick dashed line and the
dotted line show the results for a fixed mesh with no velocity boost,
comparing our code {\small AREPO} with {\small ATHENA} (the latter
using the Roe solver and second-order accurate reconstruction). The
results of the two codes for the evolution of the kinetic energy agree
qualitatively very well, showing first a maximum, followed by a rapid
decline to a nearly constant level that then declines over a much
longer timescale. This is produced by the initially very symmetric
evolution of the RT instability when a Cartesian mesh is used, and the
subsequent transition to turbulent motions at later times. The
symmetry is broken when a velocity boost is applied in the Eulerian
calculations, and even a velocity as small as $v_x=0.033$ is
sufficient for that. The other two dashed lines of
Fig.~\ref{FigRTKineticEnergy} show the evolution of the kinetic energy
(relative to the rest frame of the gas) when a boost of $v_x=1$ or
$v_x=10$ has been applied, respectively. Now the pronounced maximum is
gone, and for increasingly larger boost velocities the plateau of the
kinetic energy becomes ever smaller. For the moving mesh case on the
other hand, the results are insensitive to the velocity boost. This is
shown by the three solid lines, which give the kinetic energy for
$v_x=0$, $v_x=1$ and $v_x=10$, respectively.  These Lagrangian
calculations also do not produce the strong maximum seen in the
$v_x=0$ case for the Eulerian code, which we interpret as effectively
being an artefact of the grid symmetry, because the higher asymmetries
in the numerical round-off errors present in the moving-mesh approach
are sufficient to break the symmetry in the evolution early on.

\subsection{Moving boundaries} \label{SecCoffee}

As briefly discussed earlier, the moving mesh approach can also be quite
easily adapted to describe curved boundaries of essentially arbitrary shape,
and if desired, these boundaries can also move in complex ways. While this
feature is probably not helpful in most astrophysical problems, it has
potentially very useful applications in other areas, for example
aerodynamics. We here discuss a simple example to illustrate this
possibility.

\begin{figure} 
\bc
\resizebox{7cm}{!}{\includegraphics{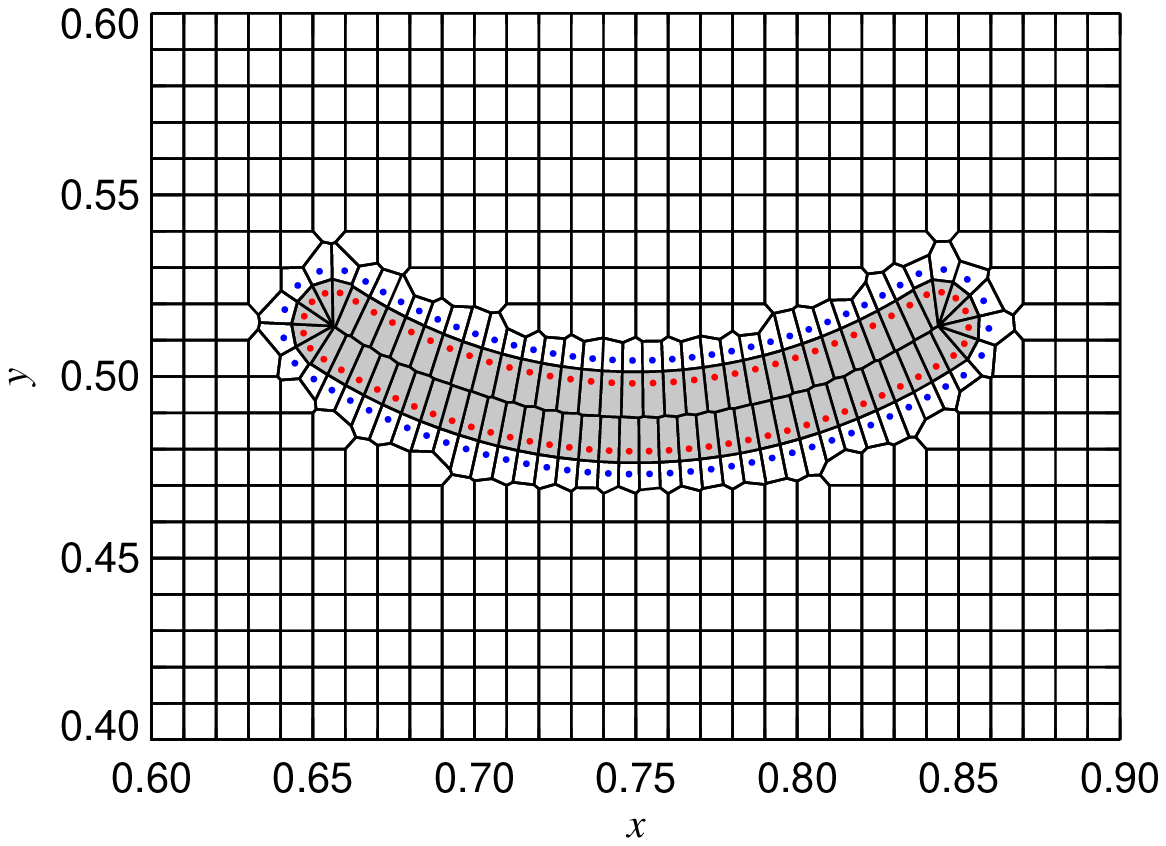}}\\%
\resizebox{7cm}{!}{\includegraphics{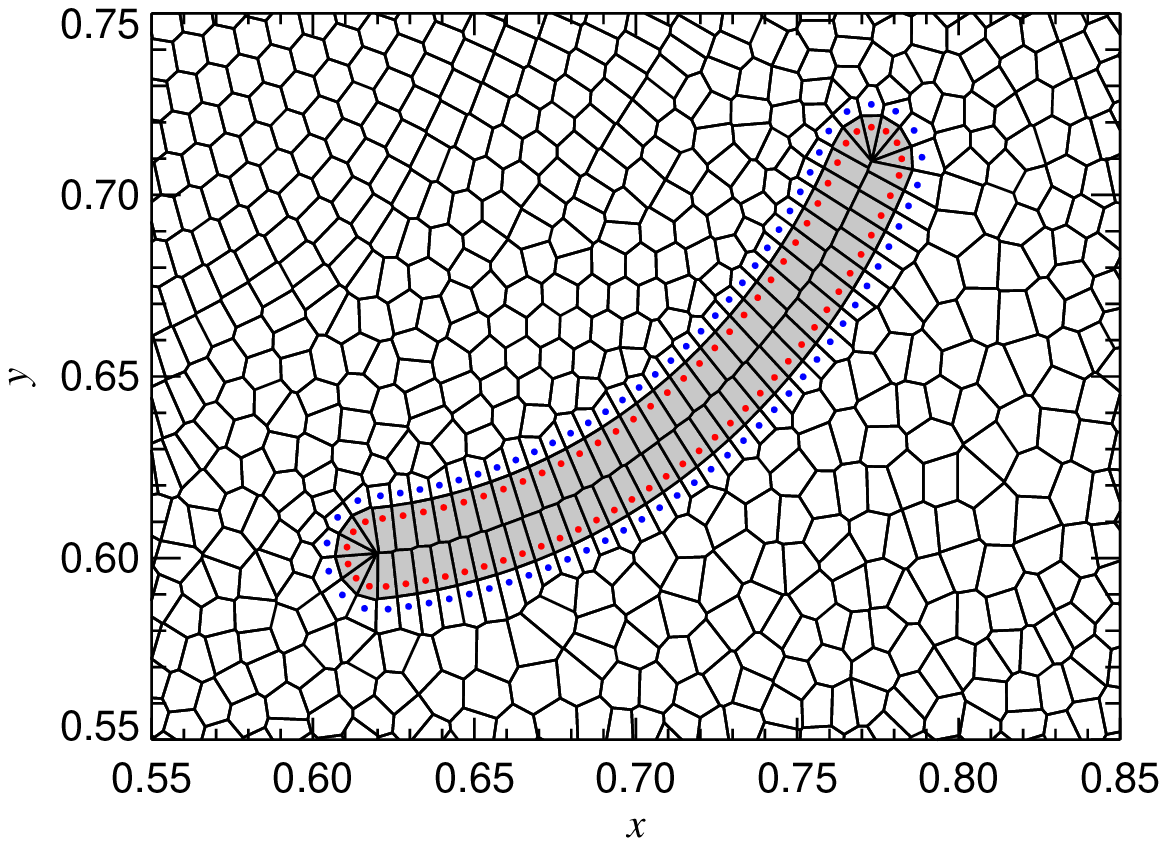}}
\caption{Example of a reflective curved boundary condition, realized with two
  adjacent strings of 60 mesh-generating points (top panel). The blue points
  are on the fluid   side, while the red points are `outside'. The Voronoi faces between these
  points are given reflective boundary conditions, while the rest of the mesh
  is treated in a regular fashion. In the bottom panel, we show the mesh
  geometry at a slightly later time, when the special points have moved as a
  solid body on a
  prescribed path, while the rest of the mesh and the surrounding fluid have
  reacted to this motion.
\label{FigCoffeeMesh}}
\ec
\end{figure}

In Figure~\ref{FigCoffeeMesh}, we show how a special curved boundary can be
constructed in terms of two parallel strings of mesh-generating points that
have an equal distance from either side of the desired contour. In our
example, we want to model a solid obstacle, where the red points are meant to
be inside the obstacle, and the blue points are outside, i.e.~on the side of
the fluid. The Voronoi faces between the points will be modelled with
reflecting boundaries by the code, and represent the surface of the solid
body. Unlike the other mesh-generating points that define the mesh, the red
and blue points are only moved together, as a rigid body. We can now impose a
motion for our solid object, and the fluid will always be forced to flow
around it. In our chosen example, we will move the object with constant
velocity according to a prescribed path, with the fluid being initially at
rest. This means we will effectively stir the fluid with the object, much like
moving a spoon in a coffee cup, except that our ``coffee cup'' is
two-dimensional here. The lower panel of Figure~\ref{FigCoffeeMesh} shows the
mesh geometry around the object after it has moved by a small amount. It is
nicely seen that the points that model the curved boundary condition have moved
together as a rigid body, while the background mesh reacted to this motion by
starting to flow around the object.

\begin{figure*} 
\bc
\resizebox{15cm}{!}{\includegraphics{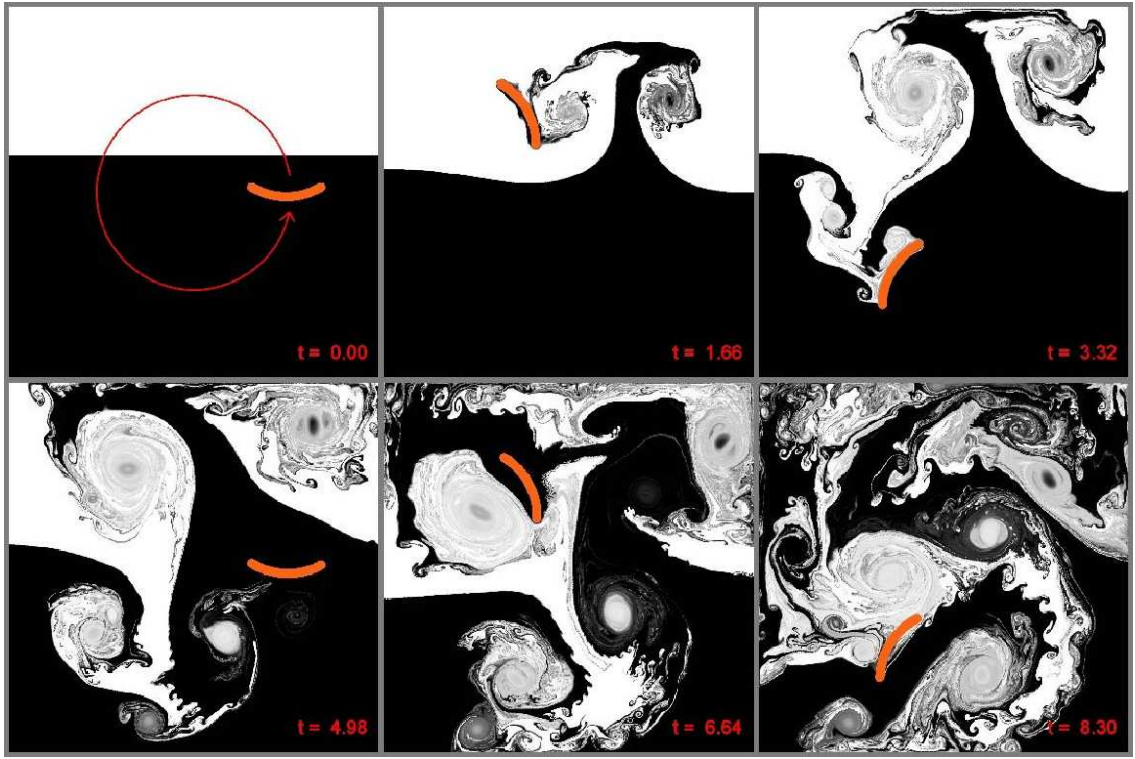}}\\%
\caption{Time evolution of the  mixing of two fluids induced by the motion of 
a solid object. This test illustrates the ability of {\small AREPO} to cope
with arbitrarily curved, moving  boundary conditions. 
As
illustrated, the orange `spoon' is moved on a circular path, through a  two-phase
gaseous medium that is initially at rest. The mixing is shown in terms of a 
 tracer dye that is advected with the flow, and which was given an initial
 value of 
1 (white) in the lower-density top phase,
and a value of 0 (black) in the higher-density lower phase. Each frame shows the value of
the dye in grey-scale, at different times as labeled. The square domain was
initially populated with a $768\times768$ mesh-generating points on a Cartesian grid, and has reflective
boundary conditions at the outer walls.
\label{FigCoffeeMixing}}
\ec
\end{figure*}

In Figure~\ref{FigCoffeeMixing} we show the time evolution of a test problem
calculated with such a moving boundary. The background fluid is represented
with $768\times 768$ points, and the solid object with 600 particles. The
domain $[0,1]\times[0.1]$ is modelled with reflecting boundaries on the
outside. The upper half for $y>0.6$ is filled with gas of density $\rho=0.5$,
the lower with gas at unit density $\rho=1$. The pressure is $P=1$ everywhere,
with $\gamma=5/3$. Our solid object is rotated around the centre in
counter-clockwise direction with an angular velocity $\omega = 2\pi/5$. In
this particular example we are especially interested in the mixing of the two
phases of the initial configuration. To this end we give each phase a `dye', a
conserved tracer variable. This is followed as a passive conserved scalar
along with the ordinary fluid variables by the code. In
Figure~\ref{FigCoffeeMixing}, we show the value of this dye as function of
time, with the solid body displayed in orange.

It is nicely seen how the motion of the object induces complex gas motions,
including the generation of vorticity and turbulence.  This eventually leads
to a complete mixing of the two phases, but for a long time partially mixed
regions survive. Thanks to the motion of the mesh with the flow, contact
discontinuities between the two media can be advected almost without numerical
errors, allowing the foliated structure of partially mixed fluid to remain
intact even while moving. Such a low level of numerical diffusivity would be
very difficult to achieve with an Eulerian treatment.

\section{Test problems with self-gravity} \label{SecGravityTests}

As the long discussion in Section~\ref{SecGravity} made clear, an accurate
treatment of self-gravity in finite volume codes is actually a surprisingly
subtle and tricky problem, more so than in SPH. In this section we will first
discuss a three-dimensional gravitational collapse problem of a cold gaseous
sphere, which is a good test for energy conservation in the presence of a
strong virialization shock, a scenario that is of direct relevance for
cosmological simulations. We then examine the collapse of Zeldovich pancakes
as a basic test of the cosmological integration in {\small AREPO}.  We finally
turn to two example applications of our new code, a colliding galaxy problem
and the `Santa Barbara cluster'. Both of these problems are primarily meant
to illustrate that the {\small AREPO} code introduced here is fully functional
and suitable for science applications in computational cosmology.

\subsection{Evrard's collapse test} \label{SecEvrard}

\citet{Evrard1988} has introduced an interesting collapse problem that has
been frequently used in the literature to test SPH simulation codes
\citep[e.g.][]{Hernquist1989,Dave1997,Springel2001gadget,Wadsley2004}, but
results for mesh codes have been rarely reported.  The initial conditions
consist of a sphere of gas with mass $M=1$ and radius $R=1$, with an initial
density profile of the form
\begin{equation}
\rho(r)=\left\{
\begin{array}{ll}
{M}/(2\pi R^2 r) & {\rm for}\; r\le R\\
0 & {\rm for} \; r>R .\\
\end{array}
\right.
\end{equation} 
The gas with adiabatic index $\gamma=5/3$ is initially at rest and has thermal
energy $u=0.05$ per unit mass, which is negligible compared with the
gravitational binding energy (assuming $G=1$).

In the beginning of the evolution, the gas is freely falling towards the
origin under self-gravity. Eventually, it bounces back in the centre, with a
strong shock propagating outwards through the still infalling outer parts of
the gas sphere. The system then virializes and settles to a spherical
distribution in hydrostatic virial equilibrium. The time evolution of the
system is hence characterized by a conversion of gravitational potential
energy first to kinetic energy, and then to heat energy. As such, it tests a
situation that is prototypical for gravitationally driven structure growth,
and also provides a sensitive test of the ability of a code to conserve the
total energy accurately in self-gravitating gaseous systems.

\begin{figure*}
\bc
\resizebox{15cm}{!}{\includegraphics{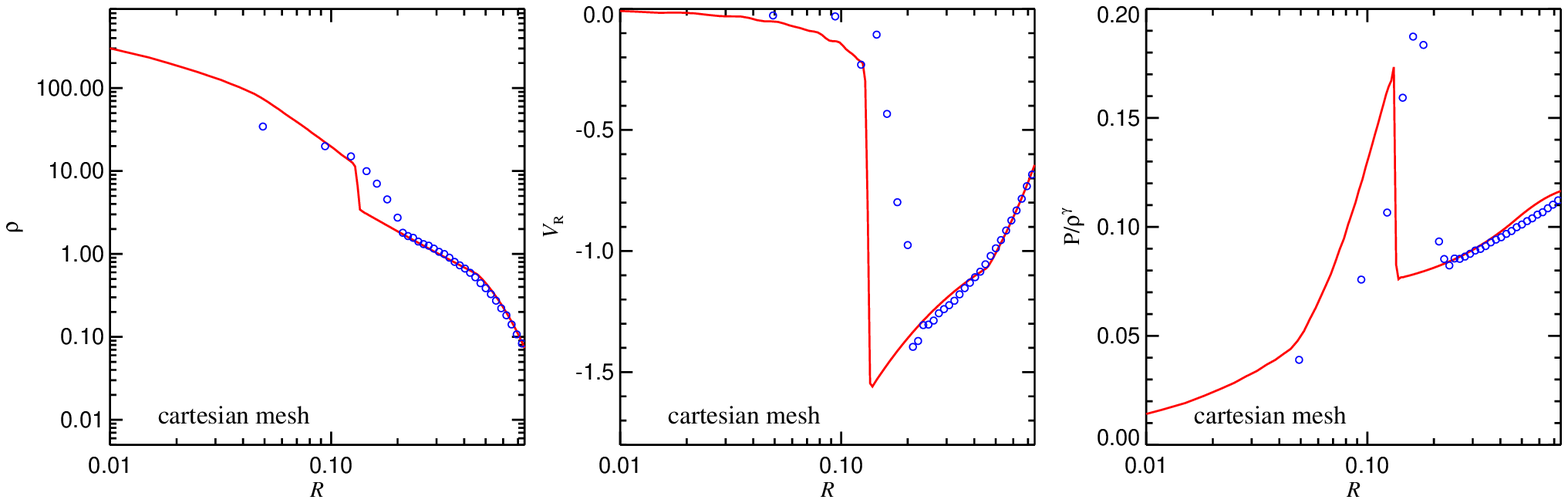}}\\
\resizebox{15cm}{!}{\includegraphics{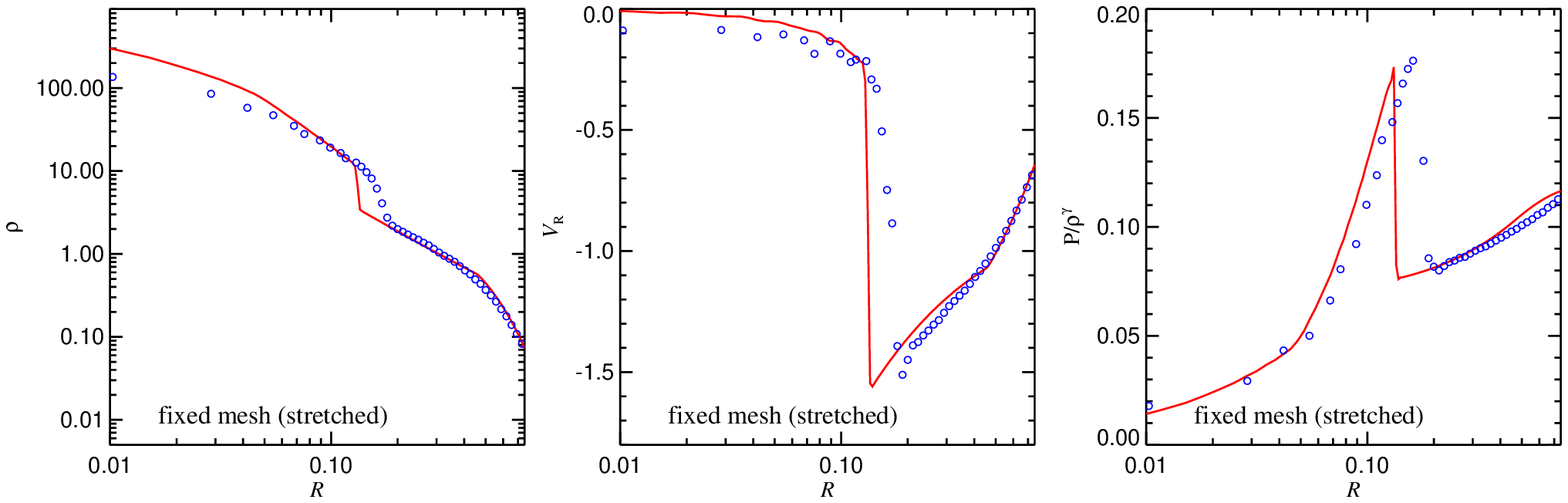}}\\
\resizebox{15cm}{!}{\includegraphics{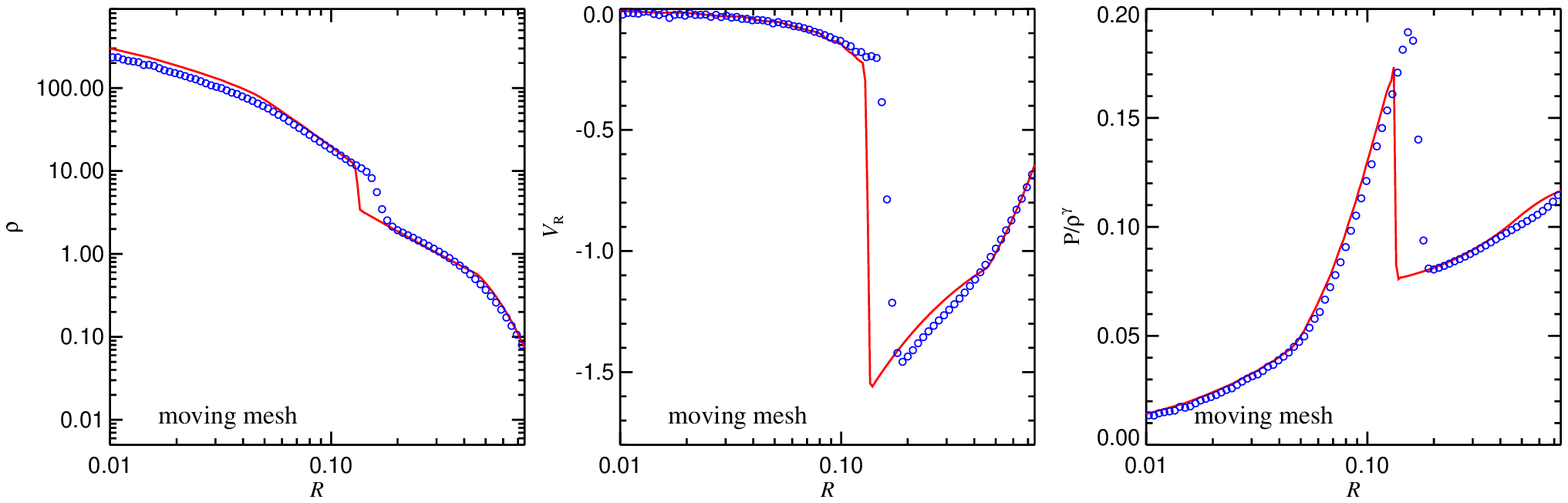}}\\
\resizebox{15cm}{!}{\includegraphics{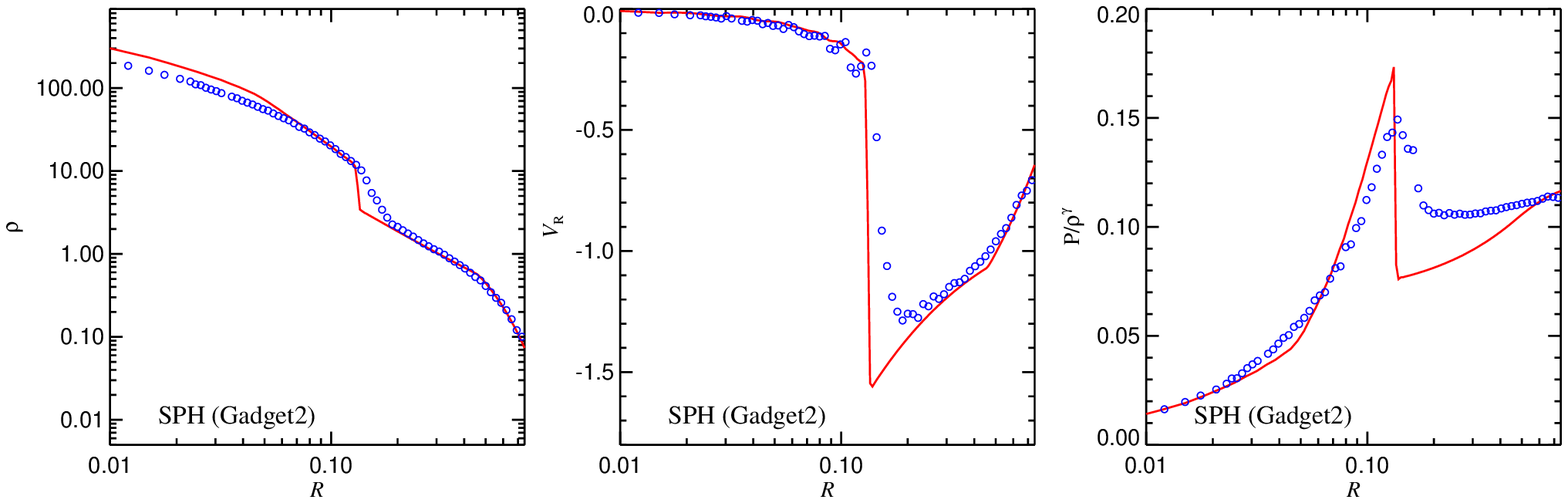}}\\
\caption{Shock profiles in the `Evrard-collapse' problem, carried out
  with different simulation techniques. In all cases, the same number
  of resolution elements inside the initial radius $R=1$ of the gas
  cloud has been used. The top panel gives the result at $t=0.8$ for a
  fixed Cartesian grid. The second row shows the result for {\small
    AREPO} when the mesh-generating points are arranged as a stretched
  grid of points such that the mass per cell is constant for the {\em
    initial} $\rho \propto 1/r$ profile, but the mesh was kept static
  in this case. This is different in the third row; here the mesh was
  allowed to move with the flow, and in addition the mesh-shaping
  scheme based on the `inverse Zeldovich' approach was enabled. 
  Finally, the bottom row gives the equivalent
  result obtained with the same particle number using SPH, as
  implemented in the {\small GADGET-2} code. The red solid line is a
  one-dimensional PPM result obtained by \citet{Steinmetz1993}.
  \label{FigEvrardProfiles}}
\ec
\end{figure*}

In Figure~\ref{FigEvrardProfiles}, we show radial profiles of density,
velocity, and entropic function $A=P/\rho^\gamma$ at time $t=0.8$, when the
strong shock has formed. We compare simulations carried out with different
calculational schemes, but all with the same number of 24464 resolution
elements in the initial radius of the sphere. The top three rows give results
calculated with our mesh-code {\small AREPO}. In the first case, we consider a
fixed Cartesian mesh, which we expect to be challenged by the large dynamic
range of this problem. In the second and third row we use a radially stretched
mesh as initial conditions, which has equal mass per cell initially. This
mesh is much better adjusted to the radial symmetry of the system, and the
increase of density towards the origin. The simulation shown in the second row
keeps this unstructured mesh fixed throughout the evolution, while in the
third row the full moving-mesh approach is applied, where the mesh-generating
points are moved with the local flow velocity. Finally, in the bottom row, an
alternative Lagrangian result is shown, this time based on SPH, using the
`entropy-formulation' of \citet{Springel2002} as implemented in {\small
  GADGET-2} \citep{Springel2005}.

\begin{figure*}
\bc
\resizebox{15cm}{!}{\includegraphics{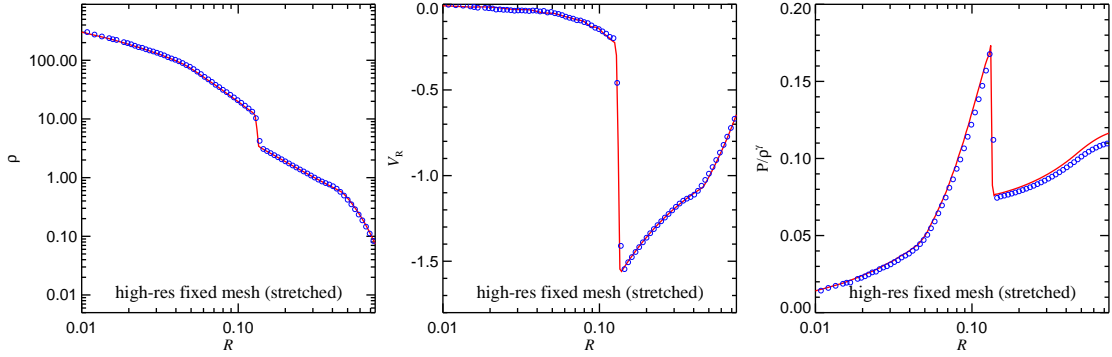}}\\
\caption{High resolution result (symbols) for the Evrard collapse problem
  calculated with {\small AREPO}, using
  $1.56\times 10^6$ resolution elements in the initial gas sphere of radius $R=1$.
An analytic solution for this problem is unavailable, but the solid 
gives the results of a one-dimensional high-resolution PPM calculation kindly
provided to us by \citet{Steinmetz1993}
which should be fairly close.
\label{FigEvrardHighRes}}
\ec
\end{figure*}

\begin{figure*}
\bc
\resizebox{8.5cm}{!}{\includegraphics{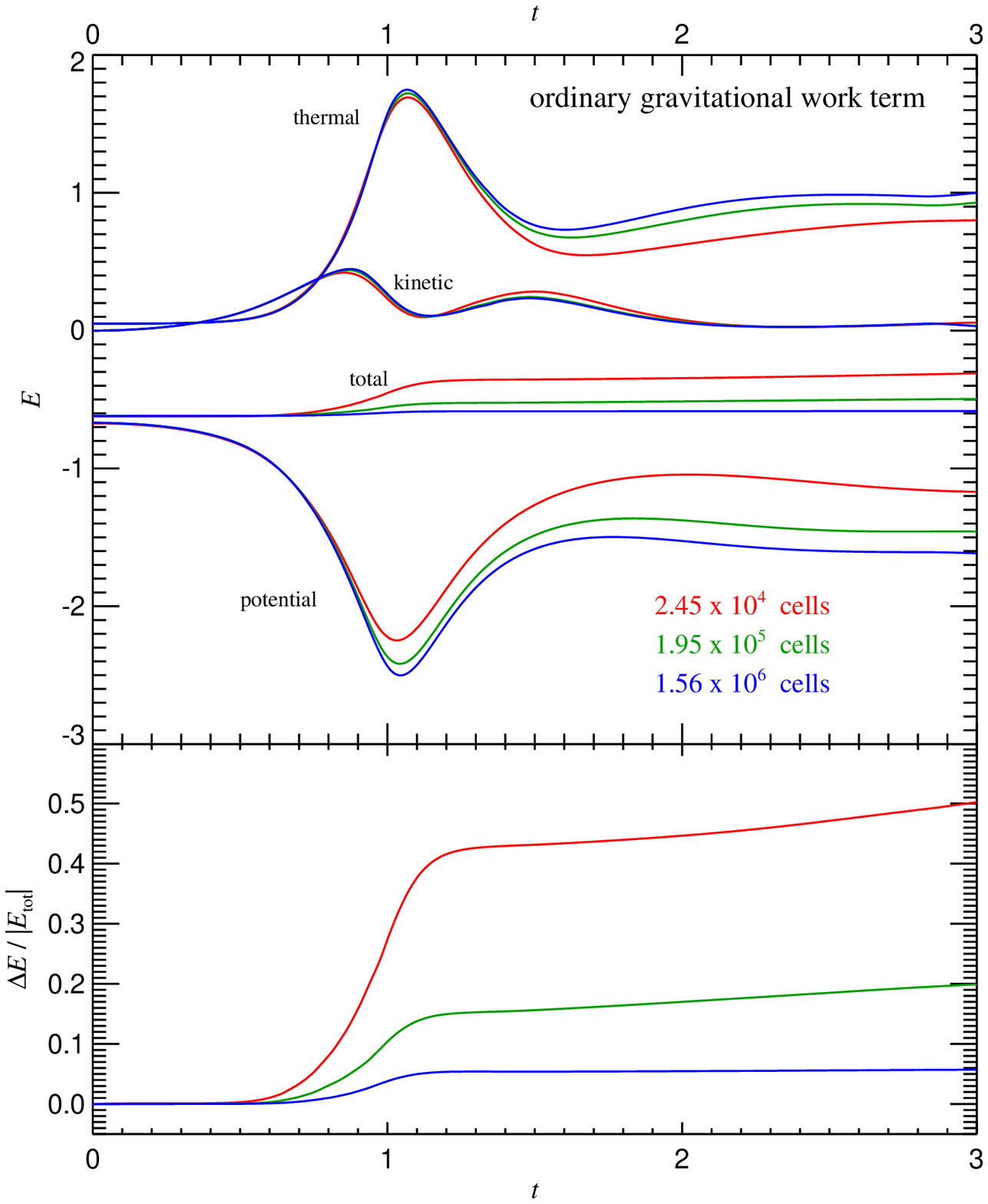}}\hspace*{0.5cm}%
\resizebox{8.5cm}{!}{\includegraphics{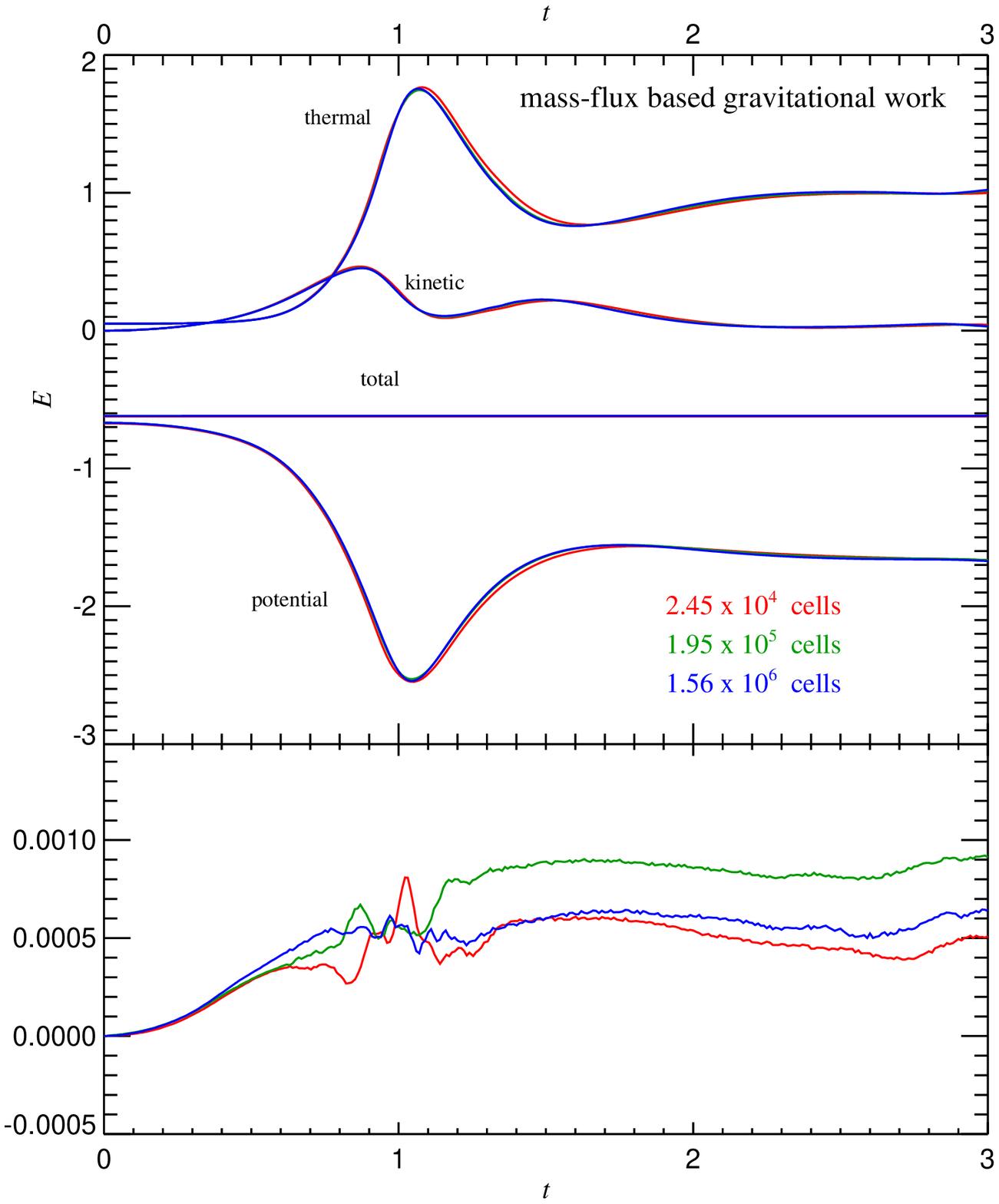}}
\caption{Energy evolution in the `Evrard-collapse' problem, calculated for
  different numerical resolutions and different calculational schemes to
  couple self-gravity to the Euler equations. In the panel on the left,
  results for a standard approach to treat self-gravity are shown. Here the
  gravitational work on a cell is estimated with cell-centred mass fluxes.
  This can however produce substantial errors in the presence of strong shock
  waves, especially when the spatial resolution is quite coarse. The plot
  on the right gives results for the same simulations, except that 
  the gravitational work term is   estimated based on the mass-fluxes determined by the Riemann solver at the
  surfaces of cells. This leads to a quite accurate conservation of the total
  energy of the system.
  \label{FigEvrardEnergy}}
\ec
\end{figure*}

Among these calculations, the least accurate result is clearly produced by the
fixed Cartesian mesh, which offers the poorest spatial resolution in the
central regions of the sphere, as a result of its lack of adaptivity. The
stretched fixed grid already gives much better results, but the central
density distribution is still significantly underestimated. However, if the
mesh is allowed to move, a significantly improved solution is obtained, even
though here also the limited spatial resolution produces a shock front that is
radially too far advanced compared to the expected solution for close to
infinite resolution. The latter is shown as a solid line and was produced by a
one-dimensional PPM calculation kindly provided by \citet{Steinmetz1993}. We
note that the SPH result shown in the bottom row also produces the main
features quite well, but its shock is significantly broader than in the moving
mesh calculation, and there is also substantial pre-shock entropy production
in the infall region ahead of the shock, as a result of the artificial
viscosity that becomes active in converging parts of the flow. 

The timing offset in the shock location appears to be a result of the low
resolution used in this test, as this vanishes for better resolution. To
illustrate this point, we show in Figure~\ref{FigEvrardHighRes} an equivalent
plot for a high-resolution simulation of the same problem using $1.56\times
10^6$ mesh-generating points, arranged in a stretched mesh that is here kept
fixed during the evolution (the moving mesh gives an essentially
indistinguishable result). The result reveals an excellent agreement with the
high-accuracy one-dimensional calculation.

Finally, we consider the conservation of total energy in this problem, as this
is not readily guaranteed in the finite volume approach with self-gravity. In
Figure~\ref{FigEvrardEnergy}, we show the time evolution of the thermal,
kinetic, and potential energy, for simulations of the Evrard collapse carried
out with different numerical resolutions and different strategies to couple
the gravitational field to the hyperbolic Euler equations. In the left panel,
results for the ``standard'' approach to treating self-gravity, discussed in
subsection \ref{SecEgyStandardApproach} are shown. We can see that there are
substantial errors in the total energy, which amount to a relative error as
large as $\sim 50\%$ for the poorest resolution considered here, where 24464
cells are inside the initial radius of the sphere. With better spatial
resolution, the size of the error progressively shrinks. However, it cannot be
made smaller by improving the time integration as it is ultimately caused by
{\em spatial} discretization errors. The cell-centred mass fluxes used to
estimate the gravitational work on a cell are not accurately balancing the
amount of energy actually extracted from the gravitational field when the
strong virialization shock propagates outwards. As a result, a substantial
energy error is produced, which, in this example corresponds to a gain of
energy of the whole system. Clearly, the energy error from this can become
quite severe, especially for poor resolution, so the first generation of
cosmic structures could be quite strongly affected by this problem.

However, rewriting the gravitational work term in terms of a surface integral,
as described in subsection~\ref{SecEgySurface}, leads to much better energy
conservation. This is shown in the right-hand panel of
Fig.~\ref{FigEvrardEnergy}, where the same simulations are shown but this time
using our improved coupling of self-gravity to the Euler equations. We see
that in this case the relative error in the total energy stays well below
$10^{-3}$ and does not show any systematic resolution dependence, which is a
dramatic improvement relative to the results above. For these results, the
gravitational work-term was calculated with the gravitational potentials, as
described in equation (\ref{EqnEgExact}). If the simpler formulation of
equation~(\ref{EqnEgApprox}) is used instead, the maximum relative errors
become considerably larger (up to $10^{-2}$ in the peak) but are still
acceptable.  All the simulations shown in Fig.~\ref{FigEvrardEnergy} were
calculated for an unstructured stretched mesh that was kept fixed; if the mesh
is allowed to move instead, the errors tend to be slightly smaller.

Lastly, we would like to examine whether the softening correction factors
discussed in subsection~\ref{secegysoftening} make a significant difference
for the energy conservation. First, note that such a difference is really only
expected if the gravitational interaction between two neighbouring points is
affected by the softening kernel. In other words, the quantities $\eta_j$
defined in equation~(\ref{eqneta}) are only different from zero if the
gravitational softening lengths are large enough so that some `overlap' with
neighbouring cells occurs at least in a fraction of the cells. This can be
guaranteed by choosing a sufficiently large value for the softening constant
$f_h$, for example $f_h=2.5$. In simulations of the Evrard collapse with this
setting, we find a maximum energy error of the same size as above, i.e.~it
stays below $10^{-3}$. However, if we disable in the code the corrective force
of equation~(\ref{EqnCorrectionForce}) that accounts for changes of the
softening lengths, the energy error goes up by more than a factor of 10, and
reaches slightly more than $1\%$ in the peak. This shows that this correction
factor should indeed be included for high-precision results.

\begin{figure*}
\bc
\resizebox{17cm}{!}{\includegraphics{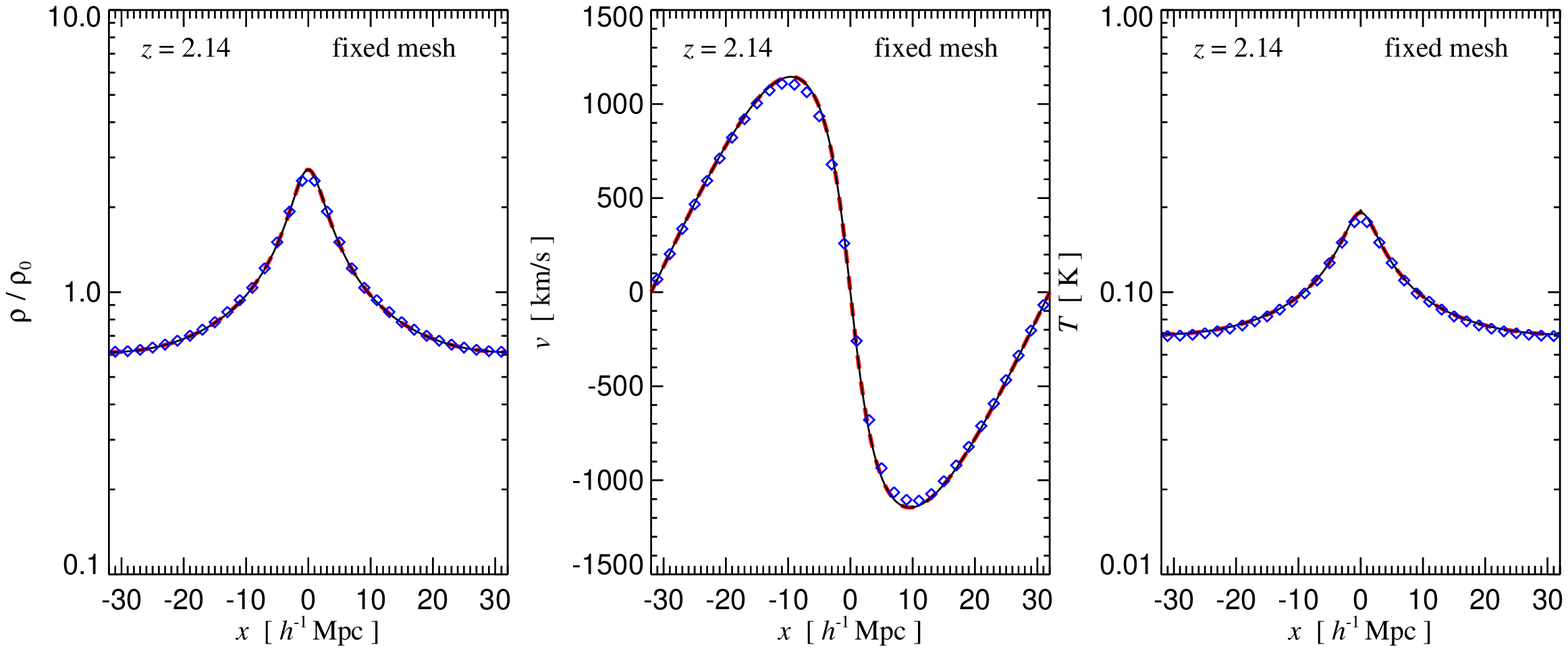}}\\
\resizebox{17cm}{!}{\includegraphics{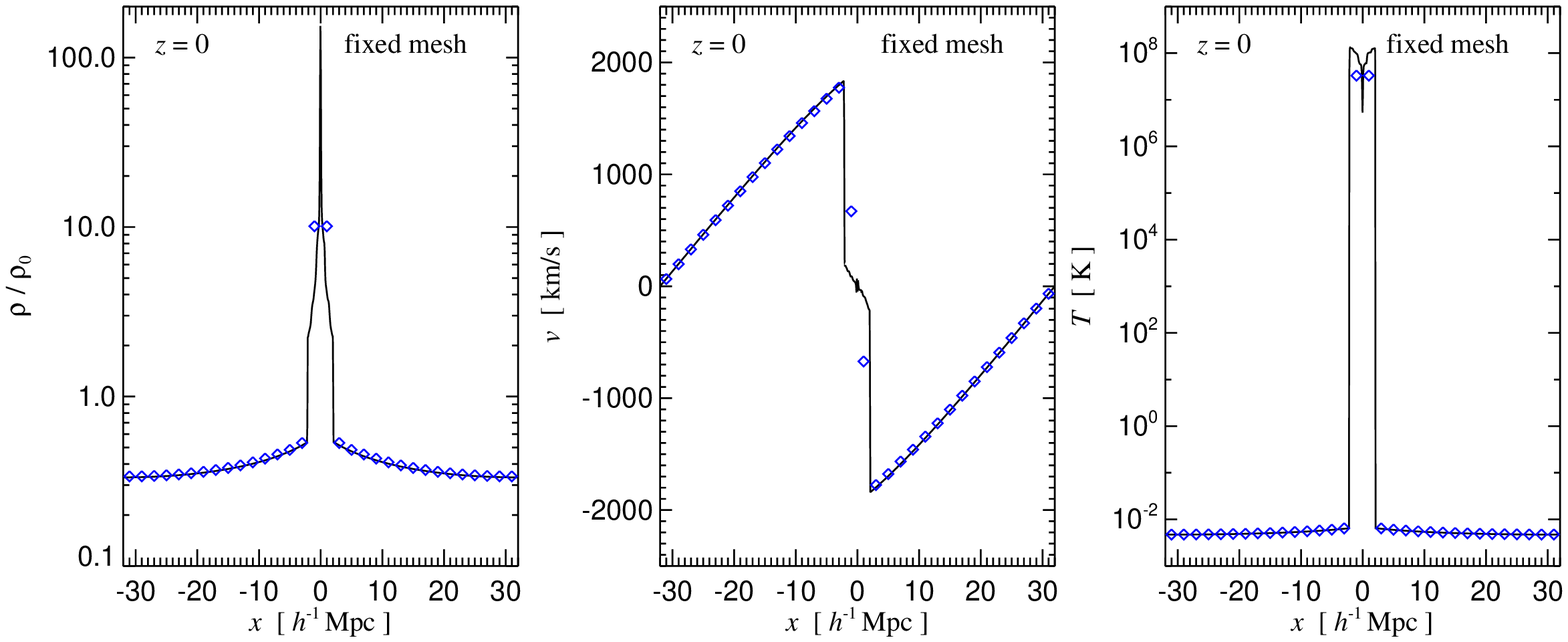}}
\resizebox{17cm}{!}{\includegraphics{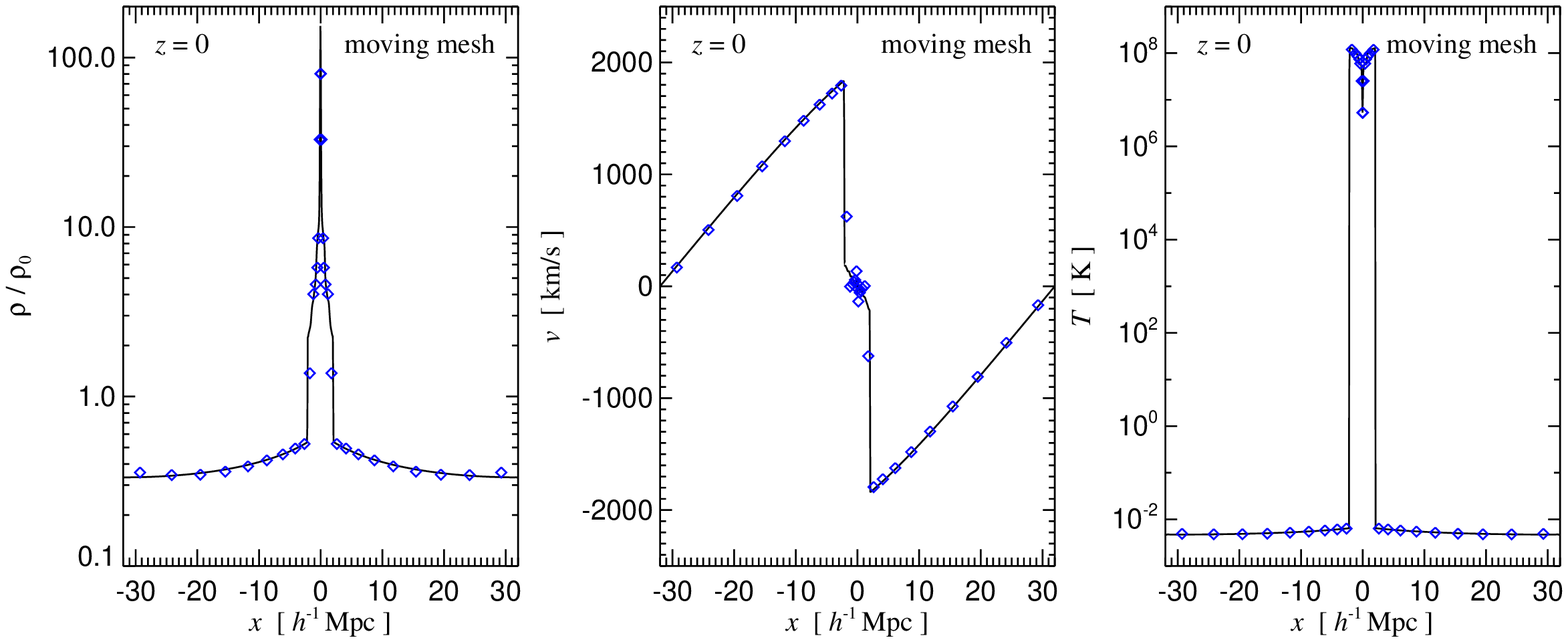}}
\caption{Zeldovich pancake test at two different times, and for two different
  ways to move the mesh. The top row of panels shows the pancake at $z=2.14$ when it is still in the
  linear regime, before the collapse occurs at $z_c=1$. We show the analytic solution
  as a thick dashed line in red, while a high-resolution calculation with 1024
  cells is shown as a solid line and the blue symbols give a low-resolution
  result with 32 fixed cells, for comparison. The middle and bottom rows of
  panels give the
  state of the pancake at $z=0$, well into the non-linear regime, once
  calculated with a fixed mesh and once with the moving mesh approach. Again, we
  compare with a high-resolution result based on 1024 fixed cells (solid line). Note in
  particular that the temperature in the unshocked parts of the flow is
   very accurate, thanks to the energy-entropy formalism that is applied by the code in very
  cold parts of the flow.
  \label{FigZeldovichNonLinear}}
\ec
\end{figure*}

\subsection{Zeldovich pancake} 

A useful standard test for cosmological codes is the evolution of a sinusoidal
density perturbation in an expanding Einstein-de-Sitter universe. After an
initial linear growth phase, the one-dimensional wave collapses to a Zeldovich
pancake, involving a pair of very strong shocks.  As this problem can be
viewed as a `single-mode' of the general cosmological structure formation
problem, it is a particularly useful test of any cosmological code.
Furthermore, it is also a useful test-bed for the dual entropy-energy
treatment described in subsection~\ref{seccoldflows}, as the initial gas
temperature is negligibly small and drops further through the cosmic expansion.

The comoving position $x$ corresponding to an initial unperturbed coordinate $q$ at
redshift $z$ is given by \citep{Zeldovich1970}
\begin{equation}
x(q,z) = q - \frac{1+z_c}{1+z}\frac{\sin(kq)}{k},
\end{equation}
where $k=2\pi/\lambda$ is the wavenumber of the perturbation of wavelength
$\lambda$. 
The comoving density corresponding to the displacement is given
by
\begin{equation}
\rho(x,z) = \frac{\rho_0}{1- \frac{1+z_c}{1+z}\cos(kq)},
\end{equation}
and the peculiar velocity is
\begin{equation}
v_{\rm pec}(x,z) = -H_0 \frac{1+z_c}{(1+z)^{1/2}}\,\frac{\sin(kq)}{k}.
\end{equation}
Here $\rho_0$ are the background density (equal to the critical density), and
$H_0$ is the Hubble constant today.  These equations describe the solution
exactly up to the redshift $z_c$ of collapse.

We follow \citet{Bryan1995} and \citet{Trac2004}, and choose
$\lambda=64\,h^{-1}{\rm Mpc}$ and $z_c=1$. Our test simulations are
started at $z_i=100$, with an initial gas temperature of $T_i =
100\,{\rm K}$. As the pressure forces are negligible up to the
formation of the pancake, the temperature should evolve adiabatically
as
\begin{equation}
T(x,z) = T_i \left[ \left(\frac{1+z}{1+z_i}\right)^3 \frac{\rho(x,z)}{\rho_0}\right]^{2/3}
\end{equation}
until collapse.  We carry out tests of the Zeldovich problem both
using a moving mesh and a fixed mesh. This in particular serves as a
useful test of the correct implementation of the cosmological time
integration, and the coupling of the gasdynamics to self-gravity in an
expanding background space.  As our code {\small AREPO} has presently
no one-dimensional gravity solver, we carry out the tests in two
dimensions instead, and for simplicity, we use only the PM solver in
two dimensions with a sufficiently large mesh.

In Figure~\ref{FigZeldovichNonLinear}, we show the density, velocity and
temperature profiles of the Zeldovich pancake at two different times, briefly
before collapse at redshift $z=2.14$, and well into the non-linear evolution
of the pancake at $z=0$. In all panels, a high resolution result (based on
1024 fixed points per dimension) is shown with solid lines, and symbols of a
low resolution calculation with initially 32 points per dimension are
overlaid.  In the high redshift result shown in the top panel, we also include
the analytic solution of Zeldovich in terms of a thick-dashed line. Before the
collapse of the pancake at $z=1$, both the fixed mesh and the moving mesh
calculation trace the analytic result with comparable accuracy, we therefore
only show one of the results. In the middle row of panels, the fixed-mesh
result at $z=0$ is shown. Outside of the shock-fronts, the solution is still
very accurate, but the lack of resolution inside the collapsed region leads to
a poor representation of the structure of the pancake, even though its
characteristic values of density and temperature are reasonably well
reproduced. The moving mesh calculation shown in the lower row of panels does
significantly better in this respect. Remarkably, even though only 32 points
were available initially, the density- and temperature structure of the
pancake, as well as the location of the two strong shocks, are represented
very accurately.

Note that the temperature evolution in this Zeldovich pancake test is
particularly difficult to get right, as there is a very large dynamic
range between the initially cold gas and the shocked heated gas in the
pancake, amounting to a difference of $\sim 10$ orders of magnitude.
Before the gas is heated by the shocks, the flow is extremely cold and
dominated by gravitational forces, meaning that the problems discussed
in subsection~\ref{seccoldflows} with respect to spurious heating of
very cold flows in finite volume methods are bound to be present in
this Zeldovich pancake test. Indeed, in the results shown in
Figure~\ref{FigZeldovichNonLinear} we have applied the entropy-energy
scheme described in subsection~\ref{seccoldflows}. The ordinary
treatment based on the total energy alone invariably leads to
significant heating of the gas well outside of the shock-front prior
to the collapse of the pancake. While this does not alter the motion
of the gas (the resulting pressure forces remain way too small), the
temperature evolution of the gas becomes inaccurate, especially when
the resolution is comparatively low. However, with the entropy scheme,
a very accurate solution is recovered in a robust way. We note that
especially with respect to the temperature evolution, our results also
compare favourably to those of \citet{Trac2004} obtained with their
moving-frame formalism. They also show much sharper shock fronts than
obtained with SPH \citep{Dave1997}.

\subsection{The Santa Barbara cluster}

In the `Santa Barbara Cluster Comparison Project' \citep{Frenk1999} a large
number of cosmological hydrodynamic codes were applied to the same initial
conditions, set-up to produce a rich cluster of galaxies in an
Einstein-de-Sitter universe. The inter-comparison of the results produced by a
this set of different codes, which included both SPH and Eulerian AMR methods,
allowed an assessment of the systematic uncertainties in such cosmological
structure formation simulations. While a fair amount of scatter between the
different results was found, there was still quite reasonable agreement in
most of the cluster bulk properties (such as total mass, temperature, etc.),
and in the radial cluster profiles (such as the radial run of density, baryon
fraction, etc.), with typical code-to-code scatter of order 10\%. The same
initial conditions have also been regularly used as hydrodynamic code test in
subsequent work \citep[e.g.][]{Wadsley2004,Springel2005,Thacker2006}.

\begin{figure}
\bc
\resizebox{8.0cm}{!}{\includegraphics{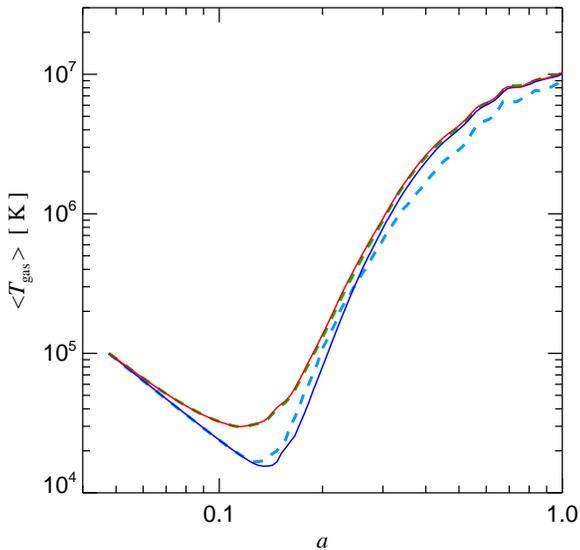}}\\
\caption{Evolution of the mean mass-weighted temperature in four
  low-resolution calculations ($2\times 32^3$) of the Santa Barbara
  cluster. The dashed green line shows the result of the moving-mesh
  approach when the ordinary total energy approach is applied. Here
  the gravitationally dominated cold flows at high redshift are
  clearly subject to some spurious heating. The red solid line shows
  the result when our entropy-energy scheme is used with a Mach-number
  threshold ${\cal M}_{\rm thresh}=1.1$, which turns out to be
  ineffective in preventing the high redshift heating, as the Mach
  numbers responsible for it are very high. If we instead use our
  alternative scheme for deciding when to preserve the entropy (with
  $\alpha_S = 0.05$) we obtain the blue solid line, where now the
  heating in very cold, low-density gas is suppressed and the expected
  adiabatic decline of the mean temperature at high redshift is
  obtained.  For comparison, the dashed lines gives an SPH result
  obtained with {\small GADGET-2} at the same resolution.
\label{FigSBTempEvolv}}
\ec
\end{figure}

It seems likely that the scatter in the results for the Santa Barbara cluster
would be smaller if the experiment was repeated today with the most recent
versions of the most commonly employed cosmological codes, thanks to the
progress made in the numerical simulation techniques in recent years.
However, at the same time there is little indication that arguably the most
important systematic difference found by \citet{Frenk1999} between the
Lagrangian SPH and the Eulerian AMR codes, namely the systematic difference in
the entropy predicted for the central cluster gas, has gone away.  This
entropy was found to be lower in SPH than in the AMR calculations, which in
turn also affects the temperature and gas density profiles in the inner parts
of the cluster. This also has an impact on cluster cooling rates if radiative
cooling is allowed, and on important observables such as the emitted X-ray
luminosity.

In SPH, entropy is accurately conserved
\citep{Springel2002,Ascasibar2003}, but it could be artificially low
due to the absence of entropy production through mixing and to SPH's
tendency to spuriously suppress fluid instabilities.  On the other
hand, the Eulerian codes may overestimate the central entropy as a
result of numerical diffusivity and overmixing. Also, they are more
prone to suffer from heating caused by the noisy gravitational field
produced by the collisionless matter. Recently, the idea that the
difference may ultimately arise from differences in the treatment of
mixing has found some support in numerical experiments
\citep{Mitchell2008}.  Presently, it remains however unclear what the
correct entropy profile for the Santa Barbara profile really is, even
though this is an important question for numerical cosmology. Note
that due to the absence of radiative cooling in this problem, the
Santa Barbara cluster represents comparatively clean and `easy'
physics. If even this case cannot be calculated fully reliably, it is
clear that the more demanding simulations that also account for
radiative cooling are fraught with numerical uncertainties.

We here give first results for the Santa Barbara Cluster with our new
moving mesh code, calculated at comparatively low resolution. All our
simulations follow the original initial conditions in a periodic box
of side-length $32\,h^{-1}{\rm Mpc}$, using homogeneous sampling of
the dark matter component, and an equal number of mesh-generating
points as dark matter particles. The simulations are started at
redshift $z=50$, and use cosmological parameters of a critical density
cosmology with dark matter content $\Omega_{\rm dm}=0.9$, baryonic
density $\Omega_b=0.1$, and Hubble constant $H_0=100\,h\,{\rm km \,
s^{-1} Mpc^{-1}}$ with $h=0.5$.

In Figure~\ref{FigSBTempEvolv}, we first show the evolution of the mean
mass-weighted temperature of the whole simulation box, from the starting
redshift to the present time. Initially, no structures have formed yet,
so that the mean mass-weighted temperature should decline as $T\propto
a^{-2}$ for a while. Eventually, the thermal energy content in the
shock-heated gas of the first forming cosmic structures starts to
dominate and the mean temperature begins to rise rapidly. This general
evolution is reflected in the four simulation results depicted in
Figure~\ref{FigSBTempEvolv}, albeit with interesting differences in
detail. The green dashed line shows the result of the moving-mesh
approach when the ordinary total energy approach is applied. The red
line gives the result when the energy-entropy formalism is used with a
Mach number threshold ${\cal M}_{\rm thresh}=1.1$, while the solid blue
line uses our alternative switch for deciding whether the entropy should
be kept instead of updating it with the total energy equation. In the
latter case, the entropy is used if the thermal energy is at most a
small fraction $\alpha_S = 0.05$ of the local kinetic energy. This
proves effective to yield the expected adiabatic decline of the mean
temperature at high redshift. On the other hand, the Mach-number based
switch does not make a difference in this regime, as the shock waves
responsible for this high-$z$ heating are typically quite
strong. However, it can still effectively act against noise-induced
heating in virialized structures at lower redshift.  For comparison, the
dashed light blue line gives an SPH result obtained with {\small
  GADGET-2} at the same resolution. It yields a high-redshift evolution
very similar to the moving-mesh code when the entropy scheme is used for
the cold gas, but at low redshifts its gas ends up being noticeably
colder on average. A substantial part of this difference in the final
temperature is probably simply caused by the lower effective resolution
of SPH, which tends to reduce the heating through shocks. Higher
resolution SPH calculations yield a mean temperature that is 5-8\%
higher, quite close to the mesh based result.

\begin{figure*}
\bc
\resizebox{7.6cm}{!}{\includegraphics{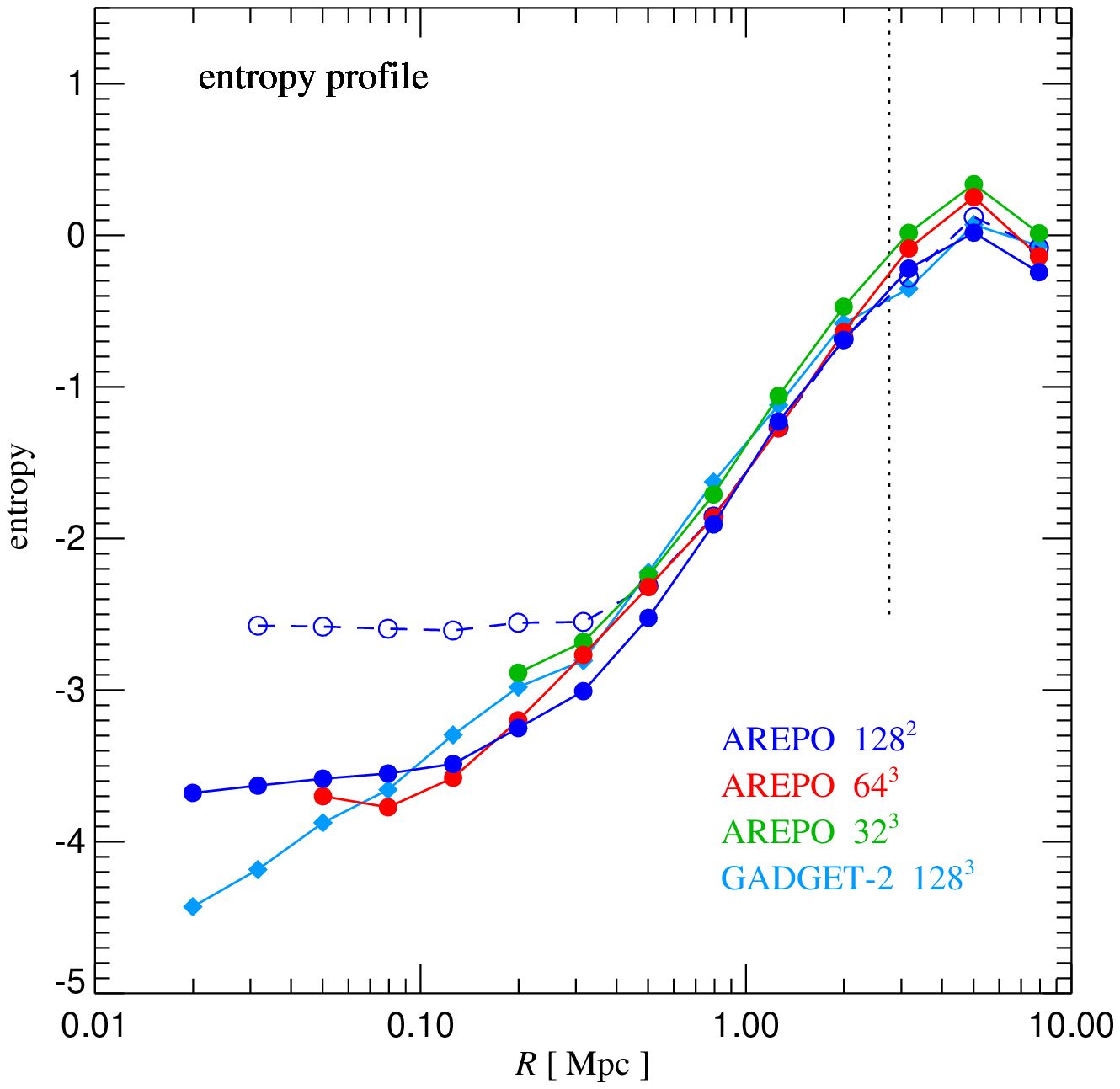}}%
\resizebox{7.6cm}{!}{\includegraphics{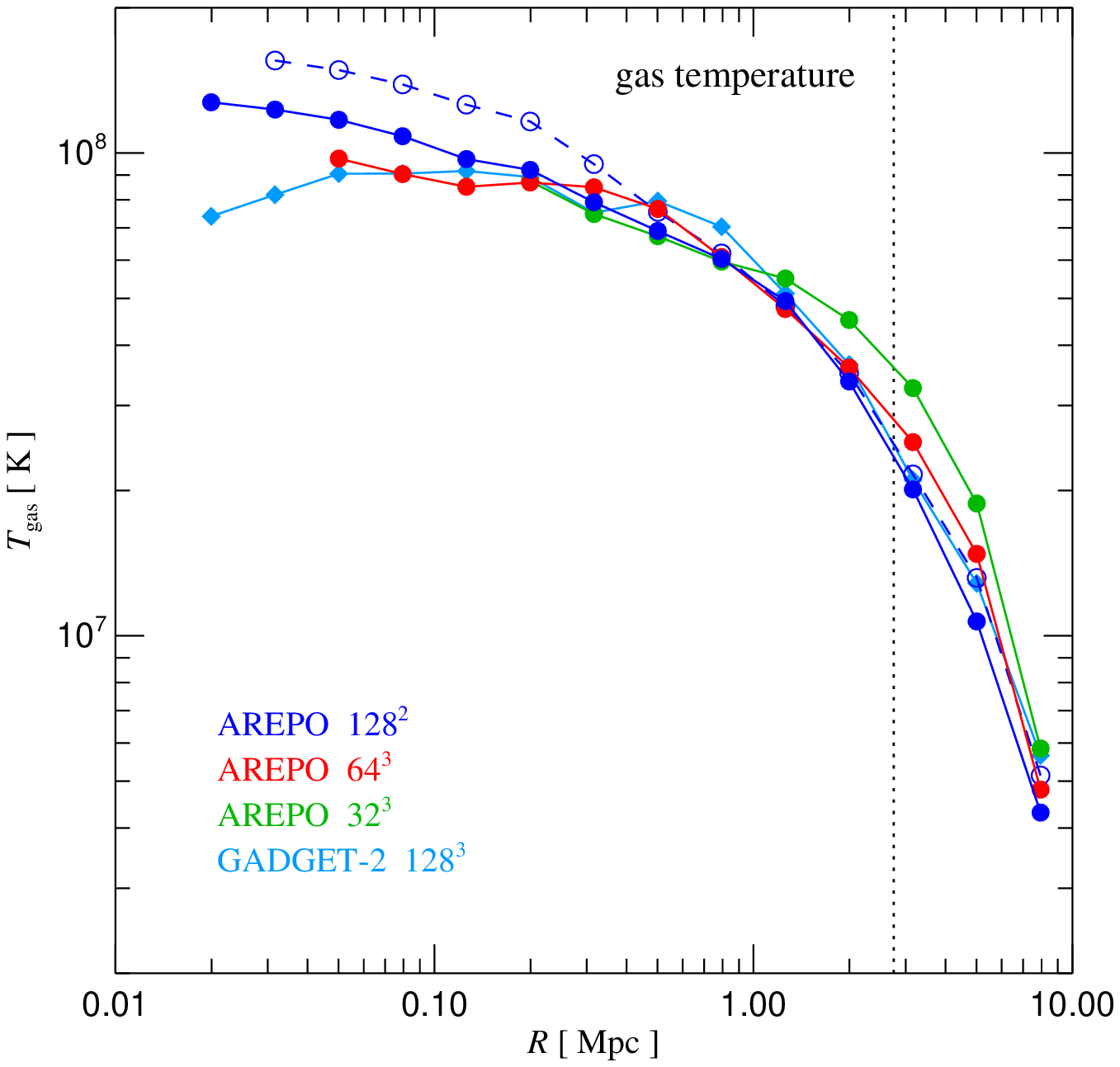}}\\
\resizebox{7.6cm}{!}{\includegraphics{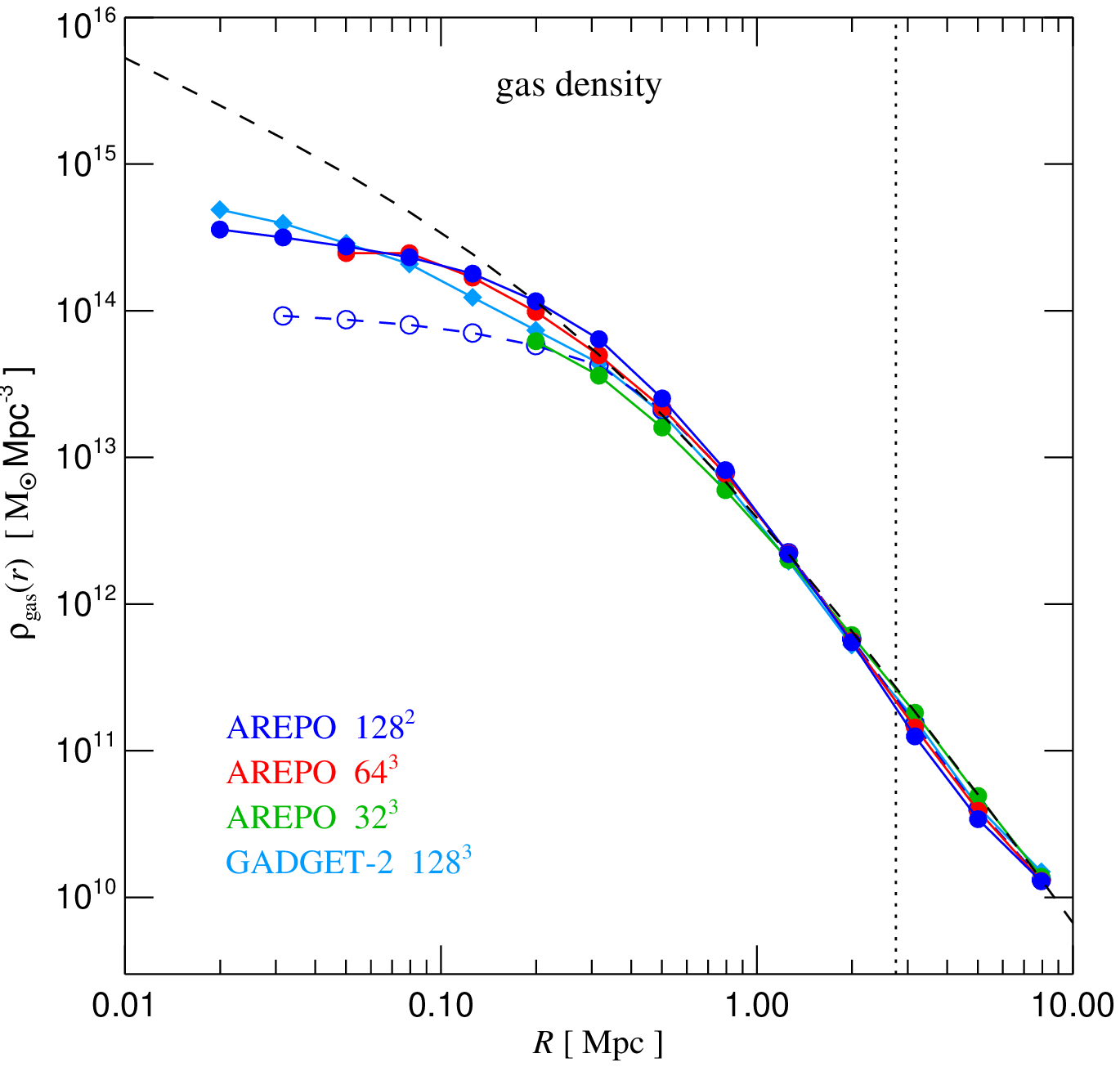}}%
\resizebox{7.6cm}{!}{\includegraphics{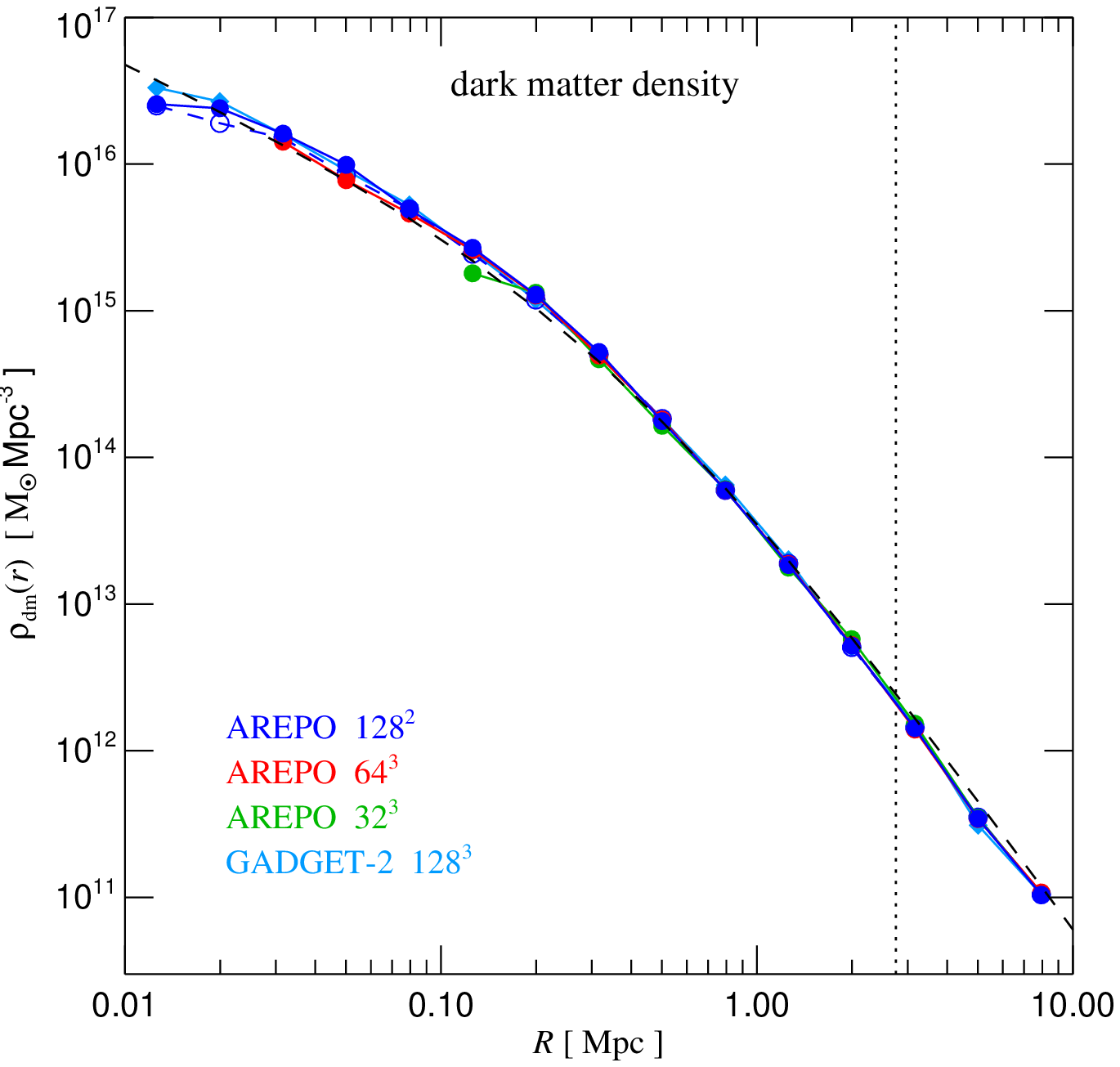}}
\caption{Radial profiles of the gas entropy (top left), gas temperature
  (top right), gas density (bottom left) and dark matter density (bottom
  right) of the Santa Barbara cluster, calculated with different
  resolutions and with different calculational methods.  For our
  moving-mesh approach, results for the three resolutions $2\times
  32^3$, $2\times 64^3$ and $2\times 128^3$ are shown with solid
  circles. Here we applied our entropy-energy scheme with a Mach number
  threshold ${\cal M}_{\rm thresh}=1.1$. The open circles show the
  result for the $2\times 128^3$ run when only the total energy equation
  is used, and the gravitational work term is estimated based on the
  actual mass fluxes. However, the results at this high resolution show
  no difference when the gravitational work term is estimated with
  cell-centred fluxes instead.  Finally, a high-resolution result
  obtained with the SPH code {\small GADGET-2} is shown with
  diamonds. The vertical dotted lines mark the virial radius of the
  cluster ($R_{\rm vir}= 2.754\,{\rm Mpc}$), while for comparison the
  dashed line in the top left panel illustrates the shape of the dark
  matter density distribution in terms of the best-fit NFW profile (with
  concentration $c=7.5$), scaled by the gas to dark matter mass ratio.
\label{FigSBProfiles}}
\ec
\end{figure*}

Radial profiles of mean gas density, gas entropy, gas temperature and
dark matter density of the final Santa Barbara cluster are given in
Figure~\ref{FigSBProfiles}. We show results for the different
numerical resolutions of $32^3$, $64^3$, and $128^3$ with solid
circles, in different colours as labelled. All these simulations use
the entropy-energy formalism with a threshold Mach number ${\cal
M}_{\rm thresh}=1.1$ in order to suppress spurious heating from the
noise in the gravitational field induced by the dark matter. The
thermodynamic profiles converge reasonably well, but not nearly as
well as the dark matter density. Interestingly, the central cluster
entropy is actually quite close to the SPH result that is shown for
comparison, but the innermost entropy profile shows a shallower slope
that produces a temperature profile that keeps slowly rising to the
very centre. If the total energy equation is applied throughout the
calculation in the $128^3$ run, we obtain the result shown with hollow
circles. It produces much higher core entropy and central gas
temperature, as well as a lowered central gas density, when compared
with our default mesh-based calculation. We think these results
clearly show that the origin of the discrepancy found first in
\citet{Frenk1999} between the central cluster entropy in SPH and AMR
codes is caused by dissipation in extremely weak shocks and the
production of mixing entropy in effectively smooth parts of the
flow. Part of this dissipation is clearly artificial and caused by
gravitational N-body noise, which has much more drastic consequence in
mesh-based calculations than in SPH. It therefore appears clear that
mesh-based results that use the energy equation alone will
overestimate the central entropy. Unfortunately, it is less clear how
much suppression of dissipation is warranted, and where hence the true
entropy level ultimately lies.  This will be investigated further in
future work.

\begin{figure}
\bc
\resizebox{8.2cm}{!}{\includegraphics{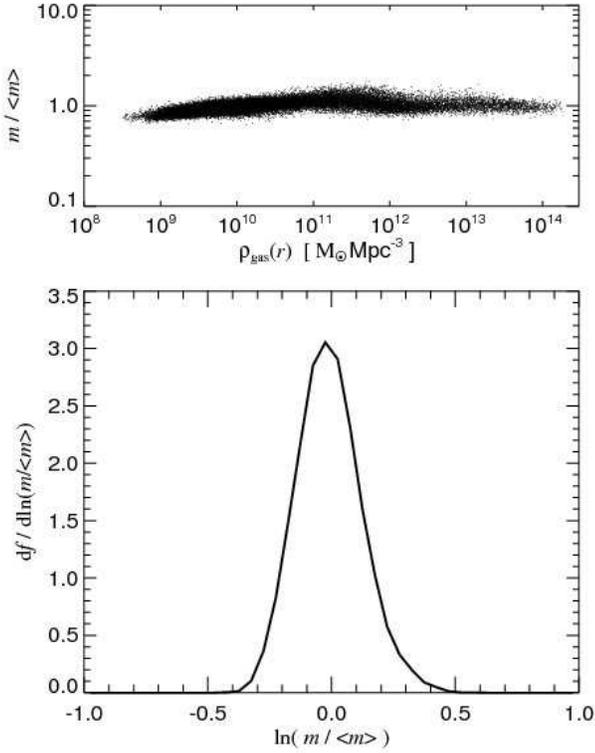}}\\
\caption{Scatter plot of the mass
 per cell as a function of local gas density (top panel). The mass
is here expressed in units of the mean mass per cell, $\left< m \right>$, and
the measurement was made at $z=0$ for our $2\times 32^3$ run of the Santa
Barbara cluster. The bottom panel shows the distribution function of $m/\left<
  m \right>$, which has roughly log-normal shape.
\label{FigSBMeshStatsMassPerCell}}
\ec
\end{figure}

\begin{figure}
\bc
\resizebox{8.2cm}{!}{\includegraphics{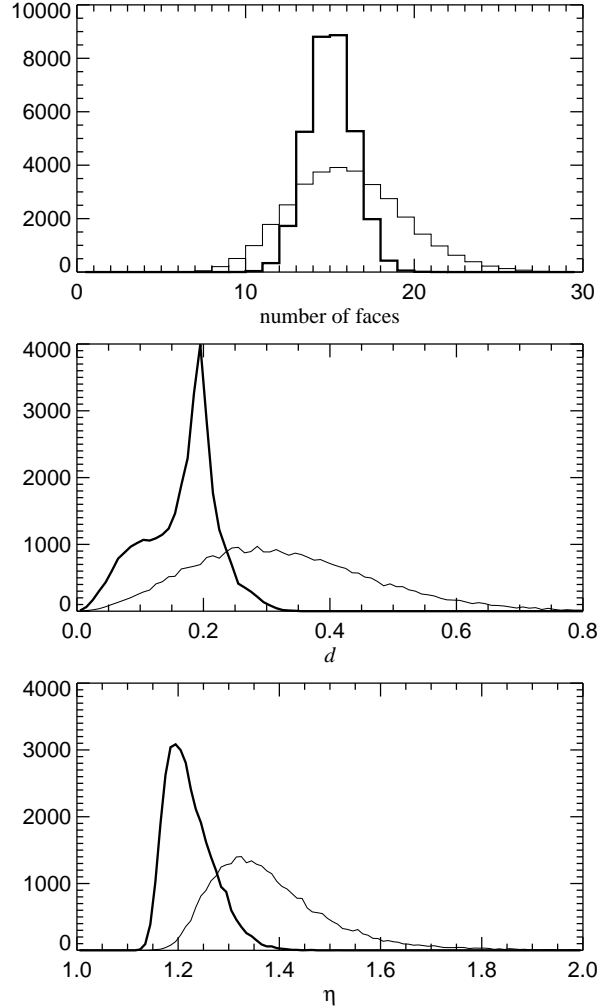}}\\
\caption{Geometry statistics of the final mesh at $z=0$ in our moving-mesh
  calculation of the Santa Barbara cluster.  The histogram in the top panel
  (thick line) counts the number of faces per cell; the average number is
  14.6. The thin line is the same statistic for a Poisson sample ($\sim 15.5$
  faces per cell on average). The middle panel gives the distribution function
  of the distance $d$ of mesh-generating points to their cell's
  centre-of-masses, in units of the cell radius. Again, we compare the
  cosmological simulation (thick line) to a Poisson sample (thin
  line). Finally, the bottom panel gives the distribution of the ratio
  $\eta=S^{3/2}/(6\sqrt{\pi}\, V)$ of the surface areas of cells to their
  volumes.
\label{FigSBMeshStats}}
\ec
\end{figure}

We note that the dark matter density profiles found with {\small
  AREPO} converge very well, and are consistent with the ones found
with {\small GADGET-2}. Also, we have found that at high resolution
($64^3$ and $128^3$) it makes essentially no difference to the results
whether the `standard' approach to treat the gravitational work term
is employed, or our alternative scheme based on the actual mass fluxes
at the surfaces of cells. Only at the low resolution of $32^3$ we have
found that the cell-centred approach gives slightly higher central
cluster entropy and temperature.

It is also interesting to examine the final state of the Santa Barbara run
with respect to statistical properties of the geometry of its Voronoi
mesh. For example, we would like to know whether the calculation was able to
maintain roughly constant mass per mesh-cell, and whether the final mesh
consists mostly of `roundish' cells, as desired. The first of these questions
is addressed by Figure~\ref{FigSBMeshStatsMassPerCell}, where we show in the
top panel a scatter plot of the gas mass per cell as a function of gas
density. It is seen that roughly constant mass per cell has been maintained
over a dynamic range of $\sim 10^5$ in density (and hence also in volume). The
distribution function of the mass per cell is shown in the bottom panel of
Figure~\ref{FigSBMeshStatsMassPerCell}. It has roughly log-normal shape, with
a typical scatter of $\sim 20\%$. This small scatter is the direct result of
our mesh-steering algorithm discussed in subsection~\ref{SecInvZel}.

In Figure~\ref{FigSBMeshStats}, we show various statistics of the geometry of
the final mesh at $z=0$ in our $2\times 32^3$ run of the Santa Barbara
cluster.  The top panel histograms the number of faces per cell and compares
it to the same statistic for a Poisson distribution with the same number of
points. The average number of faces is 14.6 per cell, meaning that on average
we need to calculate 7.3 Riemann problems per mesh-generating point. For a
structured Cartesian mesh, a factor 2.43 fewer Riemann problems per cell need
to be solved. The middle panel gives the distribution function of the distance
$d$ of a mesh-generating point to the centre-of-mass of its cell, in units of
the fiducial radius $R=(3V/4\pi)^{1/3}$ of each cell. This quantity was used
in our method to ensure reasonably roundish cells, as described in subsection
\ref{SecRoundCells}. The parameter $\chi$ was set to $\chi=0.2$, meaning that
the algorithm tries to make cells with $d> 0.2$ rounder, something that
clearly has worked well. Finally, another statistic that shows that our final
mesh is significantly more regular than a Poisson mesh is given in the bottom
panel. Here we show the distribution function of the ratio $\eta=
S^{3/2}/(6 \sqrt{\pi}\, V)$ of surface area $S$ of a cell to its volume $V$. For
a sphere, $\eta=1$ is reached, and cells with high aspect ratios will produce
larger values. For our mesh, $\eta$ peaks around $\sim 1.2$.

Finally, it is of interest to comment on the overall code speed of
{\small AREPO} for such a real-world cosmological problem in comparison
to an equivalent SPH simulation. In our present implementation, {\small
  AREPO} is a factor 1.6 slower than {\small GADGET-3} \citep[an updated
  version of {\small GADGET-2},][]{Springel2005} for the Santa Barbara
cluster, using the same number of dark matter particles and
cells/particles. This was measured for runs on 4 processors, for the
$2\times 32^3$ initial conditions. A full timestep of the moving mesh
hydrodynamics takes about $\sim 2.8$ times as much CPU-time as the SPH
calculations, and this cost is dominated by the mesh-construction, which
weighs in with twice the cost of the SPH calculations, whereas the
calculations for the finite-volume hydrodynamics itself (gradient
estimation, flux estimation with Riemann solver, etc.) is slightly
faster than SPH, as this requires no neighbour searches.  However, the
calculation of self-gravity is costly in high force-accuracy
cosmological codes, and in fact makes up nearly two thirds of the cost
in the {\small GADGET-3} calculation. This cost stays roughly equal in
{\small AREPO}, as expected, such that only a rather modest increase of
the overall run-time in the new code remains. This additional CPU-time
is well invested in our view, given that the new method yields a
substantial improvement in accuracy. Furthermore, we note that so far
comparatively little effort has been spent on optimizing the speed of
{\small AREPO}, whereas {\small GADGET-3} has been developed and tuned
over many years. There is hence certainly room for substantial
further performance improvements of {\small AREPO} in the future.

\subsection{A galaxy collision simulation}

The hierarchical bottom-up formation of structure from small building blocks
is the leading theory of galaxy formation \citep{White1978}. In this scenario,
galaxies frequently collide and merge to form bigger systems. In fact,
according to the `merger hypothesis' \citep{Toomre1972} galaxy collisions are
a primary means to form large elliptical galaxies out of disk systems,
and are hence one of the main drivers of the morphological evolution of
galaxies.

\begin{figure*}
\bc
\resizebox{8.5cm}{!}{\includegraphics{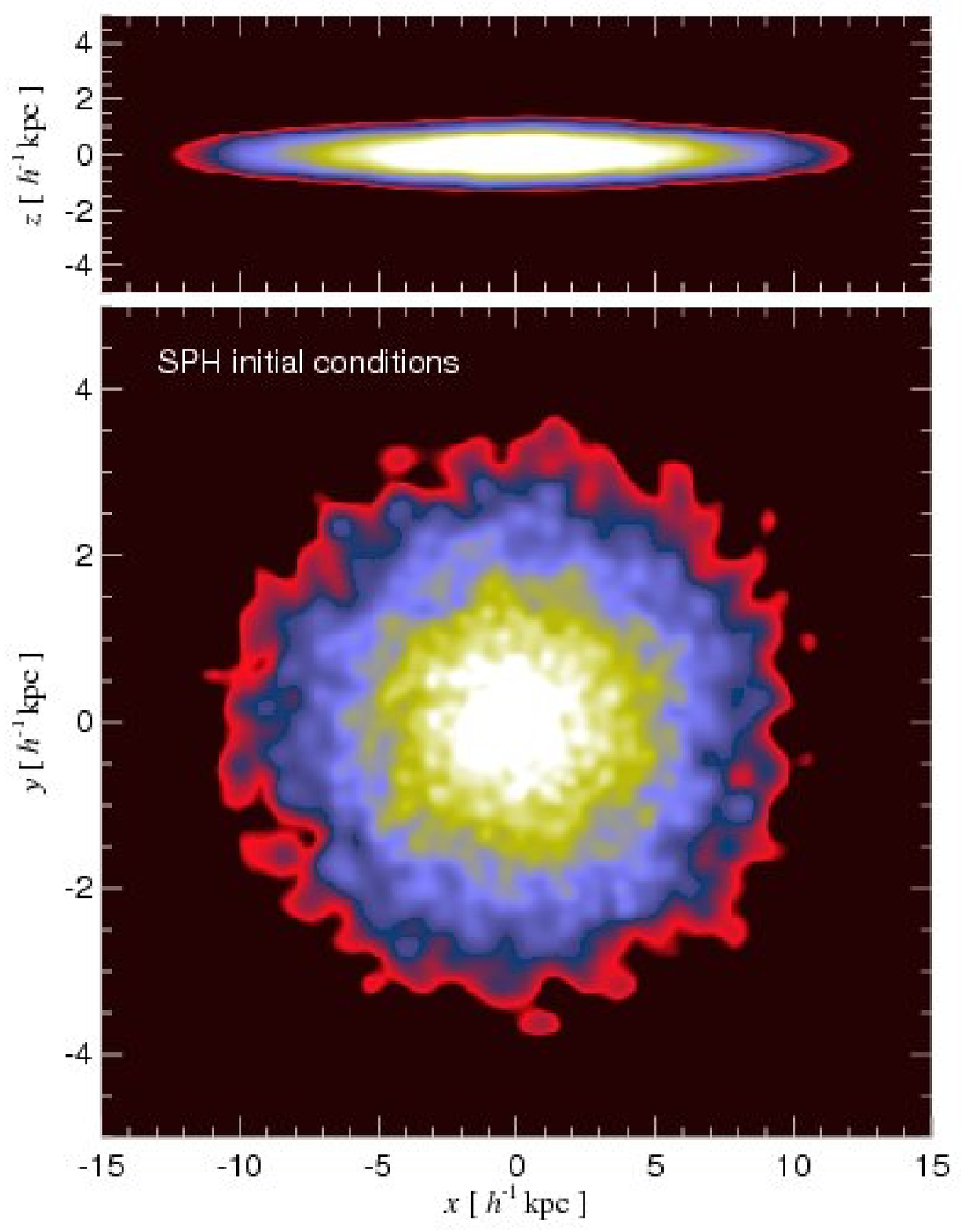}}%
\resizebox{8.5cm}{!}{\includegraphics{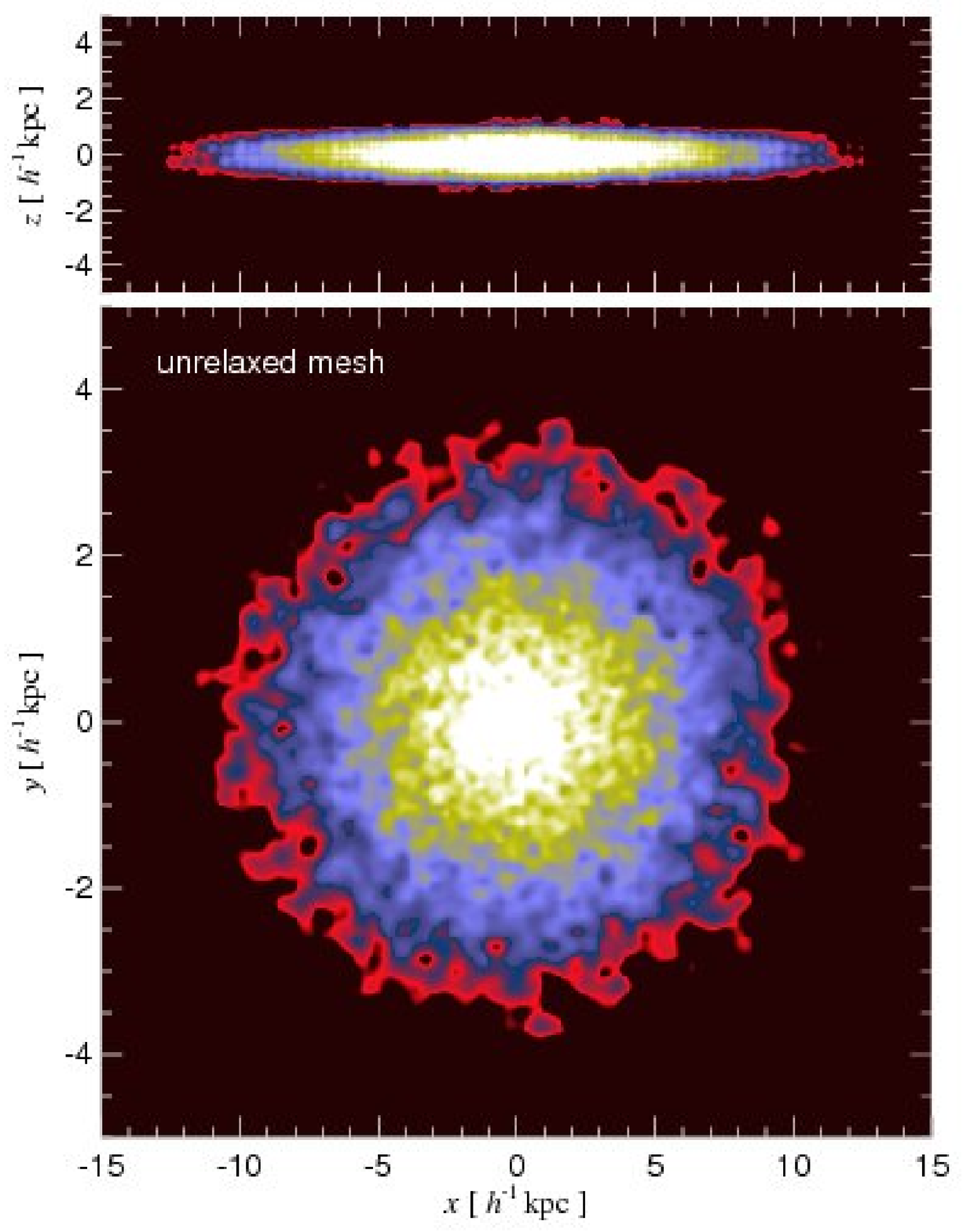}}\\
\vspace*{-1cm}\resizebox{8.5cm}{!}{\includegraphics{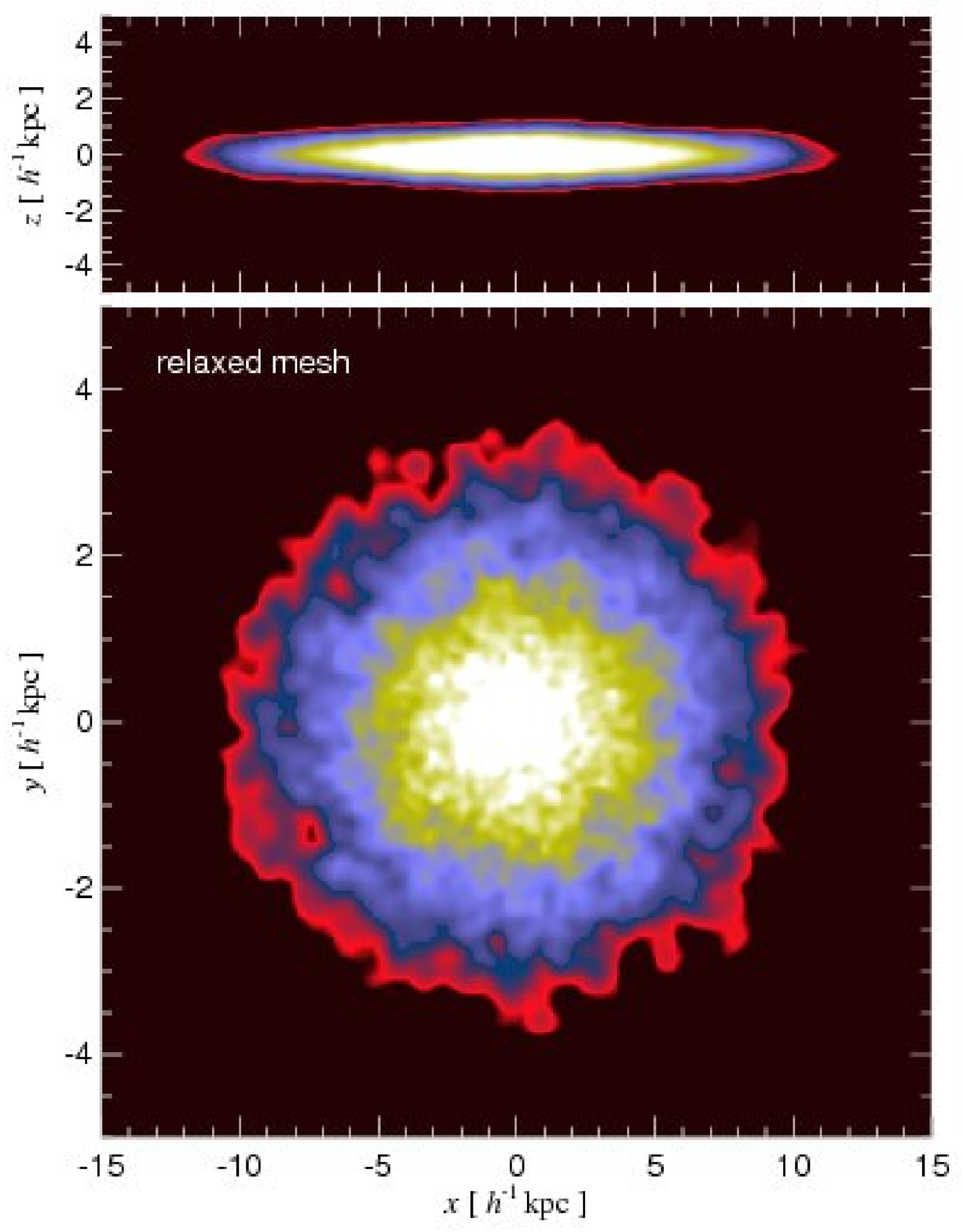}}%
\resizebox{8.5cm}{!}{\includegraphics{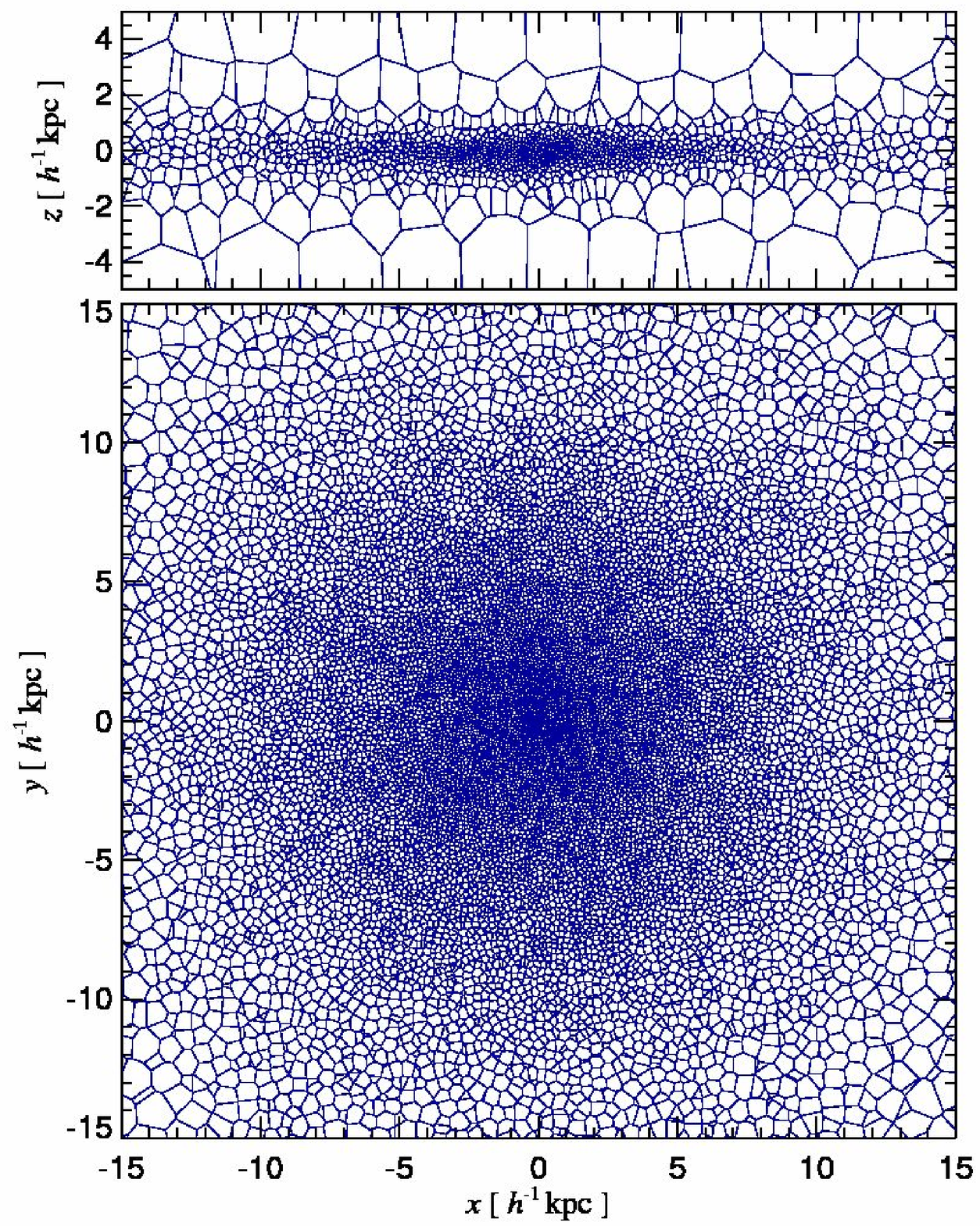}}\vspace*{-0.5cm}\\
\caption{
The top left panel shows the projected gas
density in the original SPH particle distribution, which we
adapt automatically to be used in the moving-mesh code.
In the top-right panel, we show the resulting gas distribution in the
Voronoi mesh that is created automatically based on the original SPH particle 
distribution. The lower left panel gives the density distribution
after 32 mesh relaxation steps have been applied, which improves the mesh
regularity and somewhat reduces the density fluctuations present
in the original initial conditions. Finally, the lower right shows 
two sections through the three-dimensional Voronoi mesh corresponding to the lower left distribution.
\label{FigGalaxyICs}} \ec
\end{figure*}

\begin{figure*}
\bc
\resizebox{16.0cm}{!}{\includegraphics{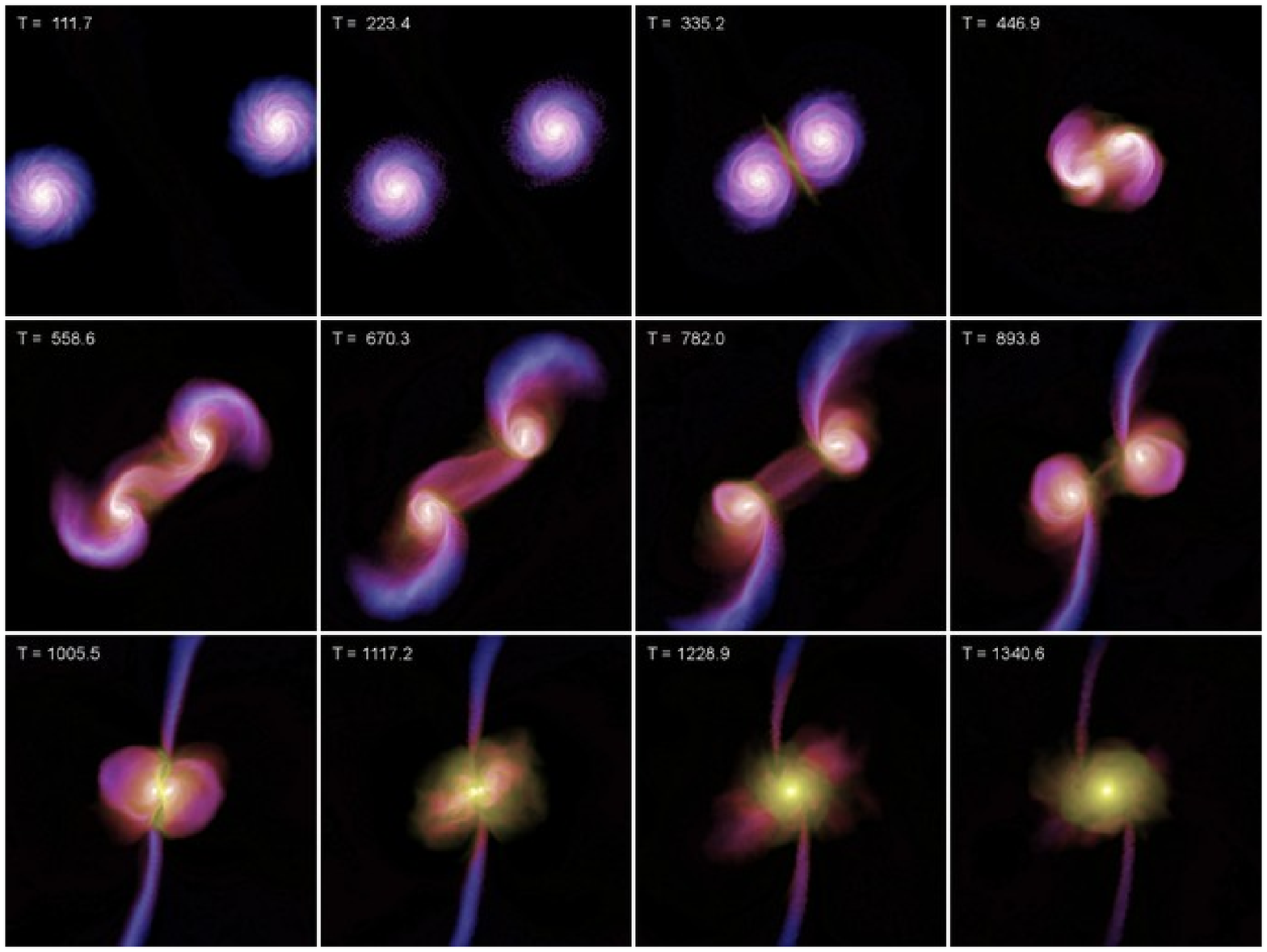}}\\%
\caption{Time evolution of the projected gas density in a 
galaxy collision with non-radiative gas, calculated with the moving-mesh
  code {\small AREPO}. Each frame has a length of $160\, h^{-1}{\rm kpc}$ on a
  side, and the elapsed time since the start of the simulation is given in
  units of $h^{-1}{\rm Myr}$. The brightness of each pixel encodes the
  projected gas surface density, and the colour hue the mass-weighted
  projected gas temperature.
  \label{FigGalaxyTimeSequence}}
\ec
\end{figure*}

Major mergers of spiral galaxies are observed in many spectacular
systems in the local Universe. They have also been extensively studied
with N-body and N-body/SPH simulations, leading to important insights
for the nature of starbursts, the formation of spheroidal galaxies,
and the secular evolution of galaxies
\citep[e.g][]{Gerhard1981,Negroponte1983,Hernquist1989b,Barnes1992,Mihos1996,Athanassoula2002}.
In recent times, galaxy merger calculations were also used to study
the growth of supermassive black holes at the centres of galaxies, and
their energy feedback on the host systems
\citep{DiMatteo2005,SpringelBH2005}. This has led to important
theoretical insights for the co-evolution of galaxies and supermassive
black holes \citep{Hopkins2006}, and for the characterization of
merger remnants and elliptical galaxies as two-component systems
\citep{Hopkins2008a}.

Interestingly, essentially all of the work thus far on isolated galaxy mergers
has been carried out with the Lagrangian SPH technique. Not without
reason. SPH can effortlessly deal with the large bulk velocities present in
the colliding galaxies before they coalesce, and the large dynamic range in
density and spatial scales that need to be resolved. At the same time, the
resolution automatically follows the mass, and is concentrated where it is
needed most, which in these calculations is naturally at the centres of the
galaxies. Achieving this same set of features with AMR is technically and
numerically substantially more challenging. This is certainly one of the
primary reasons why this method has so far not been widely applied to this
very important type of cosmological simulation. An added difficulty in the
high-speed collisions of galaxies is that there is no convenient frame of
reference where both galaxies are simultaneously at rest. In particular, the
calculational hot spots where the AMR refinements are most needed  are quickly
moving, which invokes the problems of Galilean non-invariance inherent in the
Eulerian approach. While a successful application of AMR techniques to galaxy
mergers should certainly be possible in principle, AMR does not appear
particularly well matched to the nature of the problem.

In this last subsection, we show that our new moving-mesh code can deal quite
well with the particular challenges posed by simulations of pairs of colliding
galaxies. We here focus on the technical aspects of carrying out such
simulations with {\small AREPO} and give an illustrative example,
deferring a scientific analysis of the results of moving mesh based galaxy
mergers to future work.

We begin by briefly discussing the creation of appropriate initial
conditions. SPH can easily deal with vacuum boundary conditions, and it is
hence straightforward to represent isolated gaseous disks in otherwise empty
space. In contrast, our moving-mesh code always requires a well defined total
volume that is tessellated by the mesh. We therefore enclose the isolated
galaxies with a large box that comfortably contains all the material of the
galaxies and their tidal debris. The outer walls of this box have reflective
boundary conditions for the gas, but collisionless particles are allowed
to penetrate freely. Also, the calculation of gravity is not influenced by the
presence of the box.  Next, we need some sort of background grid to fill all
of this empty space through which galaxies with their cells and gas can move.
Ideally, we would like to be able to add this background grid automatically to
existing initial conditions of SPH calculations, such that equivalent
moving-mesh initial conditions result. This allows continued use of the same
initial conditions codes and facilitates easy comparison of the results. We
have implemented the following functionality in {\small AREPO} to produce
appropriate initial conditions that fulfill these requirements:

\begin{enumerate}

\item Starting with a set of gas particles from an existing SPH initial
  conditions, we first generate a new set of points for tessellating the whole
  volume of the simulation box.  We want this set of points to produce cells
  of constant volume far away from the galaxy (or galaxies), but close to the
  original gas distribution, there should be cells of smaller size such that
  the original gas distribution stays well localized when the Voronoi mesh is
  constructed. We want to avoid that particles at the surface of the original
  SPH particle distribution end up having Voronoi cells that extend far out
  into empty space, which would cause a low-density leakage of the mass. We
  generate an appropriate set of additional points via a special
  \citet{Barnes1986} oct-tree construction. We start with a Cartesian grid
  with cells of size equal to the desired coarsest background resolution. We
  then fill in the gas particles of the SPH initial conditions one by one,
  requiring that a new set of 8 empty daughter cells is created whenever a
  particle falls into a leaf cell that already contains a particle. (Note that
  unlike in the ordinary tree construction of {\small GADGET-2}, the creation
  of empty cells is not prevented here.)  Finally, we create mesh-generating
  points at the centres of all empty leaf cells. This procedure effectively
  creates an adaptively refined grid that follows the original SPH particle
  distribution.

\item We now assign the mass, momentum, and thermal energy of the original SPH
  particles to the new set of mesh-generating points. This is done by
  distributing these quantities to the new points in a conservative fashion,
  using the SPH kernel and the original SPH smoothing lengths, and by
  weighting each new point with the volume of its associated parent tree
  node. This produces new initial conditions that faithfully represent the gas
  distribution of the original SPH simulation, and which can be directly fed
  to the {\small AREPO} code.

\item As an optional step, we may now relax the created Voronoi mesh by moving
  the mesh-generating points with the technique described in
  subsection~\ref{SecInvZel}.  If desired, this can also be used to downsample the mesh
  resolution to exactly match the particle number of the original
  SPH initial conditions. The spatial distribution of the mass density,
  momentum density and thermal energy density stays fixed in this step, only
  the mesh is moved by solving the advection equation.  The mesh relaxes to a
  distribution where the factor $m_i/\tilde m + V_i / \tilde V$ is roughly
  constant for all the cells, see equation~(\ref{eqnconstantmesh}).  Here
  $\tilde m$ is the original mean gas particle mass in the SPH initial
  conditions, and $\tilde V$ is the volume of the coarsest cell used in
  background grid. These values of $\tilde m$ and $\tilde V$ may then also
  be kept later on to steer the mesh motion during the dynamical evolution.  We
  note that the numerical diffusion from the mesh advection in this step
  reduces Poisson noise in the initial particle set, if present, which is a
  welcome effect in this case.
\end{enumerate}

In Figure~\ref{FigGalaxyICs}, we illustrate the outcome of this procedure when
applied to an isolated galaxy model. We selected SPH initial conditions
created with the methods described in \citet{Hernquist1993} and
\citet{Springel1999}. The model has circular velocity $V_c=160\,{\rm km\,
  s^{-1}}$, total mass $M_{200}=9.52\times 10^{11}\,h^{-1}{\rm M}_\odot$, and
spin parameter $\lambda=0.05$. Most of the mass is in an NFW-halo of
concentration $c=9.0$, represented with 50000 collisionless dark matter
particles.  A fraction of $m_{\rm d}=0.05$ of the mass is in a disk with an
exponential surface mass profile with a scale-length of $R_d= 3.6\,h^{-1}{\rm
  kpc}$, and a vertical scale height of $0.15\times R_d$. Half of the disk
mass is in a stellar disk, represented with 30000 collisionless particles, the
rest in a gaseous disk of 30000 SPH particles.  We enclosed the whole system
in a box of size $1200\,h^{-1}{\rm kpc}$ on a side, and applied the initial
conditions modification algorithm described above with a background grid of
$32^3$ cells. This increased the final number of mesh-generating points from
30000 to 109602.

\begin{figure}
\bc
\resizebox{8.4cm}{!}{\includegraphics{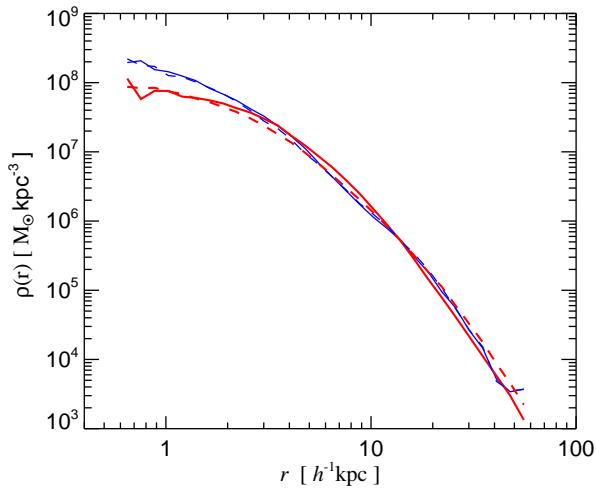}}%
\caption{Spherically averaged density profiles of the remnant galaxy
  formed in the merger simulation carried out with the moving-mesh
  code {\small AREPO}, at time $2\,h^{-1}{\rm Gyr}$ after the start of
  the simulation.  The thick red solid line is for the gas, the thin
  blue solid for the stars. Dashed lines give the result of a
  corresponding merger calculation carried out with the SPH-code
  {\small GADGET-2}. The fluctuation seen in the thick red line at small
  radii is due to counting statistics, as the number of cells there is
  small and each cell is counted in full towards the logarithmic bin in
  which its centre falls. 
  \label{FigRemnantProfiles}}
\ec
\end{figure}

In the top-left panel of Figure~\ref{FigGalaxyICs} we show the projected gas
density in the original SPH particle distribution, projected with the adaptive
SPH kernel. In the top-right panel, we show the gas distribution in the
Voronoi mesh that is created after step (ii) in the above procedure has been
completed. Here we projected the gas again with an SPH kernel, now seeking
neighbours among the mesh-generating points.  Clearly, the gas mass in the new
mesh is localized well, as desired, but there is a large amount of
high-frequency noise both in the local mesh structure and in the gas
distribution. This noise is partially eliminated in step (iii), as shown in
the bottom left panel, where we show the mass distribution after iterating the
advection equation for 32 relaxation steps. Finally, in the bottom right panel
we show planar sections through the three-dimensional Voronoi mesh
corresponding to the relaxed initial conditions. The small pieces of cells
that occasionally occur in these sections are corners from cells that are
intersected far from their centres-of-mass.

The above procedures can also be readily applied to produce initial
conditions that contain two galaxies on a collision orbit. We consider
a prograde merger of two identical copies of the above galaxy model,
placed at an initial separation of $160\,h^{-1}{\rm kpc}$, and set-up
on a zero-energy orbit with impact parameter $2\,h^{-1}{\rm kpc}$. In
Figure~\ref{FigGalaxyTimeSequence} we show the time evolution of this
galaxy merger simulated with the moving-mesh code. The baryons are
treated as a non-radiative gas in this calculation.  The galaxies
freely fall together with increasing velocity until they undergo a
first encounter.  Tidal forces and shocks largely destroy the disks
during this first passages, but the inertia of the galaxies lets them
separate again.  After sufficient braking from the dynamical friction
of their dark matter halos, the galaxies turn around and fall together
a second time. This is soon followed by complete coalescence, violent
relaxation of the collisionless components, and virialization of the
gas distribution. By time $\sim 2\,h^{-1}{\rm Gyr}$, a reasonably
relaxed spheroidal remnant galaxy has formed. We note that the images
of Figure~\ref{FigGalaxyTimeSequence} have been created by ray-tracing
through the Voronoi tessellation for each pixel, and integrating up
the linearly reconstructed density field along all ray segments cells
that intersect individual cells . This technique faithfully preserves
the full information in the three-dimensional density field and does
not rely on additional smoothing steps.

In Figure~\ref{FigRemnantProfiles}, we show the radially averaged
density profiles of gas and stars of the final merger remnant, at time
$2\,h^{-1}{\rm Gyr}$. We compare the results with the outcome of the
same merger calculation carried out with the SPH code {\small
GADGET-2}. Interestingly, both the stellar and gas density profiles
are extremely similar, even though there is a hint that the gas
density profile is slightly more concentrated in the moving-mesh
calculation in the outer parts of the halo, and the innermost gas
density profile is a bit shallower.  We have used our entropy-energy
formalism with a Mach number threshold of ${\cal M}_{\rm thresh}=1.1$
in this moving-mesh calculation, which has suppressed entropy
production in very weak shocks. If this threshold is raised to ${\cal
M}_{\rm thresh}=1.3$, the central gas density increases somewhat,
while without any such threshold it is substantially lowered, because
in this case the cold gas in the disks already experiences significant
heating from noise in the gravitational field prior to the actual
collision of the two galaxies.  Further work will be required to
better understand the dissipation in the moving-mesh code in the
presence of a collisionless particle component, and to establish the
most accurate setting of ${\cal M}_{\rm thresh}$, or to find an
alternative approach to suppress spurious dissipation in the finite
volume approach when coupled to self-gravity and a collisionless
N-body system.

\section{Discussion} \label{SecDiscussion}

We have introduced a novel moving-mesh hydrodynamical scheme that is
second-order accurate both in space and time and does not require an
artificial viscosity. The method is based on a finite volume discretization of
the Euler equations on an unstructured mesh. The mesh is constructed as the
Voronoi tessellation of a finite set of mesh-generating points, which are free
to move during the time evolution. However, unlike in many other moving-mesh
approaches, there are no mesh-tangling or mesh-twisting effects since the
motion of the mesh-generating points induces a continuous deformation of the
mesh, without the occurrence of `bow-tie' cells or other topological artefacts.
The freedom to move the mesh with a nearly arbitrary flow field adds
considerable flexibility to the method.

If the mesh-generating points are fixed, the method becomes effectively
identical to an Eulerian code, formulated with an unsplit MUSCL-Hancock scheme
on an unstructured grid. If the points are fixed and arranged on a Cartesian
grid, the method becomes identical to an ordinary Eulerian code on a regular
structured mesh.  However, the most attractive mode of operation is obtained
by tying the mesh motion to the local fluid velocity, in the simplest case
making the velocities of the mesh-generators equal to the fluid velocities of
the corresponding cells. In this Lagrangian mode, the dynamics becomes
Galilean-invariant and benefits from the automatic adaptivity of Lagrangian
approaches, which is advantageous for many problems of interest.

Conceptually, our method shares aspects of SPH and of Eulerian
hydrodynamics. From SPH, it inherits the concepts of points as carriers of
thermodynamic quantities and their spatial distribution sets the resolution
in the flow. From Eulerian codes it inherits the concept of finite volume
discretization of the Euler equations, and the Godunov approach to accurately
estimate the exchange of conserved quantities across cell faces. On the other
hand, our approach avoids some of the weaknesses of these two schemes. For
example, it does not show the same level of noise and diffusiveness as SPH,
and it avoids the Galilean non-invariance of Eulerian codes. In our view, this
new synthesis of properties makes our new approach a very attractive
technique, providing a better numerical accuracy for many problems compared
with the alternative methods available thus far.

We note that our new method is significantly different from other approaches
to improve SPH, such as, e.g. the approaches discussed by
\citet{Inutsuka2002}, where SPH is combined with a Riemann solver. The
important new concept in our method is the introduction of a well-defined
mesh, while SPH by its very definition a mesh-free technique.

In this paper, we have described in detail the numerical and algorithmic
approaches taken in our new cosmological code {\small AREPO}, ranging from
parallel mesh-construction techniques in 2D and 3D, to spatial reconstruction
and flux estimation techniques, as well as time integration with individual
and adaptive schemes.  We have shown that our new code performs very well on a
wide range of test problems. We therefore consider it to be an attractive
alternative to SPH or AMR codes used presently in cosmology, and argue that it
has the potential to become the method of choice for a number of applications.
We note, in particular, that our treatment of self-gravity should be more
accurate and better suited for the problem of cosmic structure growth than
that in the current generation of cosmological AMR codes.

We note that our new moving mesh method can also make use of many
advanced concepts that have been developed for Eulerian codes, for
example to deal with magnetic fields and radiative transfer. In
particular, if constrained transport methods for ideal
magnetohydrodynamics (MHD) can be adapted to a Voronoi mesh, this
would provide the exciting possibility of constructing a Lagrangian
MHD code. Another interesting idea is to apply the ideas outlined in
this paper in the modelling of relativistic flows, which may yield a
Lorentz invariant numerical scheme. Also, it seems possible to
increase the order of our scheme by employing more sophisticated
reconstruction steps, e.g.~those known as weighted essentially
non-oscillatory (WENO) schemes \citep[e.g.][]{Feng2004}. However, the
second-order approach followed here is probably best for cosmological
structure formation, as the integration of the collisionless component
is of second order only and involves a relatively noisy gravitational
field.

\section*{Acknowledgements}

I would like to thank Lars Hernquist and Simon White for very helpful
discussions and suggestions. Also, I appreciate useful comments from Jim
Stone, Romain Teyssier, Kees Dullemond, Dick Bond, Burton Wendroff, and
Steffen Hess. Finally, I want to thank the referee, Hy Trac, for an
insightful and constructive report that helped to improve the paper.

\bibliographystyle{mnras} \bibliography{paper}

\begin{thebibliography}{120}
\expandafter\ifx\csname natexlab\endcsname\relax\def\natexlab#1{#1}\fi

\bibitem[{Agertz} et~al.(2007){Agertz}, {Moore}, {Stadel} et~al.]{Agertz2007}
{Agertz} O., {Moore} B., {Stadel} J., et~al., 2007, \mnras, 380, 963

\bibitem[{Ascasibar} et~al.(2003){Ascasibar}, {Yepes}, {M{\"u}ller} \&
  {Gottl{\"o}ber}]{Ascasibar2003}
{Ascasibar} Y., {Yepes} G., {M{\"u}ller} V., {Gottl{\"o}ber} S., 2003, \mnras,
  346, 731

\bibitem[{Athanassoula} \& {Misiriotis}(2002)]{Athanassoula2002}
{Athanassoula} E., {Misiriotis} A., 2002, \mnras, 330, 35

\bibitem[{Balsara}(1994)]{Balsara1994}
{Balsara} D.~S., 1994, \apj, 420, 197

\bibitem[{Barnes} \& {Hut}(1986)]{Barnes1986}
{Barnes} J., {Hut} P., 1986, \nat, 324, 446

\bibitem[{Barnes} \& {Hernquist}(1992)]{Barnes1992}
{Barnes} J.~E., {Hernquist} L., 1992, \araa, 30, 705

\bibitem[Barth \& Ohlberger(2004)]{Barth2004}
Barth T., Ohlberger M., 2004, in { Encyclopedia of Computational Mechanics,
  Volume 1, Fundamentals.\/}, edited by E.~Stein, R.~de~Borst, T.~Hughes,
  439--474, John Wiley and Sons Ltd

\bibitem[Barth \& Jesperson(1989)]{Barth1989}
Barth T.~J., Jesperson D.~C., 1989, AIAA Paper, 89-0366

\bibitem[{Bate} \& {Burkert}(1997)]{Bate1997}
{Bate} M.~R., {Burkert} A., 1997, \mnras, 288, 1060

\bibitem[{Berger} \& {Colella}(1989)]{Berger1989}
{Berger} M.~J., {Colella} P., 1989, Journal of Computational Physics, 82, 64

\bibitem[{Bernardeau} \& {van de Weygaert}(1996)]{Bernardeau1996}
{Bernardeau} F., {van de Weygaert} R., 1996, \mnras, 279, 693

\bibitem[Blandford et~al.(2005)Blandford, Blelloch, Cardoze \&
  Kadow]{Blandford2005}
Blandford D.~K., Blelloch G.~E., Cardoze D.~E., Kadow C., 2005, International
  Journal of Computational Geometry \& Applications, 15, 3

\bibitem[Blandford et~al.(2006)Blandford, Blelloch \& Kadow]{Blandford2006}
Blandford D.~K., Blelloch G.~E., Kadow C., 2006, in { SCG '06: Proceedings of
  the twenty-second annual symposium on Computational geometry\/},  292--300,
  ACM, New York, NY, USA

\bibitem[Bowyer(1981)]{Bowyer1981}
Bowyer A., 1981, Comput. J., 24, 2, 162

\bibitem[{Bryan} et~al.(1995){Bryan}, {Norman}, {Stone}, {Cen} \&
  {Ostriker}]{Bryan1995}
{Bryan} G.~L., {Norman} M.~L., {Stone} J.~M., {Cen} R., {Ostriker} J.~P., 1995,
  Computer Physics Communications, 89, 149

\bibitem[Cignoni et~al.(1998)Cignoni, Montani \& Scopigno]{Cignoni1998}
Cignoni P., Montani C., Scopigno R., 1998, Computer-Aided Design, 30, 333

\bibitem[Clarkson(1992)]{Clarkson1992}
Clarkson K.~L., 1992, in { Proceedings of the 33rd Annual Symposium on
  Foundations of Computer Science\/},  387--395, IEEE Computer Society
  Washington, DC, USA

\bibitem[{Colella} \& {Woodward}(1984)]{Colella1984}
{Colella} P., {Woodward} P., 1984, Journal of Computational Physics, 54, 174

\bibitem[{Cunningham} et~al.(2009){Cunningham}, {Frank}, {Varni{\`e}re},
  {Mitran} \& {Jones}]{Cunningham2007}
{Cunningham} A.~J., {Frank} A., {Varni{\`e}re} P., {Mitran} S., {Jones} T.~W.,
  2009, \apjs, 182, 519

\bibitem[{Dave} et~al.(1997){Dave}, {Dubinski} \& {Hernquist}]{Dave1997}
{Dave} R., {Dubinski} J., {Hernquist} L., 1997, New Astronomy, 2, 277

\bibitem[{Di Matteo} et~al.(2005){Di Matteo}, {Springel} \&
  {Hernquist}]{DiMatteo2005}
{Di Matteo} T., {Springel} V., {Hernquist} L., 2005, \nat, 433, 604

\bibitem[Dobkin \& Laszlo(1989)]{Dobkin1989}
Dobkin D.~P., Laszlo M.~J., 1989, Algorithmica, 4, 3

\bibitem[{Dolag} et~al.(2005){Dolag}, {Vazza}, {Brunetti} \&
  {Tormen}]{Dolag2005}
{Dolag} K., {Vazza} F., {Brunetti} G., {Tormen} G., 2005, \mnras, 364, 753

\bibitem[Dwyer(1987)]{Dwyer1987}
Dwyer R.~A., 1987, Algorithmica, 2, 137

\bibitem[Edelsbrunner \& Mucke(1990)]{Edelsbrunner1990}
Edelsbrunner H., Mucke E.~P., 1990, ACM Trans. Graph, 9, 66

\bibitem[Edelsbrunner \& Shah(1996)]{Edelsbrunner1996}
Edelsbrunner H., Shah N.~R., 1996, Algorithmica, 15, 223

\bibitem[Evrard(1988)]{Evrard1988}
Evrard A.~E., 1988, MNRAS, 235, 911

\bibitem[{Feng} et~al.(2004){Feng}, {Shu} \& {Zhang}]{Feng2004}
{Feng} L.-L., {Shu} C.-W., {Zhang} M., 2004, \apj, 612, 1

\bibitem[{Frenk} et~al.(1999){Frenk}, {White}, {Bode} et~al.]{Frenk1999}
{Frenk} C.~S., {White} S.~D.~M., {Bode} P., et~al., 1999, \apj, 525, 554

\bibitem[{Fromang} et~al.(2006){Fromang}, {Hennebelle} \&
  {Teyssier}]{Fromang2006}
{Fromang} S., {Hennebelle} P., {Teyssier} R., 2006, \aap, 457, 371

\bibitem[{Gerhard}(1981)]{Gerhard1981}
{Gerhard} O.~E., 1981, \mnras, 197, 179

\bibitem[{Gingold} \& {Monaghan}(1977)]{Gingold1977}
{Gingold} R.~A., {Monaghan} J.~J., 1977, \mnras, 181, 375

\bibitem[{Gnedin}(1995)]{Gnedin1995}
{Gnedin} N.~Y., 1995, \apjs, 97, 231

\bibitem[{Goodman} \& {Hernquist}(1991)]{Goodman1991}
{Goodman} J., {Hernquist} L., 1991, \apj, 378, 637

\bibitem[{Gresho} \& Chan(1990)]{Gresho1990}
{Gresho} P.~M., Chan S.~T., 1990, International Journal for Numerical Methods
  in Fluids, 11, 621

\bibitem[Guibas \& Stolfi(1985)]{Guibas1985}
Guibas L.~J., Stolfi J., 1985, ACM Transactions on Graphics, 4, 74

\bibitem[{Heitmann} et~al.(2008){Heitmann}, {Luki{\'c}}, {Fasel}
  et~al.]{Heitmann2007}
{Heitmann} K., {Luki{\'c}} Z., {Fasel} P., et~al., 2008, Computational Science
  and Discovery, 1, 1, 015003

\bibitem[{Hernquist}(1989)]{Hernquist1989b}
{Hernquist} L., 1989, \nat, 340, 687

\bibitem[{Hernquist}(1993)]{Hernquist1993}
{Hernquist} L., 1993, \apjs, 86, 389

\bibitem[{Hernquist} \& {Katz}(1989)]{Hernquist1989}
{Hernquist} L., {Katz} N., 1989, \apjs, 70, 419

\bibitem[{Hopkins} et~al.(2006){Hopkins}, {Hernquist}, {Cox}, {Di Matteo},
  {Robertson} \& {Springel}]{Hopkins2006}
{Hopkins} P.~F., {Hernquist} L., {Cox} T.~J., {Di Matteo} T., {Robertson} B.,
  {Springel} V., 2006, \apjs, 163, 1

\bibitem[{Hopkins} et~al.(2008){Hopkins}, {Hernquist}, {Cox}, {Dutta} \&
  {Rothberg}]{Hopkins2008a}
{Hopkins} P.~F., {Hernquist} L., {Cox} T.~J., {Dutta} S.~N., {Rothberg} B.,
  2008, \apj, 679, 156

\bibitem[{Iapichino} et~al.(2008){Iapichino}, {Adamek}, {Schmidt} \&
  {Niemeyer}]{Iapichino2008a}
{Iapichino} L., {Adamek} J., {Schmidt} W., {Niemeyer} J.~C., 2008, \mnras, 388,
  1079

\bibitem[{Iapichino} \& {Niemeyer}(2008)]{Iapichino2008b}
{Iapichino} L., {Niemeyer} J.~C., 2008, \mnras, 388, 1089

\bibitem[{Inutsuka}(2002)]{Inutsuka2002}
{Inutsuka} S., 2002, Journal of Computational Physics, 179, 238

\bibitem[{Katz} et~al.(1996){Katz}, {Weinberg} \& {Hernquist}]{Katz1996}
{Katz} N., {Weinberg} D.~H., {Hernquist} L., 1996, \apjs, 105, 19

\bibitem[{Landau} \& {Lifshitz}(1966)]{Landau1966}
{Landau} L.~D., {Lifshitz} E.~M., 1966, {Hydrodynamik}, Lehrbuch der
  theoretischen Physik, Berlin: Akademie-Verlag, 1966

\bibitem[Lee et~al.(2001)Lee, Park \& Park]{Lee2001}
Lee S., Park C.-I., Park C.-M., 2001, Parallel Processing Letters, 11, 341

\bibitem[LeVeque(1998)]{LeVeque1998}
LeVeque R.~J., 1998, Journal of Computational Physics, 146, 346

\bibitem[{LeVeque}(2002)]{LeVeque2002}
{LeVeque} R.~J., 2002, Finite volume methods for hyperbolic systems, Cambridge
  University Press

\bibitem[Liska \& Wendroff(2003)]{Liska2003}
Liska R., Wendroff B., 2003, SIAM J. Sci. Comput., 25, 3, 995

\bibitem[Liu \& Snoeyink(2005)]{Liu2005}
Liu Y., Snoeyink J., 2005, Combinatorial and Computational Geometry, 52, 439

\bibitem[Lloyd(1982)]{Lloyd1982}
Lloyd S., 1982, IEEE Trans. Inform. Theory, 28, 129–137

\bibitem[{Lucy}(1977)]{Lucy1977}
{Lucy} L.~B., 1977, \aj, 82, 1013

\bibitem[Mavriplis(1997)]{Mavripilis1997}
Mavriplis D.~J., 1997, Annual Review of Fluid Mechanics, 29, 473

\bibitem[{Mignone} et~al.(2007){Mignone}, {Bodo}, {Massaglia}
  et~al.]{Mignone2007}
{Mignone} A., {Bodo} G., {Massaglia} S., et~al., 2007, \apjs, 170, 228

\bibitem[{Mihos} \& {Hernquist}(1996)]{Mihos1996}
{Mihos} J.~C., {Hernquist} L., 1996, \apj, 464, 641

\bibitem[{Mitchell} et~al.(2009){Mitchell}, {McCarthy}, {Bower}, {Theuns} \&
  {Crain}]{Mitchell2008}
{Mitchell} N.~L., {McCarthy} I.~G., {Bower} R.~G., {Theuns} T., {Crain} R.~A.,
  2009, \mnras, 395, 180

\bibitem[{Monaghan}(1992)]{Monaghan1992}
{Monaghan} J.~J., 1992, \araa, 30, 543

\bibitem[Monaghan(1997)]{Monaghan1997}
Monaghan J.~J., 1997, J. Comp. Phys., 136, 298

\bibitem[M{\"u}cke(1998)]{Muecke1995}
M{\"u}cke E.~P., 1998, International Journal of Computational Geometry and
  Applications, 8, 255

\bibitem[M\"{u}cke et~al.(1996)M\"{u}cke, Saias \& Zhu]{Muecke1996}
M\"{u}cke E.~P., Saias I., Zhu B., 1996, in { SCG '96: Proceedings of the
  twelfth annual symposium on Computational geometry\/},  274--283, ACM, New
  York, NY, USA

\bibitem[{M{\"u}ller} \& {Steinmetz}(1995)]{Mueller1995}
{M{\"u}ller} E., {Steinmetz} M., 1995, Computer Physics Communications, 89, 45

\bibitem[{Murphy} \& {Burrows}(2008)]{Murphy2008}
{Murphy} J.~W., {Burrows} A., 2008, \apjs, 179, 209

\bibitem[{Navarro} et~al.(2008){Navarro}, {Ludlow}, {Springel}
  et~al.]{Navarro2008}
{Navarro} J.~F., {Ludlow} A., {Springel} V., et~al., 2008, MNRAS, in press,
  ArXiv e-prints, 0810.1522

\bibitem[{Negroponte} \& {White}(1983)]{Negroponte1983}
{Negroponte} J., {White} S.~D.~M., 1983, \mnras, 205, 1009

\bibitem[{Noh}(1987)]{Noh1987}
{Noh} W.~F., 1987, Journal of Computational Physics, 72, 78

\bibitem[Okabe et~al.(2000)Okabe, Boots, Sugihara \& Nok~Chiu]{Okabe2000}
Okabe A., Boots B., Sugihara K., Nok~Chiu S., 2000, Spatial Tessellations,
  Concepts and Applications of Voronoi Diagrams, John Wiley \& Sons Ltd,
  Chichester

\bibitem[Ollivier-Gooch(1997)]{OllivierGooch1997}
Ollivier-Gooch C.~F., 1997, Journal of Computational Physics, 133, 6

\bibitem[{O'Shea} et~al.(2005){O'Shea}, {Nagamine}, {Springel}, {Hernquist} \&
  {Norman}]{Shea2005}
{O'Shea} B.~W., {Nagamine} K., {Springel} V., {Hernquist} L., {Norman} M.~L.,
  2005, \apjs, 160, 1

\bibitem[{Owen} et~al.(1998){Owen}, {Villumsen}, {Shapiro} \&
  {Martel}]{Owen1998}
{Owen} J.~M., {Villumsen} J.~V., {Shapiro} P.~R., {Martel} H., 1998, \apjs,
  116, 155

\bibitem[{Pelupessy} et~al.(2003){Pelupessy}, {Schaap} \& {van de
  Weygaert}]{Pelupessy2003}
{Pelupessy} F.~I., {Schaap} W.~E., {van de Weygaert} R., 2003, \aap, 403, 389

\bibitem[{Pen}(1998)]{Pen1998}
{Pen} U.-L., 1998, \apjs, 115, 19

\bibitem[{Pfrommer} et~al.(2006){Pfrommer}, {Springel}, {En{\ss}lin} \&
  {Jubelgas}]{Pfrommer2006}
{Pfrommer} C., {Springel} V., {En{\ss}lin} T.~A., {Jubelgas} M., 2006, \mnras,
  367, 113

\bibitem[{Press} et~al.(1992){Press}, {Teukolsky}, {Vetterling} \&
  {Flannery}]{Press1992}
{Press} W.~H., {Teukolsky} S.~A., {Vetterling} W.~T., {Flannery} B.~P., 1992,
  {Numerical recipes in C. The art of scientific computing}, Cambridge:
  University Press, 1992, 2nd ed.

\bibitem[{Price}(2008)]{Price2007KH}
{Price} D.~J., 2008, Journal of Computational Physics, 227, 10040

\bibitem[{Price} \& {Monaghan}(2007)]{Price2007}
{Price} D.~J., {Monaghan} J.~J., 2007, \mnras, 374, 1347

\bibitem[Rasio \& Shapiro(1991)]{Rasio91}
Rasio F.~A., Shapiro S.~L., 1991, ApJ, 377, 559

\bibitem[Ryder(2000)]{Ryder2000}
Ryder W.~J., 2000, Journal of Computational Physics, 162, 395

\bibitem[{Ryu} et~al.(1993){Ryu}, {Ostriker}, {Kang} \& {Cen}]{Ryu1993}
{Ryu} D., {Ostriker} J.~P., {Kang} H., {Cen} R., 1993, \apj, 414, 1

\bibitem[{Saitoh} \& {Makino}(2009)]{Saitoh2008}
{Saitoh} T.~R., {Makino} J., 2009, \apjl, 697, L99

\bibitem[{Scannapieco} \& {Br{\"u}ggen}(2008)]{Scannapieco2008}
{Scannapieco} E., {Br{\"u}ggen} M., 2008, \apj, 686, 927

\bibitem[{Schaap} \& {van de Weygaert}(2000)]{Schaap2000}
{Schaap} W.~E., {van de Weygaert} R., 2000, \aap, 363, L29

\bibitem[Schewchuk(1997)]{Shewchuk1997}
Schewchuk J.~R., 1997, Discrete \& Computational Geometry, 18, 305

\bibitem[{Serrano} \& {Espa{\~n}ol}(2001)]{Serrano2001}
{Serrano} M., {Espa{\~n}ol} P., 2001, Phys Rev E, 64, 046115

\bibitem[{Shirokov} \& {Bertschinger}(2005)]{Shirokov2}
{Shirokov} A., {Bertschinger} E., 2005, ArXiv Astrophysics e-prints,
  astro-ph/0505087

\bibitem[{Slyz} \& {Prendergast}(1999)]{Slyz1999}
{Slyz} A., {Prendergast} K.~H., 1999, \aaps, 139, 199

\bibitem[{Springel}(2005)]{Springel2005}
{Springel} V., 2005, \mnras, 364, 1105

\bibitem[{Springel} et~al.(2005){Springel}, {Di Matteo} \&
  {Hernquist}]{SpringelBH2005}
{Springel} V., {Di Matteo} T., {Hernquist} L., 2005, \mnras, 361, 776

\bibitem[{Springel} \& {Hernquist}(2002)]{Springel2002}
{Springel} V., {Hernquist} L., 2002, \mnras, 333, 649

\bibitem[{Springel} \& {White}(1999)]{Springel1999}
{Springel} V., {White} S.~D.~M., 1999, \mnras, 307, 162

\bibitem[{Springel} et~al.(2008){Springel}, {White}, {Frenk}
  et~al.]{Springel2008Nature}
{Springel} V., {White} S.~D.~M., {Frenk} C.~S., et~al., 2008, \nat, 456, 73

\bibitem[{Springel} et~al.(2001){Springel}, {Yoshida} \&
  {White}]{Springel2001gadget}
{Springel} V., {Yoshida} N., {White} S.~D.~M., 2001, New Astronomy, 6, 79

\bibitem[{Stadel} et~al.(2009){Stadel}, {Potter}, {Moore} et~al.]{Stadel2008}
{Stadel} J., {Potter} D., {Moore} B., et~al., 2009, \mnras, 398, L21

\bibitem[{Steinmetz} \& {M{\"u}ller}(1993)]{Steinmetz1993}
{Steinmetz} M., {M{\"u}ller} E., 1993, \aap, 268, 391

\bibitem[{Steinmetz} \& {White}(1997)]{Steinmetz1997}
{Steinmetz} M., {White} S.~D.~M., 1997, \mnras, 288, 545

\bibitem[{Stone} et~al.(2008){Stone}, {Gardiner}, {Teuben}, {Hawley} \&
  {Simon}]{Stone2008}
{Stone} J.~M., {Gardiner} T.~A., {Teuben} P., {Hawley} J.~F., {Simon} J.~B.,
  2008, \apjs, 178, 137

\bibitem[{Stone} \& {Norman}(1992)]{Stone1992}
{Stone} J.~M., {Norman} M.~L., 1992, \apjs, 80, 753

\bibitem[Strang(1968)]{Strang1968}
Strang G., 1968, SIAM Journal on Numerical Analysis, 5, 3, 506

\bibitem[{Tasker} et~al.(2008){Tasker}, {Brunino}, {Mitchell}
  et~al.]{Tasker2008}
{Tasker} E.~J., {Brunino} R., {Mitchell} N.~L., et~al., 2008, \mnras, 390, 1267

\bibitem[{Thacker} \& {Couchman}(2006)]{Thacker2006}
{Thacker} R.~J., {Couchman} H.~M.~P., 2006, Computer Physics Communications,
  174, 540

\bibitem[{Toomre} \& {Toomre}(1972)]{Toomre1972}
{Toomre} A., {Toomre} J., 1972, \apj, 178, 623

\bibitem[{Toro}(1997)]{Toro1997}
{Toro} E., 1997, Riemann solvers and numerical methods for fluid dynamics,
  Springer

\bibitem[{Trac} \& {Pen}(2004)]{Trac2004}
{Trac} H., {Pen} U.-L., 2004, New Astronomy, 9, 443

\bibitem[{Trac} et~al.(2007){Trac}, {Sills} \& {Pen}]{Trac2007}
{Trac} H., {Sills} A., {Pen} U.-L., 2007, \mnras, 377, 997

\bibitem[{Truelove} et~al.(1998){Truelove}, {Klein}, {McKee}
  et~al.]{Truelove1998}
{Truelove} J.~K., {Klein} R.~I., {McKee} C.~F., et~al., 1998, \apj, 495, 821

\bibitem[{van de Weygaert}(1994)]{Weygaert1994}
{van de Weygaert} R., 1994, \aap, 283, 361

\bibitem[{van de Weygaert} \& {Schaap}(2009)]{Weygaert2007}
{van de Weygaert} R., {Schaap} W., 2009, in { Lecture Notes in Physics, Berlin
  Springer Verlag\/}, edited by {V.~J.~Mart{\'{\i}}nez, E.~Saar,
  E.~Mart{\'{\i}}nez-Gonz{\'a}lez, \& M.-J.~Pons-Border{\'{\i}}a}, vol. 665 of
  { Lecture Notes in Physics, Berlin Springer Verlag\/},  291--413

\bibitem[{van Leer}(1984)]{Leer1984}
{van Leer} B., 1984, SIAM J. Sci. Stat. Comput., 5, 1

\bibitem[{van Leer}(2006)]{Leer2006}
{van Leer} B., 2006, Communications in Computational Physics, 1, 192

\bibitem[{Wadsley} et~al.(2004){Wadsley}, {Stadel} \& {Quinn}]{Wadsley2004}
{Wadsley} J.~W., {Stadel} J., {Quinn} T., 2004, New Astronomy, 9, 137

\bibitem[{Wadsley} et~al.(2008){Wadsley}, {Veeravalli} \&
  {Couchman}]{Wadsley2008}
{Wadsley} J.~W., {Veeravalli} G., {Couchman} H.~M.~P., 2008, \mnras, 387, 427

\bibitem[Watson(1981)]{Watson1981}
Watson D.~F., 1981, Comput. J., 24, 2, 167

\bibitem[White(1996)]{White1996}
White S. D.~M., 1996, in { Cosmology and Large-Scale Structure\/}, edited by
  R.~Schaefer, J.~Silk, M.~Spiro, J.~Zinn-Justin, Dordrecht: Elsevier,
  astro-ph/9410043

\bibitem[{White} \& {Rees}(1978)]{White1978}
{White} S.~D.~M., {Rees} M.~J., 1978, \mnras, 183, 341

\bibitem[{Whitehurst}(1995)]{Whitehurst1995}
{Whitehurst} R., 1995, \mnras, 277, 655

\bibitem[{Woodward} \& {Colella}(1984)]{Woodward1984}
{Woodward} P., {Colella} P., 1984, Journal of Computational Physics, 54, 115

\bibitem[{Xu}(1997)]{Xu1997}
{Xu} G., 1997, \mnras, 288, 903

\bibitem[{Zeldovich}(1970)]{Zeldovich1970}
{Zeldovich} Y.~B., 1970, \aap, 5, 84

\bibitem[{Zingale} et~al.(2002){Zingale}, {Dursi}, {ZuHone}
  et~al.]{Zingale2002}
{Zingale} M., {Dursi} L.~J., {ZuHone} J., et~al., 2002, \apjs, 143, 539

\end{thebibliography}

\end{document}